\pdfoutput=1
\documentclass[11pt]{article}
\usepackage{mathtools}
\usepackage{silence}
\usepackage{amssymb}
\usepackage{graphicx} 
\def\laq{\raise 0.4ex\hbox{$<$}\kern -0.8em\lower 0.62
ex\hbox{$\sim$}}
\def\gaq{\raise 0.4ex\hbox{$>$}\kern -0.7em\lower 0.62
ex\hbox{$\sim$}}

\begin{document}

\begin{titlepage}
\vspace*{1cm}

\begin{center}
{\LARGE {\bf Gravitational wave astronomy}}
\vskip0.5cm
{\LARGE {\bf and the expansion history of the Universe}}
\vskip2.cm
\large{Massimo Giovannini \footnote{e-mail address: massimo.giovannini@cern.ch}}
\vskip 1.cm
{\it Department of Physics, CERN, 1211 Geneva 23, Switzerland}
\vskip 0.5cm
{\it INFN, Section of Milan-Bicocca, 20126 Milan, Italy}

\end{center}
\begin{abstract}
The timeline of the expansion rate ultimately defines the interplay between high-energy physics, astrophysics and cosmology. The guiding theme of this topical review is provided by the scrutiny of the early history of the space-time curvature through the diffuse backgrounds of gravitational radiation that are sensitive to all the stages of the evolution of the plasma. Due to their broad spectrum (extending from the aHz region to the THz domain) they bridge the macroworld described by general relativity and the microworld of the fundamental constituents of matter. It is argued that during the next score year the analysis of the relic gravitons may infirm or confirm the current paradigm where a radiation plasma is assumed to dominate the whole post-inflationary epoch. The role of high frequency and ultra-high frequency signals between the MHz and the THz is emphasized in the perspective of quantum sensing. The multiparticle final state of the relic gravitons and its macroscopic quantumness  is also discussed with particular attention to the interplay between the 
entanglement entropy and the maximal frequency of the spectrum.
\end{abstract}
\end{titlepage}

\pagenumbering{arabic}

\tableofcontents

\newpage
\renewcommand{\theequation}{1.\arabic{equation}}
\setcounter{equation}{0}
\section{Introduction}
\label{sec1}
\subsection{Ten years of gravitational wave astronomy}
Gravitational waves have been predicted by Einstein in 1916 \cite{AE1} as a direct consequence of general relativity \cite{AE2}. Later on this problem has been revisited by Einstein and Rosen with somehow contradicting conclusions \cite{AE2a} suggesting that gravitational waves could be unphysical. While the  legacy of Ref. \cite{AE2a} brought eventually some late skepticism on the true physical nature of gravitational radiation (see, for instance, \cite{AE2b}) the gauge-invariant nature of gravitational waves has been well established in the 1970s \cite{AE2c}. In spite of the pioneering attempts of Weber \cite{WEB1,WEB2} and of the subsequent resonant detectors of gravitational radiation in the early 1970s, the first direct evidence of gravitational radiation dates back to the early 1980s when the orbital decay of a binary neutron star system has been originally observed \cite{AE3}.  Roughly speaking almost one century after the first speculations, the gravitational waves have been detected by the wide-band interferometers \cite{AE4,AE5,AE6}.  The signals observed so far mainly come from astrophysical processes occurring at late time in the life of the Universe and they are the result of accelerated mass distributions with non-vanishing quadrupole moment. One of the most exciting directions  is however related to the possible existence of diffuse backgrounds of gravitational radiation produced thanks to the early variation of the space-time curvature. This collection of random waves encodes a snapshot of the early expansion history of the Universe prior to the formation of light nuclei. The purpose of this topical review is to summarize what can be said on the early expansion history of the Universe from the analyses of the stochastic backgrounds of relic gravitational waves. 

\subsection{Gravitational waves in curved backgrounds}
In the 1960s and 1970s it was believed that the tensor modes 
of the geometry could not be excited in curved background geometries. Although the chain of arguments leading to such a  conjecture would be per se interesting, this misleading perspective implied that both electromagnetic and gravitational waves could be considered invariant for a Weyl rescaling of the four-dimensional background geometry; from a practical viewpoint  Weyl invariance implies that both electromagnetic and gravitational waves should obey the same equations in a Minkowski background and in curved geometries eventually obtained by Weyl rescaling from a flat space-time \cite{AA01,AA02}. This viewpoint persisted until the mid 1970s when it was challenged by a series of papers \cite{AA1,AA2} suggesting that gravitational waves can be indeed excited in curved backgrounds and, more specifically, in Friedmann-Robertson-Walker cosmologies \cite{AA3,AA4}.

Almost fifty years after these pioneering analyses the relic signals represent today a well defined (and probably unique) candidate source for typical frequencies exceeding the kHz region where wide-band detectors are currently operating. Following the formulation of the inflationary scenarios \cite{INFL1,INFL2,INFL3,INFL4} it became gradually clear that the conventional lore would predict a minute spectral energy density in the MHz region \cite{AA7,AA8,AA9}. This is ultimately the reason why the most stringent tests of the conventional lore could come, in the near future, from the largest scales \cite{AA10} where the  limits on the tensor to scalar ratio $r_{T}$ are in fact direct probes of the spectral energy density in the aHz region.  Throughout the discussions of this article  the standard prefixes of the international system of units are  systematically employed; so for instance $1 \mathrm{kHz} = 10^{3} \mathrm{Hz}$, $1\, \mathrm{aHz} = 10^{-18} \, \mathrm{Hz}$ and similarly for all the other relevant frequency domains mentioned hereunder.

 \subsection{The expansion history}
During the last thirty years cosmology astrophysics and particle physics experienced a progressive unification towards two complementary paradigms accounting for the observations at small and large distance scales. The standard model of particle interactions describes the strong and electroweak physics or, as we could say for short, the microworld; although there are various hints on its possible incompleteness (typically related to the existence of dark matter), so far the standard model has not been falsified. The so-called concordance paradigm (based on general relativity) is customarily employed to analyze the macroworld of cosmological and astrophysical observations involving, in particular, the data associated with the temperature and polarization anisotropies of the Cosmic Microwave Background, the large-scale structure data and the supernova observations. The concordance paradigm is sometimes dubbed $\Lambda$CDM where $\Lambda$ accounts for the dark energy component and CDM stands for the cold dark matter. It is fair to say that, at the moment, the standard model of particle interactions and the $\Lambda$CDM scenario seem mutually consistent but conceptually incomplete. 

In the concordance paradigm the source of large-scale inhomogeneities 
is represented by the adiabatic and Gaussian fluctuations produced during a stage 
of conventional inflationary expansion. The subsequent evolutionary history 
of the plasma assumes a long period of expansion dominated 
by radiation until the epoch of matter-radiation equality and this timeline is broadly compatible 
with the idea that all the particle species were in thermal equilibrium above 
typical temperatures of the order of $200$ GeV but  there is no direct evidence 
either in favour of this hypothesis or against it. In the 
past the radiation dominance of the primeval plasma before big bang nucleosynthesis 
has been taken as a general truism also because it was practically impossible to check 
directly the early timeline of the expansion rate by simply looking at electromagnetic effects. This 
was the viewpoint conveyed in the pioneering analyses of the hot big bang hypothesis 
formulated by Gamow, Alpher, Bethe and Herman \cite{HBB1,HBB2,HBB3} and subsequently 
confirmed with the discovery of the Cosmic Microwave Background (CMB) \cite{HBB4} 
by Penzias and Wilson also thanks to the neat theoretical interpretation formulated by Peebles \cite{HBB4a,HBB4b}. As we know 
the plasma became transparent to radiation around the time of photon decoupling. After 
that moment the slightly perturbed geodesics of the photons could be used to reconstruct 
 the temperature and polarization anisotropies of the CMB \cite{HBB5}
but the electromagnetic signals coming from the earlier expansion history 
were quickly reabsorbed by the plasma and are today completely inaccessible to any direct detection.

The sensitivities of operating detectors \cite{LIGO0,LIGO0a,LIGO0b,LIGO1} are notoriously insufficient to measure the diffuse backgrounds of relic gravitons but in the future new detectors might cover different frequencies \cite{LIGO3} even beyond the so-called audio band ranging between few Hz and $10$ kHz.  The gravitational waves produced thanks to the variation of the space-time curvature should then become an object of future empirical investigations even at high frequencies while 
at intermediate frequencies (in the nHz range) the backgrounds of relic gravitons could be observed by the pulsar timing arrays \cite{NANO1,NANO2,PPTA1,PPTA2} that are now primarily focussed  on the diffuse astrophysical signals. We actually know that every variation of the expansion rate produces shots of gravitons with given averaged multiplicities and specific statistical properties. If these spectra will ever be detected the timeline of the expansion rate might be directly tested without the need of 
 postulating a particular post-inflationary paradign before the curvature scale of big bang nucleosynthesis whose striking success is the last certain signature of radiation dominance for typical temperatures smaller than ${\mathcal O}(10)$ MeV. When considering these possibilities at face value there are at least two conceptually different issues that must 
be addressed.
\begin{itemize}
\item{} The first problem concerns the early expansion history of the current Hubble patch and its physical properties: is the conventional timeline of the $\Lambda$CDM scenario really compelling or just plausible?
\item{} The second class of questions involves the way relic gravitons could be used 
as a diagnostic of the early expansion history: how sensitive is the spectral energy density 
of the relic gravitons on the early expansion rates deviating from the $\Lambda$CDM timeline?
\end{itemize}
To address the first group of subjects we should first acknowledge that the causal structure 
of Friedmann-Robertson-Walker models provides already a number of relevant constraints 
on the expansion history. However, even admitting that, at early times, 
the particle horizon should disappear or diverge (as it happens in the case of conventional inflationary scenarios) to be replaced by an event horizon, the subsequent evolution of the space-time curvature remains undetermined. To appreciate this relevant point we should actually observe that the total number of $e$-folds does depend on the post-inflationary rate of expansion. For instance when we say that $60$ $e$-folds of accelerated expansion are necessary to suppress the spatial curvature we are actually referring to a post-inflationary evolution dominated by radiation. The same tacit assumption is systematically employed 
to confront the temperature and the polarization anisotropies of the 
CMB with the conventional inflationary scenarios \cite{RR1,RR2,RR3}. 

\subsection{The relic gravitons and the expansion history}
One of the purposes of this article is to argue that the spectra of relic gravitons provide the only direct probe of the post-inflationary evolution prior to the formation of light nuclei. This is why a detailed analysis of such a signal is mandatory even in the absence of sensitive detectors that might be available only in the 
far future. Various secondary effects may produce different backgrounds of gravitational radiation 
during a fixed post-inflationary evolution like the one endorsed in the context of the $\Lambda$CDM scenario.
These effects, however, always assume a specific knowledge that is still missing. Conversely 
the relic gravitons do represent the only conceivable direct diagnostic of the post-inflationary expansion 
history and this is the general perspective developed here. Since 
the spectrum of the relic gravitons extends from the aHz region up to the THz domain 
we can partition this broad frequency domain  into three complementary 
ranges where different stages of the early expansion rate are correspondingly probed:
\begin{itemize}
\item{} the low-frequency region (between few aHz and the fHz) is directly 
sensitive to the expansion rate during inflation; in this region the upper limits on the tensor to scalar 
ratio deduced from the temperature and polarization anisotropies of the CMB 
are in fact bounds on the early expansion rate; the CMB can be in fact considered as 
the largest electromagnetic detector of long-wavelength gravitational waves;
\item{} at intermediate frequencies various potential constraints are associated 
with the Pulsar Timing Arrays (typically operating in the nHz domain); from the 
viewpoint of the expansion history this region may set constraints both 
on the post-inflationary evolution and on the modifications introduced 
{\em during} the inflationary stage;
\item{} finally in the high frequency domain the constraints from the operating 
wide-band detectors between few Hz and $10$ kHz (as well as
 from other electromagnetic detectors operating in the MHz or GHz region) 
 will be essential for the analysis of potential peaks in the spectrum 
 of relic gravitons.
 \end{itemize}
 The first speculations suggesting that the relic gravitons could be used 
as a direct probe of the post-inflationary expansion history goes back to the 
 late 1990s and this will be the general inspiration of  this 
 article. In particular in Ref. \cite{EXP1} it has been suggested 
 that different post-inflationary stages modify the slopes of the spectral 
 energy density of the relic gravitons for frequencies larger than the 
 mHz.  It was found, quite surprisingly, that when the expansion rate is slower than radiation 
 the spectral energy density exhibits a high frequency spike \cite{EXP2,EXP3}.
 The original observation of Ref. \cite{EXP1} was that the post-inflationary 
 evolution may be modified and this would be especially true if we have to accommodate a late-time dominance 
 of the dark energy. In this case a post-inflationary evolution 
 dominated by radiation would be less likely than a long stiff stage expanding slower than radiation \cite{EXP1}. One of the first frameworks where these observations have been applied 
 are the quintessential inflationary models \cite{EXP4}. In this context 
 the late-time dominance of dark energy occurs via a quintessence 
 field that ultimately coincides with the inflaton. Later on different scenarios 
 based on different premised 
 have been proposed \cite{MGB} with the aim of accommodating an intermediate 
 stage expanding at a rate different from radiation. For the purpose 
 of this review, however, we do not want to commit ourselves to a specific 
 scenario or to a specific class of models. Indeed, as suggested in Ref. \cite{EXP1}, the spectra 
 of the relic gravitons chiefly depend on the evolution of the space-time 
 curvature and not on the particular features involving the different sources.
 
\subsection{The layout of the present article}
The interplay between the timeline of the expansion 
rate and the spectra of the relic gravitons  
promises a direct connection between cosmology, quantum field theory and the
effective description of gravitational interactions. On a more practical ground, in this 
topical investigation astrophysics and gravitational wave astronomy are seen 
as a tool for high-energy physics. In the past the common wisdom suggested instead that
high-energy physics was probably the sole tool to infer properties of the primeval plasma 
prior to big bang nucleosynthesis. This conventional viewpoint did rest on the assumption that 
the post-inflationary expansion rate had to be fixed and almost perpetually dominated 
by radiation down to the scale of matter-radiation equality. In our context the timeline of the 
post-inflationary expansion rate is only a working hypothesis subjected to the direct tests 
associated with the diffuse backgrounds of gravitational radiation.  Given the wealth of the connections between the various aspects of the problem it is impossible to analyze in detail all the relevant themes and this is why various collateral topics are swiftly mentioned but the interested readers may usefully consult a recently published book that dwells on the physics of the relic gravitons \cite{MGB} where most of the considerations omitted here are systematically addressed. The layout of this article is, in short, the following. Before elaborating on the unknowns, section \ref{sec2}  is focussed on what it is understood about the early expansion history with 
the goal of distinguishing the facts from the tacit assumptions. Section \ref{sec3} deals more directly with the interplay between the relic gravitons and the expansion history and since the various ranges of the spectra are directly sensitive to the evolutionary stages of the background geometry, it seems useful to examine separately the interplay between the relic gravitons and the expansion histories 
in the low (see section \ref{sec4}), intermediate (see section \ref{sec5}) and high frequency (see section \ref{sec6}). In section \ref{sec4} we point out that the inflationary observables are either suppressed or enhanced
depending upon the post-inflationary evolution that affects the total number of $e$-folds.
In section \ref{sec5} we present a discussion on the mutual interplay between the modified expansion 
histories and the pulsar timing arrays; towards the end of section \ref{sec5} we also argue 
that a the post-inflationary evolution may also produce signals between few $\mu$Hz and the Hz where 
usually different sources are claimed to be relevant for the (futuristic) space-borne detectors. 
Finally in section \ref{sec6} we specifically address  the direct bounds on the 
post-inflationary expansion rate coming from the high frequency and ultra-high frequency
regions where absolute bounds on the maximal frequency of the spectra can be derived. 
Some ideas related to the use of quantum sensing for the detection of the relic gravitons will also be 
analyzed.  The obtained limits on the maximal frequency are deeply rooted in the quantumness of the produced gravitons whose multiparticle final sates are macroscopic but always non-classical. As the unitary evolution preserves their coherence, the quantumness of the gravitons can be associated with an entanglement entropy that  is related with the loss of the complete information on the underlying quantum field. 
In the appendices we elaborated on some of the technical aspects that are 
often recalled in the main discussions. In particular appendix \ref{APPA} illustrates 
a number of relevant complements on the evolution of curvature inhomogeneities that are specifically needed in the discussion while appendix \ref{APPB} treats the forms of the action of the relic gravitons in different frames.

\renewcommand{\theequation}{2.\arabic{equation}}
\setcounter{equation}{0}
\section{The timeline of the expansion rate: facts and tacit assumptions}
\label{sec2}
In the last fifty years the interplay between high-energy physics, 
astrophysics and cosmology has been guided by the 
tacit assumption that prior to matter-radiation equality the primeval 
plasma was always dominated by radiation \cite{PJE,SW1} and this general truism 
is also reflected in various cartoons that are customarily employed to represent 
the timeline of the expansion rate where different moments of the life of the Universe are illustrated with the supposed matter content of the plasma. This viewpoint has been also propounded by S. Weinberg in one of the first popular accounts of the subject \cite{SW2}.  After the formulation of the inflationary paradigm in its different variants (see e.g. \cite{INFL1,INFL2,INFL3,INFL4}) the hypothesis of a post-inflationary radiation dominance remained practically unmodified and even today it is customary to assume that after an explosive stage of reheating 
the Universe should become, almost suddenly, dominated 
by radiation (see, for instance, \cite{BBK1,BBK2,BBK3}). 
Among the various conclusions that emanate from the assumption of an 
evolution dominated by radiation, the most notable one is probably that the 
 plasma as a whole is described by a single temperature for most of its history. 
An equally relevant statement is that the inflationary expansion must 
(or should) last for at least $60$ $e$-folds \cite{BBK1,BBK2,BBK3}. 
Since this tacit assumption of radiation dominance is not directly tested 
(at least for temperatures larger than few MeV) more general possibilities will be discussed. 

\subsection{What do we know about the early expansion history?}
\subsubsection{Particle horizon and causally disconnected regions}
A relevant constraint on the early expansion history  comes from the causal structure of Friedmann-Robertson-Walker (FRW) models whose line element in its canonical form is given by:
\begin{equation}
ds^2 = g_{\mu\nu} dx^{\mu} dx^{\nu} = dt^2 - a^2(t) \biggl[ \frac{dr^2}{1 - \kappa r^2} + r^2 (d\vartheta^2 +
\sin^2{\vartheta} d\varphi^2)\biggr],
\label{FRW1}
\end{equation}
where $g_{\mu\nu}$ denotes the metric tensor and $a(t)$ is the scale factor.
In the parametrization of Eq. (\ref{FRW1}), $\kappa=0$ corresponds to a spatially flat 
Universe; if  $ \kappa >0$ the Universe is spatially closed and, finally, $\kappa <0$ describes an open spatial section. In Eq. (\ref{FRW1}) the time $t$ indicates the {\em cosmic} time coordinate but 
depending upon the physical problem 
at hand, different time parametrizations can be also
adopted. A particularly useful one is the so-called 
conformal time parametrization that turns out to be particularly useful in the 
analysis of the inhomogeneities (see, in this respect, the appendices \ref{APPA} and \ref{APPB}). 
In the conformal time coordinate $\tau$ the line element of Eq. (\ref{FRW1}) becomes
\begin{equation}
ds^2 =g_{\mu\nu} dx^{\mu} d x^{\nu} = 
a^2(\tau)\biggl\{ d\tau^2 - \biggl[ \frac{dr^2}{1 - \kappa r^2} + r^2 (d\vartheta^2 +
\sin^{2}\vartheta d\varphi^2)\biggr]\biggr\}.
\label{FRW2}
\end{equation}
Since light rays follow null geodesics in Eq. (\ref{FRW1}) we may suppose 
that a signal is emitted at the time $t_{\mathrm{e}}$ (at a radial position $r_{\mathrm{e}}$) 
and received at the time $t_{\mathrm{r}}$ (at a radial position $r_{\mathrm{r}}=0$).
Then from Eq. (\ref{FRW1}) with $ds^2 =0$ and $d\vartheta = d\varphi =0$ we will have, for a 
null radial geodesic
\begin{equation}
\int_{t_{e}}^{t_{r}}\frac{d t}{a(t)}= \int_{0}^{r_{e}} \frac{d r}{\sqrt{1- \kappa \,r^2}}.
\label{FRW3}
\end{equation}
The position of the emitter is fixed in the comoving coordinate system. We can then say that the signal 
was emitted at a physical distance $d(t)$ from the origin:
\begin{equation}
d(t) = a(t) \int_{0}^{r_{e}} \frac{ d\, r}{\sqrt{ 1 - \kappa \,r^2}} = a(t) \, \int_{t_{e}}^{t} \, \frac{d t^{\prime}}{a(t^{\prime})}.
\label{FRW4}
\end{equation}
If we now introduce the concept of $t_{min}$ (corresponding to the maximal past extension of the time
coordinate on the FRW space-time), in the limit $t_{e} \to t_{min}$ we can introduce  
the particle horizon at time $t$ as: 
\begin{equation}
d_{p}(t) = a(t) \int_{t_{min}}^{t} dt'/a(t').
\label{FRW5}
\end{equation}
In short, for any given observer the particle horizon divides the regions of the space-time already
observed from the ones that have not been observed yet. In the hot big bang scenario the background expands but it is simultaneously {\em decelerated} and this means that the first derivative of the scale factor $a(t)$ with respect to the (cosmic) time coordinate $t$ is {\em positive} while its second derivative gets {\em negative}:
\begin{equation}
\dot{a}>0, \qquad \ddot{a} <0, \qquad a \, H >0,
\label{TWO1}
\end{equation}
where the overdot denotes here a derivation with respect to the cosmic time coordinate $t$
and $H= \dot{a}/a$ indicates the usual Hubble rate. If the scale factor is parametrized in terms of a power law $a(t) \simeq a_{1} (t/t_{1})^{\alpha}$ the conditions (\ref{TWO1}) imply 
$0 < \alpha < 1$ and this means that the {\em particle horizon} exists and it is finite 
\begin{equation}
d_{p}(t) = a(t) \int_{t_{min}}^{t} dt'/a(t') \to H^{-1}(t).
\label{TWO2}
\end{equation}
In the limit $t_{min}\to 0$ (when $t$ remains finite) $d_{p}(t)$ coincides (up to an irrelevant multiplicative constant factor) with $H^{-1}(t) \simeq t$. The particle horizon of Eq. (\ref{TWO2}) measures the extension of causally connected regions at time $t$ and its finiteness poses a problem if the Universe always expands in a decelerated manner. Since during a decelerated stage of expansion the extension of a causal patch is of the order of $d_{p}(t) \simeq t$,  the Hubble radius at any time preceding the current epoch must contain a finite number of causally disconnected regions. If we indicate with $H_{0}^{-1}$ the Hubble rate at the present time, at equality $H_{0}^{-1} (a_{eq}/a_{0})< H_{0}^{-1}$. In other words, the Hubble patch at equality is comparatively smaller than today but the typical size of causally connected regions at equality is $d_{p}(t_{eq}) \sim \,t_{eq}$, as suggested by Eq. (\ref{TWO2}). If we then measure $H_{0}^{-1} (a_{eq}/a_{0})$ in units of $t_{eq}$ we obtain:
\begin{equation}
\frac{H_{0}^{-1} (a_{eq}/a_{0})}{d_{p}(t_{eq})} = {\mathcal O}(50).
\label{TWO3}
\end{equation}
As $d_{p}(t_{eq})$ measures the extension of causally connected domains at $t_{eq}$, Eq. (\ref{TWO3})
suggests that, at matter-radiation equality, the region corresponding to the present Hubble patch 
contained about $50$ causally disconnected regions or $50^{3} \simeq 10^{5}$ disconnected volumes. The reference time selected in Eq. (\ref{TWO3}) can be modified but the essence of the problem remains the same. Furthermore if the typical reference time is larger than $t_{eq}$ the number of causally disconnected regions is comparatively smaller. Conversely, when $t < t_{eq}$
the number of the disconnected regions increase and quickly approaches its Planckian 
limit\footnote{Indeed, if we reach the Planck time, the blue-shifted value of the Hubble radius is of the order of  $4 \mu\mathrm{m} = 4 \times 10^{-4} \, \mathrm{cm}$. But since at the Planck time the particle horizon is $d_{p}(t_{P}) \simeq t_{P} \simeq 10^{-33}$ cm,
the ratio between $ 4 \times 10^{-4} \, \mathrm{cm}$ and $10^{-33} \,\,\mathrm{cm}$
is approximately ${\mathcal O}(10^{29})$ and the number of disconnected 
volumes is ${\mathcal O}(10^{87})$. }.
 
\subsubsection{Event horizon}
The previous discussion clarifies why the existence of the particle horizon leads necessarily to causally disconnected volumes; this occurrence clashes, among other things, with the high degree of homogeneity and isotropy of the Universe as it follows, for instance, from the analysis of the temperature and polarization anisotropies of the CMB. How come that 
regions emitting a highly homogeneous and isotropic CMB were causally disconnected in the past? To solve the causality problems of the conventional big bang scenario the idea is then to complement the standard decelerated stage of Eq. (\ref{TWO1}) with an epoch where the scale factor accelerates 
\begin{equation}
\dot{a}>0, \qquad \ddot{a} >0, \qquad a \, H >0,
\label{TWO4}
\end{equation}
and the particle horizon diverges.
If the scale factor is parametrized with a power law $a(t) \simeq a_{1} (t/t_{1})^{\alpha}$ the conditions (\ref{TWO4}) demand that $\alpha > 1$ and, in this situation, the  particle horizon 
does not exist: when $\alpha >1$ the integral of Eq. (\ref{TWO2}) is divergent in the limit 
$t_{min} \to 0$ and for any finite value of the cosmic time coordinate $t$. If the Universe expands as in Eq. (\ref{TWO4}) there exist however an {\em event horizon}
\begin{equation}
d_{e}(t) = a(t) \int_{t}^{t_{max}} dt'/a(t') \simeq H^{-1}(t),
\label{TWO5}
\end{equation}
where the second approximate equality holds for $t$ finite and $t_{max} \to +\infty$. In Eq. (\ref{TWO2})  $t_{min}$ measured the maximal past extension of the time coordinate in the given FRW space-time; in the case of the event horizon $t_{max}$ measures instead  the maximal future extension of the cosmic time
coordinate. For this reason the event horizon measures the maximal distance over which we can admit, even 
in the future, a causal connection. If $d_{e}(t)$ is finite in the limit $t_{max}\to \infty$ (for finite $t$) we can conclude that the event horizon exist. When the phase of accelerated expansion is parametrized in terms of the (expanding) branch of four-dimensional de Sitter space-time, namely $a(t) \simeq e^{H_{i} t}$ (with $H_{i} >0$)  
the particle and event horizons are, respectively, 
\begin{eqnarray}
&& d_{p}(t) = H_{i}^{-1} \biggl[ e^{H_{i}(t - t_{min})} -1\biggr],
\label{TWO5a}\\
&&  d_{e}(t) = H_{i}^{-1} \biggl[1-  e^{H_{i}(t - t_{max})}\biggr].
\label{TWO5b}
\end{eqnarray}
The cosmic time coordinate is allowed to run
from\footnote{Although this point is often ignored we like to point out that the limit $t_{min} \to - \infty$ 
is not well defined; strictly speaking an ever expanding inflationary evolution is not past geodesically complete \cite{MGB}. The limit $t_{min} \to -\infty$ can be better defined by introducing a geodesically complete extension of the de Sitter space-time. This problem has been discussed in the past but will not be specifically addressed here.} $t_{min}\to - \infty$ up to $t_{max}\to + \infty$. Consequently, 
for $t_{\mathrm{min}}\to -\infty$ (at fixed $t$) the particle horizon will diverge and the 
typical size of causally connected regions at time $t$ scales as $L_{i}(t) \simeq H_{i}^{-1} a(t)/a(t_{min})$.
While for the standard decelerated expansion the particle horizon increases faster than the scale factor, the 
typical size of causally connected regions scales exactly  as the scale factor.
In the limit $t_{max}\to \infty$ the event horizon exist and it is given, from Eq. 
(\ref{TWO5b}), by $d_{e}(t) \simeq H_{i}^{-1}$.

\subsubsection{Total number of $e$-folds?}
If the accelerated stage of expansion is sufficiently long, all the scales 
that were inside $H^{-1}$ at the onset of inflation are today comparable (or larger) than 
the Hubble radius. It is essential to appreciate that the quantitative meaning of the locution {\em sufficiently long} depends also {\em on the post-inflationary evolution} and not only on the inflationary dynamics itself. 
The duration of the accelerated stage of expansion is customarily 
parametrized in terms of the ratio between the scale factors at the end 
(i.e. $a_{f}$) and at the beginning (i.e. $a_{i}$) of inflation\footnote{Throughout this article $\ln{}$ denotes 
the natural (Neperian) logarithm, $\log{}$ indicates instead the common logarithm.}:
\begin{equation}
\exp{N} = \biggl(\frac{a_{f}}{a_{i}}\biggr) \Rightarrow N = \ln{(a_{f}/a_{i})},
\label{TWO6}
\end{equation}
where $N$ denotes the number of $e$-folds. Later on in this section 
we shall be introducing with $N_{k}$ (namely the number of $e$-folds elapsed 
since a given scale crossed the Hubble radius) as well as other notions derived from Eq. (\ref{TWO6}); 
the notion of $N_{k}$ becomes particularly relevant for the analysis of section \ref{sec4}.  Equation (\ref{TWO6a}) is clearly equivalent to 
\begin{equation}
N = \int_{t_{i}}^{t_{f}} \, H \, d t = \int_{a_{i}}^{a_{f}} \frac{d \, a}{a}.
\label{TWO6a}
\end{equation}
\begin{figure}[!ht]
\centering
\includegraphics[width=0.7\textwidth]{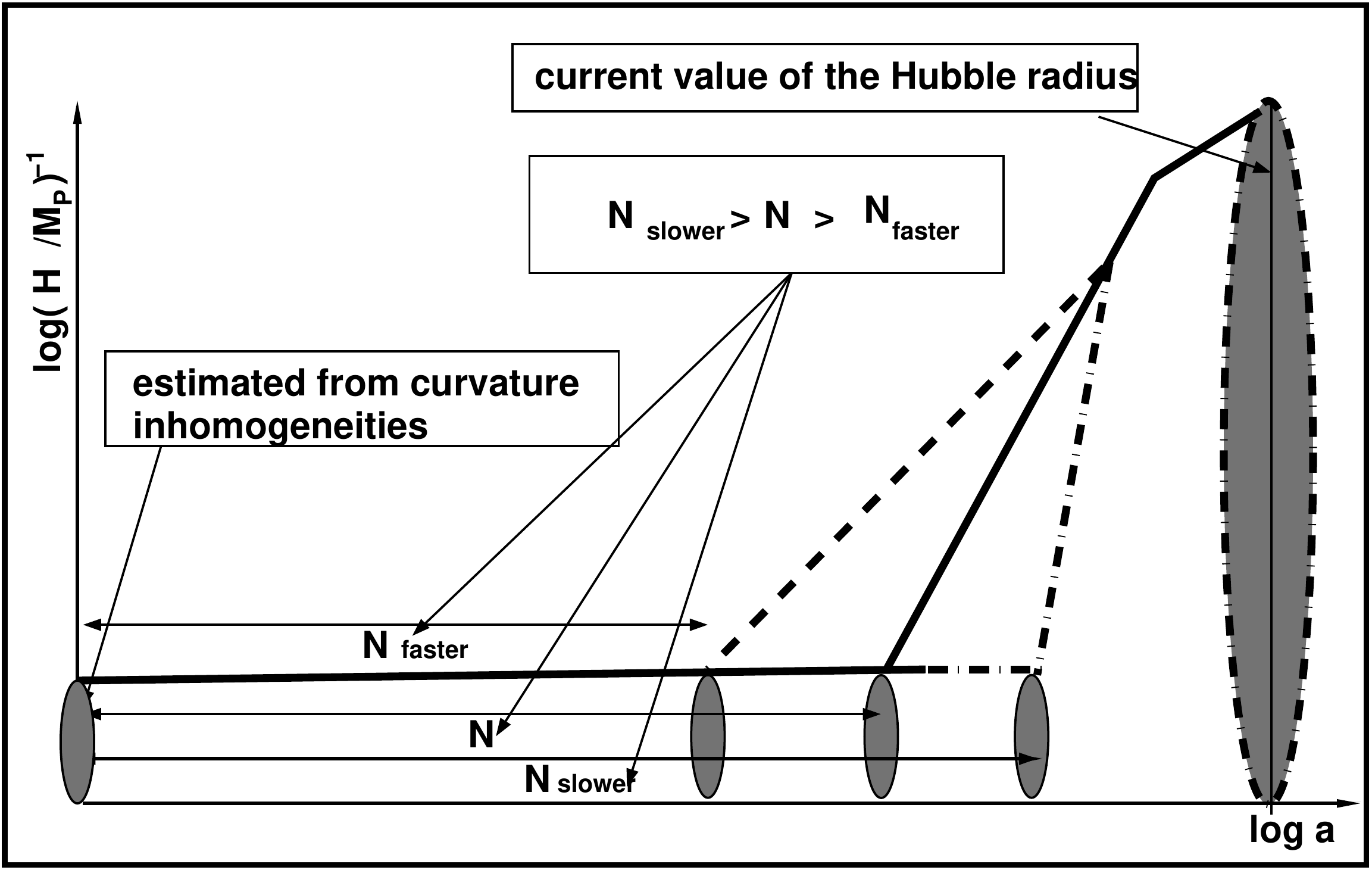}
\caption[a]{On the vertical axis the profile of $H^{-1}$ is illustrated in Planck units as a function of the logarithm of the scale factor. In this cartoon (where, for the sake of simplicity, the slow-roll corrections have been neglected) the full thick line describes the standard inflationary evolution followed by a radiation-dominated stage. The dashed and dot-dashed curves correspond instead to a post-inflationary expansion rate  that is either faster or slower than radiation, at least for some time before radiation dominance. In subsection \ref{subsec22} the early expansion rate is estimated from the large-scale curvature inhomogeneities whereas in 
subsection \ref{subsec23} we are going to present a series of quantitative estimates 
of $N$, $N_{slower}$ and $N_{faster}$. }
\label{FIGU0a} 
\end{figure}
The number of $e$-folds required for the consistency of a given inflationary scenario does not only depend on the inflationary dynamics as it might seem to follow from Eqs. (\ref{TWO6})--(\ref{TWO6a}).
In other words, while the physical features of the decelerated and of the accelerated 
expansions are per se relevant, what we want to stress here is that the indetermination of the post-inflationary evolution affects the specific value of the total number of $e$-folds. 
To clarify this point we consider the ratio between the intrinsic (spatial) and the extrinsic (Hubble) 
curvatures and recall that it is notoriously given by
\begin{equation}
\frac{\kappa}{a^2 \, H^2} = \frac{\kappa}{\dot{a}^2}.
\label{TWO7}
\end{equation}
The right-hand side of Eq. (\ref{TWO7}) gets suppressed when the Universe accelerates (see Eq. (\ref{TWO4})) while $\kappa/\dot{a}^2$ increases during a stage where $\ddot{a} <0$ (see Eq. (\ref{TWO1})). The {\em total} suppression of the ratio given 
in Eq. (\ref{TWO7}) cannot be simply attributed to the inflationary stage of expansion 
unless we artificially assume that the post-inflationary evolution is known and 
corresponds, for instance, to the dominance of radiation.  This is ultimately the reason why 
the total number of $e$-folds  suffers a 
theoretical indetermination associated with the post-inflationary evolution.
In Fig. \ref{FIGU0a} we illustrate with a cartoon the sensitivity of the number of $e$-folds
to the post-inflationary evolution. In the rightmost part of the plot we have the current value of the 
Hubble radius and the thick line denotes the standard evolution where 
the post-inflationary expansion rate is always dominated by radiation. If the blue-shifted value of the current Hubble radius must fit exactly inside the inflationary event horizon, the value of $N$ becomes, as we shall see, ${\mathcal O}(60)$. In Fig. \ref{FIGU0a} two qualitatively different possibilities are also mentioned: in the first case the post-inflationary expansion rate is faster than radiation (see the dashed line 
of  Fig. \ref{FIGU0a}) in the second case  the post-inflationary expansion rate is slower than radiation (see the dot-dashed line of  Fig. \ref{FIGU0a}). When expansion rate is faster 
than radiation the number of $e$-folds required to fit the blue-shifted Hubble radius 
inside the inflationary event horizon is comparatively smaller than $N$ (i.e. $N_{faster} < N$); the opposite is true in case the post-inflationary expansion rate is slower than radiation
(i.e. $N_{slower} > N$).  The quantitative aspects of Fig. \ref{FIGU0a} are
analyzed in subsection \ref{subsec23} after a discussion of the early expansion rate (see subsection \ref{subsec22}) which is relevant also for the determination of the number of $e$-folds.

\subsection{The early expansion rate}
\label{subsec22}
\subsubsection{Conventional inflationary stages}
The expansion history during the inflationary stage follows from the equations connecting the Hubble rate to the corresponding sources. The  single-field inflationary models
can be notoriously analyzed in terms of the following scalar-tensor action (see, for instance, \cite{BBK3})
\begin{equation}
S_{\varphi} = \int d^4 x\, \sqrt{ - g}\,\biggl[ - \frac{R}{ 2 \ell_{P}^2} \,+ \,\frac{1}{2} g^{\alpha\beta} \partial_{\alpha} \varphi \partial_{\beta} \varphi 
- V(\varphi) \biggr],
\label{SINGLE1}
\end{equation}
where $\ell_{P}$ denotes the Planck length and the following notations 
will be used throughout the whole discussion:
\begin{equation}
\ell_{P} = \sqrt{ 8 \pi G}, \qquad \overline{M}_{P} = M_{P}/\sqrt{8\pi} = 1/\ell_{P}.
\label{PLANCK}
\end{equation}
Equation  (\ref{SINGLE1}) should be regarded as the first term of an effective description where the higher derivatives are suppressed by the negative powers of a large mass $M$ associated with the fundamental theory that underlies the effective action.  The leading corrections to Eq. (\ref{SINGLE1}) consist of all possible terms containing four space-time derivatives \cite{EFFT}
and Eq. (\ref{SINGLE1}) itself can be studied in two complementary perspectives:
\begin{itemize}
\item{} if we presume, by fiat, that the post-inflationary evolution is fixed and known then the 
only sensible question and the sole concern should be somehow to reconstruct the functional dependence 
of the potential; 
\item{} conversely if post-inflationary evolution is unknown (or only partially known) 
it is less meaningful to aim at a reconstruction of the inflaton potential from the large-scale 
data since the number of $e$-folds ultimately depends on the post-inflationary evolution.
\end{itemize}
For different reasons both approaches are appealing but the former corresponds to the 
conventional lore while the latter perspective is pursued in this discussion: the ultimate goal would be
to test the post-inflationary expansion rate rather than arbitrarily postulating a specific timeline.  
The post-inflationary evolution is not going to be fixed 
and this choice has a specific impact on the remaining part of the discussion.
For this reason the properties of the expansion rate during inflation are 
technically more essential than the form of the potential although the obtained results 
can be related (at any step) to the more conventional approach. 
In the single-field case (and for the background geometry of Eqs. (\ref{FRW1})--(\ref{FRW2})) the evolution equations follow from Eq. (\ref{SINGLE1}) and they can be written as:
\begin{eqnarray}
&& 3 H^2 = \ell_{P}^2 \biggl[ \frac{\dot{\varphi}^2}{2} + V(\varphi)\biggr] - \frac{3 \kappa}{a^2},
\label{SINGLE2}\\
&& \ddot{\varphi} + 3 H \dot{\varphi} + V_{,\,\varphi} =0,
\label{SINGLE3}
\end{eqnarray}
where, as previously remarked, the overdot denotes a derivation with respect to the cosmic 
time coordinate $t$. If Eqs. (\ref{SINGLE2})--(\ref{SINGLE3}) 
are combined we obtain $  2 \dot{H} = - \ell_{P}^2 \dot{\varphi}^2 + 2 \kappa/a^2$; this 
is, in practice, the explicit form of the Raychaudhuri equation \cite{MGB} written in the case 
of a scalar field source.  In a stage where the decrease of the Hubble rate is sufficiently slow (i.e. $\dot{H} \ll - H^2$) Eqs. (\ref{SINGLE2})--(\ref{SINGLE3}) can be approximated as 
\begin{equation}
3 H^2 \, \overline{M}_{P}^2 = V(\varphi), \qquad 3 H \dot{\varphi} + V_{,\,\varphi} =0,
\label{SINGLE4}
\end{equation}
where the contribution of the spatial curvature is also neglected since it is sharply suppressed during an accelerated stage of expansion. In connection with 
Eqs. (\ref{SINGLE3})--(\ref{SINGLE4}) the last technical remark concerns the cosmic 
time parametrization that can be traded for the conformal time 
coordinate defined as $a(\tau) d \tau = d\, t$. In the conformal time coordinate  the expansion rate and its 
derivative are defined as
\begin{equation}
{\mathcal H} = a^{\prime}/a = a \, H, \qquad \dot{H} = ({\mathcal H}^{\prime} - {\mathcal H}^2)/a^2,
\label{SINGLE5}
\end{equation}
with the prime now denoting a derivation with respect to $\tau$. Both parametrizations of the time coordinate
will be used interchangeably; in the analysis of the effective action of the tensor modes of the geometry (see the appendix \ref{APPB}) the conformal time parametrization turns out to be more 
convenient, as we are going to see. 

\subsubsection{The early expansion rate}
Although the post-inflationary expansion rate modifies the number 
of $e$-folds (and consequently all the inflationary observables), the early expansion rate 
can be estimated, at least approximately,  without a detailed knowledge of the post-inflationary evolution. This happens since the early expansion rate ultimately follows from the analysis of the spectrum of curvature inhomogeneities associated with the CMB scales that left the Hubble radius during the first stages of inflation 
and reentered before matter radiation equality.
\begin{figure}[!ht]
\centering
\includegraphics[width=0.7\textwidth]{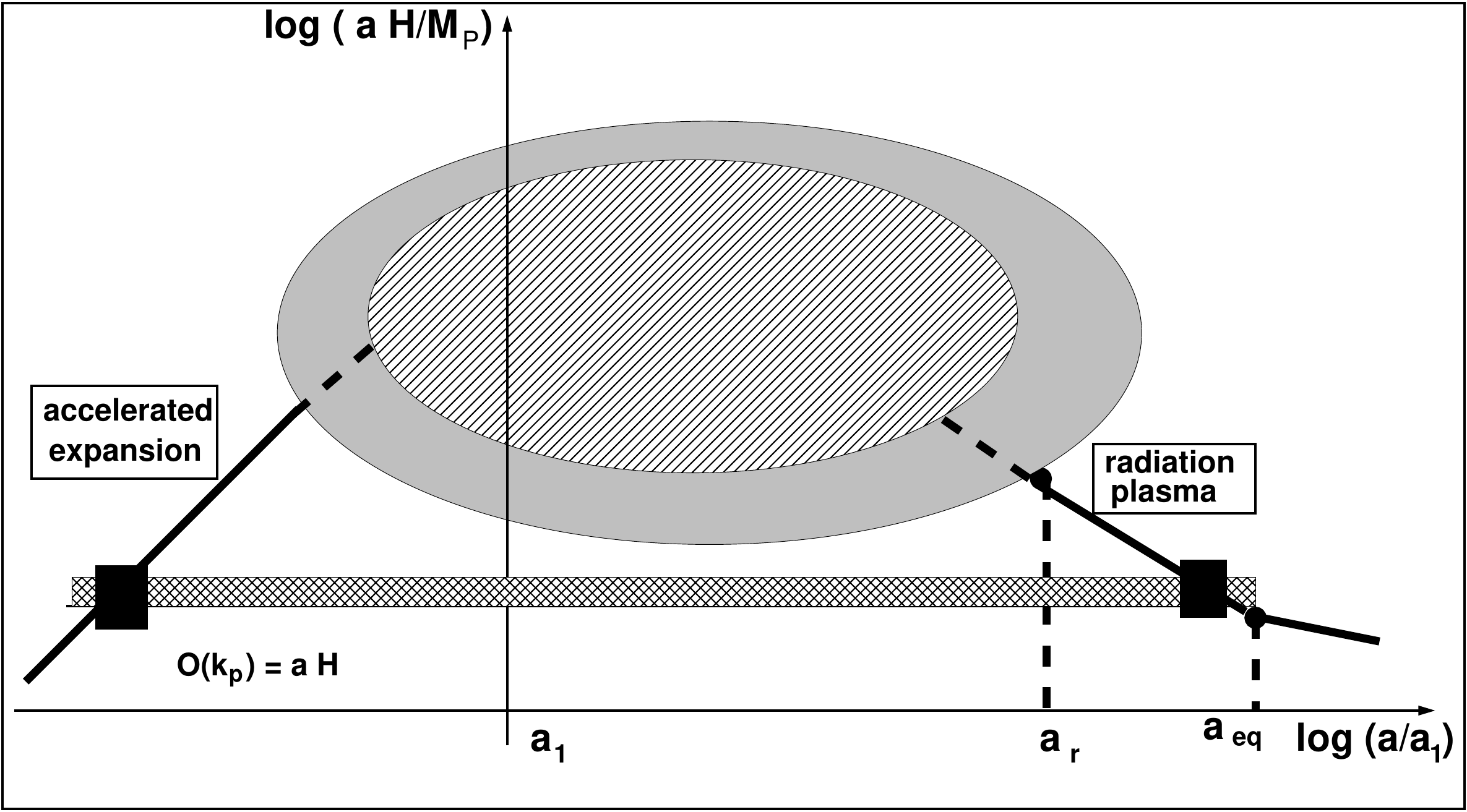}
\caption[a]{The common logarithm of $a\, H$ is illustrated as a function 
of the common logarithm of the scale factor. The two ellipses account for the indetermination 
of the post-inflationary evolution that can have different durations depending on the 
differences in the timeline of the expansion rate. In the lower part of the cartoon the CMB scales  $k = {\mathcal O}(k_{p})$ approximately cross $a\, H$ (see the two filled squares).  }
\label{FIGU0} 
\end{figure}
Since the curvature inhomogeneities are conserved when they evolve for scales larger than the Hubble radius, 
the early expansion rate does not depend upon the total 
number of $e$-folds and the rationale for this statement is illustrated in Fig. \ref{FIGU0} where the common logarithm of $a \, H$ is reported as a function of the common logarithm of the scale factor. While during inflation $a\,H \propto a$, in a radiation-dominated stage $a\, H \propto a^{-1}$;  the two ellipses of  Fig. \ref{FIGU0} parametrize the unknowns of the intermediate evolution but a detailed knowledge of that regime is not strictly necessary to set initial conditions 
for the temperature and for the polarization anisotropies. The CMB observations involve in fact a bunch of wavenumbers $k = {\mathcal O}(k_{p})$ where  $k_{p} = 0.002\,\, \mathrm{Mpc}^{-1}$ is the conventional pivot scale that is used to normalize the large-scale power spectra. These typical scales are  pictorially indicated in the lower part of Fig. \ref{FIGU0} where the two filled squares denote 
the moment where $k = {\mathcal O}(k_{p})$ gets of the order of $a\, H$. While the first crossing time occurs during inflation,  the second one  takes place prior to matter-radiation equality (see the right part of the cartoon).   The scales $k = {\mathcal O}(k_{p})$ become again of the order of $ a\, H$ when the Universe is already dominated by a radiation plasma (i.e. before matter-radiation equality) and their evolution is not affected by 
the unknowns of the post-inflationary evolution that may however modify the spectra at smaller scales; in this 
case the reentry of the fluctuations might not take place when the plasma is dominated by radiation.

\subsubsection{Adiabatic and non-adiabatic solutions}
The argument of Fig. \ref{FIGU0} holds only under the hypothesis that curvature inhomogeneities 
are conserved in the limit $k < a H $, i.e. for typical wavelengths larger than the Hubble radius. 
This is exactly what happens when the evolution of the curvature inhomogeneities  
on comoving orthogonal hypersurfaces (conventionally denoted by ${\mathcal R}$) is 
analyzed in the limit $k < a\, H$ (or $k \tau < 1$). A complementary possibility is to employ $\zeta$ which 
measures the curvature inhomogeneities on the hypersurfaces where the density contrast 
is constant (see, for instance, \cite{BBK2} and references therein). Although ${\mathcal R}$ and $\zeta$ are different variables,  the Hamiltonian constraint associated with the 
relativistic fluctuations of the geometry stipulates that 
\begin{equation}
{\mathcal R} = \zeta - \frac{2\nabla^2 \Psi}{3 \ell_{P}^2 (p_{t} + \rho_{t})},
\label{HC}
\end{equation}
where $\Psi$ is the gauge-invariant generalization of the Newtonian potential (the so-called Bardeen potential \cite{bardeen}) while $(p_{t} + \rho_{t})$ is the total enthalpy density of the sources. According to Eq. (\ref{HC}),  ${\mathcal R}$ and $\zeta$ must obey 
the same evolution for $k \ll a\, H$:
\begin{equation}
{\mathcal R}^{\prime} \simeq \zeta^{\prime} = - \frac{\delta p_{nad}}{p_{t} + \rho_{t}}, \qquad \frac{k}{a\, H} < 1.
\label{FLUC1}
\end{equation}
 In Eq. (\ref{FLUC1}) 
$\delta p_{nad}$ indicates the non-adiabatic pressure fluctuation\footnote{The inhomogeneity 
of the total pressure $\delta p_{t}$ can be decomposed as $\delta p_{t} = c_{st}^2 \delta\rho_{t} 
+ \delta p_{nad}$ where $c_{st}$ is the sound speed of the plasma.} which may  arise if different barotropic fluids are simultaneously present \cite{BBK1,BBK2,BBK3}. In the single field case $\delta p_{nad} =0$  and 
heeding observations  the temperature and polarization anisotropies 
of the CMB are consistent with adiabatic and Gaussian initial conditions  
 \cite{RR1,RR2,RR3}. The true 
 question to ask in connection with Eq. (\ref{FLUC1}) is how many adiabatic solutions and how many non-adiabatic solutions 
are compatible, for instance, with the conservation of ${\mathcal R}$ and 
$\zeta$. This question, usually approached within the
separate Universe picture \cite{bardeen2} (see also \cite{liddleetal} for a reintroduction 
of some of arguments given in \cite{bardeen2}),
has been addressed by Weinberg  in a series of papers \cite{wein1,wein2,wein3}.
In short the argument suggested in Refs. \cite{wein1,wein2,wein3} 
stipulates that the evolution equations of the relativistic fluctuations 
of the geometry have always a pair of physical solutions for which 
$\delta p_{nad} \to 0$ and ${\mathcal R}$ approaches a constant 
for $k\tau\to 0$. In other words, following the usual terminology, there will be 
always at least a pair of adiabatic solutions with $\delta p_{ nad}=0$ 
and ${\mathcal R}$ constant in the limit $ k\tau \to 0$: 
one solution with ${\mathcal R} \neq 0$; the other with ${\mathcal R}=0$. 
Although this analysis is most easily performed in the 
conformally Newtonian gauge \cite{bardeen,MB} the result is in fact gauge-invariant 
since ${\mathcal R}$ is itself gauge-invariant. All in all we have that 
the adiabatic modes are ultimately conserved and this is why 
the early expansion rate does not depend on the post-inflationary evolution.
Therefore, if the only source of large-scale inhomogeneity are the scalar and tensor modes of the geometry excited during the inflationary stage an estimate of the early expansion rate follows from the amplitude of the curvature inhomogeneities assigned for typical scales $k = {\mathcal O}(k_{p})$ as  
\begin{equation}
P_{{\mathcal R}}(k) = {\mathcal A}_{{\mathcal R}} (k/k_{p})^{n_{s}-1}, \qquad P_{T}(k) = {\mathcal A}_{T} (k/k_{p})^{n_{T}},
\label{TWO7A}
\end{equation}
where $n_{s}$ and $n_{T}$ are, respectively, the scalar and the tensor spectral indices while ${\mathcal A}_{{\mathcal R}}$ and ${\mathcal A}_{T}$ are the corresponding amplitudes at the scale $k_{p}$. The ratio between ${\mathcal A}_{T}$ and ${\mathcal A}_{{\mathcal R}}$ is the tensor to scalar ratio and it is customarily denoted by $r_{T}$. The values of $r_{T}$, ${\mathcal A}_{{\mathcal R}}$ and $n_{T}$, in various different combinations, determine the properties of the expansion rate during inflation. In the context of single-field inflationary models these quantities are related via the so-called consistency conditions, as we are now going to discuss in further 
detail.

\subsubsection{The scale-dependence of the expansion rate}
In the absence of non-adiabatic contributions the evolution of the curvature inhomogeneities obeys a source-free 
evolution equation that can be written in a decoupled form (see appendix \ref{APPA} and discussion therein); since the inflationary bound on the expansion rate follows from the large-scale evolution of ${\mathcal R}$, it is practical to recall the equation obeyed by the corresponding Fourier amplitudes:
\begin{equation}
{\mathcal R}_{k}^{\prime\prime} + 2 \frac{z_{\varphi}^{\prime}}{z_{\varphi}} {\mathcal R}_{k}^{\prime} + k^2 {\mathcal R}_{k} =0, \qquad z_{\varphi} = \frac{a \, \varphi^{\prime}}{{\mathcal H}}.
\label{ERRV1}
\end{equation}
 During an inflationary stage of expansion the evolution of $z_{\varphi}$ is proportional to the scale factor $a$ via the (time-dependent) expression of the slow-roll parameter $\epsilon(\tau)$:
\begin{equation}
z_{\varphi} = \frac{ a \, \varphi^{\prime}}{{\mathcal H}} = a \biggl(\frac{\dot{\varphi}}{H}\biggr) = 
a\,\sqrt{2\, \epsilon} \,\overline{M}_{P},\qquad \epsilon = - \dot{H}/H^2 <1.
\label{ERRV2}
\end{equation}
The role of $\epsilon(\tau)$ is here 
the most relevant since it measures the progressive suppression of the Hubble rate during the inflationary stage of expansion. Recalling now Fig. \ref{FIGU0}  Eq. (\ref{ERRV1}) may be approximately\footnote{We privilege the approximate expressions for the evolution of the mode functions but the final results coincide with more accurate (and conventional) strategies such as the ones based 
on the exact evolution of the mode functions during the inflationary stage (see, in this respect, the discussion of appendix \ref{APPA}).} solved  by iteration when $k \ll a H$:
\begin{equation}
{\mathcal R}_{k}(\tau) = {\mathcal R}_{k}(\tau_{ex}) + {\mathcal R}_{k}^{\prime}(\tau_{ex}) \int_{\tau_{ex}}^{\tau} \frac{a_{ex}^2}{a^2(\tau_{1})} \, d\tau_{1} - k^2 \int_{\tau_{ex}}^{\tau} \frac{d \tau_{2}}{a^2(\tau_{2})} \int_{\tau_{ex}}^{\tau_{2}}\, 
{\mathcal R}_{k}(\tau_{1}) \, a^2(\tau_{1}) \, d \tau_{1}.
\label{ERRV6}
\end{equation} 
The first term of Eq. (\ref{ERRV6}) is the constant adiabatic mode which obeys 
${\mathcal R}^{\prime} \simeq 0$ while the second term vanishes asymptotically and 
corresponds to the second adiabatic solution with ${\mathcal R} \to 0$ (see also Eq. (\ref{HC})  
and discussion thereafter); finally the third term of Eq. (\ref{ERRV6}) vanishes exactly  in the large-scale 
limit, i.e. for $k \to 0$. The form of the large-scale solution given by Eq. (\ref{ERRV6}) is actually a concrete example of the general argument suggested in Refs. \cite{wein1,wein2,wein3}.  Let us now go back to Fig. \ref{FIGU0} and focus on the wavenumbers $k = {\mathcal O}(k_{p})$ that are larger than $a \, H$ (or which is the same $k^2 \gg |z_{\varphi}^{\prime\prime}/z_{\varphi}|$); in this case  ${\mathcal R}_{k}(\tau) = q_{k}(\tau)/z_{\varphi}(\tau)$ where $q_{k}(\tau) = e^{- i k \tau}/\sqrt{2 k}$. Since the solution must be continuous and differentiable in $\tau_{ex}$ we can compute the approximate form of the scalar power spectrum
\begin{equation}
P_{{\mathcal R}}(k,\tau) = \frac{k^3}{2 \pi^2} \bigl| {\mathcal R}_{k}(\tau)\bigr|^2 =  \frac{k^3}{2 \pi^2} \frac{\bigl| q_{k}(\tau)\bigr|^2}{z_{\varphi}^2}.
\label{ERRV3}
\end{equation}
From Eq. (\ref{ERRV3}) recalling that $|q_{k}(\tau)|^2 = (2 k)^{-1}$ we obtain
\begin{equation}
P_{{\mathcal R}}(k,\tau) = \frac{k^2}{4\pi^2 a^2} \biggl(\frac{H^2}{\dot{\varphi}^2}\biggr),
\label{ERRV4}
\end{equation}
where the expression of $z_{\varphi}$ has been made explicit. The overall normalization of the 
scalar power spectrum is determined by the expansion rate at the typical time 
$\tau_{\ast} \simeq 1/k$; thanks to Eq. (\ref{SINGLE4}) we can use 
$3 \, H^2 \overline{M}_{P}^2 = V$ and $ 2 \overline{M}_{P}^2 \dot{H} = - \dot{\varphi}^2$ to simplify Eq. (\ref{ERRV4}):
\begin{equation}
P_{{\mathcal R}}(k,\tau) = \frac{k^2}{4 \pi^2 \, a^2(\tau) \overline{M}_{P}^4} \, \biggl( 
\frac{V}{V_{\,,\varphi}}\biggr)^2 = \frac{k^2}{\pi\, a^2(\tau) M_{P}^2 \, \epsilon(\tau)}, 
\label{ERRV5}
\end{equation}
where, as before, $\epsilon(\tau) = - \dot{H}/H^2$. During a de Sitter 
stage of expansion the scale factor is approximately given by $a(\tau) = (- H\tau)^{-1}$
so that Eq. (\ref{ERRV5}) ultimately becomes:
\begin{equation}
P_{{\mathcal R}}(k,\tau) = \frac{\bigl| k\tau\bigr|^2 }{\pi \, \epsilon(\tau)} \biggl(\frac{H^2}{M_{P}^2}\biggr).
\label{ERRVV6}
\end{equation}
For a typical time $\tau_{\ast} =1/k$ we then obtain the power spectrum that should be directly compared with 
the first of the two parametrizations of Eq. (\ref{TWO7A})
\begin{equation}
P_{{\mathcal R}}(k,\tau_{\ast}) = \frac{\bigl| k\tau_{\ast}\bigr|^2 }{\pi \, \epsilon_{\ast}} \biggl(\frac{H_{\ast}^2}{M_{P}^2}\biggr) \simeq \frac{H_{k}^2}{\pi \, \epsilon_{k} M_{P}^2}, \qquad k \tau_{\ast} = {\mathcal O}(1),
\label{ERVV7}
\end{equation}
where $H_{\ast} = H_{k}$ and $\epsilon_{\ast}= \epsilon_{k}$ are  
evaluated for $\tau_{\ast} = 1/k$. Thus, after comparing Eqs. (\ref{ERVV7}) and (\ref{TWO7A}) for $k = {\mathcal O}(k_{p})$ we obtain the wanted estimate of $H_{k}$:
\begin{equation}
\frac{H_{k}^2}{\pi \, \epsilon_{k} \,M_{P}^2} \simeq {\mathcal A}_{{\mathcal R}}, \qquad k = {\mathcal O}(k_{p}).
\label{ERVV8}
\end{equation}
The condition (\ref{ERVV8}) determines the expansion rate in Planck units which is then given by 
\begin{equation}
\frac{H_{k}}{M_{P}} \simeq \sqrt{\pi \, \epsilon_{k} \, {\mathcal A}_{{\mathcal R}} } = \frac{\sqrt{\pi \, r_{T} \, {\mathcal A}_{{\mathcal R}}}}{4}, \qquad k = {\mathcal O}(k_{p}), \qquad {\mathcal A}_{{\mathcal R}} = {\mathcal O}(10^{-9}).
\label{ERVV9}
\end{equation}
The second equality of Eq. (\ref{ERVV9}) follows from the consistency relations stipulating that $r_{T}(k) \simeq 16 \, \epsilon_{k}$; this condition is typical of single-field inlationary scenarios (see appendix \ref{APPA} and discussion therein) and in Eq. (\ref{ERVV9}) we adopted the notation\footnote{In general we have $r_{T}(k,\tau)$ but when $k = {\mathcal O}(k_{p})$ and $\tau \simeq 1/k$ we obtain the standard value of $r_{T}$ which is customarily quoted in the literature \cite{RR1,RR2,RR3}. } $r_{T}= r_{T}(k_{p})$. 
If we assume that $r_{T} \leq 0.03$ and ${\mathcal A}_{{\mathcal R}} = {\mathcal O}(10^{-9})$, Eq. (\ref{ERVV9}) 
implies  $H_{k}/M_{P} \ll 1$ and since $H$ decreases very little during inflation, the expansion rate few $e$-folds before the end of inflation is also comparable with $H_{k}$, i.e.  $H_{k} = {\mathcal O}(H)$. The ratio  $H/H_{k}$ may be estimated from the condition that defines the crossing of a given scale, i.e. $a_{k} H_{k} = k$ 
\begin{equation}
H_{k} = \frac{H_{f}}{1 - \epsilon_{k}} \biggl|\frac{k}{a_{f} \, H_{f}} \biggr|^{ - \frac{\epsilon_{k}}{1 - \epsilon_{k}}} = H  \biggl|\frac{k}{a_{f} \, H} \biggr|^{ - \epsilon_{k}}\biggl[1 + {\mathcal O}(\epsilon_{k})\biggr].
\label{ERVV10}
\end{equation}
But for typical wavenumbers $k= {\mathcal O}(k_{p})$ it turns out that $k_{p}\,\ll \, |a_{f} \, H_{f}|$; more specifically the approximate value of $k_{p}/(a_{f}\, H_{f})$ is estimated as
\begin{equation}
\frac{k_{p}}{a_{f}\, H_{f}} ={\mathcal O}(10^{-26}) \,\biggl(\frac{k_{p}}{0.002\,\, \mathrm{Mpc}^{-1}}\biggr) \biggl(\frac{r_{T}}{0.03}\biggr)^{-1/4} \biggl(\frac{{\mathcal A}_{{\mathcal R}}}{2.41\times 10^{-9}}\biggr)^{-1/4} \, \biggl(\frac{h_{0}^2 \, \Omega_{R0}}{4.15\times 10^{-5}}\biggr)^{-1/4},
\label{ERVV11}
\end{equation}
where $H_{f} \simeq H$.
If we now insert Eq. (\ref{ERVV11}) into Eq. (\ref{ERVV10}), we can conclude, as previously anticipated  that $H_{k} \simeq H$ so that 
\begin{equation}
\frac{H_{k}}{M_{P}} \simeq \frac{H}{M_{P}} = 5.32\times 10^{-6} \biggl(\frac{{\mathcal A}_{{\mathcal R}}}{2.41\times 10^{-9}}\biggr)^{1/2}\,   \biggl(\frac{r_{T}}{0.06}\biggr)^{1/2}.
\label{ERVV12}
\end{equation}
Equation (\ref{ERVV12}) estimates the expansion rate but does not imply any specific duration of the inflationary phase.  As we are going to see in the following subsection, the duration of the inflationary stage of expansion is ultimately related to the nature and to the rate of the post-inflationary evolution.

\subsection{What do we know about the late expansion history?}
\label{subsec23}
As already discussed after Eq. (\ref{TWO6}), the duration of inflation does depend on the post-inflationary evolution and this means that different expansion histories affect the number of $e$-folds required to bring all the physical scales of the model in causal contact. Different possibilities are  
examined hereunder with the purpose of quantifying the theoretical indetermination on the total number of $e$-folds. 

\subsubsection{A radiation-dominated Universe?}
One of the standard (unproven) assumptions both of the hot big bang model and of the 
conventional post-inflationary evolution is that the plasma must always 
be dominated by radiation even before the scale of big bang nucleosynthesis 
where the deviations from radiation dominance are severely constrained (see section \ref{sec4} and discussion therein). The gist of this argument is that, in the early hot and dense plasma, it is appropriate to assume an equation of state corresponding to a gas of relativistic particles; this choice is compatible with all the current data but it is neither compelling nor unique. A radiation-dominated stage of expansion extending between $H ={\mathcal O}(10^{-5})\, M_{P}$ and the equality time is one of the assumptions customarily adopted for the timeline of the expansion rate in the context $\Lambda$CDM paradigm. In terms of the cartoon of Fig. \ref{FIGU0} we would then have that $a \, H \propto a$ during inflation while in the radiation stage $a\, H \propto a^{-1}$. This means, in practice, that the energy density of the background scales approximately as $a^{-4}$ between the end of inflation and the equality time. The critical number of $e$-folds 
required to fit inside the current Hubble patch the redshifted value of $H^{-1}$ (i.e. 
the approximate size of the event horizon at the onset of inflation) follows from the condition:
\begin{equation}
H_{i}^{-1} \biggl(\frac{a_{0}}{a_{i}}\biggr) \simeq H_{0}^{-1}, \qquad H_{i} \simeq H,
\label{TWO8}
\end{equation}
where $a_{0}$ is the current value of the scale factor\footnote{Throughout the present article 
the scale factor is normalized as $a_{0} =1$. This remark is quite relevant since 
by choosing $a_{0}=1$ we will have that comoving and physical 
frequencies of the relic gravitons coincide at the present time.}.
Equation (\ref{TWO8}) can be made even more explicit by rewriting it in a slightly different manner:
\begin{equation}
\frac{a_{0} \, H_{0}}{a_{i}\, H_{i}} = \biggl(\frac{a_{0} \, H_{0}}{a_{eq}\, H_{eq}} \biggr) \, 
\biggl(\frac{a_{eq} \, H_{eq}}{a_{r}\, H_{r}} \biggr) \biggl(\frac{a_{r} \, H_{r}}{a_{f}\, H_{f}} \biggr)
\biggl(\frac{a_{f} \, H_{f}}{a_{i}\, H_{i}} \biggr) \simeq 1.
\label{TWO9}
\end{equation}
The terms appearing in the second equality of Eq. (\ref{TWO9}) can be directly evaluated
when the post-inflationary evolution is dominated by radiation; for instance, 
 by definition, $ 3 H_{eq}^2 \overline{M}_{P}^2 = 2 \rho_{M0}(a_{0}/a_{eq})^3$ where 
$\rho_{M0}$ denotes the present matter density and the factor $2$ follows since, at equality, the matter and radiation energy density coincide; furthermore the redshift to equality can be 
estimated as $(a_{0}/a_{eq}) = \Omega_{M\,0}/\Omega_{R\,0}$ where $\Omega_{M\,0}$ and $\Omega_{R\,0}$ are the critical fractions of matter and radiation in the concordance scenario. All in all we can eventually estimate 
\begin{equation}
 \biggl(\frac{a_{0} \, H_{0}}{a_{eq}\, H_{eq}} \biggr) = \frac{1}{(2 \, \Omega_{R\, 0})^{1/4}} \sqrt{\frac{H_{0}}{H_{eq}}}.
\label{TWO10}
\end{equation}
Moreover, between the equality time and $a_{r}$ the evolution is dominated by radiation, thanks to Eq. (\ref{TWO10}), Eq. (\ref{TWO9}) becomes
\begin{equation}
\biggl(\frac{a_{eq}^4 \, H_{eq}^2}{a_{r}^4 H_{r}^2}\biggr) = \frac{a_{eq}^4 \, T_{eq}^4 \, g_{\rho,\,eq}}{a_{r}^4 \, T_{r}^4 g_{\rho,\, r}} = \biggl(\frac{g_{\rho,\,eq}}{g_{\rho,\,r}}\biggr)  \biggl(\frac{g_{s,\,r}}{g_{s,\,eq}}\biggr)^{4/3},
\label{TWO11}
\end{equation} 
where $g_{\rho}$ denotes the number of effective relativistic degrees of freedom appearing 
in the {\em energy density} of the plasma while $g_{s}$ corresponds to the number of effective relativistic degrees of freedom of the {\em entropy density}. 
If the entropy density is conserved  between the $r$-stage and the equality epoch we should have that $ g_{s,\,r} \, a_{r}^3 \, T_{r}^3 = g_{s,\,eq} \, a_{eq}^3 \, T_{eq}^3$ and this observation affects the redshift between the two epochs\footnote{The difference due to $g_{s}$ and $g_{\rho}$ in the final results is actually negligible for the present purposes and it involves a 
factor $1.3$ (instead of $1$) at the level of Eq. (\ref{TWO12}). However, from the conceptual viewpoint 
this difference is certainly relevant and this is why it will be taken into account.}: 
\begin{equation}
\frac{a_{eq} \, H_{eq}}{a_{r} \, H_{r}} = \biggl(\frac{g_{\rho,\,eq}}{g_{\rho,\,r}}\biggr)^{1/4}  \biggl(\frac{g_{s,\,r}}{g_{s,\,eq}}\biggr)^{1/3} \, \sqrt{\frac{H_{eq}}{H_{r}}}.
\label{TWO12}
\end{equation}
Because during inflation the Hubble rate is nearly constant  (i.e. $H_{f} \simeq H_{i} \simeq H$), 
once Eqs. (\ref{TWO11})--(\ref{TWO12}) are inserted 
into Eq. (\ref{TWO9})  the number of $e$-folds $\overline{N}_{max} = \ln{(a_{f}/a_{i})}$ can be 
determined by requiring that Eqs. (\ref{TWO8})--(\ref{TWO9}) are satisfied. 
The final result for $\overline{N}_{max}$ becomes:
\begin{equation}
e^{\overline{N}_{max}} = (2 \, \Omega_{R\, 0})^{1/4} {\mathcal C}(g_{s}, g_{\rho}, \tau_{r}, \tau_{eq}) \sqrt{\frac{H}{H_{0}}},
\label{TWO13}
\end{equation}
where, for the sake of conciseness, we wrote ${\mathcal C}(g_{s}, g_{\rho},\tau_{r},\tau_{eq}) =(g_{\rho,\,r}/g_{\rho,\,eq})^{1/4}  (g_{s,\,eq}/g_{s,\,r})^{1/3}$.
Once more Eq. (\ref{TWO13}) determines the critical number of $e$-folds 
necessary to fit the redshifted value of $H^{-1}$  inside $H_{0}^{-1}$, as postulated in Eq. (\ref{TWO8}). It should be stressed that  $\overline{N}_{max}$ corresponds to {\em the maximal number of $e$-folds currently accessible to large-scale 
observations}. In other words the conditions  (\ref{TWO8})--(\ref{TWO9}) fix $\overline{N}_{max}$ by requiring that all the physical scales inside the inflationary (event) horizon 
are all contained inside the current Hubble patch $H_{0}^{-1}$.
It is of course possible that the total number of $e$-folds {\em exceeds} $\overline{N}_{max}$ and this happens if we require 
\begin{equation}
e^{\overline{N}} > (2 \, \Omega_{R\, 0})^{1/4} {\mathcal C}(g_{s}, g_{\rho},\tau_{r},\tau_{eq}) \sqrt{\frac{H}{H_{0}}}.
\label{TWO13a}
\end{equation}
The condition (\ref{TWO13a}) implies that some of the scales originally contained inside the inflationary (event) horizon are today larger than the current value of the Hubble patch; in this case the causal connection 
is realized on a region possibly larger than $H_{0}^{-1}$.
The overlines appearing both in $\overline{N}$ and $\overline{N}_{max}$ remind that 
the corresponding quantities have been deduced for a post-inflationary 
evolution dominated by radiation. From Eq. (\ref{TWO13}) the explicit value of $\overline{N}_{max}$ becomes
\begin{eqnarray}
\overline{N}_{max}  &=& 61.9 - \ln{(h_{0}/0.7)} + \frac{1}{4} \ln{\biggl(\frac{r_{T}}{0.06}\biggr)} +\frac{1}{4} \ln{\biggl(\frac{{\mathcal A}_{{\mathcal R}}}{2.41\times 10^{-9}}\biggr)} 
\nonumber\\
&+& \ln{{\mathcal C}(g_{s}, \, g_{\rho}, \tau_{r}, \tau_{eq})}+   \frac{1}{4} \ln{\biggl(\frac{h_{0}^2 \, \Omega_{R0}}{4.15\times 10^{-5}}\biggr)}, 
\label{TWO14}
\end{eqnarray}
and it is, as anticipated, ${\mathcal O}(60)$. For the actual estimates relating Eqs. (\ref{TWO13}) and (\ref{TWO14}) the following three observations
should be emphasized:
\begin{itemize}
\item{} the inflationary expansion rate is estimated from the amplitude of the scalar power 
spectrum and, more specifically, from Eq. (\ref{ERVV12});
\item{} it is assumed that, in practice, there is no energy loss between the inflationary phase 
and the post-inflationary evolution (i.e. $H_{r} \simeq H$);
\item{} in the standard situation where $g_{s,\, r}= g_{\rho,\, r} = 106.75$ and $g_{s,\, eq}= g_{\rho,\, eq} = 3.94$ the value of ${\mathcal C}(g_{s}, \, g_{\rho},\tau_{r}, \tau_{eq})$ is given by  $0.75$; the contribution of ${\mathcal C}(g_{s}, \, g_{\rho}, \tau_{r}, \tau_{eq})$ to Eq. (\ref{TWO14}) is numerically not essential for the determination of $\overline{N}_{max}$.
\end{itemize}
The approximation $H_{r} \simeq H$ is customarily enforced by  CMB experiments when setting bounds, for instance, on the total number of $e$-folds \cite{RR1,RR2,RR3} and although 
energy is lost during reheating,  in the case 
of single-field inflationary models this approximation is rather plausible since 
the combined action of the reheating and of the preheating dynamics leads to 
a process that is almost sudden \cite{BBK1,BBK2,BBK3}. In this sense, if $H_{last}$ denotes the expansion rate 
during the last few $e$-folds of inflationary expansion, it is true that $H_{r} < H_{last}$; however, even for a difference of few orders of magnitude the quantitative arguments illustrated here 
will not be crucially affected. We recall that, conventionally, the reheating is the period where the entropy observed in the present Universe is produced and it typically takes place when all the large-scale inhomogeneities of observational interest are outside the horizon. The different approaches to the reheating dynamics are not expected to affect the large-scale power spectra \cite{MB}.

The number of inflationary $e$-folds introduced in Eqs. (\ref{TWO6})--(\ref{TWO7}) depends on the post-inflationary evolution but it also scale-dependent. This happens because the actual observations always probe a typical scale so that this dependence also enters the number of $e$-folds and the expansion rate. In what follows $N_{k}$ and $H_{k}$ are associated, respectively, with the number of $e$-folds and with the expansion rate at the crossing of the CMB scales $k = {\mathcal O}(k_{p})$. Even though $H_{k}$ and $H$ are conceptually different, $H_{k}/H = {\mathcal O}(1)$ since the curvature scale decreases very slowly during the inflationary stage.
Most of the previous estimates can the be repeated in the case of  $\overline{N}_{k} = \ln{(a_{f}/a_{k})}$.
As in the case of $\overline{N}_{max}$, also $\overline{N}_{k}$ is estimated in the present 
section for a post-inflationary thermal history dominated by a radiation background and this is why the overline is included; the values of $\overline{N}_{k}$ are implicitly determined from:
\begin{equation}
\frac{k}{a_{k}\, H_{k}} = e^{\overline{N}_{k}} \, \biggl(\frac{H_{f}}{H_{k}}\biggr) \frac{k}{a_{f}\, H_{f}},
\label{TWO14a}
\end{equation} 
and when the given wavenumber is of the order of the comoving expansion rate $k \simeq a_{k}\, H_{k}$. The latter condition fixes the value of $\overline{N}_{k}$ not only in terms of $H_{f}$ (the expansion rate at the end of inflation) but also as a function of the subsequent expansion history, exactly as in the case of $\overline{N}_{max}$. By then repeating all all the different steps in the case of Eq. (\ref{TWO14a}) we deduce 
\begin{equation}
e^{\overline{N}_{k}} = ( 2 \, \Omega_{R\,0})^{1/4} \frac{H_{k}}{\sqrt{H_{0} \, H_{f}}}\, {\mathcal C}(g_{s}, g_{\rho},\tau_{r},\tau_{eq})\biggl(\frac{a_{0} \, H_{0}}{k}\biggr).
\label{TWO14b}
\end{equation}
If  the determinations of $\overline{N}_{k}$ and $\overline{N}_{max}$ are compared in the case of a post-inflationary 
evolution dominated by radiation we obtain:
\begin{equation}
\overline{N}_{k} = \overline{N}_{max} - \ln{\biggl(\frac{k}{a_{0}\, H_{0}}\biggr)} -\ln{(H_{k}/H_{f})}.
\label{TWO14c}
\end{equation}
As already mentioned in Eq. (\ref{ERVV12}), $H_{k} = {\mathcal O}( H_{f})$ so that $\overline{N}_{k}$ and $\overline{N}_{max}$ are of the same order as long as $k \simeq a_{0}\, H_{0}$. The explicit value of $\overline{N}_{k}$ can then be written as\footnote{The result of Eq. (\ref{TWO14d}) is in fact obtained from Eq. (\ref{TWO14c}) by recalling that $H_{k}/M_{P} = \sqrt{ \pi \epsilon_{k} {\mathcal A}_{{\mathcal R}}}$. 
From the consistency relations we also have that $r_{T} \simeq 16 \epsilon_{k}$ so that, for $r_{T} =0.06$ Eq. (\ref{TWO14d}) demands that the 
value of $\overline{N}_{k}$ is given by 
$\overline{N}_{k} = 59.7384$ (while all the other parameters are kept fixed at their typical values).} 
\begin{eqnarray}
\overline{N}_{k} &=& 59.408  + \frac{1}{4} \ln{\biggl(\frac{\epsilon_{k}}{0.001}\biggr)} +\frac{1}{4} \ln{\biggl(\frac{{\mathcal A}_{{\mathcal R}}}{2.41\times 10^{-9}}\biggr)} + \ln{{\mathcal C}(g_{s}, \, g_{\rho}, \tau_{r}, \tau_{eq})}   
\nonumber\\
&-& \ln{\biggl(\frac{k}{0.002\,\,\mathrm{Mpc}^{-1}}\biggr)} + \frac{1}{4} \ln{\biggl(\frac{h_{0}^2 \, \Omega_{R0}}{4.15\times 10^{-5}}\biggr)} - \frac{1}{2} \ln{\biggl(\frac{H_{1}}{H_{k}}\biggr)}.
\label{TWO14d}
\end{eqnarray}
\subsubsection{An extra phase preceding big bang nucleosynthesis}
\begin{figure}[!ht]
\centering
\includegraphics[width=0.7\textwidth]{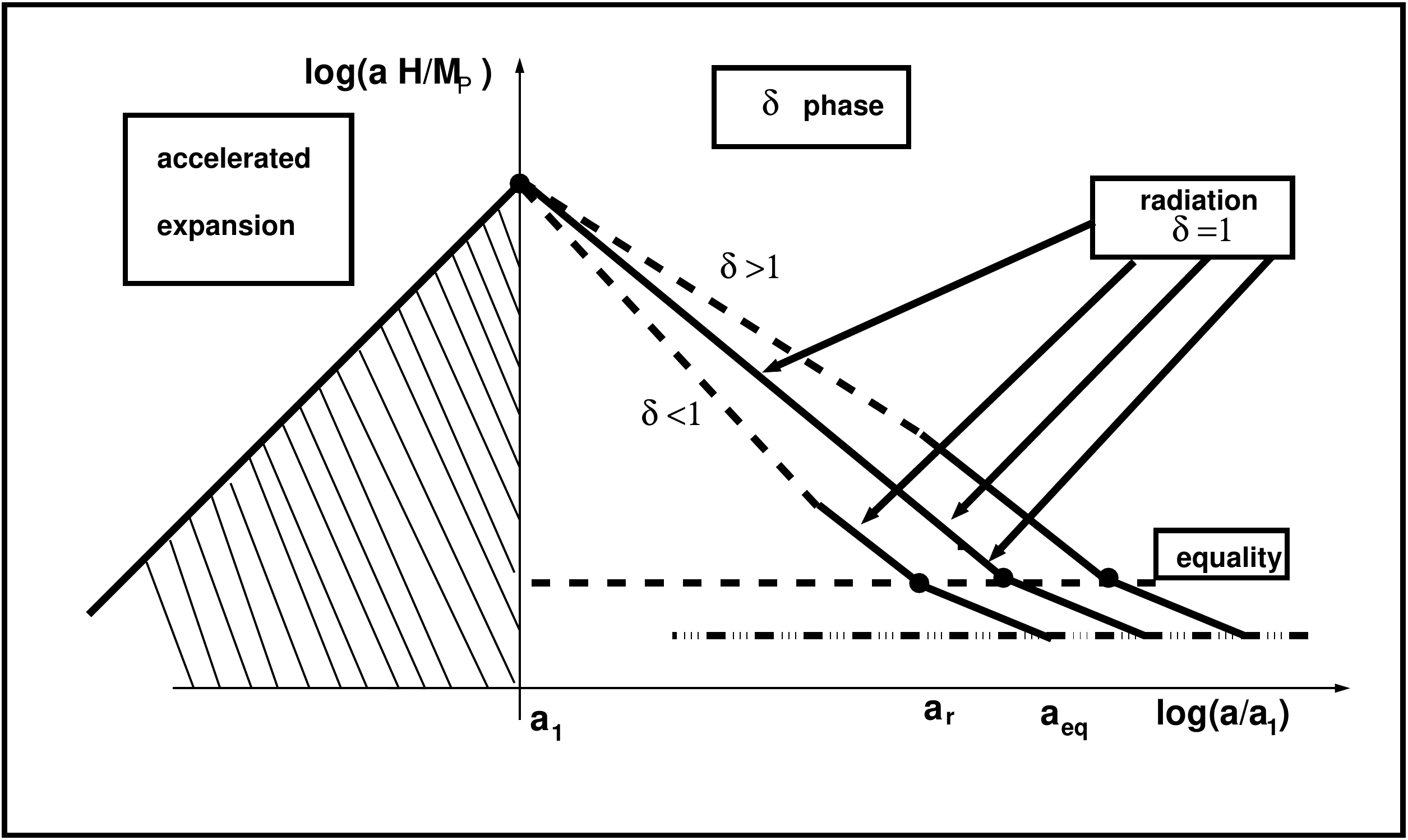}
\caption[a]{The conventional radiation-dominated epoch (taking place for $a> a_{r}$) is preceded by an intermediate phase parametrized by the value of $\delta$. If $\delta\to 1$ we recover the case of a single post-inflationary radiation epoch. When $\delta < 1$ the expansion rate is slower than radiation; conversely if $\delta > 1$ the expansion rate is faster than radiation.}
\label{FIGU0b} 
\end{figure}

In the previous subsection we considered a timeline dominated 
by radiation between the end of inflation and the equality 
epoch. We are now going to suppose that, prior to radiation dominance, 
the expansion rate is modified for a sufficiently long period where the 
expansion rate can be either faster or slower than radiation. 
Probably the simplest example along this perspective consists 
in adding a further stage of expansion between the end of inflation and the 
onset of the radiation-dominated phase. The ellipses of Fig. \ref{FIGU0} are now replaced by the cartoon of Fig. \ref{FIGU0b} and the following comments are in order:
\begin{itemize}
\item{} as before during inflation we have that $H \,a \propto a$ 
while in a radiation stage we would get $a \, H \propto a^{-1}$: the simplest 
timeline is then the one illustrated with the full thick line;
\item{}  prior to the onset
of the radiation stage and after inflation we have instead that $a\, H \propto a^{-1/\delta}$
where now $\delta$ parametrizes the expansion rate in the intermediate 
regime;
\item{} if $\delta >1$ the expansion rate is faster than radiation; conversely 
when $\delta < 1$ the expansion rate is slower than radiation (see, in this respect,
the dashed timelines of Fig. \ref{FIGU0b}).
\end{itemize}
According to Fig. \ref{FIGU0b} the condition
imposed by Eq. (\ref{TWO9}) becomes different and its modification depends 
on $\delta$. Indeed, if the estimate of Eq. (\ref{TWO9}) is repeated, the value of $\overline{N}_{max}$ gets shifted \cite{EXP1,EXP2,EXP3} (see also \cite{LID1,MGshift})
\begin{equation}
\overline{N}_{max} \to N_{max} = \overline{N}_{max} + \frac{(\delta -1)}{2 (\delta + 1)} \ln{(H_{r}/H)},
\label{TWO15}
\end{equation}
where we now denote with $N_{max}$ the maximal number of $e$-folds for a {\em generic} 
post-inflationary evolution while $\overline{N}_{max}$ corresponds to the case 
of a timeline dominated by radiation right after the end of the inflationary expansion.
This is why, as anticipated, in the case $\delta\to 1$  the timeline of Fig. \ref{FIGU0b} 
reproduces a (single) radiation-dominated stage of expansion and 
 $N_{max} \to \overline{N}_{max}$ (see Eqs. (\ref{TWO13})--(\ref{TWO15}) and discussion therein). Because 
 $H_{r} < H < H_{i}$ in Eq. (\ref{TWO15}), for arbitrary values of $\delta$ the following two remarks are in order: 
\begin{itemize}
\item{} when the background expands faster than radiation (i.e. $\delta > 1$) the value 
of $N_{max}$ gets smaller than in the case of radiation dominance (i.e. 
$N_{max} < \overline{N}_{max}$);
\item{} conversely when the expansion rate is slower than radiation (i.e. $\delta < 1$) we have that $N_{max} > \overline{N}_{max}$. 
\end{itemize}
The orders of magnitude involved in Eq. (\ref{TWO15}) 
are estimated by considering that the typical expansion scale of big bang 
nucleosynthesis (BBN) is approximately $H_{bbn} = {\mathcal O}(10^{-44}) \, M_{P}$ whereas 
the inflationary expansion rate follows from Eq. (\ref{ERVV12})  (i.e. $ H \simeq \sqrt{ \pi {\mathcal A}_{{\mathcal R}} \, r_{T}}/4$). This means that the relation between $N_{max}$ 
and $\overline{N}_{max}$ is approximately given by:
\begin{equation}
N_{max} = \overline{N}_{max} - {\mathcal O}(45) \biggl(\frac{\delta -1}{\delta+1}\biggr),\qquad 
H_{r} = {\mathcal O}(H_{bbn}).
\label{TWO16}
\end{equation}
Let us now suppose, for instance, that $\delta >1$. If the sources for the evolution of the geometry are parametrized in terms of perfect fluids with barotropic equation of state  $\delta =2/(3 w +1)$ so that $\delta_{max} \to 2$ corresponds\footnote{To avoid confusions  $w_{min}$ and $w_{max}$ indicate, respectively, the minimal and the maximal values of $w$.} to $w \to w_{min} =0$. In this case, from Eqs. (\ref{TWO15})--(\ref{TWO16}), $N_{max} = \overline{N}_{max} 
- {\mathcal O}(15)$. In case $\delta < 1$ the post-inflationary expansion rate between 
$H$ and $H_{r}$ is instead slower than radiation so that we would have $N_{max} > \overline{N}_{max}$. Again, assuming  the background is driven by 
perfect barotropic fluids, $\delta_{min} =1/2$ and it corresponds to a plasma 
$w_{max} = 1$ where the sound speed and the speed of light coincide. Therefore 
 for $\delta \to delta_{min} = 1/2$ Eqs. (\ref{TWO15})--(\ref{TWO16}) imply that $N_{max} = \overline{N}_{max} + {\mathcal O}(15)$.
In summary the critical number of $e$-folds required 
to fit the redshifted event horizon inside the current value of the Hubble radius
does depend on the post-inflationary expansion rate; thanks to the 
results of Eqs. (\ref{TWO15})--(\ref{TWO16}) we can then estimate the theoretical error
associated with the unknown post-inflationary expansion rate as
\begin{equation}
N_{max} = \overline{N}_{max} \pm {\mathcal O}(15), \qquad {\mathcal O}(60) < \overline{N}_{max} = {\mathcal O}(62),
\label{TWO17}
\end{equation}
where, we remind, the value of $\overline{N}_{max}$ is determined in the case $\delta \to 1$ corresponding to a radiation dominated stage of expansion.The same kind of evaluation 
leading to Eq. (\ref{TWO17}) can be repeated for different classes of sources driving the background evolution.
For instance the post-inflationary expansion rate might correspond to a stage dominated by 
an oscillating scalar field with an approximate potential $\varphi^{2 q}$ near 
the origin \cite{turn1} (see also \cite{turn2,turn3,turn4}). In this case $\delta = (q+1)/(2 q-1)$ 
and the condition $\delta \geq 1$ implies that $q \leq 2$; this means, once more, that 
$\delta_{max} = 2$ while $\delta_{min} \to 1/2$ corresponding either to the asymptote $q \gg 1$ or to the absence of the potential. Thus the case $\delta_{min} \to 1/2$  may be realized in a number of physically different situations \cite{MGshift2}.

All in all, if  the total number of $e$-folds 
is ${\mathcal O}(60)$  in the case of a radiation-dominated universe, Eq. (\ref{TWO17}) suggests that the potential indetermination due to a modified expansion rate ranges\footnote{We stress, in this respect, that the indetermination 
on $N_{max}$ is not related to the considerations discussed in Eq. (\ref{TWO13a}): in that 
context $N$ denoted the {\em total number of $e$-folds} which may be, for different reasons, 
larger than $N_{max}$. } 
between $45$ and $75$.  The same indetermination affecting $N_{max}$ also enters the value of $N_{k}$. Indeed even in the presence of an intermediate stage preceding the conventional 
radiation-dominated epoch  Eqs. (\ref{TWO14b})--(\ref{TWO14c}) remain fully valid.
The value of $N_{k}$ is relevant for various phenomenological aspects of the problem since it affects the inflationary observables that are specifically discussed later on\footnote{For the moment it is sufficient to note that, for monomial inflationary potentials, the tensor to scalar ratio scales as $N_{k}^{-1}$ whereas for plateau-like potential the same 
quantity scales as $N_{k}^{-2}$. Both values may get eventually larger or smaller than in the 
radiation phase depending on the post-inflationary expansion rate. Moreover, as we shall see, the value of $N_{k}$ ultimately affects the value of the maximal frequency of the relic graviton spectrum.} in section \ref{sec4}. We finally recall that Eq. (\ref{ERVV11}) has been correctly deduced in the case of radiation dominance (i.e. $\delta \to 1$ in the language of this subsection) and the same indetermination affecting the number of $e$-folds may also modify the value of the pivot scale in units of the inflationary expansion rate. In the presence of the $\delta$-phase illustrated 
in Fig. \ref{FIGU0b} we have that Eq. (\ref{ERVV11}) gets modified as
\begin{equation}
\frac{k_{p}}{a_{f}\, H_{f}} ={\mathcal O}(10^{-26}) ( H_{r}/H_{f})^{(\delta -1)/[2 (\delta +1)]}.
\label{TWO7Gnew}
\end{equation}
Depending on the value of $H_{r}$, when the expansion rate is faster than radiation the value of $k_{p}/(a_{f} H_{f})$ may get smaller than $10^{-26}$. The opposite is true when the background expands at a rate slower than radiation since, in this second instance, $k_{p}/(a_{f} H_{f})$ gets larger than $10^{-26}$. In both situations, however, it is fully justified to assume $H_{f} \simeq H_{k} \simeq H$, as already established in Eq. (\ref{ERVV10}).

\subsubsection{Multiple stages preceding big bang nucleosynthesis}
\begin{figure}[!ht]
\centering
\includegraphics[width=0.7\textwidth]{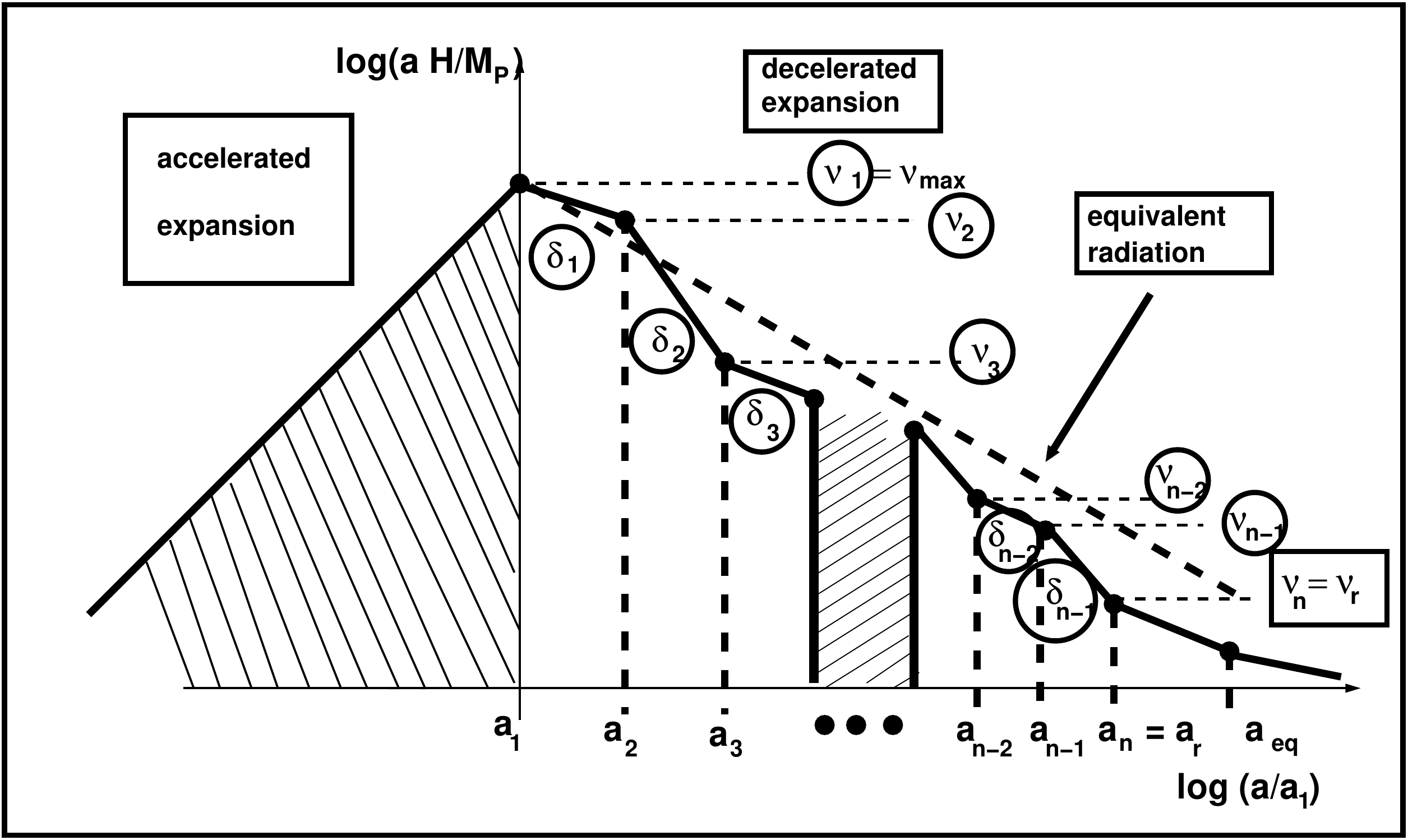}
\caption[a]{As in the previous cartoons of this section, on the vertical axis the common logarithm of $a\, H$ is reported as a function 
of the common logarithm of the scale factor. The region at the left corresponds, as usual,  to the inflationary evolution while for $a> a_{1}$ the background is decelerated. It is also understood throughout the discussion that the post-inflationary epoch is bounded by the curvature scale of big bang nucleosynthesis so that $H_{r} \geq 10^{-44} \, M_{P}$. In this plot we adopt the convention that $a_{n}=a_{r}$ and $H_{n} =H_{r}$ implying that 
the end of the sequence of intermediate stages coincides with the onset of the radiation-dominated evolution.}
\label{FIGU0c} 
\end{figure}

\begin{figure}[!ht]
\centering
\includegraphics[height=7cm]{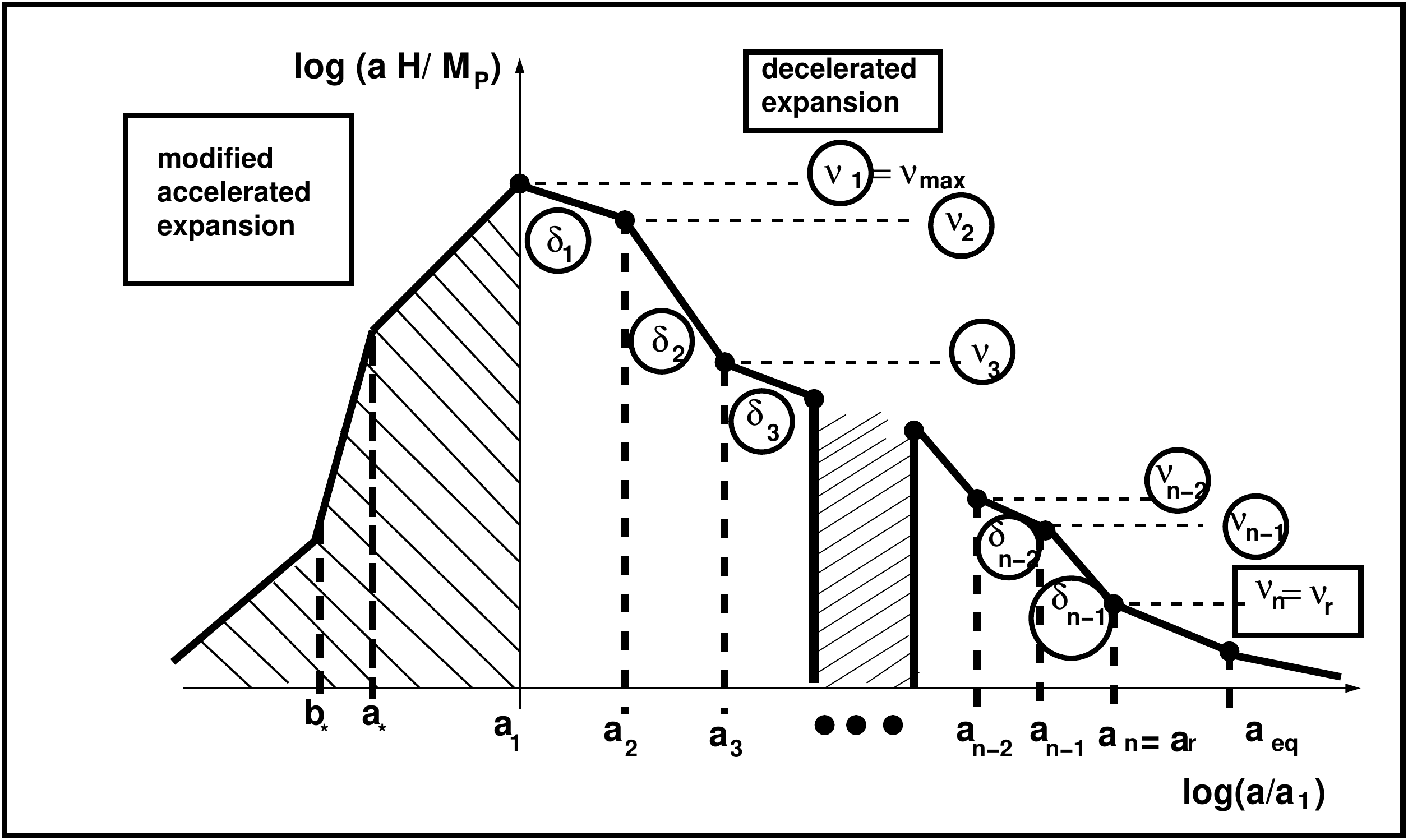}
\caption[a]{As in the previous figure the common logarithm of the comoving expansion rate is illustrated as 
a function of the common logarithm of the scale factor. Prior to the onset of the standard inflationary stage of expansion the evolution 
is however modified.}
\label{FIGU3}      
\end{figure}

A natural extension of the results obtained in Eqs. (\ref{TWO15})--(\ref{TWO17}) involves the presence of multiple post-inflationary stages parametrized by different values of the 
expansion rate conventionally denoted by $\delta_{i}$ with $i =1,\,.\,.\,n$. It is actually 
plausible to generalize the previous considerations by replacing the single $\delta$ stage with $n$ intermediate phases of expansion preceding the epoch of radiation dominance, as illustrated in Fig. \ref{FIGU0c}.
The cartoon of Fig. \ref{FIGU0b} is then substituted by the timeline of Fig. \ref{FIGU0c} where 
the initial stage of the post-inflationary evolution begins after the end of inflation 
(i.e.  $H \simeq H_{f} = H_{1}$) while the $n$-th stage conventionally coincides with the standard radiation-dominated evolution i.e. $a_{r} = a_{n}$ and $\delta_{n} = 1$. As already explained before, we should always require $H_{r} >10^{-44} \, M_{P}$ implying that the big bang nucleosynthesis takes place when radiation is already dominant. 
During the $i$-th stage of the sequence $a \, H \propto a^{- 1/\delta_{i}}$ and the expression of 
$N_{max}$ given in Eq. (\ref{TWO15}) can be generalized to the 
timeline of Fig. \ref{FIGU0c}:
\begin{equation}
N_{max} = \overline{N}_{max} + \frac{1}{2}\sum_{i}^{n-1} \, \biggl(\frac{\delta_{i} -1}{\delta_{i} + 1}\biggr) \, \ln{\xi_{i}}.
\label{TWO15N}
\end{equation}
The various $\xi_{i}$ appearing in Eq. (\ref{TWO15N}) measure the duration of each post-inflationary stage of expansion and since the rate is always decreasing we may conclude that
\begin{equation}
\xi_{i} = \frac{H_{i+1}}{H_{i}} < 1. 
\label{TWO15O}
\end{equation}
Since, by construction, $a_{n} = a_{r}$ we also have that $H_{n}= H_{r}$; this means that 
$\xi_{n-1} = H_{n}/H_{n-1} = H_{r}/H_{n-1}$. It finally follows from Fig. \ref{FIGU0c} that the 
product of all the $\xi_{i}$ coincides with $H_{r}/H$, namely
\begin{equation}
\prod_{i=1}^{n-1} \xi_{i} = \xi_{1}\,\,\xi_{2}\,\,.\,.\,.\,\,\xi_{n-2} \,\, \xi_{n-1} = \xi_{r} = H_{r}/H < 1.
\label{TWO15P}
\end{equation}
This also means that if all the $\delta_{i}$ are equal the result of Eq. (\ref{TWO15N}) coincides with the one of Eq. (\ref{TWO15}) obtained for a single $\delta$-phase. 
If the post-inflationary plasma is only dominated by radiation then in Eq. (\ref{TWO15N}) all the $\delta_{i}$ go to $1$ and the whole contribution disappears. Conversely when some of the $\delta_{i}$ are smaller than $1$ both $N_{max}$ and $N_{k}$ increase. For $\delta_{i} > 1$ we may have the opposite effect suggesting an overall reduction of $N_{max}$ and $N_{k}$. Both effects are relevant in low-frequency region of the relic graviton spectrum, as we are going to see more specifically in section \ref{sec4}. Indeed 
the result of Eq. (\ref{TWO14c}) remains valid also for the timeline of Fig. \ref{FIGU0c}; this 
means that not only $N_{max}$ but also $N_{k}$ gets reduced or enhanced depending 
on the values of the various $\delta_{i}$, as it follows from the explicit 
expression of $N_{k}$:
\begin{equation}
N_{k} = \overline{N}_{k} + \frac{1}{2}\sum_{i}^{n-1} \, \biggl(\frac{\delta_{i} -1}{\delta_{i} + 1}\biggr) \, \ln{\xi_{i}}.
\label{NK1}
\end{equation}
Besides the case of Fig.  \ref{FIGU0c} we can also take into account a further possibility 
that is  illustrated in Fig. \ref{FIGU3}. Prior to a conventional stage of inflationary expansion there could be a stage where the expansion rate is different. This may happen for various reasons and, in the most conservative perspective, it could be that the evolution 
of the relic gravitons develops a refractive index even thought 
the dynamics of the background is always inflationary\footnote{ It can 
also happen that the background evolution at early times is genuinely 
different from a stage of inflationary expansion. Both possibilities 
will be swiftly mentioned later on in section \ref{sec5}.} (see, in this respect, appendix \ref{APPB} and section \ref{sec5}).

\begin{figure}[!ht]
\centering
\includegraphics[width=0.7\textwidth]{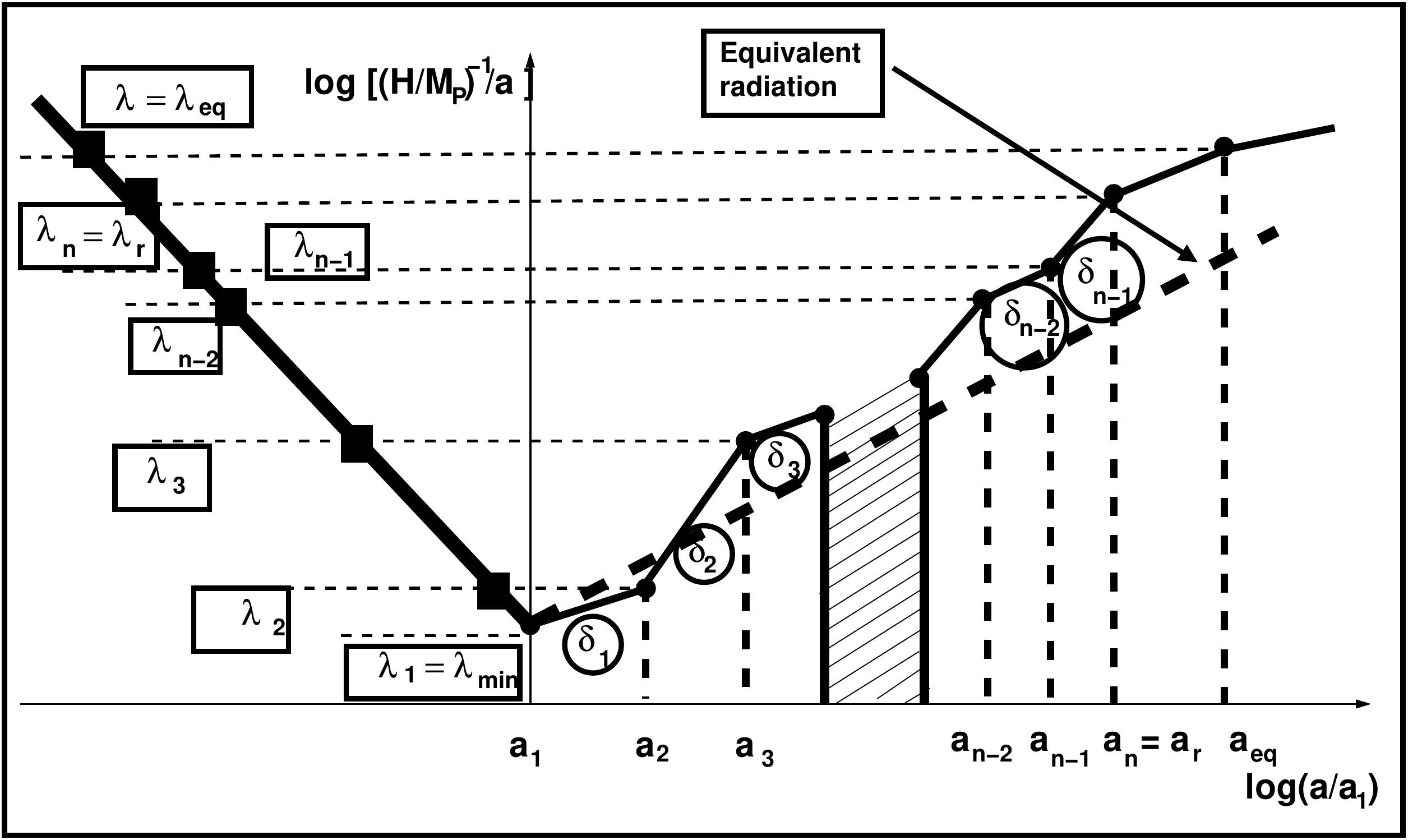}
\caption[a]{We illustrate the inverse of the comoving expansion rate (i.e. the comoving Hubble radius) in the case of the timeline already introduced in Fig. \ref{FIGU0c}. The  terminology followed here is the one commonly employed in te literature. When a given wavelength crosses the comoving Hubble radius for the first time we say that it {\em exits} the horizon (see the filled squares). When the wavelengths crosses the comoving Hubble radius for the second time we say that it {\em reenters} the horizon. This is why in the text we indicated these moments as $\tau_{ex}$ and $\tau_{re}$ respectively. We stress however that, in this context, the terminology ``horizon'' is actually a misnomer since the evolution of the Hubble radius is just a way to illustrate the dynamics of large-scale inhomogeneities and has has nothing to do with the causal structure of the underlying space-time. }
\label{FIGU0d} 
\end{figure}

In the previous cartoons of this section we illustrated the effective rate of expansion in Planck units even though, in various cases, it is also useful to reason in terms of the inverse of $a\, H$. For this reason in Fig. \ref{FIGU0d} we now plot $ (a \,H)^{-1}$. Sometimes in the literature $(a\, H)^{-1}$ is referred to as the horizon or simply the Hubble radius. 
According to this terminology the different wavelengths of the gravitational waves 
and of the scalar modes of the geometry cross the Hubble radius 
at different times. The first crossing typically occurs during inflation (see 
the left part of Fig. \ref{FIGU0d}); after the first crossing the wavelength gets 
 larger than the Hubble radius. This moment is then referred to as 
the {\em exit} of the given wavelength. The second crossing 
(see the right part of Fig. \ref{FIGU0d}) occurs in the decelerated 
stage of expansion and it is conventionally referred to as the {\em reentry} 
of the given wavelength since after this typical time the wavelength 
gets again smaller than the Hubble radius. The  filled squares in Fig. 
\ref{FIGU0d} define the exit of a given (comoving) wavelength while the dots in the right portion of the plot denote reentry of the 
selected scale. According to Fig. \ref{FIGU0d} the wavelengths smaller than $\lambda_{r}$ reenter before radiation dominance while the wavelengths $\lambda> \lambda_{r}$ reenter between the onset of radiation dominance and the epoch of matter--radiation equality. 
For $\lambda < \lambda_{r}$ the wavelength ${\mathcal O}(\lambda_{min})$ corresponds to comoving frequencies close to the maximal (i.e. $\nu = {\mathcal O}(\nu_{max})$). The scales  $\lambda_{r} < \lambda <\lambda_{eq}$ were still larger than the comoving horizon prior to matter--radiation equality and exited about $N_{k}$ $e$-folds before the end of inflation; the corresponding wavenumbers range therefore between $0.05\,\, \mathrm{Mpc}^{-1}$ and $0.002\,\, \mathrm{Mpc}^{-1}$.

\renewcommand{\theequation}{3.\arabic{equation}}
\setcounter{equation}{0}
\section{The relic gravitons and the expansion history}
\label{sec3}
During the last fifty years a recurrent viewpoint has been that, ultimately, 
high-energy physics is a tool for cosmology and astrophysics. This argument rests on the 
observation that the plasma became transparent to electromagnetic 
radiation only rather late (i.e. after the last scattering of photons). 
Therefore there cannot be direct signals coming, for instance, from an expanding stage with 
a typical temperature of the order of few TeV. However, since these energy scales 
are reachable by colliders, particle physics is the only 
tool that we might have to scrutinize the early Universe. 
This perspective (implicitly assuming the dominance of radiation and the existence 
of a prolonged stage of local thermal equilibrium) should be probably revamped in the light 
of the direct detection of gravitational radiation. Indeed we do know that every variation of the space-time curvature produces shots of relic gravitons with given multiplicities 
and specific spectra \cite{MGB}. Since the sensitivities 
of gravitational wave detectors greatly improved in the last thirty years, it is plausible to assume 
that the direct observations might hit the thresholds of the cosmological signals during the next score year or so. 
Under this hypothesis the timeline of the expansion rate illustrated in the 
previous section may be one day testable in practice as it is already scrutinized in principle. Along this
revamped perspective gravitational wave astronomy could become a tool for high-energy physics by conveying a more specific knowledge of energy scales that might not be accessible to colliders in the future. 

The relic gravitational waves produced by the early variation of the space-time curvature \cite{AA1,AA2,AA3,AA4} lead to a late-time background of diffuse radiation. In the simplest situation the relic gravitons are produced in pairs of opposite three-momenta from the inflationary vacuum and this is why they appear as a collection of standing (random) waves which are the tensor analog of the so-called Sakharov oscillations \cite{BB1}; this phenomenon has been also independently discussed in the classic paper of Peebles and Yu \cite{BB2} (see also \cite{BB3}). The late-time properties of the signal not only rest on the features of the inflationary vacuum but also on the post-inflationary evolution. It is well established that in the concordance paradigm the spectral energy density at late times is quasi-flat \cite{AA7,AA8,AA9} and it gets larger at smaller frequencies 
of the order of the aHz \cite{AA10}. This happens because, in the concordance scenario, the spectral energy density scales as $\nu^{-2}$ between few aHz and $100$ aHz in the region where the current Cosmic Microwave Background (CMB) observations are now setting stringent limits on the contribution of the relic gravitons to the temperature and polarization anisotropies \cite{RR1,RR2,RR3}. Along this perspective 
the low-frequency constraints translate into direct bounds on the tensor to scalar ratio $r_{T}$ and seem to suggest that at higher frequencies (i.e. in the audio band and beyond) the spectral energy density in critical units should be ${\mathcal O}(10^{-17})$ or even smaller. This result has been realized, at a different level of accuracy, in various
papers starting from Refs. \cite{AA7,AA8,AA9} (see also \cite{FLS1,FLS2}).
 The minuteness of the spectral energy density follows from the presumption that radiation dominates (almost) right after the end of inflation and it is otherwise invalid. As we argued in section \ref{sec2}, the post-inflationary evolution prior to BBN nucleosynthesis is not probed by any direct observation and may deviate from the radiation dominated timeline; if this is the case, the high frequency spectrum of the relic gravitons can be much larger \cite{EXP1,EXP2,EXP3}. In what follows we are going to discuss first the statistical properties of the gravitons produced by the variation of the space-time curvature; in the second 
part of the section the discussion is focussed on the slopes of the spectral energy density of the relic gravitons and on their connection with  the expansion rate of the Universe.

\subsection{Random backgrounds and quantum correlations}
The random backgrounds associated with the relic gravitons are homogeneous but not stationary and this property is ultimately related with their quantum mechanical origin. Conversely the homogeneity of the background does not directly follow from the properties of the  quantum mechanical correlations. In what follows we shall try to clarify the analogies and the differences between these two aspects of the problem by swiftly summarizing the main conclusions of a recent analysis \cite{MGSTOC} that follows previous attempts along 
similar directions \cite{CORR6}.

\subsubsection{The energy density of random backgrounds}
We start by considering a tensor random field $h_{i\, j}(\vec{x}, \tau)$ and its Fourier transform\footnote{ Since the tensor amplitude $h_{i\,j}(\vec{x},\tau)$ is real the corresponding Fourier amplitude must obey $h_{i\,j}^{\ast}(\vec{k},\tau)= h_{i\,j}(-\vec{k},\tau)$. Moreover $h_{i\,j}(\vec{x},\tau)$ is also solenoidal and traceless;  
thus $h_{i\,j}^{\ast}(\vec{k},\tau)$ must obey 
$\hat{k}^{i} \, h_{i\, j}(\vec{k},\tau) = \hat{k}^{j} \, h_{i\, j}(\vec{k},\tau)=0$ and 
$h_{i}^{\,\,i} =0$. See also the considerations developed in appendix \ref{APPB}.}:
\begin{equation}
h_{i\, j}(\vec{x}, \tau) = \frac{1}{(2 \pi)^{3/2}} \int d^{3} k \, e^{- i \vec{k}\cdot\vec{x}} \, h_{i\, j}(\vec{k}, \tau),
\label{RB1}
\end{equation}
where $\vec{k}$ is the comoving three-momentum. The Fourier amplitude $h_{i\, j}(\vec{k},\tau)$ can be decomposed in terms of the tensor polarizations as
\begin{equation}
h_{i\, j}(\vec{k},\tau) = \sum_{\lambda} \,e^{(\lambda)}_{i\,j}(\hat{k}) h_{\lambda}(k,\tau),
\label{RB1A}
\end{equation}
where the sum over $\lambda$ runs over $\oplus$ and $\otimes$. If we introduce a triplet of mutually 
orthogonal unit vectors $\widehat{m}$, $\widehat{n}$ and $\widehat{k}$ (where $\widehat{m} \times \widehat{n} = \widehat{k}$) the two tensor polarizations are:
\begin{equation}
e^{(\oplus)} = \widehat{m}_{i} \, \widehat{m}_{j} - \widehat{n}_{i} \, \widehat{n}_{j}, \qquad \qquad e^{(\otimes)} = \widehat{m}_{i} \, \widehat{n}_{j} + \widehat{n}_{i} \, \widehat{m}_{j}.
\label{RB1B}
\end{equation}
If background is isotropic and unpolarized
the corresponding ensemble averages of the Fourier amplitudes (and of their first derivatives) can be expressed as
\begin{eqnarray}
\langle h_{ij}(\vec{k},\tau) \, h_{mn}(\vec{k}^{\prime}, \tau) \rangle &=& \frac{2 \pi^2}{k^3}\, P_{T}(k,\tau) \, \delta^{(3)}(\vec{k} + \vec{k}^{\prime}) \, 
{\mathcal S}_{i \,j \,m \,n}(\widehat{k}),
\label{RB2}\\
\langle \partial_{\tau}\, h_{ij}(\vec{k},\tau) \, \partial_{\tau} h_{mn}(\vec{k}^{\prime}, \tau) \rangle &=& \frac{2 \pi^2}{k^3}\, Q_{T}(k,\tau) \, \delta^{(3)}(\vec{k} + \vec{k}^{\prime}) \, 
{\mathcal S}_{i\,j\,m\,n}(\widehat{k}),
\label{RB3}
\end{eqnarray}
where the two tensor power spectra $P_{T}(k,\tau)$ and $Q_{T}(k,\tau)$ fully describe the tensor random field; 
 $\langle .\,.\,.\rangle$ denotes an average over an ergodic ensemble of random functions. 
  In Eq. (\ref{RB3}) the tensor ${\mathcal S}_{i\,j\,m\,n}(\widehat{k})$ arises from the sum over the two 
tensor polarizations:
\begin{equation}
{\mathcal S}_{i\,j\,m\,n}(\hat{k}) = \biggl[p_{i\,m}(\hat{k}) \,\,
p_{j\,n}(\hat{k}) + p_{i\,n}(\hat{k}) \,\, p_{j\,m}(\hat{k}) - p_{i\,j}(\hat{k})\,\, p_{m\,n}(\hat{k})\biggr]/4,
\label{RB4}
\end{equation}
where $p_{i\,j} = (\delta_{i\,j} - \hat{k}_{i}\, \hat{k}_{j})$. From the $(00)$ component of the energy-momentum pseudo-tensor discussed in the appendix \ref{APPB} (see in particular Eq. (\ref{APPA14})) the energy density of the relic gravitons becomes:
\begin{equation}
\rho_{gw} = \frac{1}{8 \ell_{P}^2 a^2} \biggl( \partial_{\tau} h_{k \ell}\, \partial_{\tau}h^{k \ell} + \partial_{m} h_{k\ell} \partial^{m} h^{k\ell}\biggr).
\label{RB5}
\end{equation}
If we now insert Eq. (\ref{RB1}) inside Eq. (\ref{RB5}) 
and average the obtained result according to Eqs. (\ref{RB2})--(\ref{RB3})
we obtain
\begin{equation}
\overline{\rho}_{gw} \,  = \frac{1}{8 \ell_{P}^2 a^2} \biggl(\langle \partial_{\tau} h_{k \ell}\, \partial_{\tau}h^{k \ell} \rangle + \langle \partial_{m} h_{k\ell} \partial^{m} h^{k\ell} \rangle\biggr)
= \frac{1}{8\, \ell_{P}^2\, a^2} \int_{0}^{\infty} \, \frac{d k}{k} \biggl[ k^2 P_{T}(k, \tau) + Q_{T}(k,\tau) \biggr].
\label{RB6}
\end{equation}
In Eq. (\ref{RB6}) $\overline{\,\rho\,}_{gw} = \langle \rho_{gw} \rangle$ represents the ensemble average of the energy density; the second equality in Eq. (\ref{RB6}) directly follows from Eq. (\ref{RB1}) after taking the ensemble average of each term according to 
Eqs. (\ref{RB2})--(\ref{RB3}). From Eq. (\ref{RB6}) we can always introduce  the spectral energy density in critical units:
\begin{equation}
\Omega_{gw}(k,\tau) = \frac{1}{\rho_{crit}} \,  \frac{d\, \overline{{\,\rho\,}}_{gw}}{d \ln{k}} = \frac{1}{24 H^2 a^2 } \biggl[ k^2 P_{T}(k, \tau) + Q_{T}(k,\tau) \biggr],
\label{RB7}
\end{equation}
where, as before, $\rho_{crit} = 3\, \overline{M}_{P}^2 \,H^2$.
The value of $\Omega_{gw}(k,\tau)$ depends {\em both} on  $P_{T}(k,\tau)$ and $Q_{T}(k,\tau)$. Sometimes $\Omega_{gw}(k,\tau)$ is swiftly referred to as the energy density (in critical units) of the random background but this 
terminology is incorrect: the energy density does not depend on the 
frequency (or on the momentum); $\Omega_{gw}(k,\tau)$ represents the energy density (in critical units) and per logarithmic interval of momentum (or frequency) since, in our units, $\omega = k = 2 \pi\, \nu$. Since $Q_{T}(k,\tau) \to k^2 P_{T}(k,\tau) $ for $k \gg a H$, the spectral energy density  for typical wavelengths shorter that the Hubble radius can also be expressed as\footnote{Sometimes in the literature Eq. (\ref{RB8}) is taken as definition 
of $\Omega_{gw}(k,\tau)$. This is also incorrect since Eq. (\ref{RB8})
is only an approximation that holds for wavelengths that are sufficiently 
small in comparison with the effective horizon (or, in equivalent terms, 
wavenumbers much larger than the expansion rate).}: 
\begin{equation}
\Omega_{gw}(k,\tau) =\frac{k^2}{12 H^2 a^2 } P_{T}(k,\tau), \qquad k \gg a\, H.
\label{RB8}
\end{equation}

\subsubsection{Homogeneity in space}
The results of Eqs. (\ref{RB7})--(\ref{RB8}) follow  by considering the basic features of traceless and solenoidal tensor random fields supplemented by the notion of stochastic average introduced in Eqs. (\ref{RB2})--(\ref{RB3}). 
A relevant result following from the previous considerations 
is that the two-point function of the tensor modes is {\em homogeneous 
in space}. By this we mean that the two-point function only depends on the 
distance between two spatial locations. If we compute
the correlation functions of $h_{i\,j}(\vec{x},\tau)$ and of its derivative at equal times (but for two different spatial locations) we obtain
\begin{eqnarray}
&& \langle \,h_{i\,j}(\vec{x}, \tau)\, h^{i \, j}(\vec{x} +\vec{r},\tau) \,\rangle =  \int_{0}^{\infty} \frac{d\, k}{k} P_{T}(k,\tau) \, j_{0}(k\,r),
\label{RB9}\\
&& \langle \,\partial_{\tau} h_{i\,j}(\vec{x}, \tau)\,\,\partial_{\tau} h^{i \, j}(\vec{x} +\vec{r},\tau) \,\rangle =  \int_{0}^{\infty} \frac{d\, k}{k} Q_{T}(k,\tau) \, j_{0}(k\,r),
\label{RB10}
\end{eqnarray}
where $j_{0}(k\,r)$ is the spherical Bessel function of zeroth order \cite{abr1,abr2}. 
We remark that Eqs. (\ref{RB9})--(\ref{RB10}) follow directly from the definition
of Eq. (\ref{RB1}) and from the averages 
of Eqs.  (\ref{RB2})--(\ref{RB3}). We note that the homogeneity 
in space implies that both correlators are evaluated  at the same 
values of the conformal time coordinate $\tau$.
The results of Eqs. (\ref{RB9})--(\ref{RB10}) demonstrate that the tensor 
random fields, heuristically defined by Eqs. (\ref{RB1}) and (\ref{RB2})--(\ref{RB3}), 
can be described by stochastic processes that are homogeneous in space. 
This means also that the two-point function computed at equal times (but for different locations) is invariant under spatial translations.

\subsubsection{Homogeneity in time (stationarity)} 
It would now seem that the same kind of invariance should also hold when the spatial location is fixed but the time coordinates are shifted shifted. In this case the two-point function of the tensor fluctuations would also be {\em stationary}, i.e. invariant under 
time translations. The stationarity is actually more restrictive than homogeneity if the random background is defined by Eqs. (\ref{RB1}) and (\ref{RB2})--(\ref{RB3}). Indeed, as we are going to see, the stationarity ultimately restricts the time-dependence of the power 
spectra $P_{T}(k,\tau)$ and $Q_{T}(k,\tau)$. If we then avoid  
the complication of the spatial dependence and directly discuss a single tensor polarization $h(\tau)$, instead of an ensemble or random fields we deal an ensemble of real random functions $h(\tau)$. We then introduce the autocorrelation function $\Gamma_{h}( \Delta\tau )$ defined in the context of the generalized harmonic analysis and associated with the finiteness of the integral \cite{STOC1,STOC2}
\begin{equation}
\Gamma_{h}( \Delta\tau ) = \lim_{T \to \infty}\,\, \frac{1}{2 \, T}\, \int_{-T}^{T} \, h(\tau) h(\tau + \Delta\tau) \, d \tau.
\label{RB11}
\end{equation}
Wiener considered the
class of functions (all measurable in a Lebesgue sense) for which the integral 
(\ref{RB11}) exists and demonstrated that 
the spectral density exists \cite{WK1}. In the case of a stationary and  ergodic ensemble of random functions, the autocorrelation of Eq. (\ref{RB11}) can be replaced by 
\begin{equation}
\Gamma_{h}(|\tau_{1}- \tau_{2}|) = \langle \, h(\tau_{1}) \, h(\tau_{2})\, \rangle,
\label{RB12}
\end{equation}
where $\langle \,.\,.\,.\rangle$ now denotes an ensemble average and the results of Eqs. (\ref{RB11}) and (\ref{RB12}) must ultimately coincide under the hypotheses of ergodicity. The property expressed by Eq. (\ref{RB12}) is characteristic of a stationary process whose 
autocorrelation function is {\em invariant under a shift of the time coordinate}. This is why the  
 Fourier transform of the autocorrelation function is associated with a well defined spectral amplitude \cite{WK1,WK2}.
Recalling Eq. (\ref{RB12}) we can Fourier transform $h(\tau)$, obtain $h(\nu)$ as a function of the frequency $\nu$ and eventually evaluate the corresponding ensemble average; the result is  
\begin{equation}
h(\tau) = \int_{-\infty}^{+\infty} e^{ 2 \, i\, \pi\, \nu \, \tau} h(\nu) \, d\nu, \qquad \langle h(\nu)\, h(\nu^{\prime}) \rangle = \delta(\nu + \nu^{\prime}) \, S_{h}(\nu).
\label{RB13}
\end{equation}
From Eqs. (\ref{RB12})--(\ref{RB13}) the autocorrelation function and the spectral amplitudes are then related as 
\begin{eqnarray}
\Gamma_{h}(\tau_{1} - \tau_{2}) = \frac{1}{2\pi} \int_{-\infty}^{\infty} e^{i \, \omega (\tau_{1} - \tau_{2})} S_{h}(\omega) d \omega 
= \int_{-\infty}^{\infty} e^{2 i \,\pi \, \nu (\tau_{1} - \tau_{2})} \, S_{h}(\nu) \, d \nu.
\label{RB14}
\end{eqnarray}
According to Eqs. (\ref{RB13})--(\ref{RB14}) the spectral amplitude and the autocorrelation function of the process 
form a Fourier transform pair; this statement is often referred to as Wiener-Khintchine 
theorem (see e.g. \cite{STOC2}) and was originally developed in the framework of the  generalized harmonic analysis  that establishes a rigorous connection between Eqs. (\ref{RB11}) and (\ref{RB12}) \cite{WK1,WK2}. The possibility of defining a spectral amplitude relies then on the stationary nature of the underlying random process. The spectral amplitude is actually measured in units of inverse frequencies and can also be assigned in the case of a generic spatial dependence.  Both stationarity and homogeneity play an important role when 
analyzing the correlation between gravitational wave detectors of arbitrary geometry\footnote{ In particular the intrinsic noises of the instruments are customarily assumed to be  stationary, Gaussian, uncorrelated, much larger in amplitude than the gravitational strain, and statistically independent on the strain itself.  The stationarity and the homogeneity are also conjectured for the signals associated with the diffuse background of gravitational radiation \cite{CORR7}. So far we demonstrated that the diffuse backgrounds of relic gravitons are homogeneous in space but  to address the stationarity it is instead essential to take into account the quantum mechanical aspects of the problem. } \cite{CORR1,CORR2,CORR3,CORR4}.

\subsection{Random backgrounds and quantum mechanics}
For a quantum description of the relic gravitons the first step is to recall 
the second-order action for the tensor inhomogeneities deduced in appendix \ref{APPB}
(see, in particular, Eq. (\ref{APPA13})). The canonical momentum deduced from Eq. (\ref{APPA13}) is in fact given by $\pi_{i\, j} = a^2 \partial_{\tau} h_{i\, j}/(8 \ell_{P}^2)$
and the resulting classical Hamiltonian is:
\begin{equation}
H_{g}(\tau) = \int d^3x \biggl[ \pi_{i\, j} \partial_{\tau} h^{i\,j} + \pi^{i\, j} \partial_{\tau} h_{i\,j} - 
{\mathcal L}_{g}(\vec{x},\tau)\biggr].
\label{QB1}
\end{equation}
By promoting the classical fields 
to the status of quantum operators (i.e. $ h_{i\, j}(\vec{x}, \tau) \to \widehat{h}_{i\,j}(\vec{x}, \tau)$ and $ \pi_{i\, j}(\vec{x}, \tau) \to\widehat{\pi}_{i\,j}(\vec{x}, \tau)$) the quantum Hamiltonian $\widehat{H}_{g}(\tau)$ becomes
\begin{equation}
\widehat{H}_{g}(\tau) = \int d^{3} x \biggl[\frac{8 \,\ell_{P}^2}{a^2} \, \widehat{\pi}_{i\, j} \,\, \widehat{\pi}^{i\,j} + \frac{a^2}{8 \, \ell_{P}^2} \, \partial_{k} \widehat{h}_{i\, j} \, \partial^{k} \widehat{h}^{i\, j} \biggr],
\label{QB2}
\end{equation}
where $\widehat{h}_{i\,j}^{\, \dagger}(\vec{x}, \tau) = \widehat{h}_{i\,j}(\vec{x}, \tau)$ and $\widehat{\pi}_{i\,j}^{\, \dagger}(\vec{x}, \tau) = \widehat{\pi}_{i\,j}(\vec{x}, \tau)$ are both Hermitian; the dagger denotes, as usual, the Hermitian conjugation. 
From Eq. (\ref{QB2}) the evolution equations of the field operators in the Heisenberg description are:
\begin{equation}
\partial_{\tau} \, \widehat{\pi}_{i\, j} = i\, [\widehat{H}_{g}, \widehat{\pi}_{i\, j}]=  \frac{a^2}{8 \, \ell_{P}^2} \nabla^2 \widehat{h}_{i\, j}, \qquad \partial_{\tau} \, \widehat{h}_{i\, j} = i\, [\widehat{H}_{g}, \widehat{h}_{i\, j}] = \frac{8 \ell_{P}^2}{a^2} \, \widehat{\pi}^{i\, j},
\label{QB3}
\end{equation}
and their explicit form in the Heisenberg representation is
\begin{eqnarray}
\widehat{h}_{i\, j}( \vec{x}, \tau) &=& \frac{\sqrt{2} \, \ell_{P}}{(2 \pi)^{3/2}} \sum_{\alpha = \oplus, \otimes} \int d^{3} k \,\, e^{(\alpha)}_{i\,j}(\hat{k}) \,\,
\biggl[ F_{k,\alpha}(\tau) \, \widehat{b}_{\vec{k}, \, \alpha} e^{- i \vec{k}\cdot \vec{x}} + \mathrm{H.\,c.}\biggr],
\label{QB4}\\
\widehat{\pi}_{i\, j}( \vec{x}, \tau) &=& \frac{a^2(\tau)}{4 \sqrt{2} \, \ell_{P}\, (2 \pi)^{3/2}} \sum_{\beta = \oplus, \otimes} \int d^{3} k \,\, e^{(\beta)}_{i\,j}(\hat{k}) \,\,
\biggl[ G_{k,\beta}(\tau) \, \widehat{b}_{\vec{k}, \, \beta} e^{- i \vec{k}\cdot \vec{x}} + \mathrm{H.\,c.}\biggr].
\label{QB5}
\end{eqnarray}
In Eqs. (\ref{QB4})--(\ref{QB5}) the second term inside the square bracket denotes the Hermitian conjugate of the preceding one. As before the sum runs over the two tensor polarizations defined in Eq. (\ref{RB1B}); because of Eq. (\ref{QB3}) the mode functions $F_{k}(\tau)$ and $G_{k}(\tau)$ 
obey: 
\begin{equation}
G_{k}^{\, \prime} + 2 {\mathcal H} \, G_{k} = - k^2 \, F_{k}, \qquad\qquad G_{k} = F_{k}^{\, \prime},
\label{QB6}
\end{equation}
where, as usual, ${\mathcal H} = a^{\prime}/a$.  The Fourier transforms of the Hermitian field operators of Eqs. (\ref{QB4})--(\ref{QB5}) are
\begin{eqnarray}
\widehat{h}_{i\,j}(\vec{q},\tau) &=& \sqrt{2}\, \ell_{P} \, \sum_{\alpha} \biggl[ e^{(\alpha)}_{i\,j}(\hat{q})\,\widehat{b}_{\vec{q},\, \alpha} \, F_{q,\alpha}(\tau) +  e^{(\alpha)}_{i\,j}(-\hat{q})\widehat{b}_{-\vec{q}, \alpha}^{\dagger} F_{q,\alpha}^{\ast}(\tau)\biggr],
\label{QB7}\\
\widehat{\pi}_{m\,n}(\vec{p},\tau) &=& \frac{a^2}{4\,\sqrt{2}\, \ell_{P}}\, \sum_{\beta} \biggl[ e^{(\beta)}_{m\,n}(\hat{p})\,\widehat{b}_{\vec{p},\, \beta} \, G_{p,\beta}(\tau) +  e^{(\beta)}_{m\,n}(-\hat{p})\widehat{b}_{-\vec{p}, \beta}^{\dagger} G_{p,\beta}^{\ast}(\tau)\biggr].
\label{QB8}
\end{eqnarray}
The field operators of Eqs. (\ref{QB7})--(\ref{QB8}) obey the canonical commutation 
relations 
\begin{equation}
\biggl[ \widehat{h}_{i\, j}(\vec{q},\tau), \, \widehat{\pi}_{m\, n}(\vec{p}, \tau) \biggr] = i\,\, {\mathcal S}_{i\,j\, m\, n}(\widehat{q}) \,\, \delta^{(3)}(\vec{q} + \vec{p}),
\label{QB9}
\end{equation}
provided the mode functions obey the Wronskian normalization
\begin{equation}
F_{k}(\tau) \, G_{k}^{\ast}(\tau) - F_{k}^{\ast}(\tau) G_{k}(\tau) = i/a^2(\tau).
\label{QB10}
\end{equation}
The condition expressed by Eq. (\ref{QB10}) is essential to obtain the correct form of the commutation relations that must be preserved throughout all the stages of the dynamical evolution. The mode functions can also be rescaled 
as $F_{k}(\tau) = a \, f_{k}(\tau)$ and $G_{k}(\tau) = a \, g_{k}(\tau)$; in this case Eq. (\ref{QB10}) becomes $f_{k}(\tau) \, g_{k}^{\ast}(\tau) - f_{k}^{\ast}(\tau) g_{k}(\tau) = i$.

\subsubsection{Quantum mechanics and non-stationary processes}
To analyze the stationarity of the process we need to introduce 
the autocorrelation functions depending on two different times $\tau_{1}$ and $\tau_{2}$:
\begin{eqnarray}
\Gamma_{i\,j\,m\,n}(\vec{k}, \vec{p}, \tau_{1}, \tau_{2}) &=& \frac{1}{2}\biggl[\langle \widehat{h}_{i\,j}(\vec{k}, \tau_{1}) \, \widehat{h}_{m\,n}(\vec{p}, \tau_{2}) \rangle  +  \langle \widehat{h}_{i\,j}(\vec{p}, \tau_{2}) \, \widehat{h}_{m\,n}(\vec{k}, \tau_{1}) \rangle\biggr],
\label{QB12}\\
\overline{\Gamma}_{i\,j\,m\,n}(\vec{k}, \vec{p}, \tau_{1}, \tau_{2}) &=& \frac{1}{2}\biggl[\langle \partial_{\tau_{1}} \widehat{h}_{i\,j}(\vec{k}, \tau_{1}) \,\, \partial_{\tau_{2}} \widehat{h}_{m\,n}(\vec{p}, \tau_{2}) \rangle  +   \langle \partial_{\tau_{2}}\, \widehat{h}_{i\,j}(\vec{p}, \tau_{2}) \,\partial_{\tau_{1}} \widehat{h}_{m\,n}(\vec{k}, \tau_{1}) \rangle\biggr].
\label{QB13}
\end{eqnarray}
The explicit form of Eqs. (\ref{QB12})--(\ref{QB13}) follows from the actual expressions of the field operators in Fourier space (see Eqs. (\ref{QB7})--(\ref{QB8})) and the final result becomes:
\begin{eqnarray}
\Gamma_{i\,j\,m\,n}(\vec{k}, \vec{p}, \tau_{1}, \tau_{2}) &=& {\mathcal S}_{i\, j \, m\,n}(\hat{k})\, \delta^{(3)}(\vec{k} + \vec{p}) \, \Delta_{k}(\tau_{1},\, \tau_{2}), 
\label{QB14}\\
\overline{\Gamma}_{i\,j\,m\,n}(\vec{k}, \vec{p}, \tau_{1}, \tau_{2}) &=&  {\mathcal S}_{i\, j \, m\,n}(\hat{k}) \, \delta^{(3)}(\vec{k} + \vec{p}) \,\overline{\Delta}_{k}(\tau_{1},\, \tau_{2}) ,
\label{QB15}
\end{eqnarray}
where $\Delta_{k}( \tau_{1},\, \tau_{2})$ and $\overline{\Delta}_{k}(\tau_{1},\, \tau_{2})$ are now given by:
\begin{eqnarray}
\Delta_{k}(\tau_{1},\, \tau_{2}) &=& 4 \ell_{P}^2 \biggl[ F_{k}(\tau_{1}) \,\, F_{k}^{\ast}(\tau_{2}) + F_{k}(\tau_{2}) \,\, F_{k}^{\ast}(\tau_{1}) \biggr],
\label{QB16}\\
\overline{\Delta}_{k}(\tau_{1},\, \tau_{2}) &=& 4 \ell_{P}^2 \biggl[ G_{k}(\tau_{1}) \,\, G_{k}^{\ast}(\tau_{2}) + G_{k}(\tau_{2}) \,\, G_{k}^{\ast}(\tau_{1}) \biggr].
\label{QB17}
\end{eqnarray}
If we now consider the limit $\tau_{1} \to \tau_{2}$, we have, from Eqs. (\ref{QB14})--(\ref{QB15}) and (\ref{QB16})--(\ref{QB17}), that 
\begin{eqnarray} 
&& \langle \widehat{h}_{i\,j}(\vec{k}, \tau) \, \widehat{h}_{m\,n}(\vec{p}, \tau) \rangle 
= \frac{2 \pi^2}{k^3} {\mathcal S}_{i\,j\,k\,m\,n}(\hat{k}) P_{T}(k,\tau) \delta^{(3)}(\vec{k} + \vec{p}),
\label{TEST1}\\
&& \langle\partial_{\tau} \widehat{h}_{i\,j}(\vec{k}, \tau) \, \partial_{\tau}\widehat{h}_{m\,n}(\vec{p}, \tau) \rangle 
= \frac{2 \pi^2}{k^3} {\mathcal S}_{i\,j\,k\,m\,n}(\hat{k}) Q_{T}(k,\tau) \delta^{(3)}(\vec{k} + \vec{p}).
\label{TEST2}
\end{eqnarray}
Equations (\ref{TEST1})--(\ref{TEST2}) reproduce exactly Eqs. (\ref{RB9})--(\ref{RB10})
with the difference that  random fields are replaced by the field operators and  the 
ensemble averages are now quantum mechanical expectation 
values. Furthermore the two power spectra $P_{T}(k,\tau)$ and $Q_{T}(k,\tau)$ depend on the specific form of the 
mode functions 
\begin{equation}
P_{T}(k,\tau) = \frac{4 \ell_{P}^2}{\pi^2}\, k^3\, \bigl| F_{k}(\tau) \bigr|^2, \qquad Q_{T}(k,\tau) = \frac{4 \ell_{P}^2}{\pi^2}\, k^3\, \bigl| G_{k}(\tau) \bigr|^2.
\label{QB11}
\end{equation}
Since  Eqs. (\ref{RB9})--(\ref{RB10}) hold also in the quantum case, the  production of relic gravitons can be certainly mimicked with a homogeneous process.
To analyze of the stationarity we need the explicit form of the autocorrelation functions, at late times \cite{MGSTOC}. For this purpose the phases must be correctly determined from the continuity of the mode functions and from the Wronskian normalization condition of Eq. (\ref{QB10}). These requirements 
ultimately lead to the standing oscillations\footnote{These standing oscillations are in fact related to the tensor analog of the Sakharov 
oscillations \cite{BB1,BB2,BB3} (see also \cite{MGB}). Both during the radiation phase and in the matter epoch the standing oscillations appearing in the power spectrum lead to non-stationary features in the autocorrelation function\cite{MGSTOC}. } both in the mode 
functions and in the autocorrelation functions. This means that the autocorrelation functions at late times do not only depend on the time-difference $|\tau_{1} - \tau_{2}|$ (as it would happen in the case of a stationary process)
but also on $(\tau_{1} + \tau_{2})$.
The non-stationary features are simpler to illustrate in the radiation epoch since the explicit expressions 
are less cumbersome; during the radiation stage the mode functions can be expressed in terms of a $2\times2$ unitary matrix
\begin{eqnarray}
f_{k}(\tau) &=& A^{(r)}_{f\, f}(u,\, u_{r}) \overline{f}_{k}
\, +\, A^{(r)}_{f\,g}(u, \, u_{r}) \, \overline{g}_{k}/k,
\nonumber\\
g_{k}(\tau) &=& A^{(r)}_{g\,f}(u,\, u_{r}) k \, \overline{f}_{k} \, + \, A^{(r)}_{g\,g}(u,\, u_{r}) \overline{g}_{k},
\label{QB18}
\end{eqnarray}
where, as already mentioned after Eq. (\ref{QB10}), the mode functions have been rescaled as $F_{k}(\tau) = f_{k}(\tau)/a(\tau)$ and $G_{k}(\tau) = g_{k}(\tau)/a(\tau)$; in Eq. (\ref{QB18}) $\overline{f}_{k}$ and $\overline{g}_{k}$ indicate the initial conditions of the mode functions as they emerge from the inflationary stage of expansion. To describe the smooth evolution of the background and of the mode functions in Eq. (\ref{QB18}) it is appropriate to introduce the variable $u(\tau)$ 
\begin{equation}
u(\tau) = k \,[\tau + (2 - \epsilon) \tau_{r}], \qquad \tau > - \tau_{r},\qquad \epsilon = - \dot{H}/H^2,
\label{QB19}
\end{equation}
 where $\tau_{r}$ conventionally marks the onset of the radiation-dominated stage and $\epsilon$ is the standard slow-roll parameter; by definition $u_{r} = u(-\tau_{r}) = k ( 1 -\epsilon) \tau_{r}$. The inflationary initial conditions determine the amplitude of the tensor power spectrum during inflation for typical wavelengths larger than the Hubble radius and also the normalization of the autocorrelation function. In particular we can introduce 
\begin{eqnarray}
\overline{P}_{T}^{(r)}&=& \frac{4 \ell_{P}^2}{\pi^2} k^{3} \,\bigl| \overline{f}_{k}\bigr|^2= \frac{16}{\pi} \biggl(\frac{H_{r}}{M_{P}}\biggr)^2 \biggl(\frac{k}{a_{r} \, H_{r}}\biggr)^{n_{T}}
= \frac{128}{3} \biggl(\frac{V}{M_{P}^4}\biggr)_{k \simeq H_{r} a_{r}}  \biggl(\frac{k}{a_{r} \, H_{r}}\biggr)^{n_{T}}, 
\label{QB20}
\end{eqnarray}
where we defined, for the sake of conciseness, $\overline{P}_{T}^{(r)}=\overline{P}_{T}(k,\tau_{r})$ and $n_{T} = - 2 \epsilon = - r_{T}/8$. In Eq. (\ref{QB20}) $V$ denotes the inflationary potential which is related to the expansion rate in the slow-roll approximation $3\, H^2 \overline{M}_{P}^2 \simeq V$; as already mentioned prior to Eq. and in the related footnote $H_{r}$ is, roughly speaking, the expansion rat at the end of inflation. It is relevant to point out that 
in the limit $n_{T} \to 0$ the last equality in Eq. (\ref{QB20}) is ill defined even though 
it is still true that $\overline{P}_{T}^{(r)}= (16/\pi) (H_{r}/M_{P})^2$.
In summary, during the radiation stage the autocorrelation functions of Eqs. (\ref{QB16})--(\ref{QB17}) become:
\begin{eqnarray}
\Delta_{k}(\tau_{1},\, \tau_{2}) &=& \frac{\pi^2 \, \overline{P}_{T}^{(r)}}{k^3} \, \frac{[\cos{(u_{1} - u_{2})} - \cos{(u_{1} + u_{2})}]}{u_{1}\, u_{2}},
\label{QB21}\\
\overline{\Delta}_{k}(\tau_{1},\, \tau_{2}) &=& \frac{\pi^2 \, \overline{P}_{T}^{(r)}}{k}\biggl[ \frac{\cos{(u_{1} - u_{2})}}{u_{1}\, u_{2}} \biggl( 1 + \frac{1}{u_{1}\, u_{2}}\biggr) + \frac{\sin{(u_{1} - u_{2})}}{u_{1} \, u_{2}} \biggl(\frac{1}{u_{2}} - \frac{1}{u_{1}}\biggr)
\nonumber\\
&+& \frac{\cos{(u_{1} + u_{2})}}{u_{1}\, u_{2}} \biggl( 1 - \frac{1}{u_{1}\, u_{2}}\biggr) -  \frac{\sin{(u_{1} + u_{2})}}{u_{1} \, u_{2}} \biggl(\frac{1}{u_{2}} + \frac{1}{u_{1}}\biggr)\biggr],
\label{QB22}
\end{eqnarray}
where, by definition, $u_{1} = u(\tau_{1})$ and $u_{2} = u(\tau_{2})$; at late-times $u_{1} \simeq k \tau_{1}$ and 
$u_{2} = k\tau_{2}$. The autocorrelation functions of Eqs. (\ref{QB21})--(\ref{QB22}) do not only depend on the time difference (as implied in the case of stationary processes); on the  contrary both $\Delta_{k}(\tau_{1},\, \tau_{2})$ and $\overline{\Delta}_{k}(\tau_{1},\, \tau_{2})$ include distinct functions of   $(\tau_{1} - \tau_{2})$ and of $(\tau_{1}  + \tau_{2})$. In Eqs. (\ref{QB21})--(\ref{QB22}) we can also see a number of corrections going as inverse powers of $u_{1}$ and $u_{2}$; some of these corrections are suppressed when the wavelengths of the gravitons are much smaller than the Hubble radius during the radiation stage. In the matter stage the discussion is technically more involved but the final result is the same \cite{MGSTOC}. This means that the backgrounds of relic gravitons are homogeneous in space but they cannot be reduced to  
 a stationary stochastic process. This conclusion has various implications that are not analyzed here (see, however, \cite{MGSTOC}). It is appropriate to remark, however, 
that the use of the spectral amplitude should be limited to the signals that are describable in terms of homogeneous {\em and} stationary processes\footnote{While it is debatable if  the non-stationary features associated with the diffuse backgrounds of relic gravitons are (or will be) directly detectable, the spectral amplitude following from the Wiener-Khintchine theorem is generally inappropriate for a consistent description of the relic signal. }.  

\subsubsection{The averaged multiplicity}
In a quantum mechanical perspective the amplification of the field amplitudes corresponds to the creation of gravitons either from the vacuum or from some other initial state. Since the production of particles of various spin in cosmological backgrounds is a unitary process \cite{AA01,AA02} (see also \cite{PARKTH,BIRREL,PTOMS}) which is closely analog to the ones arising in the context of the quantum theory of parametric amplification \cite{MANDL,HBT4,louisell,mollow1,mollow2,RGS0,HBTG1}, the relation between the creation and the annihilation operators in the asymptotic states is given by: 
\begin{eqnarray}
\widehat{a}_{\vec{p}, \lambda}(\tau) &=& \alpha_{p,\, \lambda}(\tau)\,\, \widehat{b}_{\vec{p}}\,- \,
\beta_{p,\,\lambda}(\tau)\,\, \widehat{b}_{-\vec{p},\,\lambda}^{\,\dagger},  
\label{QB23}\\
\widehat{a}_{-\vec{p},\lambda}^{\,\dagger}(\tau) &=& \alpha_{p,\,\lambda}^{\ast}(\tau) \,\,\widehat{b}_{-\vec{p},\,\lambda}^{\,\dagger}\,  - \,\beta_{p,\,\lambda}^{\ast}(\tau)\,\, \widehat{b}_{\vec{p},\,\lambda}.
\label{QB24}
\end{eqnarray}
The time-dependent (complex) functions $\alpha_{p,\,\lambda}(\tau)$ and $\beta_{p,\,\lambda}(\tau)$ appearing in Eq. (\ref{QB24}) satisfy $\bigl|\alpha_{p,\,\lambda}(\tau)|^2 - \bigl|\beta_{p,\,\lambda}(\tau) \bigr|^2 = 1$;
because of the unitary evolution $[\widehat{a}_{\vec{p},\,\lambda},\, \widehat{a}_{\vec{k},\,\lambda^{\prime}}^{\,\dagger}] = \delta^{(3)}(\vec{k} - \vec{p})\, \delta_{\lambda \, \lambda^{\prime}}$ and $[\widehat{b}_{\vec{p},\,\lambda},\, \widehat{b}_{\vec{k},\,\lambda^{\prime}}^{\,\dagger}] = \delta^{(3)}(\vec{k} - \vec{p})\,\delta_{\lambda \, \lambda^{\prime}}$.
Since the total three-momentum must be conserved, Eqs. (\ref{QB23})--(\ref{QB24}) describe the production of graviton pairs with opposite momenta and the averaged multiplicity is obtained by computing the mean number of gravitons for with momentum $\vec{k}$ and $- \vec{k}$, i.e. 
\begin{equation}
\langle \hat{N}_{k} \rangle = \sum_{\lambda=\oplus, \, \otimes} \, \langle \widehat{a}_{\vec{k},\,\lambda}^{\dagger} \widehat{a}_{\vec{k},\, \lambda} + \widehat{a}_{-\vec{k},\,\lambda}^{\dagger} \widehat{a}_{-\vec{k},\,\lambda} \rangle = 4 \, \,\overline{n}(k,\tau),
\label{QB25}
\end{equation}
where $\overline{n}(k,\tau)= |v_{k}(\tau)|^2$ denotes the multiplicity of the pairs of relic gravitons and the further factor of $2$ counts the polarizations.  From Eq. (\ref{QB25}) the spectral energy density in critical units is expressed in terms of the averaged multiplicity of the produced gravitons with opposite three-momenta as: 
\begin{equation}
\Omega_{gw}(\nu, \tau) = \frac{1}{\rho_{crit}} \, \frac{ d \langle \rho_{gw} \rangle }{d \ln{\nu} } = \frac{128\, \pi^3}{3} \,\, \frac{\nu^{4}}{H^2 \, M_{P}^2 \, a^{4}}\,\, \overline{n}(\nu, \tau).
\label{QB26}
\end{equation}
The result of Eq. (\ref{QB26}) does not show any specific dependence on 
$\hbar$ and $c$ just because we are  working here in the natural system of units where 
$\hbar = c = \kappa_{B}= 1$. The $\hbar$ dependence can be however 
restored by recalling that the energy of a single graviton is given by $\hbar\, \omega$ (which we simply wrote as $k$ in units 
 $\hbar = c=1$); moreover another $\hbar$ is present in the definition of Planck mass squared. Thus, as suggested in \cite{WK} $\Omega_{gw}(\nu,\tau_{0}) \propto \hbar^2$ and this means that the relic gravitons have a truly quantum mechanical origin since their energy density goes to zero in the limit $\hbar \to 0$. 

\subsubsection{Upper bound on the maximal frequency of the spectrum}
Since we are here normalizing the scale factor as $a_{0} =1$, the 
physical and the comoving frequencies coincide at the present time  
and from Eq. (\ref{QB26}) the spectral energy density in critical units becomes:
\begin{equation}
\Omega_{gw}(\nu, \tau_{0}) = \frac{128\, \pi^3}{3} \,\, \frac{\nu^{4}}{H_{0}^2 \, M_{P}^2}\,\, \overline{n}(\nu, \tau_{0}),
\label{QB27}
\end{equation}
where $\tau_{0}$ denotes the current value of the conformal 
time coordinate. Equation (\ref{QB27}) suggests that the maximal frequency of the spectrum corresponds to the production of a single pair of gravitons with opposite three-momenta. For this reason we can always refer Eq. (\ref{QB27}) to a putative maximal frequency of the spectrum (be it $\nu_{max}$) and obtain \cite{MGF,MGQ}
\begin{equation}
\Omega_{gw}(\nu, \tau_{0}) =   \frac{128 \pi^3}{3} \frac{\nu_{max}^{4}}{H_{0}^2 \, M_{P}^2}
 \biggl(\frac{\nu}{\nu_{max}}\biggr)^{4}  \, \overline{n}(\nu/\nu_{max}, \tau_{0}).
 \label{QB28}
 \end{equation} 
While in Eqs. (\ref{QB27})--(\ref{QB28}) the single graviton limit \cite{MGHF} is achieved when $\overline{n}(\nu_{max}, \tau_{0}) \to 1$, in a classical perspective the maximal frequencies correspond to the bunch of wavenumbers that experience the minimal amplification.
All the wavelengths reentering the Hubble radius between the end of inflation and BBN must  comply with the bound\footnote{The rationale for the bound of Eq. (\ref{HH2}) is  discussed 
in section \ref{sec5}.}\cite{bbn1,bbn2,bbn3,bbn5}
\begin{equation}
h_{0}^2 \, \int_{\nu_{bbn}}^{\nu_{max}} \,\Omega_{gw}(\nu,\tau_{0}) \,\,d\ln{\nu} < 5.61\times 10^{-6} \biggl(\frac{h_{0}^2 \,\Omega_{\gamma0}}{2.47 \times 10^{-5}}\biggr) \, \Delta N_{\nu},
\label{HH2}
\end{equation}
where $\Omega_{\gamma\,0}$ is the (present) critical fraction of CMB photons. 
Equation (\ref{HH2}) sets an indirect constraint  on the extra-relativistic species possibly present at the time of nucleosynthesis. Since Eq. (\ref{HH2}) is also relevant in the context of neutrino physics the limit is  often expressed in terms of $\Delta N_{\nu}$ (i.e. the contribution of supplementary neutrino species). The actual bounds on $\Delta N_{\nu}$ range from $\Delta N_{\nu} \leq 0.2$ to $\Delta N_{\nu} \leq 1$ so that the integrated spectral density in Eq. (\ref{HH2}) must range, at most, between  $10^{-6}$ and $10^{-5}$.  The averaged multiplicity for $\nu \ll \nu_{max}$ corresponds to the mean number of produced pairs and it is approximately given by
\begin{equation}
\overline{n}(\nu/\nu_{max}, \tau_{0}) = \bigl|\beta(\nu,\tau_{0}) \bigr|^2 = \bigl(\nu/\nu_{max}\bigr)^{n_{T} -4}, \qquad \nu< \nu_{max},
\label{HH2aa}
\end{equation}
where $n_{T}$ denotes, in practice, to the spectral slope of $\Omega_{gw}(\nu,\tau_{0})$ in a given frequency interval. In the conventional lore where the consistency relations are enforced $n_{T} = n_{T}^{(low)}\simeq - r_{T}/8 + {\mathcal O}(r_{T}^2)$. There are, however, different physical situations where $n_{T}> 0$ 
\cite{EXP1,EXP2,EXP3} or even $n_{T}>1$ (see for instance \cite{MGB}). In all these situations $\Omega_{gw}(\nu, \tau_{0})$ increase at high frequencies while the averaged multiplicity for $\nu\ll \nu_{max}$ is comparatively less suppressed than in the standard lore where $n_{T} \simeq n_{T}^{(low)}\to 0$.  
The pair production process implies that for $\nu > \nu_{max}$ the averaged multiplicity is suppressed  \cite{AA01,AA02,BIRREL}:
\begin{equation}
\frac{\bigl|\overline{n}(\nu,\tau_{0}) \bigr|^2 }{1 + \bigl|\overline{n}(\nu,\tau_{0}) \bigr|^2 } = e^{- \gamma (\nu/\nu_{max})}, \qquad\qquad \nu> \nu_{max}.
\label{FFF5}
\end{equation}
The degree of suppression of Eq. (\ref{FFF5}) depends on $\gamma$, i.e. a numerical factor ${\mathcal O}(1)$ controlled by the smoothness of the transition between the 
inflationary and the post-inflationary phase; the value of $\gamma$ can be numerically estimated if the evolution of the mode functions is carefully integrated frequency by frequency \cite{ST3,ST3a}. The mean number of pairs produced from the vacuum can be written in the following suggestive form:
\begin{equation}
\overline{n}(\nu, \tau_{0}) = \gamma  \,\, x^{n_{T} - 3}/\bigl[e^{\gamma\,x} -1\bigr],\qquad\qquad x = (\nu/\nu_{max}),
\label{FFF6}
\end{equation}
where $n_{T}$ coincides, in practice, withe high frequency spectral index $n_{T}^{(high)}$. Equation (\ref{FFF6})
interpolates between the power-law behaviour of Eq. (\ref{HH2aa}) and the exponential suppression of Eq. (\ref{FFF5}) and 
suggests that the spectrum of relic gravitons should be
represented by a distorted thermal spectrum as argued 
long ago \cite{PARKTH}. There are furthermore situations 
where $n_{T}^{(high)} \to 3$ and this happens 
in some bouncing scenarios (see e.g. \cite{MGB} and discussions therein). It is therefore possible that a thermal spectrum of relic gravitons is produced purely from geometric effects, as originally suggested in Ref. \cite{PARKTH}.
We should also stress, incidentally, that the existence of the exponential suppression for $\nu>\nu_{max}$  
guarantees the convergence of the integral (\ref{HH2}) also in the case when the integration is performed up to $\nu \to \infty$.  If, for some reason, the diffuse background of relic gravitons has been formed  {\em after},  BBN $\Omega_{gw}(\nu, \tau_{0})$ must always be smaller than the current fraction of relativistic species to avoid an observable impact on the CMB and matter power spectra \cite{DD5,DD6}. Inserting therefore Eq. (\ref{FFF6}) 
into Eq. (\ref{QB28}) we can directly use Eq. (\ref{HH2}) 
(or its analog at even later times) to obtain a general bound 
on $\nu_{max}$:
\begin{equation}
\nu_{max} \leq 0.165 \,\, \Omega_{R0}^{1/4}\,\, \sqrt{H_{0}\, M_{P}} < \,\mathrm{THz},
\label{HH3}
\end{equation}
where the inequality follows in the limit $\overline{n}(\nu_{max}, \tau_{0}) \to {\mathcal O}(1)$, i.e. in the case where a single pair of gravitons is produced \cite{MGQ,MGF}. The argument leading to the bound of Eq. (\ref{HH3}) follows directly from the quantum mechanical result of Eq. (\ref{QB26})
that vanishes in the limit $\hbar \to 0$; in short we can 
therefore say that, according to a purely quantum perspective, 
$\nu_{max}$ coincides with the frequency where only  a graviton pair is produced. Since $h_{0}^2 \Omega_{gw}(\nu, \tau_{0})$ scales as $\nu^4$  when $\overline{n}(\nu,\tau_{0})$ is (approximately) frequency-independent single gravitons (or bunches of coherent gravitons) could be preferentially detected in the high frequency range: this observation has been eventually pointed out in Ref. \cite{MGHF} and a similar argument is in fact due to Dyson \cite{FA} who suggested that only at high frequencies it might be eventually possible to detect single gravitons. 
The existence of $\nu_{max}$ rules out the possibility of considering gravitons of arbitrary large 
frequencies as sometimes assumed by those who prefer to ignore the physical implications 
of high frequency gravitons. On the contrary, for frequencies $\nu = {\mathcal O}(\nu_{max})$, as we shall see in the second part of this section, $h_{0}^2 \Omega_{gw}(\nu, \tau_{0})$ may exceed 
the signal of the concordance scenario and may even exceed 
the contribution of the single-graviton line at lower frequencies \cite{MGHF}. 
As already suggested in the past \cite{EXP1,EXP2,EXP3} 
high frequency detectors might resolve single-gravitons \cite{MGHF,FA}.  In the perspective of Ref. \cite{FA} this could happen by conversion to photons in a strong magnetic field \cite{FB} with experimental techniques very similar to the ones employed for the scrutiny of axion-like particles\cite{FC} (see also \cite{FA1,FA2,FA3,FA4,FD} for some other papers with similar inspiration). In the case of Eq. (\ref{HH3}) the relic gravitons 
cannot exceed the THz domain where  coupled microwave cavities with superconducting walls \cite{CAV1,CAV2,CAV3,CAV4,CAV5,CAV6}, waveguides \cite{WG1,WG2,WG3,WG4,WG5} or even small interferometers \cite{SI1,SI2,SI3} could be used for direct detection even if the current sensitivities should not be overestimated as often done in recent times \cite{MGHF}. The observation of Ref. \cite{MGHF} raised a debate on the possibility of assessing the quantumness of the relic gravitons by looking at the analysis of the Bose-Einstein correlations \cite{MGF,MGQ}. In this second perspective the quantumness of the relic gravitons 
does not rest on the possibility of literally detecting a single gravitons (as sometimes 
misunderstood) but rather on the correlation properties of the underlying macroscopic quantum state\footnote{The analyses of the Bose-Einstein correlations, however, cannot be pursued in spite of the properties 
of the sources; this is why to  overlook the physical properties of the cosmic gravitons leads to conclusions that are ambiguous and ultimately superficial. It makes actually little sense 
to consider potentials signals coming from diffuse backgrounds for arbitrarily large frequencies (possibly much larger than the THz) without bothering about the underlying physical constraints. This approach is probably motivated by the 
need of claiming large sensitivities for potential instruments but has no physical 
basis unless the class of bounds related to Eq. (\ref{HH3}) is understood and acknowledged. We shall get back to the 
quantumness of the relic gravitons at the end of section \ref{sec6}.} \cite{MGQ}.

\subsection{The expansion history and the spectral energy density}
\subsubsection{The maximal frequencies}
While the bound on $\nu_{max}$ deduced in Eq. (\ref{HH3}) follows from quantum mechanical 
considerations, in a classical perspective the maximal frequency is computed from the 
smallest wavelength that crosses the Hubble radius of \ref{FIGU0c} and immediately 
reenters; this is why Eq. (\ref{APA8a}) depends upon $H_{1}\simeq H$ and also 
upon the timeline of the post-inflationary expansion rate discussed in section \ref{sec2}.
Let us therefore start from the simplest situation where the post-inflationary evolution 
is dominated by radiation. In terms of the cartoons of  Figs. \ref{FIGU0c} and \ref{FIGU0d} this 
means that all the $\delta_{i} \to 1$. Since in this case we already denoted the number of $e$-folds 
with an overline (e.g. $\overline{N}_{max}$, $\overline{N}_{k}$ and do on) we are now going to indicate the maximal frequency deduced in this case by $\overline{\nu}_{max}$:
\begin{eqnarray}
\overline{\nu}_{max} &=& \frac{M_{P}}{2 \pi} \,\bigl(2\, \Omega_{R0}\bigr)^{1/4} \, \sqrt{\frac{H_{0}}{M_{P}}}\, \sqrt{\frac{H_{1}}{M_{P}}}\, \, {\mathcal C}(g_{s}, g_{\rho}, \tau_{r}, \tau_{eq})
\label{APA8a}\\
&=& 195.38 \, {\mathcal C}(g_{s}, g_{\rho},\tau_{r},\tau_{eq}) \, \biggl(\frac{{\mathcal A}_{{\mathcal R}}}{2.41\times 10^{-9}}\biggr)^{1/4}\,\,
\biggl(\frac{\epsilon_{k}}{0.001}\biggr)^{1/4} \,\, \biggl(\frac{h_{0}^2 \, \Omega_{R\,0}}{4.15\times 10^{-5}}\biggr)^{1/4} \,\,\mathrm{MHz}.
\label{NK5}
\end{eqnarray}
If the transition to radiation dominance is almost sudden,  $\tau_{r}$ coincides approximately with $\tau_{1}$; for notational convenience we also use the convention $H \simeq H_{1}$ where $H$ indicates the inflationary expansion 
rate estimated from the power spectrum of the curvature inhomogeneities. Equation (\ref{NK5}) does not assume a specific relation between $\epsilon_{k}$ and $r_{T}$ however, 
if the consistency relations are enforced, we can always trade $\epsilon_{k}$ for $r_{T}$ and the value of $\overline{\nu}_{max}$ becomes: 
\begin{equation}
\overline{\nu}_{max} = 271.93 \, {\mathcal C}(g_{s}, g_{\rho},\tau_{r},\tau_{eq}) \, \biggl(\frac{{\mathcal A}_{{\mathcal R}}}{2.41\times 10^{-9}}\biggr)^{1/4}\,\,
\biggl(\frac{r_{T}}{0.06}\biggr)^{1/4} \,\, \biggl(\frac{h_{0}^2 \, \Omega_{R\,0}}{4.15\times 10^{-5}}\biggr)^{1/4} \,\,\mathrm{MHz}.
\label{NK6}
\end{equation}
The impact of ${\mathcal C}(g_{s}, g_{\rho},\tau_{r},\tau_{eq})$  on $\overline{\nu}_{max}$ is minor;
for typical values of the late-time parameters (i.e. $g_{\rho, \, r} = g_{s,\, r} = 106.75$ 
and  $g_{\rho, \, eq} = g_{s,\, eq} = 3.94$) ${\mathcal C}(g_{s}, g_{\rho},\tau_{r}, \tau_{eq}) =0.7596$ and the determination of $\overline{\nu}_{max}$ of Eq. (\ref{NK6}) moves from $\overline{\nu}_{max} = 271.93$ MHz to $206.53$ MHz. When the timeline of the post-inflationary evolution is not dominated by radiation 
but by a generic sequence of stages expanding either faster or slower than radiation 
(see  Figs. \ref{FIGU0c} and \ref{FIGU0d} and discussions therein) 
the maximal frequency can be related to $\overline{\nu}_{max}$ and is given by
\begin{equation}
\nu_{max} =  \prod_{i\,=\,1}^{n-1} \, \xi_{i}^{\frac{\delta_{i} -1}{2 (\delta_{i}+1})}\,\, \overline{\nu}_{max}.
\label{APA8}
\end{equation}
When all the $\delta_{i} \to 1$ the value of $\nu_{max}$
coincides with the $\overline{\nu}_{max}$ of Eq. (\ref{APA8a}). In case all the $\delta_{i}$ 
are equal (i.e. $ \delta_{i} = \delta$) the post-inflationary evolution 
consists of a single stage. The product of all the $\xi_{i}$ then coincides 
with $\xi_{r} = H_{r}/H$, as explained in Eqs. (\ref{TWO15O})--(\ref{TWO15P}).
Provided $\delta < 1$ (i.e. when the expansion rate is slower than radiation) $\nu_{max} > \overline{\nu}_{max}$; conversely, when $\delta >1$ (and the expansion rate is faster than radiation) 
$\nu_{max} < \overline{\nu}_{max}$. According to Eq. (\ref{NK6}) the value 
of $\overline{\nu}_{max}$ is ${\mathcal O}(300)$ MHz. This means that 
if the post-inflationary evolution is dominated by an expanding stage with 
$\delta \to 1/2$ with $H_{r} = {\mathcal O}(10^{-30})\,\,M_{P}$ 
the value of $\nu_{max}$ is going to be ${\mathcal O}(10)$ GHz. Similarly 
if $\delta \to 2$ (and with the same choice of $H_{r}$) $\nu_{max} = {\mathcal O}(100)$ kHz. 
In summary we can say that:
\begin{itemize}
\item{} in a model-dependent perspective the maximal frequency of the relic gravitons obeying the bound (\ref{HH3})
is sensitive to the timeline of the post-inflationary expansion rate;
\item{}  in the case of radiation-dominated evolution extending throughout the post-inflationary evolution 
the maximal frequency is of the order of $300$ MHz;
\item{} if the post-inflationary expansion rate is smaller than radiation for some time $\nu_{max} > {\mathcal O}(300)$ MHz;
\item{} if the expansion rate is instead faster than radiation $\nu_{max} < {\mathcal O}(300)$ MHz.
\item{} in general terms, recalling the considerations of section \ref{sec2}, we have that 
${\mathcal O}(100)\,\, \mathrm{kHz} < \nu_{max} < \mathrm{THz}$.
\end{itemize}
Although the maximal frequency alone cannot be used to determine observationally the 
timeline of the expansion rate, Eqs. (\ref{NK5})--(\ref{APA8}) suggest nonetheless that 
$\nu_{max}$ of the spectrum is sensitive to all the aspects of the 
 post-inflationary evolution\footnote{This also means that the maximal frequency, the intermediate frequencies and the shape of 
 $\Omega_{gw}(\nu, \tau_{0})$ can all be employed, in different combinations, to infer timeline of the expansion rate 
 as we are going to see in the following sections.}. 
 
\subsubsection{The intermediate frequencies}
From Figs. \ref{FIGU0c} and \ref{FIGU0d} we have that the bunch of frequencies $\nu = {\mathcal O}(\nu_{max})$ corresponds to the wavelengths that left the horizon at the end of inflation and reentered immediately after. Depending on the timeline of the post-inflationary evolution 
there are other typical frequencies that can be explicitly computed. Moreover, since
$\overline{\nu}_{max}$ depends on $r_{T}$, also all the other frequencies are sensitive 
to the specific value of the tensor-to-scalar-ratio. Rather than starting from the general considerations 
it is better to consider a specific example. Let us then suppose that, before the onset of radiation dominance,
the post-inflationary epoch consists of {\em thee separate phases}. This means, according to Figs. \ref{FIGU0c} and \ref{FIGU0d},  that the final spectrum 
is going to be characterized by the three typical frequencies $\nu_{1} = \nu_{max}$, $\nu_{2}$ and $\nu_{3} = \nu_{r}$. As already  stressed after Eq. (\ref{TWO15O}) we actually recall that we can always assume that $a_{n} = a_{r}$ and $\nu_{r} = \nu_{n}$ so that the $n$-th stage of expansion corresponds (by construction) to the radiation phase. 
In the case $n=3$ the expression of $\nu_{max}$ follows from Eq. (\ref{APA8}) and it is 
\begin{equation}
\nu_{max} = \nu_{1} = \prod_{i\,=\,1}^{2} \, \xi_{i}^{\frac{\delta_{i} -1}{2 (\delta_{i}+1})}\,\, \overline{\nu}_{max} = \xi_{1}^{\frac{\delta_{1} -1}{2 (\delta_{1}+1})}\, \xi_{2}^{\frac{\delta_{2} -1}{2 (\delta_{2}+1})}\,\, \overline{\nu}_{max},
\label{APA9ca}
\end{equation}
where $\xi_{1} = H_{2}/H_{1}$ and $\xi_{2} = H_{r}/H_{2}$.
The intermediate frequencies $\nu_{2}$ and $\nu_{r}$ are related to $\overline{\nu}_{max}$ and 
they are
\begin{eqnarray}
\nu_{2} &=& \sqrt{\xi_{1}} \,\, \xi_{2}^{(\delta_{2} -1)/[2 (\delta_{2}+1)]} \, \,\overline{\nu}_{max},
\nonumber\\
\nu_{r} &=&\nu_{3} = \sqrt{\xi_{1}}\, \sqrt{\xi_{2}}\, \overline{\nu}_{max} = \sqrt{\xi_{r}} \,\overline{\nu}_{max},
\label{APA9cb}
\end{eqnarray}
where, by definition, $\xi_{r} = \, \xi_{1} \, \xi_{2} = H_{r}/H_{1}$. Equation (\ref{APA9cb}) can be easily generalized so that when $n$ intermediate phases are present prior to $a_{r}$ the generic intermediate frequencies $\nu_{m}$  and $\nu_{r}$ are:
\begin{eqnarray}
\nu_{m} &=& \sqrt{\xi_{1}} \, .\,.\,.\, \sqrt{\xi_{m-1}}\,  \prod_{m\,=\,1}^{n-1} \xi_{i}^{\frac{\delta_{i} -1}{2 (\delta_{i}+1})}\,\, \overline{\nu}_{max},
\label{APA9d}\\
\nu_{r} &=& \nu_{n} = \sqrt{\xi_{1}}\,\sqrt{\xi_{2}} \, .\,.\,.\, \sqrt{\xi_{n-2}}\,\sqrt{\xi_{n-1}} \,\,\overline{\nu}_{max}.
\label{APA9e}
\end{eqnarray}
Recalling the remarks presented before, since 
the different phases must not last below $H_{r}$,  the product of all the $\xi_{i}$ equals $H_{r}/H_{1}$, i.e. by definition $\xi_{1}\, \xi_{2} \,.\,.\,. \xi_{n-1} \,\xi_{n} =\xi_{r} = H_{r}/H_{1}$.
Therefore, in case the consistency relations are enforced, Eqs. (\ref{APA9ca})--(\ref{APA9cb}) and (\ref{APA9d})--(\ref{APA9e}) show that both the maximal and the intermediate frequencies of the spectrum depend on $r_{T}$ through $\xi_{r}$. Since $m =1,\,.\,.\,., n -2$, if there are $n$ different stages 
there are $(n-2)$ intermediate frequencies between $\nu_{1}$ and $\nu_{r}$. If $n=3$, as exemplified above, the only intermediate frequency is $\nu_{2}$ and it is given in Eq. (\ref{APA9cb}).

\subsubsection{The slopes of the spectra}
In the previous subsection we derived the typical frequencies 
of the spectrum in the case of a generic sequence of post-inflationary stages
with expansion rates that can be either faster or slower than radiation.
Within the same framework we could now discuss the slopes 
of $\Omega_{gw}(k,\tau)$ within the various frequency domains.
The calculation of the spectral energy density can be sometimes 
carried on in analytic terms but more often with appropriate numerical 
techniques. Here we shall not review all these aspects but just 
remark that, for a sound estimate of the spectral slopes, it is 
sufficient to employ an approximate description that is
based on the Wentzel–Kramers–Brillouin (WKB) solution of the mode functions (see, for instance, \cite{LIGO3} and discussion therein). 
If the power spectra $P_{T}(k,\tau)$ and $Q_{T}(k,\tau)$ of Eq. (\ref{QB11}) 
are inserted into Eq. (\ref{RB7}) $\Omega_{gw}(k,\tau)$
can be directly expressed in terms of the mode functions 
\begin{equation}
\Omega_{gw}(k,\tau) = \frac{k^{5}}{6 \pi^2 H^2 \, a^2 \overline{M}_{P}^2} \biggl[ \bigl| F_{k}(\tau)\bigr|^2 + \frac{\bigl| G_{k}(\tau) \bigr|^2}{k^2} \biggr].
\label{SL1}
\end{equation}
We note that $F_{k}(\tau)$ and $G_{k}(\tau)$ can also be rescaled, i.e. $a(\tau) F_{k}(\tau) = f_{k}(\tau)$ and $a(\tau) G_{k}(\tau) = g_{k}(\tau)$; in this way 
Eq. (\ref{SL1}) becomes
\begin{equation}
 \Omega_{gw}(k,\tau) = \frac{k^{5}}{6 \pi^2 H^2 \, a^4 \overline{M}_{P}^2} \biggl[ \bigl| f_{k}(\tau)\bigr|^2 + \frac{\bigl| g_{k}(\tau) \bigr|^2}{k^2} \biggr].
\label{SL2}
\end{equation}
Before a given wavelength exits the Hubble radius 
(see Fig. \ref{FIGU0d}) the mode functions are simple plane waves normalized 
as in Eq. (\ref{QB10}) to preserve the canonical commutation relations 
between field operators. After the wavelengths reenter the Hubble radius  $f_{k}(\tau)$ and $g_{k}(\tau)$ are
\begin{eqnarray}
f_{k}(\tau) &=& \frac{e^{- i k \tau_{ex}}}{\sqrt{ 2 k}} \biggl[ A_{k}(\tau_{ex},\tau_{re}) \, \cos{k \,\Delta\tau} + B_{k}(\tau_{ex},\tau_{re}) \, \sin{k \,\Delta\tau} \biggr],
\label{SL3}\\
g_{k}(\tau) &=& e^{- i k \tau_{ex}}\, \sqrt{\frac{k}{2}}\, \biggl[ - A_{k}(\tau_{ex},\tau_{re}) \, \sin{k \,\Delta\tau} + B_{k}(\tau_{ex},\tau_{re}) \, \cos{k \,\Delta\tau} \biggr],
\label{SL4}
\end{eqnarray}
where $\Delta\tau = (\tau - \tau_{re})$. In Eqs. (\ref{SL3})--(\ref{SL4}) $\tau_{ex}$ and $\tau_{re}$ denote, respectively, the moments where a given scale
exits and reenters the Hubble radius; the two coefficients $A_{k}(\tau_{ex},\tau_{re})$ and $B_{k}(\tau_{ex},\tau_{re})$ are given by:
\begin{eqnarray}
A_{k}(\tau_{ex},\tau_{re}) &=& \biggl(\frac{a_{re}}{a_{ex}}\biggr) \, {\mathcal J}^{(t)}_{k}(\tau_{ex}, \tau_{re}),
\label{SL5}\\
B_{k}(\tau_{ex}, \tau_{re}) &=& \biggl(\frac{{\mathcal H}_{re}}{k} \biggr) \,  \biggl(\frac{a_{re}}{a_{ex}}\biggr) \, {\mathcal J}^{(t)}_{k}(\tau_{ex}, \tau_{re})
- \biggl(\frac{a_{ex}}{a_{re}}\biggr) \biggl(\frac{{\mathcal H}_{ex} + i k}{k}\biggr),
\label{SL6}\\
{\mathcal J}^{(t)}_{k}(\tau_{ex}, \tau_{re}) &=& 1 - ({\mathcal H}_{ex} + i k) \int_{\tau_{ex}}^{\tau_{re}} \frac{a_{ex}^2}{a^2(\tau^{\prime})} \,\, d \tau^{\prime}.
\label{SL7}
\end{eqnarray}
In Eqs. (\ref{SL5})--(\ref{SL7}), with obvious notations, ${\mathcal H}_{ex} = {\mathcal H}(\tau_{ex})$, 
and ${\mathcal H}_{re} = {\mathcal H}(\tau_{re})$. It can be immediately checked that Eqs. (\ref{SL5})--(\ref{SL7}) together with Eqs. (\ref{SL3})--(\ref{SL4}) imply the Wronskian normalization condition
$f_{k}(\tau) g_{k}^{\ast}(\tau) - f_{k}^{\ast}(\tau) \, g_{k}(\tau) = i$. If the background expands between $a_{ex}$ and $a_{re}$ we have that all the terms containing the ratio $(a_{re}/a_{ex}) \gg 1$ superficially dominate 
against those proportional to $(a_{ex}/a_{re}) \ll 1$.  If we use this logic in a simplified 
manner we would  keep the dominant terms and completely neglect the subdominant ones; by following this logic the 
approximate expression of the mode functions becomes:
\begin{eqnarray}
f_{k}(\tau) &\simeq & \frac{e^{- i k \, \tau_{ex}}}{ \sqrt{2 \, k}} \, {\mathcal J}^{(t)}_{k}(\tau_{ex}, \tau_{re}) \, \biggl(\frac{a_{re}}{a_{ex}}\biggr) \biggl\{ \frac{{\mathcal H}_{re}}{k} \sin{k \Delta \tau}
+ \cos{k \Delta \tau} \biggr\},
\label{SL8}\\
g_{k}(\tau) &\simeq & e^{- i k \, \tau_{ex}}\,\, \sqrt{\frac{k}{2}}\,\, {\mathcal J}^{(t)}_{k}(\tau_{ex}, \tau_{re}) \, \biggl(\frac{a_{re}}{a_{ex}}\biggr) \biggl\{ \frac{{\mathcal H}_{re}}{k} \cos{k\Delta\tau}
- \sin{k \, \Delta \tau}\biggr\}.
\label{SL9}
\end{eqnarray}
We stress that Eqs. (\ref{SL8})--(\ref{SL9}) are quantitatively correct but they do not faithfully 
account for the unitary evolution of the field operators since the all the subleading terms have been 
neglected. These terms are essential if the unitarity is to be restored order by order in the perturbative expansion (see 
in this respect that discussion at the end of this section).
 Inserting finally Eqs. (\ref{SL8})--(\ref{SL9}) in Eq. (\ref{SL2}) we can deduce 
 the explicit expression of the spectral energy density in critical units:
\begin{equation}
\Omega_{gw}(k,\tau) = \frac{k^4}{12\, \pi^2 \,a^4\, H^2 \, \overline{M}_{P}^2} \bigl| {\mathcal J}_{k}(\tau_{ex}, \tau_{re})\bigr|^2 \biggl(\frac{a_{re}}{a_{ex}}\biggr)^2 \biggl( 1 + \frac{{\mathcal H}_{re}^2}{k^2}\biggr) \biggl[ 1 + {\mathcal O}\biggl(\frac{{\mathcal H}}{k}\biggr)\biggr].
\label{SL10}
\end{equation}
Equations (\ref{SL8})--(\ref{SL9}) are  valid for $k \gg a\, H$ (when all the corresponding wavelengths are shorter than the Hubble radius) but they do not satisfy the Wronskian normalization condition owing to their approximate form. Equation (\ref{SL10}) holds within the same approximations which are adequate for the estimates we are presenting here and in the following sections. 

\subsubsection{Spectral energy density, exit and reentry}
According to Eq. (\ref{SL10}) the slopes of  $\Omega_{gw}(k, \tau)$ in a given 
range of wavenumbers chiefly depend on the dynamics of the expansion rate at
$\tau_{ex}$ and $\tau_{re}$. For illustrative purposes we can consider that all 
the wavelengths of spectrum exited the Hubble radius during a conventional 
inflationary stage; this is the viewpoint of Figs. \ref{FIGU0c} and \ref{FIGU0d}. 
The exit may also occur as in Fig. \ref{FIGU3} but this possibility is going to be separately examined in section \ref{sec5} . Since the exit of all wavelengths of the spectrum occurs during inflation, 
\begin{equation}
a_{ex} \, H_{ex}  = - \frac{1}{(1- \epsilon) \tau_{ex} }, \qquad k \tau_{ex} \simeq {\mathcal O}(1).
\label{EXsc}
\end{equation}
the scale factor in this regime can be deduced from Eq. (\ref{EXsc}) and it is 
approximately given by $a(\tau) = ( - \tau/\tau_{1})^{-\mu}$. 
We shall assume that the reentry 
takes place during a decelerated stage of expansion not necessarily coinciding with a radiation epoch.
If the reentry takes place in a generic $\delta$ phase for $\tau \geq -\tau_{1}$ we have that 
\begin{equation}
a_{re}= a(\tau_{re}) = \biggl[ \frac{\mu}{\delta} \biggl( \frac{\tau_{re}}{\tau_{1}} +1 \biggr) +1\biggr]^{\delta}, \qquad\qquad k \tau_{re} \simeq {\mathcal O}(1),
\label{SSTT2}
\end{equation}
where the continuity of the scale factor and of the Hubble rate has been enforced.
The condition $k \tau_{re} = {\mathcal O}(1)$ holds provided $\delta \neq 1$. This can be 
understood by appreciating that the condition of the crossing of a given wavelength is not 
simply given by $k \simeq a\, H$ but, more precisely, by
\begin{equation}
k^2 = a^2 \, H^2 [ 2 - \epsilon(a)], \qquad\qquad \epsilon(a) = - \dot{H}/H^2.
\label{TS18}
\end{equation}
This condition ultimately 
follows from the observation that the evolution of the mode functions 
can be decoupled in terms of $f_{k}(\tau)$ as $f_{k}^{\prime\prime} + [ k^2 - a^{\prime\prime}/a] f_{k} =0$;
it is easy to show that the crossing condition obtained from these considerations is 
$k^2 \simeq |a^{\prime\prime}/a|$ which can also be rewritten as $k^2 \simeq a^2 \, H^2 ( 2 - \epsilon)$.
When $\epsilon \neq 2$ both turning points are regular and this implies that the two 
solutions of Eq. (\ref{TS18}) are given, respectively, by  $k \tau_{ex} = {\mathcal O}(1)$ and by $k \tau_{re} = {\mathcal O}(1)$. For instance when a given wavelength crosses the Hubble radius during inflation we have that $\epsilon \ll 1$ and $k \simeq a_{ex} \, H_{ex}$ 
that also means, by definition, $k \tau_{ex} \simeq 1$. Similarly 
if the given wavelength reenters in a decelerated stage of expansion different from radiation $k \simeq a_{re} \, H_{re}$. However, if the reentry occurs in the radiation stage (or close to it) we have that $\epsilon_{re}\to 2$ (i.e. $\delta = \delta_{re} \to 1$) and the condition (\ref{TS18}) implies that $k \tau_{re} \ll 1$. Equation (\ref{SL10}) can be further simplified in the following form
 \begin{equation}
\Omega_{gw}(k,\tau) = \frac{k^4}{6 H^2 \overline{M}_{P}^2 \pi^2 a^4} \biggl(\frac{a_{re}}{a_{ex}}\biggr)^2  \biggl[ 1 + {\mathcal O} \biggl(\frac{1}{k^2 \tau^2}\biggr) \biggr],
\label{SSTT2a}
\end{equation}
which is valid for $k\tau > 1$, $\tau > \tau_{re}$. Recalling then Eqs. (\ref{EXsc}) and (\ref{SSTT2}) 
 the WKB estimate of the spectral energy density becomes\footnote{In Eq. (\ref{SSTT1}) we restored $M_{P}$ by recalling its relation with $\overline{M}_{P}$, i.e. 
 $\overline{M}_{P} = M_{P}/\sqrt{8\,\pi}$.}:
\begin{equation}
\Omega_{gw}(k,\tau) = \frac{4}{3\, \pi} \biggl(\frac{H_{1}}{M_{P}}\biggr)^2 \,
\biggl(\frac{a_{1}^2 \, H_{1}}{a^2 \, H}\biggr)^2 \biggl| 
\frac{k}{a_{1}\, H_{1}}\biggr|^{\overline{n}_{T}}, 
\label{SSTT1}
\end{equation}
where the spectral index $\overline{n}_{T}$ determines the slope of $h_{0}^2 \Omega_{gw}(k,\tau)$. If the consistency relations are enforced the slow-roll parameter can be traded for the tensor-to-scalar ratio $r_{T}$ so that the slope is ultimately given by\footnote{The result of Eq. (\ref{SSTTnew1}) holds, strictly speaking, in the case $\delta \geq 1/2$. When $\delta < 1/2$ the contribution of ${\mathcal J}^{(t)}(\tau_{ex},\tau_{re})$ must be carefully 
evaluated and contributes to the slope which becomes $\overline{n}_{T} = 2 \delta + {\mathcal O}(r_{T})$. } :
\begin{equation}
\overline{n}_{T}(\delta,\,r_{T}) = \frac{32 - 4 r_{T}}{16 - r_{T}} - 2 \delta.
\label{SSTTnew1}
\end{equation}
Equation (\ref{SSTTnew1}) implies that the high frequency spectral slope is, respectively, increasing or decreasing depending if the expansion rate is either 
 slower or faster than radiation:
\begin{eqnarray}
&& \overline{n}_{T}(\delta, r_{T}) >0 \qquad \mathrm{for}\qquad \delta < 1 - \frac{r_{T}}{16} + {\mathcal O}(r_{T}^2),
\nonumber\\
&& \overline{n}_{T}(\delta, r_{T}) <0 \qquad \mathrm{for}\qquad \delta > 1 - \frac{r_{T}}{16} + {\mathcal O}(r_{T}^2).
\label{SSTTnew2}
\end{eqnarray}
The  analysis leading to Eq. (\ref{SSTT1}) determines the conventional low-frequency slope which is applicable for the frequencies $\nu< \nu_{r} \simeq a_{r} \, H_{r} $ and which is given by 
Eq. (\ref{SSTTnew1}) evaluated in the limit $\delta \to 1$: 
\begin{equation}
n^{(low)}_{T} = \lim_{\delta\to 1} \, \overline{n}_{T}(\delta,\,r_{T}) = - \frac{2 \, r_{T}}{16 - r_{T}} = - \frac{r_{T}}{8}  + {\mathcal O}(r_{T}^2).
\label{SSTTnew3}
\end{equation}
Equation (\ref{SSTTnew3}) corresponds to the slope of the
spectral energy density obtained for the transition between a conventional 
inflationary stage of expansion and a radiation phase. Thanks to the consistency relations, $r_{T} \simeq 16 \, \epsilon$ 
so that the result of Eq. (\ref{SSTTnew3}) also implies that $\overline{n}_{T} = - \,2 \, \epsilon$. 
We can similarly have that the high frequency slope is  given, in practice, by  
$\overline{n}_{T}(\delta,r_{T})$ in the limit $r_{T} \to 0$:
\begin{equation}
n^{(high)}_{T} = \lim_{r_{T}\to 0} \, \overline{n}_{T}(\delta,\,r_{T}) = 2 - 2 \delta.
\label{SSTTnew3a}
\end{equation}
 The spectral index $\overline{n}_{T}(\delta, r_{T})$ can be expressed also in different ways; for instance if we consider that the post-inflationary stage is governed by a perfect barotropic fluid we can also write the spectral index as a function of $w$ (the barotropic index) and $\epsilon$ (the slow-roll parameter):
\begin{equation}
\overline{n}_{T} = \frac{12\, w \, (1- \epsilon) - 2 ( 3 w +1 )}{(3 w+1 ) ( 1 - \epsilon)}.
\label{SSTT1aa}
\end{equation}
where the slow-roll parameter denoted by $\epsilon$ coincides with $\epsilon_{ex} = \epsilon(\tau_{ex})$ and it should be a slowly varying function of the conformal time coordinate; in the limit $\epsilon \to 0$ and $w\to 1$ 
the spectral index in the high frequency region is blue, i.e. $\overline{n}_{T} \to 1$.
 Different values of $w$ slightly reduce the slope (e.g. for $ w \to 2/3$ we have $\overline{n}_{T} \to 2/3$) so that  $0< \overline{n}_{T} \leq 1 $ as long as  $1/3 < w \leq 1$; as expected, the results expressed by Eqs. (\ref{SSTT1})--(\ref{SSTT1aa}) cannot be applied for $\epsilon \to 1$ since, in this limit, the slow-roll approximation breaks down\footnote{ The analysis leading to the results discussed above can be generalized to the situation where there are many post-inflationary stages characterized by 
 different rates $\delta_{i}$. It is also possible to use different approximation schemes that 
 will not be specifically discussed here (see however \cite{MGB}).}.
 
\subsubsection{Approximate forms of the averaged multiplicities and unitarity}
In the past there have been various attempts to justify the loss of quantum coherence 
of the relic gravitons by claiming that when particles are copiously produced 
the averaged multiplicities are very large (see e.g. \cite{ST11} and references therein). The averaged multiplicity $\overline{n}(k,\tau)$ accounting for the pairs of gravitons with opposite three-momenta for each tensor polarization follows then from Eqs. (\ref{QB23})--(\ref{QB24}) 
\begin{equation}
\langle \hat{N}_{k} \rangle = \langle \widehat{a}_{\vec{k}}^{\dagger} \widehat{a}_{\vec{k}} + \widehat{a}_{-\vec{k}}^{\dagger} \widehat{a}_{-\vec{k}} \rangle = 2 \, \overline{n}(k,\tau), \qquad\qquad  \overline{n}(k,\tau) =  |\beta_{k}(\tau)|^2.
\label{NNNew1}
\end{equation}
The largeness of the averaged multiplicity is (incorrectly) used to argue that the final state of the evolution of the relic gravitons is classical and the argument is, in short, the following. Since $|\beta_{k}(\tau)|^2\gg 1$ we also have that $|\alpha_{k}(\tau)|^2 \gg 1$ and this means that the field operators approximately commute\footnote{This conclusion would follow by appreciating that $\alpha_{k,\,\alpha}(\tau) = [k \, f_{k,\, \alpha}(\tau)\, + \, i g_{k,\,\alpha}(\tau)]/\sqrt{2 k}$ and also that 
$\beta_{k,\,\alpha}^{\,\ast}(\tau) = - [k \, f_{k,\, \alpha}(\tau)\, - \, i g_{k,\,\alpha}(\tau)]/\sqrt{2 k}$.}:
\begin{equation}
|\alpha_{k}(\tau)|^2 \simeq |\beta_{k}(\tau)|^2, \qquad\qquad f_{k}(\tau) \, g_{k}^{\ast}(\tau) \simeq g_{k}(\tau) f_{k}^{\ast}(\tau) ,\qquad\qquad [\widehat{\mu}_{i\,j}(\vec{k},\tau), \, \widehat{\pi}_{m\,n}(\vec{p},\tau)] \simeq 0.
\label{NNNew2}
\end{equation}
The heuristic argument of Eq. (\ref{NNNew2}) is  self-contradictory since it suggests 
that unitarity is approximately lost every time a large number of pairs is produced. This is 
markedly false. On the contrary when the approximation scheme is accurate the violations of unitarity are neither explicit nor implicit even if the averaged multiplicity of the produced pairs is very large. 
 To clarify this point we first note that, in the WKB approach of Eqs. (\ref{SL5})--(\ref{SL7}) the values of
 $\alpha_{k}(\tau)$ and $\beta_{k}(\tau)$ are given by:
 \begin{eqnarray}
\alpha_{k}(\tau) =  \frac{1}{2} \bigl[A_{k}(\tau_{ex},\tau_{re}) + i\, B_{k}(\tau_{ex},\tau_{re})\bigr] e^{- i \, k (\Delta \tau + \tau_{ex})},
\label{NNNew3}\\
\beta_{k}^{\ast}(\tau) = - \frac{1}{2}  \bigl[A_{k}(\tau_{ex},\tau_{re}) - i\, B_{k}(\tau_{ex},\tau_{re})\bigr] e^{i \, k (\Delta \tau + \tau_{ex})}.
\label{NNNew4}
\end{eqnarray}
Following the logic of the argument (\ref{NNNew2}) we could now naively argue that
 all the terms containing the ratio $(a_{re}/a_{ex}) \gg 1$ superficially dominate 
against those proportional to $(a_{ex}/a_{re}) \ll 1$, always in the approximation that the background expands between $a_{ex}$ and $a_{re}$. If we would use this logic we would simply keep the dominant terms and discard the subdominant ones. This approach violates unitarity and a the correct strategy is instead to expand systematically in a Laurent series $\alpha_{k}(\tau)$ 
and $\beta_{k}(\tau)$ with the constraint that $|\alpha_{k} (\tau) |^2 - |\beta_{k}(\tau)|^2 =1$.
The result of this strategy is 
\begin{eqnarray}
\alpha_{k}(\tau) = \frac{e^{- i k\tau}}{2} \biggl[ i +  q_{ex}(1 - i) \,  {\mathcal I}(\tau_{ex}, \tau_{re}) \, \biggr] (q_{re} - \, i) \biggl(\frac{a_{re}}{a_{ex}}\biggr) + ( 1 - i \, q_{ex}) \biggl(\frac{a_{ex}}{a_{re}}\biggr)+ {\mathcal O}\biggl[ \biggl(\frac{a_{ex}}{a_{re}}\biggr)^{5}\biggr],
\nonumber\\
\beta_{k}(\tau) = \frac{e^{- k \tau}}{2} \biggl[ - 1 +  (q_{ex} - i){\mathcal I}(\tau_{ex}, \tau_{re}) \biggr] (q_{re} - \, i) \biggl(\frac{a_{re}}{a_{ex}}\biggr) + ( 1 + i \, q_{ex}) \biggl(\frac{a_{ex}}{a_{re}}\biggr)+ {\mathcal O}\biggl[ \biggl(\frac{a_{ex}}{a_{re}}\biggr)^{5}\biggr],
\label{NNNew5}
\end{eqnarray}
where $q_{ex} = a_{ex} \, H_{ex}/k$ and $q_{re} = a_{re} \, H_{re}/k$. We can immediately verify that, to the given order in the perturbative expansion (i.e. $(a_{re}/a_{ex})^{5}$), Eq. (\ref{NNNew5}) implies that $|\alpha_{k}(\tau)|^2 - |\beta_{k}(\tau)|^2=1$ so that unitarity is not lost while keeping the leading terms of the expansion. The result of Eq. (\ref{NNNew5}) can be further simplified by neglecting various factors so that, ultimately, $\alpha_{k}(\tau_{ex}, \tau_{re})$ and 
$\beta_{k}(\tau_{ex}, \tau_{re})$ appearing in Eqs. (\ref{QB23})--(\ref{QB24}) can be evaluated as
\begin{equation}
\alpha_{k}(\tau_{ex},\tau_{re}) \,\simeq\,  \frac{1}{2}\biggl[ \biggl(\frac{a_{re}}{a_{ex}}\biggr) + \biggl(\frac{a_{ex}}{a_{re}}\biggr)\biggl], \qquad 
\beta_{k}(\tau_{ex},\tau_{re}) \,\simeq\, \frac{1}{2}\biggl[ \biggl(\frac{a_{ex}}{a_{re}}\biggr) - \biggl(\frac{a_{re}}{a_{ex}}\biggr)\biggl],
\label{NNNew6}
\end{equation}
where, again, $|\alpha_{k}(\tau_{ex},\tau_{re})|^2 - |\alpha_{k}(\tau_{ex},\tau_{re})|^2 =1$. 
Equation (\ref{NNNew6}) should be still used with some attention since, in the derivation, we artificially 
neglected some phases just for the benefit of a simple expression. However the approximation 
of Eq. (\ref{NNNew6}), unlike other approaches, is at least consistent with unitarity. If the 
limit $a_{re} \gg a_{ex}$ is then taken at the very end of the calculation, the previous results 
for the spectral energy density are explicitly obtained without any violation of the unitary evolution.

\renewcommand{\theequation}{4.\arabic{equation}}
\setcounter{equation}{0}
\section{The expansion history and the low-frequency gravitons}
\label{sec4}
\subsection{General considerations}
\subsubsection{Enhancements and suppressions of the inflationary 
observables}
The low-frequency range of the relic gravitons falls in the aHz domain 
and it corresponds to the CMB wavelengths that 
left the Hubble radius $N_{k}$ $e$-folds before the end of inflation.  As already mentioned in section 
\ref{sec2}, these wavelengths are ${\mathcal O}(\lambda_{p})$ where $\lambda_{p} = 2 \pi/k_{p}$ and $k_{p} = 0.002\, \mathrm{Mpc}^{-1}$ is the pivot scale 
at which the spectra of the scalar and tensor modes of the geometry are normalized within the present conventions; note in fact that $\nu_{p} = k_{p}/(2\pi)= {\mathcal O}(3)$ aHz. The timeline of the post-inflationary evolution
directly affects the values of the tensor to scalar ratio and of the 
other inflationary observables\footnote{For a generic quantity
that is both scale-dependent and time-dependent (be it for instance 
$W(k,\tau)$) we the have that its value is given by $W_{k}= W(k,\tau) = W(k,1/k)$ when the CMB wavelengths cross the Hubble radius during 
inflation. } through their dependence upon $N_{k}$ which can be substantially different 
from ${\mathcal O}(60)$. For instance a stage expanding faster than radiation has been suggested in the past with the purpose of enhancing  the values of $r_{T}$ (see for instance \cite{RR1a,RR2a,RR3a}). Indeed, if the expansion rate is faster than radiation $N_{k}$ gets eventually smaller than the value it would have when the post-inflationary evolution is dominated by radiation (see Eq. (\ref{NK1}) and discussion therein). But since the inflationary observables and the tensor to scalar ratio are all suppressed by different powers of $N_{k}$, they might all experience a certain level of enhancement as long as the post-inflationary expansion rate is faster than radiation\footnote{For instance the BICEP2 observations \cite{REF3a} suggested $r_{T} = {\mathcal O}(0.2)$ that 
looked rather large for single-field inflationary models with monomial potentials. If $N_{k} \ll 60$ 
the value of $r_{T}(N_{k})$ is comparatively larger than in the case $N_{k} = \overline{N}_{k} = {\mathcal O}(60)$.} and this is why this possibility has been employed to account for the BICEP2 excess \cite{REF3a}. Different mechanisms have been suggested for the same purpose like the violation of the consistency relations caused, for instance, by the quantum initial conditions in the case of a short inflationary stage \cite{fluid1,REF3b}. A post-inflationary stage expanding faster than radiation efficiently enhances the value of $r_{T}$ especially in the case of single-field scenarios with monomial potentials.  We now know that the BICEP2 measurements were seriously affected by foreground contaminations so that, at the moment, the current bounds suggests $r_{T} \leq 0.06$ \cite{RR1,RR2,RR3}; this also 
means that the observational evidence would suggest that $r_{T}(N_{k})$ 
is comparatively more suppressed than in the case $N_{k} = \overline{N}_{k} = {\mathcal O}(60)$.
In this respect an even earlier suggestion \cite{EXP1,EXP2,EXP3} (discussed well before the controversial BICEP2 observations \cite{REF3a}) implies that the values of the inflationary observables can be further reduced (rather than enhanced) thanks to a stage expanding more slowly than radiation  \cite{ST3,ST3a} (see also \cite{LID1,MGshift}); this is ultimately the punchline of the considerations of section \ref{sec2} 
(see in particular Eq. (\ref{NK1}) and discussions therein).  As pointed out in Ref. \cite{MGshift2}  a reduction of $r_{T}$ (such as the one suggested by current determinations \cite{RR1,RR2,RR3}) implies that the spectral energy density of relic gravitons is enhanced for frequencies larger than the kHz. This conclusion is particularly interesting since two widely separated frequency domains (i.e. the aHz and the MHz regions) may eventually cooperate in the actual determination of the post-inflationary expansion history, as originally pointed out in \cite{ST3,ST3a}.
 
\subsubsection{The number of $e$-folds and the potential} 
When the pivot scales cross the comoving Hubble radius the values of the inflationary observables can be 
directly expressed as a function of  $N_{k}$ for $k = {\mathcal O}(k_{p})$. 
For this purpose the values of the slow-roll parameters 
(and their dependence on $N_{k}$) must be evaluated not simply for the 
conventional post-inflationary evolution dominated by radiation but in the case of different expansion rates. The total number of $e$-folds elapsed since the crossing of the CMB wavelengths follows from Eq. (\ref{TWO6a}) and it is given by
\begin{eqnarray}
N_{k} =\int_{t_{i}(k)}^{t_{f}} \, H \, dt  = \int_{t_{i}(k)}^{t_{f}} \biggl(\frac{H}{\dot{\varphi}}\biggr) \,\, d\varphi = \frac{1}{\overline{M}_{P}^2} \int_{t_{f}}^{t_{i}(k)} \biggl(\frac{V}{V_{,\, \varphi}}\biggr) \,\,d \varphi,
\label{NEF1}
\end{eqnarray}
where $t_{i}(k)$ coincides with the crossing of the CMB wavelengths and $t_{f}$ marks the end 
of the inflationary stage. In the case of plateau-like potentials it makes sense to normalize $V(\varphi)$  in terms of its scale of variation conventionally denoted hereunder by $M$:
\begin{equation}
V(\varphi) = M^4 \, v(\Phi), \qquad\qquad \Phi = \varphi/\overline{M}_{P}.
\label{NEF2}
\end{equation}
Once Eq. (\ref{NEF2}) is inserted into Eq. (\ref{NEF1}) the number of $e$-folds elapsed since the crossing of the CMB wavelengths becomes:
\begin{equation}
N_{k} = \int_{\Phi_{f}}^{\Phi_{k}} \biggl(\frac{v}{\partial_{\Phi} v }\biggr) \, d \Phi, \qquad \Phi_{k}= \Phi(1/k),
\label{NEF3}
\end{equation}
where $\Phi_{k}$ is now the value of the field when the scale $k$ crosses the comoving Hubble radius while  $\Phi_{f}$ coincides with the end of inflation. Even if different approaches can be envisaged we are here suggesting that the end of inflation effectively occurs when 
\begin{equation}
\epsilon(\Phi_{f}) = \epsilon_{f} \to 1 \,\,\,\Rightarrow\,\,\, H^2 \epsilon_{f} = - \dot{H} \,\,\,\Rightarrow \,\,\, \dot{\varphi}_{f}^2 = \frac{2 \epsilon_{f}}{3 - \epsilon_{f}} \, V_{f}.
\label{NEF4}
\end{equation}
From Eq. (\ref{NEF4}) it also follows that
\begin{equation}
\dot{\varphi}_{f}^2 = \frac{2 \epsilon_{f}}{3 - \epsilon_{f}} \, V_{f}, \qquad \rho_{\varphi}^{(f)} = \biggl[1 + \frac{\epsilon_{f}}{3 - \epsilon_{f}}\biggr] \, V_{f},
\label{NEF5}
\end{equation}
  where, by definition, $\rho_{\varphi} = V + \dot{\varphi}^2/2$. If $\epsilon_{f} \to 1$ we have that 
  $\dot{\varphi}_{f}^2 =  V_{f}$ and  $\rho_{\varphi}^{(f)} = 3 V_{f}/2$.
 Equations (\ref{NEF3})--(\ref{NEF4})  imply a direct connection between 
$\Phi_{k}$ and $N_{k}$ even if the scaling of the various inflationary observables 
and of the slow-roll parameters may be different. For instance the slow-roll parameter $\epsilon(\tau)$ evaluated at $\tau \simeq 1/k$
(i.e. $\epsilon(\tau) = \epsilon(1/k) = \epsilon_{k}$) scales differently as a function of $N_{k}$: we could have $\epsilon_{k} \propto 1/N_{k}$ (as it happens in the case of monomial potentials \cite{RR1a,RR2a,RR3a}) or $\epsilon_{k} \propto 1/N_{k}^2$ (a typical scaling of plateau-like potentials) or even other more complicated scalings like the ones of hill-top potentials \cite{HT3,HT1,HT2,HT2aa} (see also \cite{FR1,FR2,FR3}).  As suggested in Eq. (\ref{NK1}) the value of $N_{k}$ ultimately depends on the post-inflationary evolution
and it cannot be reliably fixed without a specific knowledge of the early expansion 
rate right after the end of inflation. In particular, if the evolution of the background prior to nucleosynthesis is faster than radiation, some of the $\delta_{i}$ in Eq. (\ref{NK1}) will be larger than $1$ and $N_{k}$ gets comparatively smaller than $\overline{N}_{k}$. 
The opposite is true if the expansion rate is slower 
than radiation: in this case some of the $\delta_{i}$ in Eq. (\ref{NK1}) will be smaller than $1$ so that $\epsilon_{k}$ will be more suppressed than in the case of radiation dominance where $N_{k} \to \overline{N}_{k}$.  This perspective is further scrutinized and more concretely illustrated hereunder.

\subsubsection{Illustrative examples and physical considerations}
In the case of plateau-like potentials $v(\Phi)$ may be written as the ratio of two functions approximately scaling with the  same power for $\Phi \gg 1$; for instance we can have:
\begin{equation}
v(\Phi) = \frac{\beta^{p} \Phi^{2 q}}{[ 1 + \beta^2\Phi^{\frac{4 q}{p}}]^{\frac{p}{2}}},\qquad 4 q > p,\qquad \beta>0.
\label{POTEX1}
\end{equation}
In Eq. (\ref{POTEX1}) $\beta$, $p$ and $q$ are the parameters of the potential and, for technical reasons related with the limit $\Phi \gg 1$, it is practical to require $4\, q>p$. Under the conditions of spelled out in Eq. (\ref{POTEX1}), an  oscillating stage may arise 
in the limit $\Phi \ll 1$ where $v(\Phi) = \beta^{p} \Phi^{2 \, q}$. With the same strategy different concrete examples can be 
explicitly constructed:
\begin{equation}
v(\Phi) = \frac{\bigl(e^{\gamma \Phi} -1)^{2 q}}{\bigl(e^{\frac{4 \gamma q}{p} \,\Phi} +1 \bigr)^{\frac{p}{2}}},\qquad 4 q > p, \qquad \gamma >0.
\label{POTEX2}
\end{equation}
The potentials of Eqs. (\ref{POTEX1})--(\ref{POTEX2}) depend upon 
three parameters (i.e. $p$, $q$ and $\beta$). The examples of Eqs. (\ref{POTEX1})--(\ref{POTEX2})
could be further simplified by choosing, for instance, 
\begin{equation}  
v(\Phi) = \bigl(1 - e^{-\beta\Phi})^{2 q}, \qquad \beta > 0, \qquad q >0.
\label{POTEX3}
\end{equation} 
With similar logic we may also consider 
\begin{equation}  
v(\Phi) = \tanh^{2 q}{(\beta\, \Phi)}, \qquad \beta > 0, \qquad q >0.
\label{POTEX4}
\end{equation} 
The examples of plateau-like potentials can be (obviously) multiplied but 
instead of focussing on the peculiar features 
of various potentials it is more interesting to consider the effects of different post-inflationary 
evolutions on the interplay between the large-scale and small-scale 
constraints, as originally pointed out in \cite{EXP1,EXP2,EXP3}; along a more technical 
perspective the same suggestion has been reinstated also in Refs. \cite{ST3,ST3a}.
The form of the potential is then useful for illustration but from the physical viewpoint the production 
of the relic gravitons is determined by the evolution of the space-time 
curvature: while the scalar inhomogeneities (see appendix \ref{APPA}) 
may depend on the features of the potential, the production of the relic gravitons is 
directly sensitive to the expansion rate.  It is nonetheless useful 
to consider which potential might lead to an invisible $r_{T}$ in the aHz 
range together with a very large signal in the MHz and GHz regions \cite{MGshift2}.
The potentials suggested above go along this perspective even though one can find other classes of potentials that may suppress $r_{T}$ for the CMB wavelengths without leading to a large signal at higher frequencies (see \cite{HT2,HT2a} and references therein).

\subsection{The tensor to scalar ratio}
The amplitudes of the tensor and scalar power spectrum are related via $r_{T}$ which is, in general terms, both scale-dependent and time-dependent:
\begin{equation}
r_{T}(k, \tau) = P_{T}(k,\,\tau)/P_{{\mathcal R}}(k,\tau), \qquad P_{{\mathcal R}}(k,\tau) = \frac{k^3}{2\pi^2} |F^{(s)}_{k}(\tau)|^2,
\label{TSR1}
\end{equation}
where $F^{(t)}_{k}(\tau)$ and $F^{(s)}_{k}(\tau)$ denote, respectively, the tensor and the scalar mode functions, i.e. 
\begin{eqnarray}
&& F^{(t)\prime\prime}_{k} + 2 {\mathcal H} F^{(t)\prime}_{k} + k^2 F^{(t)}_{k} =0, \qquad F^{(t)\prime}_{k} = G^{(t)}_{k},\qquad {\mathcal H} = a^{\prime}/a,
\label{MFT11}\\ 
&& F^{(s)\prime\prime}_{k} + 2 {\mathcal F} F^{(s)\prime}_{k} + k^2 F^{(s)}_{k} =0, \qquad  F^{(s)\prime}_{k} = G^{(t)}_{k},\qquad {\mathcal F} = z^{\prime}/z,
\label{MFS11}
\end{eqnarray}
where, notational convenience, in the following discussion we posit $z_{\varphi}(\tau) = z(\tau)$.
 The definition of $P_{T}(k,\,\tau)$ has been already introduced in Eq. (\ref{QB11}) whereas the scalar power spectrum is also discussed in appendix \ref{APPA}. 
To avoid confusions the tensor mode functions (denoted by $F_{k}(\tau)$ in Eq. (\ref{QB11})) are distinguished from their scalar counterpart by a superscript (i.e. $F^{(t)}_{k}(\tau)$).
With these notations, the tensor to scalar ratio (\ref{TSR1}) becomes:
\begin{equation}
r_{T}(k, \tau) = 8 \ell_{P}^2 \bigl| F^{(t)}_{k}(\tau) \bigr|^2/\bigl| F^{(s)}_{k}(\tau) \bigr|^2. 
\label{TSR2}
\end{equation}
We are now going to evaluate $r_{T}(k,\tau)$ before and after reentry. In appendix \ref{APPA} the 
tensor to scalar ratio is discussed by using the exact solutions for the evolution of the mode functions 
during the inflationary stage; by construction the analysis of appendix \ref{APPA} applies for wavelengths larger than the Hubble radius during inflation.

\subsubsection{The tensor to scalar ratio before reentry}
The  initial conditions of the temperature and polarization anisotropies of the CMB are set 
prior to matter-radiation equality (i.e. $\tau < \tau_{eq}$) when the relevant wavelengths are larger than the Hubble radius. This means that Eq. (\ref{TSR2}) should be evaluated for for $\tau_{ex} \leq \tau < \tau_{re}$ and $k \ll a\, H$; as before $\tau_{ex}(k)$ and $\tau_{re}(k)$ denote, respectively, the moments at which a given wavelength either exits or reenters the Hubble radius (see Fig. \ref{FIGU0d} and discussions therein). In this approximation Eqs. (\ref{MFT11})--(\ref{MFS11}) can be independently solved: 
\begin{equation}
F^{(s)}_{k}(\tau) = \frac{e^{- i k\tau_{ex}}}{ z_{ex}\sqrt{2 \, k}} \,{\mathcal J}^{(s)}_{k}(\tau_{ex}, \tau), 
\qquad\qquad F_{k}^{(t)}(\tau) = \frac{e^{- i k\tau_{ex}}}{ a_{ex} \sqrt{2 \, k}} \, {\mathcal J}^{(t)}_{k}(\tau_{ex}, \tau),
\label{TSR3}
\end{equation}
where the shorthand notations $z_{ex} = z_{\varphi}(\tau_{ex})$ and $a_{ex} = a(\tau_{ex})$ have been used; note that $z_{\varphi}(\tau)$ already appears right after Eq. (\ref{APPB1}) and its definition will not be repeated. In the cartoon of Fig. \ref{FIGU0d}  the solutions of Eq. (\ref{TSR3}) hold 
for $ \tau_{ex} < \tau < \tau_{re}$, i.e. for frequencies that are larger than the Hubble radius; both solutions of Eq. (\ref{TSR3}) have been correctly normalized to their respective quantum mechanical initial data. The functions 
${\mathcal J}^{(s)}_{k}(\tau_{ex}, \tau)$ and ${\mathcal J}^{(t)}_{k}(\tau_{ex}, \tau)$ 
are defined as:
\begin{eqnarray}
{\mathcal J}_{k}^{(s)}(\tau_{ex}, \tau) &=& 1 - ( i k + {\mathcal F}_{ex}) \, z_{ex}^2\,\int_{\tau_{ex}}^{\tau} 
\frac{d \,\tau_{1}}{z_{\varphi}^2(\tau_{1})},
\nonumber\\
{\mathcal J}_{k}^{(t)}(\tau_{ex}, \tau) &=& 1 - ( i k + {\mathcal H}_{ex}) \, a_{ex}^2\,\int_{\tau_{ex}}^{\tau} 
\frac{d\,\tau_{1}}{a^2(\tau_{1})},
\label{TSR4}
\end{eqnarray}
where, we remind, ${\mathcal F}= z_{\varphi}^{\prime}/z_{\varphi}= z^{\prime}/z$ and ${\mathcal H} = a^{\prime}/a$; the integral ${\mathcal J}_{k}^{(t)}(\tau_{ex}, \tau)$ already appears in Eq. (\ref{SL7}).
When Eqs. (\ref{TSR3})--(\ref{TSR4}) are inserted into Eq. (\ref{TSR1}), the explicit form of $r_{T}(k,\tau)$ is obtained 
\begin{equation}
r_{T}(k,\tau) = 8 \, \ell_{P}^2 \biggl(\frac{z_{ex}}{a_{ex}}\biggr)^2 \, \frac{|{\mathcal J}_{k}^{(t)}(\tau_{ex}, \tau) |^2}{|{\mathcal J}_{k}^{(s)}(\tau_{ex}, \tau) |^2} \to 8 \, \ell_{P}^2 \biggl(\frac{z_{ex}}{a_{ex}}\biggr)^2,
\label{TSR5}
\end{equation}
and it is valid in the regime $k < a\, H$ and $\tau_{ex} \leq \tau < \tau_{re}$.
In the case of single field inflationary models 
and for the timeline of the comoving horizon illustrated in Fig. \ref{FIGU0d} we deduce:
\begin{equation}
r_{T}(k,\tau) = 8 \, \ell_{P}^2 \biggl(\frac{\dot{\varphi}^2}{H^2}\biggr)_{ex} \simeq 16 \, \epsilon_{k}, \qquad 
\qquad \epsilon_{k} = - \biggl(\frac{\dot{H}}{H^2}\biggr)_{ex},
\label{TSR6}
\end{equation}
since ${\mathcal J}^{(t)}_{k}(\tau_{ex}, \tau) \simeq {\mathcal J}^{(s)}_{k}(\tau_{ex}, \tau) \to 1$; in Eq. (\ref{TSR6}) the second equality follows from $ 2 \dot{H} = - \ell_{P}^2 \dot{\varphi}^2$ (see Eqs. (\ref{SINGLE2})--(\ref{SINGLE3}) and discussion thereafter).

\subsubsection{The tensor to scalar ratio after reentry}
The expression of the scalar and tensor mode functions after reentry can be 
directly obtained from the previous results of Eqs. (\ref{SL3})--(\ref{SL4}) and from the subsequent discussion. In particular, within the same 
approximation leading to Eqs. (\ref{SL8})--(\ref{SL9}), the evolution of the tensor mode functions is approximately given by:
\begin{equation}
F^{(t)}_{k}(\tau) = \frac{e^{- i k \, \tau_{ex}}}{a \sqrt{2 \, k}} \, {\mathcal J}^{(t)}_{k}(\tau_{ex}, \tau_{re}) \, \biggl(\frac{a_{re}}{a_{ex}}\biggr) \biggl\{ \frac{{\mathcal H}_{re}}{k} \sin{(k \Delta\tau)}
+ \cos(k\Delta\tau) \biggr\}, 
\label{TSR7}
\end{equation}
where, as in Eqs. (\ref{SL3})--(\ref{SL4})  $\Delta \tau =( \tau- \tau_{re})$; ${\mathcal J}^{(t)}_{k}(\tau_{ex}, \tau_{re})$ has been already defined in Eq. (\ref{TSR4}) and it is now evaluated for $\tau \to \tau_{re}$. Equation (\ref{TSR7}) holds when all the comoving frequencies are larger than the expansion rate (i.e. for $k  \gg a \, H$) and in the same approximation $G^{(t)}_{k}(\tau)$ becomes:
\begin{equation}
G^{(t)}_{k}(\tau) = \frac{e^{- i k \, \tau_{ex}}}{a}\,\, \sqrt{\frac{k}{2}}\,\, {\mathcal J}^{(t)}_{k}(\tau_{ex}, \tau_{re}) \, \biggl(\frac{a_{re}}{a_{ex}}\biggr) \biggl\{ \frac{{\mathcal H}_{re}}{k} \cos{(k\Delta\tau)}
- \sin(k\Delta\tau) \biggr\}.
\label{TSR8}
\end{equation}
Equations (\ref{TSR7})--(\ref{TSR8}) assume an expanding background (i.e.  $a_{re} \gg a_{ex}$) but they are otherwise general since the  rates at $\tau_{ex}(k)$ and $\tau_{re}(k)$ have not been specified. Always in the limit of short wavelengths, the mode function for the curvature inhomogeneities becomes
\begin{equation}
F^{(s)}_{k}(\tau) = \frac{e^{- i k \, \tau_{ex}}}{z_{\varphi}(\tau) \sqrt{2 \, k}} \, {\mathcal J}^{(s)}_{k}(\tau_{ex}, \tau_{re}) \, \biggl(\frac{z_{re}}{z_{ex}}\biggr) \biggl\{ \frac{{\mathcal F}_{re}}{k} \sin{(k \Delta\tau)} + \cos(k\Delta\tau) \biggr\}.
\label{TS19}
\end{equation}
Equations (\ref{TSR7}) and (\ref{TS19}) can be finally inserted into the definition
of Eq. (\ref{TSR2})  to obtain the wanted form of $r_{T}(k, \tau)$ valid for $ \tau \geq \tau_{re}$ in the short-wavelength limit (i.e. for $k \tau> 1$):
\begin{eqnarray}
r_{T}(k,\tau) &=& 8 \ell_{P}^2\, \biggl[\frac{z(\tau)}{a(\tau)}\biggr]^2 \, \biggl(\frac{a_{re}}{a_{ex}}\biggr)^2 \biggl(\frac{z_{ex}}{z_{re}}\biggr)^2 {\mathcal G}^2(k \Delta\tau),
\nonumber\\
{\mathcal G}(k \Delta\tau) &=& \frac{{\mathcal F}_{re}\sin{(k \Delta\tau)} + k\,\cos(k\Delta\tau)}{{\mathcal H}_{re}\sin{(k \Delta\tau)} + k\cos(k\Delta\tau)}.
\label{TS20}
\end{eqnarray}
Recalling the discussion of Eq. (\ref{TS18}), in the limit $\epsilon_{re} \to 2$ we also have ${\mathcal H}_{re}/k \simeq {\mathcal F}_{re}/k \gg 1$ and, in this case, ${\mathcal G}(k \Delta \tau) \to 1$. Conversely, when $\epsilon_{re} \neq 2$ we have instead that 
${\mathcal H}_{re}/k \simeq {\mathcal F}_{re}/k = {\mathcal O}(1)$; also in this situation 
${\mathcal G}(k\Delta\tau) = {\mathcal O}(1)$. From Eq. (\ref{TS20}) we can therefore obtain:
\begin{equation}
r_{T}(k, \tau) = 16 \, \epsilon_{k} \, \frac{\epsilon(\tau)}{\epsilon_{re}}, \qquad \tau \geq \tau_{re}, \qquad k\tau > 1.
\label{TS21}
\end{equation}
After inflation $\epsilon(\tau)$ is in practice piecewise constant and it is of the order of $\epsilon_{re}$ so that, ultimately, $r_{T}(k,\tau) \to 16 \epsilon_{k}$ even for short wavelengths. 

\subsubsection{Oscillating potentials}
If the background expands as simple power-law $\epsilon(\tau)$ is constant; similarly, if  the reentry of the given wavelength takes place when the inflaton potential is still dominant (and oscillating) $\epsilon(\tau)$ remains approximately constant. To analyze this situation we can first write $\epsilon(\tau)$ in terms of the inflaton potential $V(\varphi)$, i.e. 
\begin{equation}
\epsilon(\tau) = - \dot{H}/H^2 = 3 \dot{\varphi}^2/(\dot{\varphi}^2 + 2 V).
\label{TS22}
\end{equation}
As suggested long ago the coherent oscillations of the inflaton imply the approximate constancy of the corresponding energy density \cite{turn1,turn2,turn3,turn4}. Indeed we have that the evolution 
of the inflaton energy density $\rho_{\varphi}$ can be rephrased as 
\begin{equation}
\dot{\rho}_{\varphi} + 3 H \dot{\varphi}^2 =0, \qquad \rho_{\varphi} = \dot{\varphi}^2/2 + V.
\label{TS22a}
\end{equation}
During a stage driven by the inflatoon oscillations we have that $3 H\dot{\varphi}^2 \ll \dot{\rho}_{\varphi}$
so that, approximately, the energy density is conserved, i.e. $\dot{\rho}_{\varphi} \simeq 0$, i.e.
\begin{equation}
\dot{\rho}_{\varphi} \simeq 0 \qquad\Rightarrow\qquad  \dot{\varphi}^2 = 2 (V_{max} - V),
\label{TS22b}
\end{equation}
 $V_{max} = V(\varphi_{max})$. Equation (\ref{TS22b}) can be integrated further since the inflaton potential around its minimum can be parametrized as: 
\begin{equation}
V(\varphi) = V_{0} (\varphi/\overline{M}_{P})^{2 q},\qquad \rightarrow \qquad \dot{\varphi} = \pm \sqrt{2 V_{max}}\sqrt{1 - x^{2 q}},
\label{TS23}
\end{equation}
where $x = \varphi/\varphi_{max}$. We can now go back to  
Eq. (\ref{TS22}); when the numerator and the denominator are averaged over one period of oscillations (say between $\varphi=0$ and $\varphi= \varphi_{max}$)  $\epsilon(\tau)$ becomes
\begin{equation}
\epsilon(\tau) = \frac{3 \int_{0}^{1} \sqrt{ 1 - x^{2q}} dx}{ \int_{0}^{1} dx/\sqrt{ 1 - x^{2q}}} \to \frac{3 \,q }{q +1}.
\label{TS24}
\end{equation}
Thus, from Eqs. (\ref{TS21})--(\ref{TS23}) and  Eq. (\ref{TS24}), $\epsilon(\tau)/\epsilon_{re} \to 1$ also when the reentry occurs in a stage driven by the coherent inflaton oscillations. With the same approach the average expansion rate can be computed; in particular we can obtain:
\begin{equation}
{\mathcal H}^{\prime}/{\mathcal H}^2  = (1- 2 q)/(q+1) \qquad \Rightarrow \qquad \delta = (q+1)/(2 q -1),
\label{TS25}
\end{equation}
where $\delta$ denotes the expansion rate in the conformal time coordinate 
(i.e. $a(\tau) = (\tau/\tau_{1})^{\delta}$).  The evolution of the comoving horizon 
in Fig. \ref{FIGU0d} assumes a sequence of different expanding stages characterized by the constancy of the expansion rate. A fully equivalent strategy is to consider the continuous variation of $\delta$ implying 
\begin{equation}
\frac{1}{\delta(\tau)} = - 1 - \frac{1}{2} \frac{\partial \ln{\rho_{t}}}{\partial\ln{a}} = -1 + \epsilon(\tau),
\label{TS26}
\end{equation}
where $\rho_{t}(a)$ denotes the total energy density governing the 
post-inflationary evolution prior to radiation.  In the case of inflaton-dominated 
oscillations $\rho_{t}(a) = \rho_{\varphi}$ and 
\begin{equation}
 \delta(a)  = 1/[\epsilon(a)  -1]= (q+1)/( 2 q -1).
\label{TS26a}
\end{equation}
By going back to Fig. \ref{FIGU0d} we therefore have that when the given wavelength crosses the Hubble radius prior to radiation dominance the value of $\delta$ is scale-dependent $\delta_{k}= \delta(\tau_{re}) = \delta(1/k)$. This conclusion follows by recalling that, during the post-inflationary stage illustrated in the cartoon of Fig. \ref{FIGU0d},
$\delta(a) \neq 1$ which also implies $\epsilon(a)\neq 2$.

\subsection{Consistency relations and inflationary observables}
In this final subsection we are going to analyze the dependence of the 
observables upon the post-inflationary timeline encoded in the value of $N_{k}$. We first introduce the standard form of the slow-roll parameters
\begin{equation}
\epsilon(\tau) = - \frac{\dot{H}}{H^2} = \frac{\overline{M}_{P}^2}{2} \biggl(\frac{V_{\,\,, \varphi}}{V} \biggr)^2, 
\qquad 
\eta(\tau) = \frac{\ddot{\varphi}}{H \, \dot{\varphi}} = \epsilon(\tau) - \overline{\eta}(\tau), \qquad \qquad \overline{\eta}(\tau) = \overline{M}_{P}^2 \biggl(\frac{V_{\,,\varphi\varphi}}{V}\biggr),
\label{TWO7H1}
\end{equation}
and recall that in terms of the dimensionless variables of Eq. (\ref{NEF2}) $\epsilon(\tau)$ and $\overline{\eta}(\tau)$ become:
\begin{equation}
\epsilon(\tau) = \frac{1}{2}\biggl(\frac{v_{\,\,, \Phi}}{v} \biggr)^2,\qquad 
\overline{\eta}(\tau) = \biggl(\frac{v_{\,,\Phi\Phi}}{v}\biggr)
\label{TWO7H1a}
\end{equation}
During inflation all the slow-roll parameters are much smaller than $1$ and the corresponding observables at the crossing time become: 
\begin{equation}
n_{s}(k) = 1 - 6 \epsilon_{k} + 2 \overline{\eta}_{k}, \qquad r_{T}(k) = 16 \, \epsilon_{k},\qquad n^{(low)}_{T}(k) = - 2 \, \epsilon_{k},
\label{TWO7I}
\end{equation}
where $\epsilon_{k}= \epsilon(1/k)$ and $\overline{\eta}_{k} = \overline{\eta}(1/k)$ denote the slow-roll parameters evaluated when the bunch of wavelengths corresponding to the CMB scales exited the comoving horizon approximately $N_{k}$ $e$-folds prior to the end of inflation. 
According to the current limits, the tensor-to-scalar-ratio and the scalar spectral index are determined as\footnote{We stress once more that  $r_{T}(k,\tau_{ex}) = r_{T}(k,1/k) = r_{T}(k)$ and similarly  $n_{s}(k,\tau_{ex}) = n_{s}(k,1/k) = n_{s}(k)$.}
\begin{equation}
r_{T}(k,\tau_{ex}) < \overline{r}_{T}, \qquad\qquad n_{s}(k,\tau_{ex}) = \overline{n}_{s},
\label{TWO7L}
\end{equation}
where $\overline{r}_{T}$ ranges between ${\mathcal O}(0.06)$ and ${\mathcal O}(0.03)$ while $0.96448 < \overline{n}_{s} < 0.96532$ with a central value corresponding to $0.9649$ \cite{RR1,RR2,RR3}.  
For the monomial potentials $\epsilon_{k}$ and $\overline{\eta}_{k}$ are of the same order and these scenarios are practically excluded by current data. 
Let us then consider, as an example, the potential given in Eq. (\ref{POTEX3}). In this case from the expression of the number of $e$-folds we obtain 
\begin{equation}
N_{k} = \frac{e^{\beta\Phi_{k}}}{ 2 \, q\, \beta^2}, \qquad \epsilon_{k} = \frac{ 1}{ 2 \beta^2 N_{k}^2}, \qquad \overline{\eta}_{k} = - \frac{1}{N_{k}},
\label{TWO7La}
\end{equation}
which also implies that 
\begin{equation}
r_{T}(k) = \frac{8}{\beta^2 \, N_{k}^2}, \qquad n_{s}(k) = 1 - \frac{3}{\beta^2 \, N_{k}^2} - \frac{2}{N_{k}}.
\label{TWO7b}
\end{equation}
\begin{figure}[!ht]
\centering
\includegraphics[height=6cm]{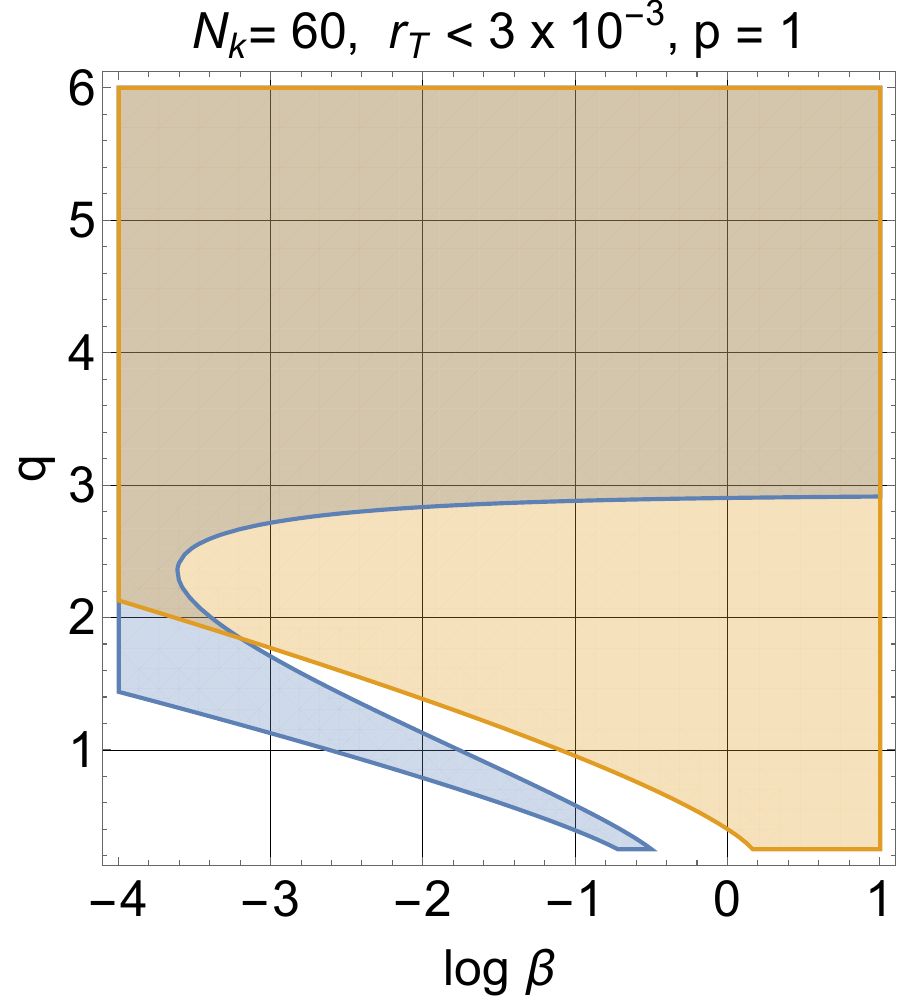}
\includegraphics[height=6cm]{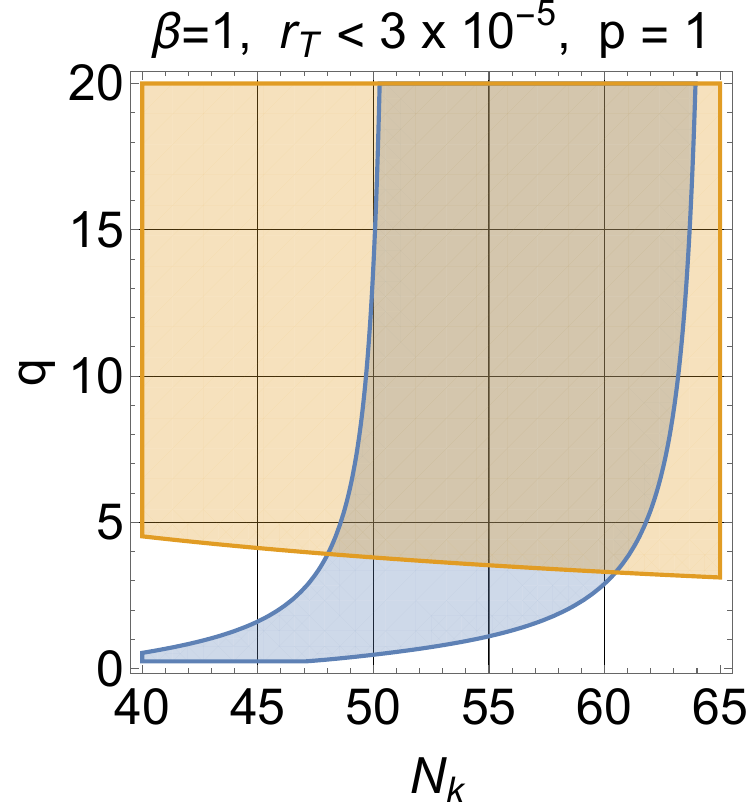}
\caption[a]{We illustrate Eqs. (\ref{PPS1})--(\ref{PPS3}) in the case 
$p =1$. In the plot at the left we consider the $(\beta,\, q)$ plane while in the 
right plot we discuss the plane $(q,N_{k})$. In both plots there are two overlapping 
regions: the wider area corresponds to the condition $r_{T}(k,\tau_{ex}) < \overline{r}_{T}$ 
while the narrower region illustrates the bounds on $n_{s}(k)$ (see Eq. (\ref{TWO7L}) and discussion thereafter). In the two plots we illustrated 
different values of $\overline{r}_{T}$.}
\label{FIGUpot1}      
\end{figure}

\subsubsection{Scaling of the spectral indices with the number of $e$-folds}
When the consistency relations are enforced the tensor to scalar 
ratio cannot be equally small for all the classes of inflationary potentials and while the monomials are clearly excluded, the plateau-like and the hill-top 
potentials may lead to $r_{T}$ that are comparatively smaller.  In the case of Eq. (\ref{POTEX1}) the explicit expressions of the slow-roll parameters 
follow from $\epsilon(\Phi)$ and $\overline{\eta}(\Phi)$ are given by:
\begin{equation}
\epsilon(\Phi) = \frac{2 \, q^2 }{\Phi^2 ( 1 + \beta^2 \Phi^{\frac{4 \,q}{p}})^2}, \qquad\qquad
\overline{\eta}(\Phi) = \frac{2 q\,[2 \, p\, q- p - \beta^2 (p + 4 q) \Phi^{\frac{4 q}{p}}]}{ p \Phi^2 ( 1 +\beta^2 \, \Phi^{\frac{4 q}{p}})^2}.
\label{PP6}
\end{equation}
In this case, according to Eq. (\ref{PP6}), the tensor-to-scalar ratio and the scalar spectral index are given by:
\begin{equation}
r_{T}(\Phi)= \frac{ 32 \, q^2}{\Phi^2 (1 + \beta^2 \Phi^{\frac{4 q}{p}})^2}, \qquad\qquad n_{s}(\Phi) = 1 - \frac{4\, p \, q( 1+ q) + 4 q ( q + 4 p) \beta^2 \Phi^{\frac{4 q}{p}}}{p \Phi^2 ( 1 + \beta^2 \Phi^{\frac{4 q}{p}})^2}.
\label{PP7}
\end{equation}
The number of $e$-folds is ultimately given, in this case, by: 
\begin{equation}
N_{k} = \frac{\Phi_{k}^2 -1}{4 q} + \frac{p \, \beta^2 \, \bigl( \Phi_{k}^{2 + \frac{4 q}{p}} -1\bigr)}{4 q ( p + 2 q)},
\label{PP10}
\end{equation}
where we simply assumed $\Phi_{f} \to 1$.
Since the field value at $\Phi_{k}$ is defined at the time of the crossing during inflation we can 
take the limit $\Phi_{k} \gg 1$ in Eq. (\ref{PP10}) and eventually determine the connection between 
$\Phi_{k}$ and $N_{k}$:
\begin{equation}
N_{k} = \frac{p \, \beta^2}{4 q\, ( p + 2 q)} \, \, \Phi_{k}^{2 + \frac{4 q}{p}}
\qquad \Rightarrow \qquad \Phi_{k} = \biggl[ \frac{4 q (p + 2 q) \, \, N_{k}}{p \, \beta^2}\biggr]^{ \frac{p}{2 (p + 2 q)}}.
\label{PP11}
\end{equation}
Thanks to Eq. (\ref{PP11})  Eqs. (\ref{PP6})--(\ref{PP7}) can be directly expressed in terms of $\Phi_{k}> 1$
\begin{equation}
\epsilon_{k} = \frac{2 q^2}{\beta^4 \, \Phi_{k}^{8 q/p +2}}, \qquad \qquad \overline{\eta}_{k} = - 
\frac{2 q (p + 4 q)}{p \,\beta^2 \, \Phi_{k}^{8 q/p+2}}.
\label{PP12}
\end{equation}
Finally using Eq. (\ref{PP11}) into eq. (\ref{PP12}) we have: 
\begin{equation}
\epsilon_{k} = \frac{ 2 q^2 \, \beta^{- \frac{2 p}{p + 2 q}}}{[ 4 q ( p + 2 q) N_{k}/p]^{\frac{p + 4 q}{p + 2 q}}}, \qquad \qquad \overline{\eta}_{k} = - \frac{ p + 4 q}{2 ( p + 2 q) \, N_{k}}.
\label{PP13}
\end{equation}
In what follows the scaling of the inflationary observables 
with the number of $e$-folds and with the other parameters 
will be swiftly discussed in the case of the example of Eqs. (\ref{PPS1})--(\ref{PPS3}).
\begin{figure}[!ht]
\centering
\includegraphics[height=6cm]{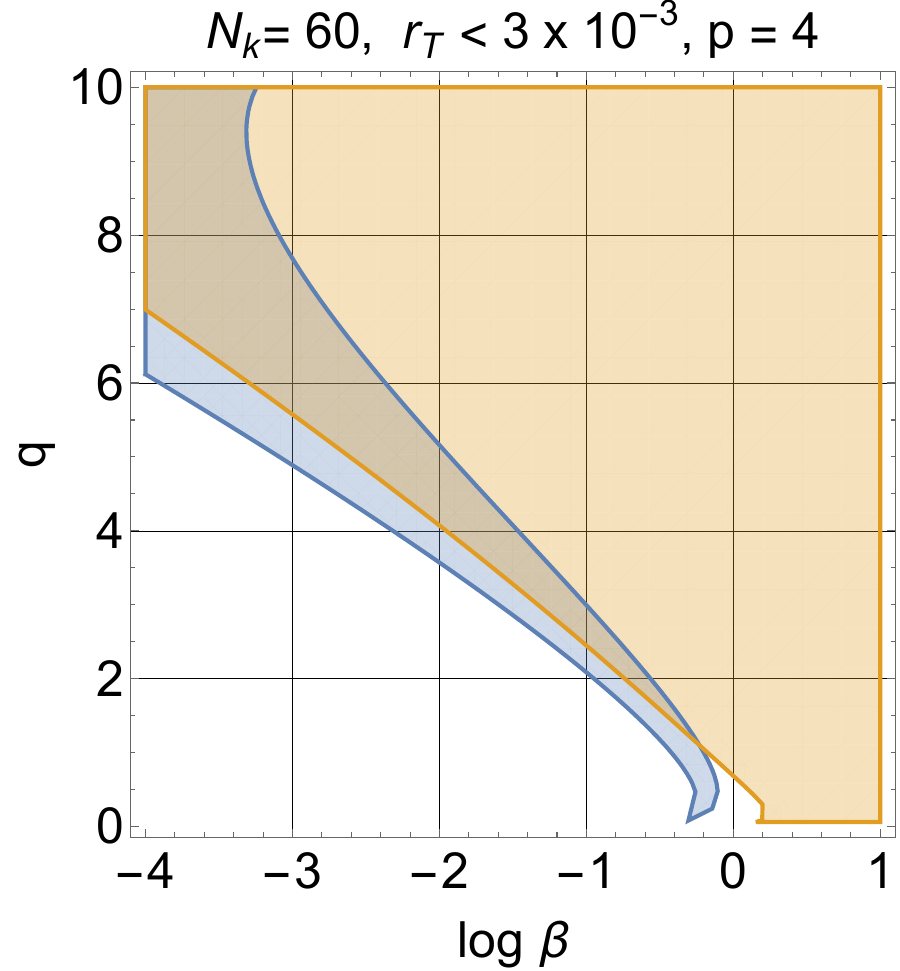}
\includegraphics[height=6cm]{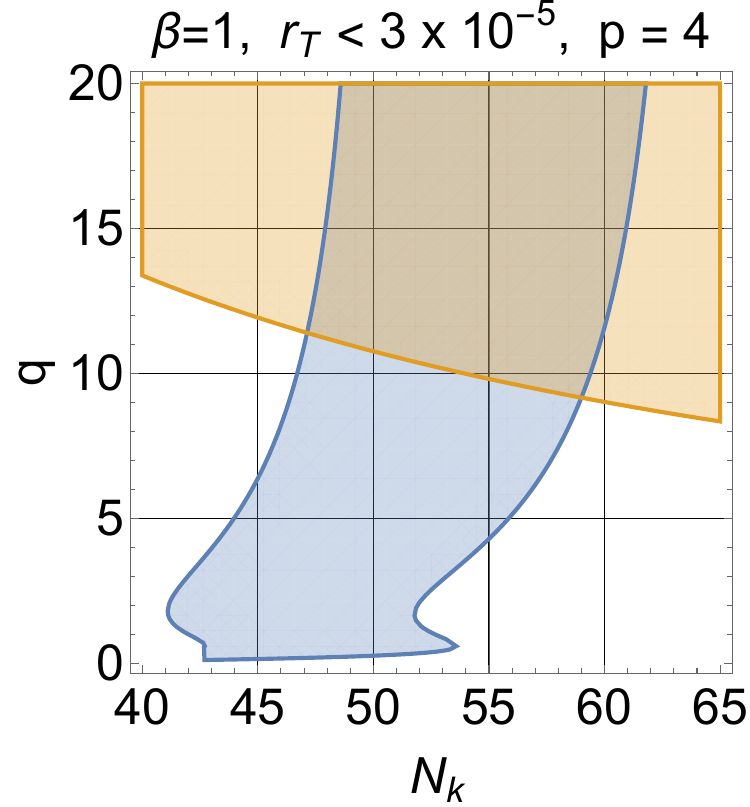}
\caption[a]{As in Fig. \ref{FIGUpot1}  we consider the example of Eqs. (\ref{PPS1})--(\ref{PPS3}) but in the case $p=4$. The same qualitative 
features already discussed in the case of Fig. \ref{FIGUpot1} can be observed. }
\label{FIGUpot2}      
\end{figure}
\subsubsection{An illustrative example}
While the examples along these lines can be multiplied, for the present purposes, different functional forms of the potential do not radically modify the scaling of 
$r_{T}(k)$ and of $\overline{\eta}_{k}$. 
From Eq. (\ref{PP13})  $n_{s}(k)$ and $r_{T}(k)$ becomes:
\begin{eqnarray}
n_{s}(k) &=& n_{s}(N_{k}) =1 - \frac{12 q^2 \, \beta^{ - 2/(1+ 2q/p)}}{[ 4\, q\, ( p + 2 q)\, N_{k}/p]^{(p+ 4 q)/(p+ 2 q)}}- \frac{p +4 q}{(p + 2\, q) \, N_{k}}.
\label{PPS1}\\
r_{T}(k) &=& r_{T}(N_{k}) = \frac{32 \, q^2 \,  \beta^{ - 2/(1+ 2q/p)}}{[ 4\, q\, ( p + 2 q)\, N_{k}/p]^{(p+ 4 q)/(p+ 2 q)}}, 
\label{PPS2}\\
n^{(low)}_{T}(k) &=& n^{(low)}_{T}(N_{k})
= - \frac{4 \, q^2 \, \beta^{ - 2/(1+ 2q/p)}}{[ 4\, q\, ( p + 2 q)\, N_{k}/p]^{(p+ 4 q)/(p+ 2 q)}},
\label{PPS3}
\end{eqnarray}
where $n^{(low)}(k) = - 2\epsilon_{k}$ denotes the low-frequency 
spectral index deduced from the enforcement of the consistency 
relations. Unlike $n^{(low)}_{T}(k)$, the high frequency spectral index may not 
be unique and it can also increase, as already discussed in section \ref{sec3}.
In Fig. \ref{FIGUpot1} we discuss the case of Eqs. (\ref{PPS1})--(\ref{PPS3}) 
for $p=1$. In each plot there are two shaded regions: the first one corresponds 
to the bounds on the scalar spectral index (i.e. $0.96448 < \overline{n}_{s} < 0.96532$) and it resembles to a vertical stripe that gets wider as $q$ increases; 
the second shaded area illustrates the bound on $r_{T}$. When the two shaded
regions overlap the constraints on $r_{T}$ and on $n_{s}$ are 
concurrently satisfied. In Fig. \ref{FIGUpot1} we artificially 
lowered the bounds on $r_{T}$ (typically $r_{T} < 0.03$) and also considered 
$N_{k}$ as a free parameter. A reduction of $r_{T}$ always entails 
large values of $q$; in this case the high frequency slope of the spectral 
energy density is increasing, as already discussed in section \ref{sec3}.
In Fig. \ref{FIGUpot2} we always illustrate  Eqs. (\ref{PPS1})--(\ref{PPS3}) but
for $p =4$. According to Figs. \ref{FIGUpot1} and \ref{FIGUpot2} 
the region where the constraints are simultaneously satisfied moves towards large
$q$-values where the inflaton oscillations effectively lead to a phase 
expanding at a rate that is slower than radiation. In this case the high frequency 
bound are therefore essential \cite{MGshift2} (see also \cite{ST3,ST3a}).

In summary we have that the low-frequency region is sensitive to the post-inflationary 
expansion rate through the number of $e$-folds which can be either larger or smaller 
than ${\mathcal O}(60)$. If the timeline of the expansion rate is 
faster than radiation $N_{k}$ gets smaller and therefore all the inflationary observables 
are comparatively less suppressed than in the radiation-dominated case. Thanks 
to the current measurements \cite{RR1,RR2,RR3} we are however in the opposite situation 
 and $N_{k}$ must comparatively larger than ${\mathcal O}(60)$. In this case the inflationary observables are more suppressed than in the standard radiation-dominated case. Probably the most economical way of achieving this goal is to consider inflationary scenarios where the post-inflationary expansion rate is slower than radiation. In this case, following the considerations of section \ref{sec4}, a high frequency background of relic gravitons must be expected between the MHz and the THz.

\renewcommand{\theequation}{5.\arabic{equation}}
\setcounter{equation}{0}
\section{The expansion history and the intermediate frequencies}
\label{sec5}
In the  intermediate region of the spectrum (extending, approximately, between few pHz and the Hz) two important scales are related, respectively, to the big bang nucleosynthesis epoch (i.e. $\nu_{bbn}$) and to the electroweak time.  (i.e. $\nu_{ew}$). While $\nu_{bbn}$ is three orders of magnitude smaller than the observational region of the pulsar timing arrays (PTA), $\nu_{ew}$ is comparable with the window where space-borne interferometers might eventually operate a score year from now. During the last four years the PTA  reported a series of evidences of gravitational radiation in the nHz range;  it is then interesting to understand if these claimed signals 
are truly primordial or are just coming from diffuse backgrounds of gravitational radiation formed 
after matter-radiation equality. In any case the PTA set already an indirect constraint on the expansion history of the Universe. Within a similar perspective, the lack of detection between few 
$\mu$Hz and the Hz (i.e. $\nu\geq \nu_{ew}$) sets an essential limit on the post-inflationary expansion rate. 
\subsection{The theoretical frequencies}
\subsubsection{Neutrino free-streaming}
Given the expansion rate at the big bang nucleosynthesis time (when the temperature of the plasma was 
approximately ${\mathcal O}(1)$ MeV), the general expression of $\nu_{bbn}$ is  
\begin{equation}
\nu_{bbn} =  \frac{H_{bbn}}{2\pi} \biggl(\frac{a_{bbn}}{a_{0}}\biggr) = \biggl( \frac{g_{\rho,\,bbn} \, \Omega_{R0}}{90 \pi}\biggr)^{1/4} \, 
\sqrt{\frac{H_{0}}{M_{P}}} \, T_{bbn},
\label{FIVE1}
\end{equation}
where $g_{\rho,\, bbn}$ denotes the effective number of relativistic species at the nucleosynthesis epoch. Since $H_{0}= 1.742\times h_{0}\,\,10^{-61} \, M_{P}$, Eq. (\ref{FIVE1}) becomes  
\begin{eqnarray}
\nu_{bbn} &=& 8.17 \times 10^{-33} g_{\rho,\, bbn}^{1/4} \, T_{bbn} \biggl(\frac{h_{0}^2 \Omega_{R0}}{4.15\times 10^{-5}} \biggr)^{1/4}
\nonumber\\
&=& {\mathcal O}(2) \times 10^{-2} \biggl(\frac{g_{\rho,\, bbn}}{10.75}\biggr)^{1/4} \biggl(\frac{T_{bbn}}{\,\,\mathrm{MeV}}\biggr) 
\biggl(\frac{h_{0}^2 \Omega_{R0}}{4.15 \times 10^{-5}}\biggr)^{1/4}\,\,\mathrm{nHz}.
\label{FIVE3}
\end{eqnarray}
Between the nHz domain and the audio band (with  $\mathrm{Hz} < \nu_{audio} < 10\,  \mathrm{kHz}$)  the spectral energy density of inflationary origin is, at most, ${\mathcal O}(10^{-17})$ in critical units and  the deviations from scale-invariance in the direction of blue spectral indices are excluded at least in the conventional situation where the corrections to $h_{0}^2\,\Omega_{gw}(\nu,\tau_{0})$ always lead to decreasing spectral slope\footnote{This happens, for instance, in the single-field case where, thanks to the consistency relations, the tensor spectral index $n^{(low)}_{T}$ is related to the tensor to scalar ratio $r_{T}$ as $n^{(low)}_{T} \simeq - r_{T}/8$. Since $r_{T}$ is currently assessed from the analysis of the temperature and polarization anisotropies of the CMB
\cite{RR1,RR2,RR3} $n^{(low)}_{T}$ cannot be positive.}. 
In the case of the concordance paradigm the spectral energy density is further reduced by various sources of damping and, most notably, by the free-streaming of neutrinos \cite{STRNU0,STRNU1,STRNU2,STRNU3,STRNU4} exactly for frequencies below the nHz. The same phenomenon also affects the spectral energy density when the corresponding slopes are increasing \cite{ST3,ST3a}; in both situations, however, the suppression due to the neutrinos operates for $\nu < \nu_{bbn}$ and when the 
expansion rate is dominated by radiation.

\subsubsection{Big bang nucleosynthesis bound}
The frequency range associated with $\nu_{bbn}$ is related to a set of direct limits on the expansion rate of the plasma at the big bang nucleosynthesis epoch when the expansion rate was $H_{bbn} = {\mathcal O}(10^{-44})\, M_{P}$. Any excess in the energy density of the massless species at the BBN time increases the value of $H_{bbn}$. The additional massless species may be either bosonic or fermionic; however they are theoretical traditionally parametrized in terms of the effective number of neutrino species as $N_{\nu} = 3 + \Delta N_{\nu}$.  The standard BBN  results are in agreement with the observed abundances for $\Delta N_{\nu} \leq 1$ \cite{Dn1,Dn2,Dn3,Dn4}. The most constraining bound for the intermediate and  high frequency branches of the relic graviton spectrum is represented by big bang nucleosynthesis as argued long ago by Schwartzman \cite{bbn1}. The increase in the expansion rate affects, in particular, the synthesis of $^{4}\mathrm{He}$ and to avoid its overproduction  the expansion and rate the number of relativistic species 
must be bounded from above. All in all, if the additional species are relic gravitons \cite{bbn1,bbn2,bbn3,bbn4,bbn5} the integral of the spectral energy density 
over the whole spectrum must satisfy the following bound: 
\begin{equation}
h_{0}^2  \int_{\nu_{bbn}}^{\nu_{max}}
  \Omega_{gw}(\nu,\tau_{0}) d\ln{\nu} = 5.61 \times 10^{-6} \Delta N_{\nu} 
  \biggl(\frac{h_{0}^2 \Omega_{\gamma\,0}}{2.47 \times 10^{-5}}\biggr),
\label{BBN1}
\end{equation}
where $\nu_{bbn}$ is given by Eq. (\ref{FIVE3}) and $\nu_{max}$ corresponds instead to the maximal frequency of the spectrum. For the relic gravitons produced within the concordance scenario $\nu_{max}$ is given by Eq. (\ref{FIVE3}). As discussed in section \ref{sec3} the maximal frequency is model 
 dependent but it is possible to deduce an absolute bound on $\nu_{max}$ 
 and, according to this bound, $\nu_{max} < \mathrm{THz}$.  Depending on the combined data sets (i.e. various light elements abundances and different combinations of CMB observations), the standard BBN scenario implies that the bounds on $\Delta N_{\nu}$ range from $\Delta N_{\nu} \leq 0.2$ 
to $\Delta N_{\nu} \leq 1$.  All the relativistic species present inside the 
Hubble radius at the BBN contribute to the potential increase in the expansion rate and this  explains why the integral in Eq. (\ref{BBN1}) must be performed from $\nu_{bbn}$ to $\nu_{max}$.  
The constraint of Eq. (\ref{BBN1})  can be relaxed in some 
non-standard nucleosynthesis scenarios, but, in what follows, the 
validity of Eq. (\ref{BBN1}) will be enforced by adopting 
$\Delta N_{\nu} \simeq 1$. The considerations discussed so far can be complemented by other bounds which are, however,  less stringent. In particular the same logic employed for the derivation of Eq. (\ref{BBN1}) can be applied at the decoupling of matter and radiation \cite{DD5,DD6}
when the typical lower extremum of integration  becomes $\nu_{dec} ={\mathcal O}(100)$ aHz:
\begin{equation}
h_{0}^2  \int_{\nu_{dec}}^{\nu_{max}}
  \Omega_{gw}(\nu,\tau_{0}) d\ln{\nu} \leq 8.7 \times 10^{-6}.
 \label{CMBlim} 
  \end{equation}
The BBN limits examined so far can be relaxed in nonstandard BBN scenarios \cite{bbn2} (see also \cite{MGB}). In particular this may happen in the presence of matter-anti-matter domains; instead 
of being ${\mathcal O}(10^{-5})$ the integral of Eq. (\ref{BBN1}) may get ${\mathcal O}(10^{-4})$.

\subsubsection{The electroweak frequency}
The standard model of particle interactions (based on the $SU_{L}(2) \otimes U_{Y}(1)\otimes SU_{c}(3)$ gauge group) appears to be successful at least up to 
energy scales ${\mathcal O}(\mathrm{TeV})$ and its basic correctness ultimately suggests  
that the electroweak phase transition cannot produce a detectable background of gravitational
radiation for typical frequencies smaller than the Hz. To explain 
this viewpoint we start by remarking that the dynamics of the electroweak phase transition has been studied since the early 1970s and while it is plausible that spontaneously broken symmetries are restored at high-temperatures, the order of the electroweak phase transition determines the physical features of the purported 
gravitational signal. The symmetry breaking phase transitions 
may cause departures from local thermal equilibrium (and from 
homogeneity) but, according to the current experimental evidence, 
the electroweak phase transition does not lead to large 
anisotropic stresses that could eventually produce a diffuse 
background of gravitational radiation. A large anisotropic stress 
can only be produced if the electroweak phase transition is of 
first-order and proceeds through the formation of bubbles 
of the new phase. It was clear already from the first 
(perturbative) estimates that the electroweak phase transition 
cannot be strongly first-order \cite{PTFIRST1,PTFIRST2,PTFIRST3};
however a definite conclusion on this issue was delayed because of
the hope that, by using non-perturbative techniques \cite{PTFIRST4}, the essence of the perturbative result could be somehow disproved. The phase diagram 
of the electroweak theory at high-temperature has been first analyzed by reducing the 
theory from $4$ to $3$ dimensions and by subsequently simulating on the lattice the lower dimensional theory with compactified time coordinate \cite{PTFIRST5,PTFIRST6,PTFIRST7}. These analyses have been later
corroborated by genuine $4$-dimensional lattice simulations discussing the 
$SU(2)$-Higgs system \cite{PTFIRST8,PTFIRST9}. The main results relevant for the present 
discussion can be summarized, in short, as follows. For approximate values of the Higgs mass $m_{H}$ smaller than the $W$-boson mass the phase diagram of the electroweak theory contains a line 
of first-order phase transitions but for $m_{H} \geq {\mathcal O}(75)$ GeV the phase transition if of higher order and when $m_{H} \gg m_{W}$ (as it is the case from an experimental viewpoint) the phase transition disappears since we can pass from the symmetric 
to the broken phase in continuous manner. In this cross-over 
regime there large deviations from homogeneity do not arise
and diffuse backgrounds of gravitational radiation are absent. 

Although the electroweak phase transition is of higher order, 
strongly first-order phase transitions may anyway lead to bursts of gravitational radiation 
 and, for this reason, the production of gravitational waves has been investigated in a number of hypothetical first-order phase transitions. Provided the phase transition proceeds thanks to the collision of bubbles of the new phase, the lower frequency scale of the burst is (at most) comparable with the Hubble radius at the corresponding epoch. Denoting by $\nu_{b}$ the 
frequency of the purported burst, we should always require that  $\nu_{b} \geq  {\mathcal O}(\nu_{ew})$ where $\nu_{ew}$ is the typical frequency corresponding to the electroweak horizon. This 
condition follows directly from the observation that gravitational waves should be 
formed inside the Hubble radius when the expansion rate of the Universe was approximately ${\mathcal O}(H_{ew})$. Assuming the electroweak plasma is dominated by radiation between $H_{ew}$ and $H_{bbn}$ the electroweak frequency is given by
\begin{equation}
\nu_{ew} = \frac{H_{ew}}{2 \pi} \biggl(\frac{a_{ew}}{a_{eq}}\biggr) \, \biggl(\frac{a_{eq}}{a_{0}}\biggr)= \frac{H_{ew}}{2 \pi} \sqrt{\frac{H_{eq}}{H_{ew}}} \, \biggl(\frac{a_{eq}}{a_{0}}\biggr),
\label{EWF1}
\end{equation}
where the second equality follows from the approximate 
expansion history and, as usual, $H_{0}$ denotes the current value 
of the Hubble rate. Since during the radiation stage $a^2 \, H$ 
is just constant,  Eq. (\ref{EWF1}) implies
\begin{equation}
\nu_{ew} = \frac{M_{P}}{2\pi} \, \sqrt{\frac{H_{ew}}{M_{P}}} \, \sqrt{\frac{H_{0}}{M_{P}}} \,\,
\sqrt{\frac{a_{eq}^2 \,\, H_{eq}}{a_{0}^2 \,\,H_{0}}}.
\label{EWF2}
\end{equation}
The result of Eq. (\ref{EWF2}) can be further simplified by recalling that 
\begin{equation}
\frac{H_{eq}\, a_{eq}^2}{H_{0}\, a_{0}^2} = \sqrt{ 2 \, \Omega_{R\, 0}}, \qquad\qquad \biggl(\frac{H_{ew}}{M_{P}}\biggr)= \sqrt{\frac{4 \pi^3}{45}} 
\sqrt{g_{\rho}} \biggl(\frac{T_{ew}}{M_{P}}\biggr)^2,
\label{EWF3}
\end{equation}
If Eq. (\ref{EWF3}) is now inserted into Eq. (\ref{EWF2}) we get the following estimate: 
\begin{equation}
\nu_{ew} = 7.98  \biggl(\frac{g_{\rho,\, ew}}{106.75}\biggr)^{1/4} \biggl(\frac{T_{ew}}{200\, \mathrm{GeV}} \biggr) \, \mu\mathrm{Hz},
\label{FF5a}
\end{equation}
where the value of $T_{ew}$ 
has been chosen to be slightly above the value of the top quark mass just to make sure that all the species of the Standard Model are in local thermal equilibrium.  Strictly 
speaking the adiabatic evolution only implies the constancy 
of  $a^3 \, T^3 \, g_{s}(T)$ so that result of Eq. (\ref{FF5a}) 
should be slightly corrected:
\begin{equation}
\frac{H_{ew}^2 \, a_{ew}^4}{H_{eq}^2 \, a_{eq}^4} ={\mathcal C}(g_{s}, g_{\rho},\tau_{ew},\tau_{eq}),
\label{EWF4}
\end{equation}
where ${\mathcal C}(g_{s}, g_{\rho},\tau_{r},\tau_{eq})$ has been already introduced in Eq. (\ref{TWO13}).
Equation (\ref{EWF4}) also implies that 
\begin{eqnarray}
\biggl(\frac{a_{ew}}{a_{eq}}\biggr) = \sqrt{\frac{H_{eq}}{H_{ew}}} \,\, {\mathcal C}(g_{s}, g_{\rho},\tau_{ew},\tau_{eq}) = 0.76 \sqrt{\frac{H_{eq}}{H_{ew}}},
\label{EWF5}
\end{eqnarray}
where the explicit estimate follows by recalling that, for $T_{ew} > m_{t}$,  $g_{s,\,ew} = g_{\rho,\,ew} = 106.75$. Moreover, since for $T= T_{eq}$  the values of $g_{s,\,eq}$ and $g_{\rho,\, eq}$  are slightly different (i.e. $3.94$ and $3.36$ respectively)
the typical value of $\nu_{ew}$ given in Eq. (\ref{FF5a}) passes 
from $7.98\,\,\mu\mathrm{Hz}$ to $6.06 \,\,\mu\mathrm{Hz}$.

\subsection{Pulsar timing arrays and the expansion history}
In the last few years a set of direct observations potentially related with the diffuse 
backgrounds of gravitational radiation have been reported for a typical benchmark frequency ${\mathcal O}(30)$ nHz. This range of frequencies is between $3$ and $4$ orders of magnitude larger than $\nu_{bbn}$ and it is currently probed by the pulsar timing arrays (PTA in what follows). As recently pointed out \cite{MGpuls} the signals possibly observed by the PTA may be the result of the pristine variation of the space-time curvature. The specific features of the current observations seem to suggest, however, that $h_{0}^2 \Omega_{gw}(\nu, \tau_{0})$ in the nHz domain may only depend on the evolution of the comoving horizon at late, intermediate and early times. This is also, in a nutshell, the systematic perspective swiftly outlined hereunder. 

\subsubsection{Basic terminology and current evidences}
A pulsar timing array  is just a series of millisecond pulsars that are monitored with a specific cadence  that ultimately depends on the choices of the given experiment.  We refer here, in particular, {\it (i)} to the NANOgrav collaboration \cite{NANO1,NANO2}, {\it (ii)} to the Parkes Pulsar Timing array (PPTA) \cite{PPTA1,PPTA2} and {\it (iii)} o the European Pulsar Timing array (EPTA) \cite{EPTA1,EPTA2}. The  PTA data have been also combined in the consortium named International Pulsar Timing array (IPTA) \cite{IPTA1}.  The data of the PTA collaborations have been released \cite{NANO2,PPTA2,EPTA2} together with the first determinations of the Chinese Pulsar Timing array (CPTA) \cite{CPTA}.  As suggested long ago the millisecond pulsars can be employed as effective detectors of random gravitational waves for a typical domain that corresponds to the inverse of the observation time during which the pulsar timing has been monitored  \cite{PP1a,PP1b,PP1c}. The signal coming from diffuse backgrounds of gravitational radiation, unlike other noises, should be correlated across the baselines.  The effect depends on the angle between a pair of Earth-pulsars baselines and it is often dubbed by saying that the correlation signature of an isotropic and random gravitational wave background follows the so-called Hellings-Downs curve \cite{PP1c}. If the gravitational waves are not characterized by stochastically distributed Fourier amplitudes the corresponding signal does not necessarily follow the Hellings-Downs correlation. In the past various upper limits on the spectral energy density of the relic gravitons in the nHz range have been obtained \cite{PP2a,PP2b,PP2c,PP2d} and during the last five years the PTA reported an evidence that could be attributed to isotropic backgrounds of gravitational radiation. The observational collaborations customarily assign the chirp amplitude at a reference frequency $\nu_{ref} = 31.68\,\, \mathrm{nHz}$ that corresponds to $\mathrm{yr}^{-1}$:
\begin{equation}
h_{c}(\nu,\tau_{0}) = \,Q \,\bigl(\nu/\nu_{ref}\bigr)^{\beta}, \qquad \qquad \nu_{ref} = 1/\mathrm{yr}=   31.68\,\, \mathrm{nHz}.
\label{PTAR1}
\end{equation}
To avoid confusions we stress that the $\beta$ appearing in Eq. (\ref{PTAR1}) 
has nothing to do with the quantity characterizing the inflaton potential (see Eq. (\ref{POTEX1}) 
and discussion thereafter). Recalling now the relation between the spectral energy density and the chirp amplitude, we have: 
\begin{equation}
\Omega_{gw}(\nu,\tau_{0}) = \frac{2 \pi^2 \nu^2}{3 \, H_{0}^2 } \,h_{c}^2(\nu, \tau_{0}) =  \frac{2 \pi^2}{3} \, Q^2 \, \biggl(\frac{\nu_{ref}}{H_{0}}\biggr)^2 \,\, \biggl(\frac{\nu}{\nu_{ref}}\biggr)^{2 + 2 \beta},
\label{PTAR2}
\end{equation}
where the second equality follows from Eq. (\ref{PTAR1}). If we now multiply Eq. (\ref{PTAR2}) by $h_{0}^2$ (where $h_{0}$ denotes the indetermination in the present value of the Hubble rate) and take into account the explicit value of $\nu_{ref}$, the expression for the spectral energy density becomes \cite{LIGO3}:
\begin{equation}
h_{0}^2 \, \Omega_{gw}(\nu,\tau_{0}) = 6.287 \times 10^{-10} \, \, q_{0}^2\,\,\bigl(\nu/\nu_{ref}\bigr)^{2 + 2 \beta}.
\label{PTAR3}
\end{equation}
In Eq. (\ref{PTAR3}) $Q$ is parametrized as $Q= q_{0} \times 10^{-15}$ 
(where $q_{0}$ is a number of order $1$) since this is basically 
the observational evidence. Clearly, for $\nu \to \nu_{ref}$ 
\begin{equation}
h_{0}^2 \, \Omega_{gw}(\nu_{ref},\tau_{0}) =  6.287 \times 10^{-10} \, \, q_{0}^2,
\label{PTAR5}
\end{equation}
implying $h_{0}^2 \, \Omega_{gw}(\nu_{ref},\tau_{0}) = {\mathcal O}(2.57) \times 10^{-8}$ in the case of Ref. \cite{NANO2} (for $q_{0} =6.4$) and $h_{0}^2 \, \Omega_{gw}(\nu_{ref},\tau_{0}) = {\mathcal O}(6.04)\times 10^{-9}$ for Ref. \cite{PPTA2} (for $q_{0} =3.1$).  With the same logic we can also deduce 
the explicit relation between the spectral and the chirp amplitudes:
\begin{equation}
S_{h}(\nu, \tau_{0}) = 3.15 \, \times 10^{-23} \, \,  q_{0}^2\,\,\bigl(\nu/\nu_{ref}\bigr)^{2 \beta-1} \, \, \mathrm{Hz}^{-1}.
\label{PTAR4}
\end{equation}
It is also customary to employ  $\sqrt{S_{h}(\nu, \tau_{0})} =5.61\times10^{-12} \, \, q_{0} (\nu/\nu_{ref})^{\beta -1/2} \,\, \mathrm{Hz}^{-1/2}$ for a direct comparison it with the spectral 
amplitude of the signal. Recalling however the considerations of section \ref{sec3} the 
use of a spectral amplitude implicitly assumes that the signal 
can be mimicked by a stationary and homogeneous stochastic process; this 
is not the case for the relic gravitons \cite{MGSTOC}. Bearing in mind the results and the notations of Eqs. (\ref{PTAR2})--(\ref{PTAR4}) the main statements of the observers can be 
summarized, in short, as follows.
\begin{itemize}
\item{} The pivotal class of models analyzed in Refs. \cite{NANO1,NANO2,PPTA1,PPTA2,EPTA1,EPTA2,IPTA1,CPTA} always assume $\beta = -2/3$ (i.e. $\overline{\gamma} = 13/3$); recall, in this respect, that the relation between $\overline{\gamma}$ and $\beta$ is simply given by $\beta = (3-\overline{\gamma})/2$.
\item{} In the former data releases the  $q_{0}$ ranged between $1.92$ and $5.13$ depending on the values of $\beta$ \cite{NANO1,PPTA1,EPTA1,IPTA1}.
\item{} The latest data releases of the Parkes and the NANOgrav collaborations \cite{PPTA2,NANO2,EPTA2} seem to suggest different origins of the diffuse background of gravitational radiation.
\item{} In particular, after considering $30$ millisecond pulsars spanning $18$ years of observations, the Parkes PTA collaboration estimates 
$q_{0}= 3.1^{1.3}_{-0.9}$ with a spectral index $\beta = -0.45\pm 0.20$ \cite{PPTA2}; for a spectral the pivotal model $\beta =-2/3$ the collaboration suggests instead $q_{0} =2.04_{-0.22}^{0.25}$ which is compatible with 
the determinations of the previous data releases \cite{PPTA1}; the Parkes PTA collaboration does not clearly claim the detection of the Hellings-Downs correlation \cite{PPTA2} and carefully considers possible issues related to time-dependence of the common noise. 
\item{} The conclusions of the Parkes PTA  seem significantly more conservative than the one of the NANOgrav collaboration examining $67$ millisecond pulsars in the last $15$ years. 
\item{} The NANOgrav experiment  claims the detection of the Hellings-Downs correlation \cite{NANO2} but the inferred values of the spectral 
parameters are slightly different from the ones of PPTA since $q_{0} = 6.4^{+4.2}_{-2.7}$
and $\beta = -0.10 \pm 0.30$ \cite{NANO2}. 
\end{itemize}

\subsubsection{The comoving horizon after inflation}
The measurements of the PTA set a number of relevant constraints on the spectrum of the relic gravitons and on the expansion rate of the Universe. If the observed excess in the nHz range  is just a consequence of the primeval variation of the space-time curvature the spectral energy density of the relic gravitons in the nHz domain only depends on the evolution of the comoving horizon at {\em late}, {\em intermediate} and {\em early} times \cite{MGpuls}. Two complementary aspects of the problem will now be addressed. In the 
first part of the discussion we are going to see if a post-inflationary modification of the expansion rate can account for the nHz excess. In the second part of the analysis we consider instead the possibility of explaining the observed PTA excess through the evolution of the effective horizon at early times. 

A first general observation is that, in the concordance paradigm, the  PTA results do not set any further constraint besides the ones of the aHz region already discussed in section \ref{sec4}. This happens because the spectral energy density of Eqs. (\ref{PTAR4})--(\ref{PTAR5}) always exceeds the 
the one of the concordance paradigm in the nHz region. Indeed, if the expansion rate is dominated by radiation after inflation,   $h_{0}^2 \, \Omega_{gw}(\nu,\tau_{0}) < {\mathcal O}(10^{-17})$ for typical frequencies larger than $\nu_{bbn}$. Furthermore, in the concordance paradigm, $h_{0}^2 \, \Omega_{gw}(\nu,\tau_{0})$ is a monotonically decreasing function of the comoving frequency between the aHz and the MHz domain. This means that in the nHz range the signal of the relic gravitons produced within the conventional lore is always ten orders of magnitude smaller than the one  suggested by Eqs. (\ref{PTAR4})--(\ref{PTAR5}). If the expansion history is modified in comparison with the concordance paradigm the relevant time-scale of the problem must coincide with $\tau_{k}$, i.e. the moment at which the wavelength associated with $\nu_{PTA} \simeq \nu_{ref}= {\mathcal O}(30) \, \mathrm{nHz}$ crossed the comoving Hubble radius after the end of inflation (see Fig. \ref{FIGU0d}). The actual value of  $\tau_{k}$ represents in fact a fraction of the time-scale associated with big bang nucleosynthesis:
\begin{eqnarray}
\frac{\tau_{k}}{\tau_{bbn}} &=& (4 \pi\, \Omega_{R\,0})^{1/4} \biggl(\frac{g_{\rho,\,bbn}}{g_{\rho,\, eq}}\biggr)^{1/4}  \,\,
\biggl(\frac{g_{\rho,\,eq}}{g_{\rho,\, bbn}}\biggr)^{1/3}  \, \sqrt{\frac{H_{bbn}}{H_{0}}} \biggl(\frac{a_{0} \, H_{0}}{\nu_{ref}}\biggr) 
\nonumber\\
&=& {\mathcal O}(3) \times 10^{-2} \biggl(\frac{\nu_{ref}}{31.68 \, \mathrm{nHz}}\biggr)^{-1}\, \biggl(\frac{T_{bbn}}{\mathrm{MeV}}\biggr) \, \biggl(\frac{h_{0}^2 \Omega_{R0}}{4.15 \times 10^{-5}}\biggr)^{1/4}.
\label{PTAR6}
\end{eqnarray}
In Eq. (\ref{PTAR6}) we are actually assuming, for simplicity, that $\nu_{PTA} = {\mathcal O}(\nu_{ref})$ and if  $\nu_{PTA} > \nu_{ref}$  the corresponding wavelength crossed the comoving horizon even earlier. Besides 
Eq. (\ref{PTAR6}), the second relevant scale of the problem follows from the ratio 
between $\nu_{PTA}$ and the expansion rate at the end of inflation:
\begin{equation}
\frac{\nu_{PTA}}{a_{1} \, H_{1}} = {\mathcal O}(2) \times 10^{-17} \biggl(\frac{\nu_{PTA}}{31.68 \, \mathrm{nHz}}\biggr)\,\biggl(\frac{h_{0}^2 \Omega_{R0}}{4.15 \times 10^{-5}}\biggr)^{-1/4} \, \biggl(\frac{r_{T}}{0.03}\biggr)^{-1/4} \, \biggl(\frac{{\mathcal A}_{{\mathcal R}}}{2.41\times 10^{-9}}\biggr)^{-1/4},
\label{PTAR7}
\end{equation}
where ${\mathcal A}_{{\mathcal R}}$ denotes, as usual, the amplitude of the curvature inhomogeneities at the pivot scale $k_{p}$ ( see Eq. (\ref{TWO7A}) and discussion thereafter).  Already from Eqs. (\ref{PTAR6})--(\ref{PTAR7}) it follows that any modification of the post-inflationary evolution is unlikely to produce a hump for frequencies ${\mathcal O}(\nu_{PTA})$: the value of $\nu_{PTA}$ in units of the expansion rate is too small.
 It is on the contrary more likely that a hump will be produced over larger frequencies $\nu > \mu\mathrm{Hz}$, as we are going to see later on in this section. 

To substantiate the previous statement we now consider a generic post-inflationary expanding stage (i.e. 
a single $\delta$-phase in the language of section \ref{sec2}). When the wavelengths $\lambda= {\mathcal O}(\lambda_{PTA})$ cross the comoving Hubble radius during the $\delta$-phase we have 
\begin{equation}
\frac{\nu_{PTA}}{a_{1} \, H_{1}} = 2.05 \times 10^{-17} \, \, (H_{r}/H_{1})^{(1 - \delta)/[ 2 (\delta +1)]}, 
\label{PTAR7a}
\end{equation}
where, once more, $H_{r}$ and $H_{1}$ denote, respectively, the Hubble rates at the onset of the radiation stage and at the end of inflation. As $H_{r} < H_{1}$ the comoving horizon at its minimum is comparatively larger for $\delta > 1 $ than for $\delta \to 1$; for the same reason the opposite is true when $\delta < 1$.  
Since the post-inflationary evolution is modified the spectral energy density of the relic gravitons 
gets larger, as it follows from Eq. (\ref{SSTTnew1}). When the PTA wavelength crosses the Hubble radius during the $\delta$-phase $h_{0}^2 \Omega_{gw}(\nu, \tau_{0})$
exhibits a twofold slope:
\begin{itemize}
\item{} in the low-frequency regime the slope is simply 
given by $n_{T}^{(low)} = - r_{T}/8$; this is true when 
the consistency relations are enforced as we are 
assuming throughout;
\item{} if the wavelength corresponding to $\nu_{PTA}$ 
 reenters the Hubble radius when $\delta \neq 1$
 the high frequency slope follows from Eqs. (\ref{SSTTnew1})--(\ref{SSTTnew3}) and it is $n^{(high)}_{T} = 2 ( 1 - \delta) + {\mathcal O}(r_{T})$.  
 \end{itemize}
 To compare $n_{T}^{(high)}$ with the potential excesses 
suggested by the PTA we may recall Eq. (\ref{PTAR3}) and then 
consider the theoretical estimate of the spectral energy density in critical units \cite{MGshift}
\begin{equation}
h_{0}^2 \, \Omega_{gw}(\nu, \tau_{0}) = {\mathcal N}(r_{T}, \nu)\, \biggl(\frac{\nu}{\nu_{r}}\biggr)^{n^{(high)}_{T}}, \qquad \nu> \nu_{r},
\label{STWO8}
\end{equation}
where ${\mathcal N}(r_{T}, \nu)$ includes the effects of the low-frequency suppressions associated with the transfer function, with the neutrino free-streaming \cite{STRNU1,STRNU2,STRNU3} and with the other late-time sources of damping (like the one related with the dark-energy dominance \cite{LIGO3}). In connection 
with ${\mathcal N}(r_{T},\nu)$  see also Eqs. (\ref{SING9a})--(\ref{SING9b}) and the discussion therein. For $r_{T} =0.03$ we can numerically estimate that ${\mathcal N}(r_{T},\nu)= 10^{-16.8}$ and since above $\nu_{bbn}$ the value of ${\mathcal N}(r_{T},\nu)$ has a mild frequency dependence controlled by the value of the low-frequency slope (i.e. $n_{T}^{(low)} \simeq - r_{T}/8$) and by the low-frequency transfer function, for the present ends we can assume ${\mathcal N}(r_{T}, \nu) \simeq {\mathcal N}  = {\mathcal O}(10^{-17})$ which is the value already quoted before.   According to Eq. (\ref{STWO8}) the theoretical amplitude at $\nu_{PTA} \simeq \nu_{ref}$ ultimately depends upon $\nu_{r}= \overline{\nu}_{max} \sqrt{\xi}$ where $\overline{\nu}_{max} = {\mathcal O}(300)\, \mathrm{MHz}$ has been already computed in Eq. (\ref{NK6}). In spite of the specific values of $\overline{\nu}_{max}$ we have that  $\nu_{r}$ cannot be smaller than $\nu_{bbn}$.  In the general case\footnote{Unless the relic gravitons would lead {\em exactly} to the same slope of the astrophysical foregrounds associated with black-hole binary systems, the value $\beta =-2/3$ is not particularly compelling in a cosmological setting.} (i.e. when the special value $\beta = -2/3$ is not preliminarily selected) the Parkes PTA collaboration \cite{PPTA2} suggests that $\beta = - 0.45 \pm 0.20$. This determination is marginally compatible with the value of Eq. (\ref{STWO7}) in the limit $\delta \geq 1/2$ and the discrepancy between the observational determination of $\beta$ and the values predicted by Eq. (\ref{STWO7}) becomes even more significant if we look at that NANOgrav data suggesting \cite{NANO2} $\beta = -0.10 \pm 0.30$. All in all, when $\delta \geq 1/2$, the relation between $\delta$, $\beta$ and $r_{T}$ is:
 \begin{equation}
\delta = - \beta - \frac{r_{T}}{r_{T} +1} >\frac{1}{2}, \qquad \beta < 0,
\label{STWO7}
\end{equation}
so that, from Eq. (\ref{STWO7}),  $\delta = -\beta + {\mathcal O}(r_{T})$.
Both in the previous \cite{NANO1,PPTA1} and in the most recent \cite{NANO2,PPTA2} data releases the value of $2 (1 + \beta)$ is always positive definite (i.e. $1 +  \beta >0$). In the special case $\beta \to -2/3$, Eq. (\ref{STWO7}) 
implies $ \delta = 2/3 + {\mathcal O}(r_{T})$. If we would now assume that the post-inflationary evolution is driven by a relativistic and irrotational fluid we would have $\delta= 2/(3 w+1)$ implying that $\beta \to- 2/3$ for $w \to 2/3$. Another possibility would be that the effective expansion rate is dictated by an oscillating scalar field (like the inflaton) with potential $V(\varphi) = V_{0} (\varphi/\overline{M}_{P})^{2 q}$; in this case the expansion rate during the oscillating phase would be given by $\delta = (q+1)/(2q -1)$  suggesting that $q = {\mathcal O}(5)$ for $\beta = {\mathcal O}(-2/3)$.  Although, for specific values of $\delta$, the theoretical and the observed slopes can be compatible the corresponding amplitudes involve orders of magnitude that are grossly different; to analyze this aspect we then impose that Eqs. (\ref{PTAR3}) and (\ref{STWO8}) should coincide at $\nu_{ref}$
\begin{equation}
{\mathcal N}(r_{T}, \nu_{ref}) \bigl(\nu_{ref}/\nu_{r}\bigr)^{n_{T}^{(high)}}= 6.287\,\times 10^{-10}\, q_{0}^2.
\label{STWO11}
\end{equation}
If the two sides of Eq. (\ref{STWO11}) would be mutually consistent the post-inflationary modification of the comoving horizon might indeed explain the observed PTA excess. But unfortunately the left-hand side of Eq. (\ref{STWO11}) is systematically smaller than the right-hand side;  the two 
contributions are of the same order only when $\nu_{r}\ll \nu_{ref}$ while, at the same time, $n^{(high)}_{T}= 2 + 2 \beta$ is sufficiently {\em large and positive}. A large (and positive) value of $n^{(high)}$ guarantees a sharp increase of the spectral energy density while the condition $\nu_{r}\ll \nu_{ref}$ widens the frequency range for a potential growth of $h_{0}^2 \, \Omega_{gw}(\nu, \tau_{0})$. Since the minimal value of $\nu_{r}$ is provided by $\nu_{bbn}$ we can select the most favourable situation and posit $\nu_{r} = {\mathcal O}(\nu_{bbn})$. 
For different values of ${\mathcal N}(r_{T},\nu_{ref})$ (see Eq. (\ref{STWO8}) and discussion thereafter) Eq. (\ref{STWO11}) leads therefore to a specific relation between $\beta$ and $\log{q_{0}}$: 
\begin{equation}
\beta = -1 + \frac{ 2 \log{q_{0}} - \log{{\mathcal N}(r_{T},\nu_{ref})} - 9.201}{2 \log{(\nu_{ref}/\nu_{bbn})}}.
\label{STWO13}
\end{equation}
\begin{figure}[!ht]
\centering
\includegraphics[height=6cm]{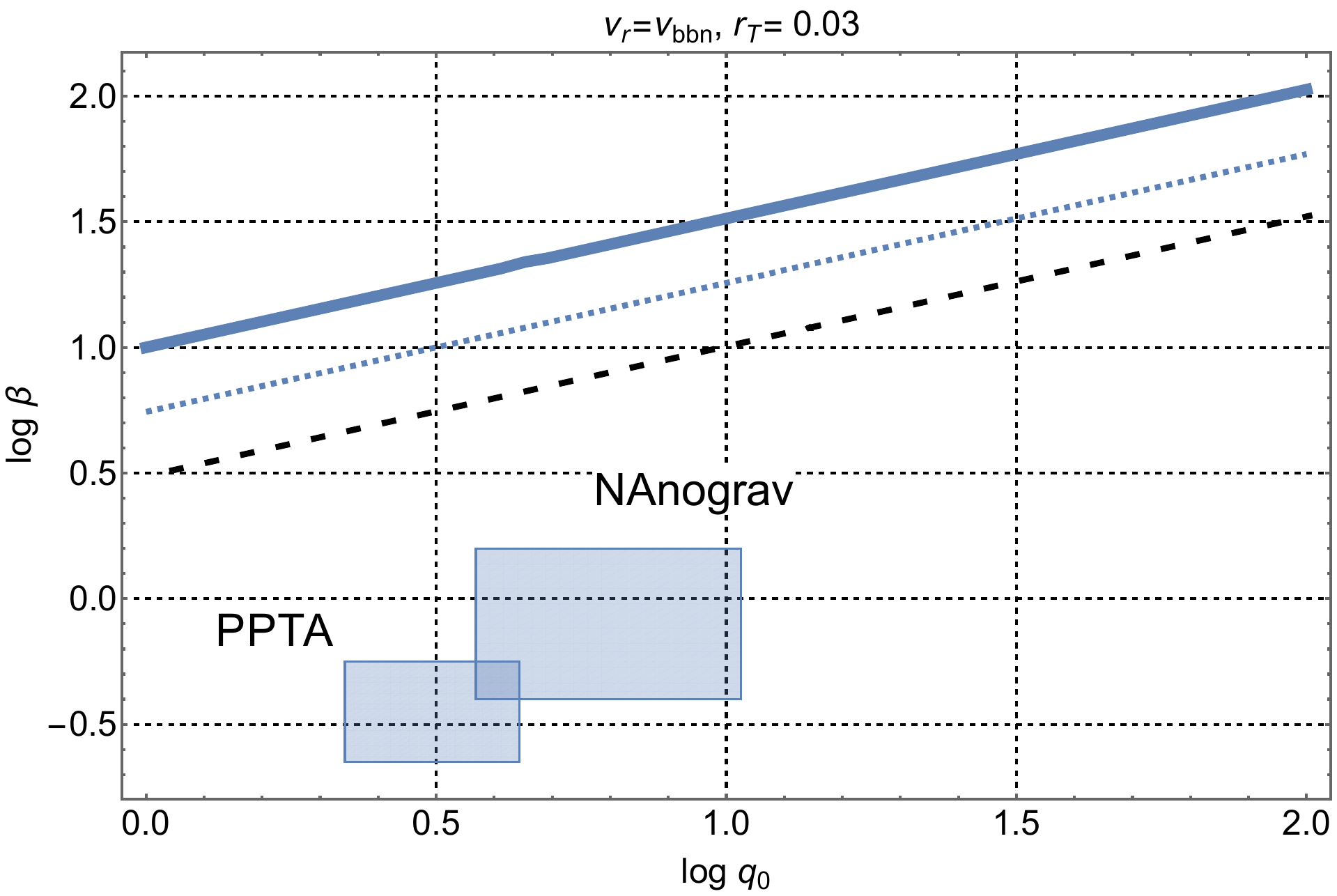}
\caption[a]{The three straight lines illustrate Eq. (\ref{STWO13}) for ${\mathcal N}(r_{T},\nu_{ref}) =10^{-17}$ 
(full line), for ${\mathcal N}(r_{T},\nu_{ref}) =10^{-16}$ (dotted line) and for ${\mathcal N}(r_{T},\nu_{ref}) =10^{-15}$ (dashed line). The two filled rectangles define the regions probed by the Parkes PTA and by NANOgrav in the plane $(\log{q_{0}}, \, \beta)$. The two diagonal lines do not overlap with the shaded areas appearing in the lower 
portion of the plot and this means that the amplitudes and the slopes of the theoretical signal cannot be simultaneously matched with the corresponding observational determinations. Common logarithms are employed on the horizontal axis.}
\label{FIGPTA}      
\end{figure}
The result of Eq. (\ref{STWO13}) must then be compared in the plane $(\log{q_{0}}, \, \beta)$ with the ranges of $\beta$ and $q_{0}$ determined by the PTA collaborations. The two filled rectangles in Fig. \ref{FIGPTA} correspond to the 
observational ranges of $q_{0}$ and $\beta$; in the same plot the relation between $\beta$ and $\log{q_{0}}$ has been illustrated as it follows from Eq. (\ref{STWO13}) for two neighbouring values of ${\mathcal N}(r_{T})$. The three diagonal lines of  Fig. \ref{FIGPTA}  imply that the values of $\beta$ required to obtain $h_{0}^2 \Omega_{gw}(\nu_{ref}, \tau_{0})$ of the order of $10^{-8}$ or $10^{-9}$ should be much larger than the ones determined observationally and represented by the two shaded regions. Since the full and dashed lines of Fig. \ref{FIGPTA} do not overlap with the two rectangles in the lower part of the cartoon, we can conclude that the excess observed by the PTA collaborations cannot be explained by the modified post-inflationary evolution suggested of Fig. \ref{FIGPTA}. 
For the specific case $\beta= -2/3$ Eq. (\ref{STWO11}) becomes
\begin{equation}
{\mathcal N}(r_{T},\nu_{ref}) \bigl(\nu_{ref}/\nu_{r}\bigr)^{2/3}= 6.287\,\times 10^{-10}\, q_{0}^2.
\label{STWO11b}
\end{equation}
Again to maximize the potential growth of the spectral energy density we set $\nu_{r} = {\mathcal O}(\nu_{bbn})$ and obtain that the left-hand side 
of Eq. (\ref{STWO11b}) is $2.015\times 10^{-15}$ whereas the right-hand side 
is always larger than ${\mathcal O}(10^{-9})$. As in the previous case, larger values of $\nu_{r}$
only reduce the left-hand side of Eq. (\ref{STWO11b}) and ultimately increase the mismatch between Eqs. (\ref{PTAR3}) and (\ref{STWO8}).  The argument based on a single $\delta$-phase can be 
generalized to include different stages expanding either 
faster or slower than radiation. In this case there is the possibility 
of developing a hump in the spectrum which is however always 
much smaller than the excess observed by the PTA \cite{MGpuls}.

\subsubsection{The comoving horizon during inflation}

The previous analysis demonstrated that the PTA excess cannot be 
explained by a post-inflationary modification of the expansion rate. 
However, if the effective comoving horizon is modified during inflation 
(as suggested in  \ref{FIGU3}) it is possible 
to explain the PTA excess in terms of a relic signal \cite{MGpuls}.
To implement an effective modification of the comoving horizon 
without obliterating the inflationary expansion we actually consider an effect 
suggested almost $10$ years ago:  
a dynamical refractive index associated with the propagation of the tensor modes 
of the geometry in curved backgrounds naturally leads to an increasing spectral energy density 
at intermediate frequencies \cite{CC2}.
The tensor modes of the geometry may indeed acquire an effective index of refraction when they travel in curved space-times \cite{CC1,CC1a} and the blue spectral slopes (compatible with the PTA excesses) arise from the variation of the refractive index even if the background geometry evolves according to a conventional stage of expansion possibly supplemented by a standard decelerated epoch \cite{CC2} (see also \cite{CC3,CC4,CC5}). When the refractive index  of the relic gravitons is dynamical ($n(a)$ in what follows)
the conditions associated with the crossing of a given wavelength are different;  the action of the tensor modes of the geometry in the case of a dynamical refractive index \cite{CC4,CC5} is given by:
\begin{equation}
S = \frac{\overline{M}_{P}^2}{8} \int \, d^{3} x\, \int d\tau \,\, a^2(\tau) \,\biggl[ \partial_{\tau} h_{i\, j} \partial_{\tau} h^{i\, j} - \frac{1}{n^2(\tau)} 
\partial_{k} h_{i\, j} \partial^{k} h^{i\, j} \biggr],
\label{STHR1}
\end{equation}
see also Eq. (\ref{APPG1}) and discussion therein. The analysis of Eq. (\ref{STHR1}) simplifies if the conformal time coordinate is redefined from $\tau$ to $\eta$ where the relation between the new and the old time parametrizations follows from  $n(\eta) \, d\eta = d \tau$. Equation (\ref{STHR1}) becomes then canonical in terms of a redefined scale factor conventionally denoted hereunder by $b(\eta)$: 
\begin{equation}
S = \frac{\overline{M}_{P}^2}{8} \int \, d^{3} x\, \int d\eta \, b^2(\eta)\, \biggl[ \partial_{\eta} h_{i\, j} \partial_{\eta} h^{i\, j} - 
\partial_{k} h_{i\, j} \partial^{k} h^{i\, j} \biggr], \qquad b(\eta) = a(\eta)/\sqrt{n(\eta)}.
\label{STHR2}
\end{equation}
The result of Eq. (\ref{STHR2})  explains how and why the evolution of the tensor modes is modified even during a conventional stage of inflationary expansion. The evolution of the
 tensor amplitude can be directly deduced from Eq. (\ref{STHR2}) and it is
\begin{equation}
\ddot{\mu}_{i\, j} - \nabla^2 \mu_{i\, j} - \frac{\ddot{b}}{b} \mu_{i\, j} =0, \qquad \mu_{i\, j}(\vec{x},\eta) = b(\eta) h_{i\, j}(\vec{x}, \eta),
\label{STHR2a}
\end{equation}
where the overdot now denotes a derivation with respect to the $\eta$-time. Equation 
(\ref{STHR2a}) also implies that the standard crossing condition $k^2 = a^{\prime\prime}/a$ is now 
replaced by $k^2 = \ddot{b}/b$. Between these twi conditions the former seems superficially equivalent to the latter but this is not the case \cite{CC2}. 
The evolution dictated by Eq. (\ref{STHR2a}) ultimately
leads to a spectral energy density that increases over intermediate 
frequencies provided the effective phase velocity\footnote{After Ref. \cite{CC2} appeared in the form of a preprint, some authors made exactly the same speculation and talked about the sound speed (or sound velocity) of the relic gravitons. While this terminology makes little sense in the context of the propagation of massless particles, the idea is exactly the same (see \cite{CC3,CC4} and references therein). In the present context we prefer to discuss this class of phenomena in terms of an effective refractive index, as originally suggested in Ref. \cite{CC2,CC1,CC1a}.} of the relic gravitons remains sub-luminal. Although the phase velocity of the relic gravitons is not required to be sub-luminal we impose, for consistency, that $n(a) \geq 1$; in particular we consider an appreciable change of the refractive index during inflation with the concurrent requirement that $n(a)$ reaches $1$ in the standard decelerated stage of expansion:
\begin{equation}
n(a) = n_{\ast} \frac{ (a/a_{\ast})^{\alpha} \,\,e^{- \overline{s}\,(a/a_{1})}}{(a/a_{*})^{\alpha} + 1} + 1, \qquad\qquad
n_{\ast} = n_{i} (a_{\ast}/a_{i})^{\alpha} = n_{i} e^{\alpha \, N_{\ast}},
\label{STHR3}
\end{equation}
where $a_{i}$ and $a_{1}$ denote, respectively, the beginning and the end of the 
inflationary epoch; $a_{*}$ indicates the boundary of the refractive stage. Equation (\ref{STHR3}) also implies the refractive index is not dynamical in the post-inflationary stage. Some other possibilities have been considered in Refs. \cite{CC2,CC3,CC4} but they will not be examined here. In what follows the parametrization of Eq. (\ref{STHR3}) is regarded as the minimal example that successfully produces a nHz excess. Equation (\ref{STHR3}) can be analyzed in three relevant  physical limits:
{\it (i)} for $a\gg a_{1}$ we have that $n(a) \to 1$  and the sharpness of the transition depends on the parameter $\overline{s} \geq 1$; {\it (ii)} in the range $a_{*} < a < a_{1}$ $n(a)$ is constant but still larger than $1$ (i.e. $n(a)\simeq n_{\ast} > 1$) and, finally, when{\it (iii)}  $a< a_{\ast}$ the refractive index is truly dynamical since $n (a) \simeq n_{\ast} (a/a_{\ast})^{\alpha}$. 
When $a < a_{\ast}$ we can directly compute $b(\eta)$ in case  the dynamics of the geometry 
is given by a conventional inflationary stage where $ a \, H = - 1/[(1 -\epsilon)\tau]$.
By direct integration from $n(\eta) d\eta = d\tau$ we obtain the relation between $\eta$ and $\tau$ and then compute $b(\eta) = a(\eta)/\sqrt{n(\eta)}$; the result is:
\begin{equation}
b(\eta) = b_{\ast} (- \eta/\eta_{\ast})^{-\zeta}, \qquad b_{\ast} = a_{\ast}/\sqrt{n_{\ast}}, \qquad 
\zeta = \frac{ 2 - \alpha}{2 ( 1 - \epsilon + \alpha)}.
 \label{STHR6}
\end{equation}
In the $\eta$-time coordinate the evolution of the tensor modes can be 
quantized in the standard manner as 
\begin{equation}
\widehat{h}_{i\, j}(\vec{x}, \eta) = \frac{ \sqrt{2} \,\ell_{P}}{ (2\pi)^{3/2} \, b^2(\eta)} \sum_{\lambda} \int d^{3} k\, e_{i\,j}^{(\lambda)}(\hat{k}) \biggl[ \widehat{a}_{\vec{k},\, \lambda} \, f_{k,\, \lambda}(\eta) e^{- i \vec{k}\cdot\vec{x}} + \mathrm{H.\, c.} \biggr],
\label{STHR6a}
\end{equation}
where, as usual, the sum over $\lambda$ runs over the two tensor polarizations. From Eq. (\ref{STHR2a}) the mode functions 
obey $\ddot{f}_{k,\,\lambda} + (k^2 - \ddot{b}/b)f_{k\,\lambda}=0$ and recalling the expression of Eq. (\ref{STHR6}),  for each tensor polarization the solution of the mode function reads 
\begin{equation} 
f_{k}(\eta) = \frac{{\mathcal N}}{\sqrt{2 k}} \sqrt{ - k \, \eta} \, H_{\nu}^{(1)}(- k \eta), \qquad \nu = \zeta +1/2,
\label{STHR6b}
\end{equation}
where $|{\mathcal N}| = \sqrt{\pi/2}$ and $H_{\nu}^{(1)}(-k\eta)$ is the Hankel function of first kind \cite{abr1,abr2}. Once more, the solution of Eq. (\ref{STHR6b}) is relatively simple in terms of the $\eta$-time but gets 
more cumbersome in the conformal time coordinate. While in the case of the concordance paradigm 
the low-frequency spectral index is red (i.e. $n_{T}^{(low)}<0$) when the refractive 
index is dynamical $n_{T}^{(low)}> 0$. Indeed if we compute the tensor power spectrum 
in the long wavelength limit (i.e. $| k \eta |< 1$) from Eq. (\ref{STHR6b}) we obtain 
\begin{equation}
P_{T}(k,\eta) = \frac{4 \,\ell_{P}^2\, k^3}{ \pi^2 \, b^2(\eta)} \bigl| f_{k}(\eta)\bigr|^2 =  \ell_{P}^2\, \overline{{\mathcal H}}^2_{\ast} \,{\mathcal C}(\nu) \, \biggl(\frac{k}{k_{\ast}}\biggr)^{n_{T}^{(low)}},\qquad {\mathcal C}(\nu)= \frac{ 2^{2\nu+1}}{\pi^3} \Gamma^2(\nu),
\label{STHR6c}
\end{equation}
where $\overline{{\mathcal H}}^2_{\ast} = 1/\eta_{\ast}^2$ and 
$n_{T}^{(low)} = 3 - 2\nu = 2( 1 -\zeta)$ is, by definition, the low-frequency spectral index
which is generically blue as long as $\alpha>0$ in Eq. (\ref{STHR3}):
\begin{eqnarray}
n_{T} = 2 - 2 \zeta &=& \frac{3 \alpha - 2\epsilon_{k}}{(1 + \alpha - \epsilon_{k})}
\nonumber\\
&=& \frac{3 \alpha}{1 + \alpha} + 
 \frac{\epsilon_{k}\, (\alpha - 2)}{(1 + \alpha)^2} + {\mathcal O}(\epsilon_{k}^2).
\label{STHR9}
\end{eqnarray}
The spectral energy density for typical wavenumbers $ k < a_{\ast}\, H_{\ast}$ can also be computed 
 once the expression of $b(\eta)$ is known; within the WKB approach already 
 outlined in section \ref{sec3} is given by 
\begin{equation}
\Omega_{gw}(k,\tau) = \frac{k^4}{12 \,\pi^2\, H^2\, \overline{M}_{P}^2\, a^{4}}\,\, \bigl|{\mathcal Q}_{k}(\eta_{ex}, \eta_{re})\bigr|^2 \,\, \biggl(\frac{b_{re}}{b_{ex}}\biggr)^2 \biggl(
1 + \frac{1}{k^2 \tau_{re}^2} \biggr).
\label{STHR5}
\end{equation}
Equation (\ref{STHR5}) has been computed by assuming that the reentry 
of the relevant wavelengths occurs when the refractive index is not dynamical and this implies 
that when the relevant wavelength reenters the $\eta$-time and the conformal time 
coordinates coincide, i.e. $\eta_{re} = \tau_{re}$. Furthermore in the simplest 
situation $\tau_{re}$ falls within the radiation phase (i.e. $a^{\prime\prime} \to 0$) so that $k \tau_{re}\ll 1$ in Eq. (\ref{STHR5}). Since any wavelength  exiting for $\eta < - \eta_{\ast}$ does its first crossing during the inflationary phase, the corresponding refractive index is  $n = n_{\ast} (a/a_{\ast})^{\alpha}$;  the explicit expression of ${\mathcal Q}_{k}(\eta_{ex}, \eta_{re})$ is 
\begin{equation}
 {\mathcal Q}_{k}(\eta_{ex}, \eta_{re}) = 1 - (\overline{{\mathcal H}}_{ex} + i k) \int_{\eta_{ex}}^{\eta_{re}} \frac{b_{ex}^2}{b^2(\tau)} \, d\eta, \qquad\qquad \overline{{\mathcal H}}= \dot{b}/b,
 \label{STHR5a}
 \end{equation}
which is the analog of the expression already obtained in section \ref{sec3} when the  refractive index is not dynamical. Finally, using Eq. (\ref{STHR6}) an even more explicit expression of the spectral energy density can be deduced:
\begin{eqnarray}
h_{0}^2 \,\Omega_{gw}(\nu,\tau_{0}) &=&  \biggl(\frac{H_{1}}{M_{P}}\biggr)^2  \,{\mathcal D}_{\ast}(\alpha,n^{(low)}_{T}) \biggl(\frac{\nu}{\nu_{\ast}}\biggr)^{n^{(low)}_{T}}, \qquad \qquad \nu_{eq} < \nu< \nu_{\ast},
 \label{STHR7}\\
{\mathcal D}_{\ast}(\alpha,n_{T}) &=& \frac{4 n_{\ast}^3\, h_{0}^2 \Omega_{R0}  }{3 \pi} \, \biggl(1 + \frac{\alpha}{1-\epsilon_{k}}\biggr)^2 \, \biggl(\frac{g_{\rho, \, r}}{g_{\rho, \, eq}}\biggr) \biggl(\frac{g_{s,\, eq}}{g_{s,\, r}}\biggr)^{4/3} \, \biggl(\frac{\Omega_{M0}}{\Omega_{\Lambda}}\biggr)^2.\, \,
\label{dstar}
\end{eqnarray}
 As usual $\Omega_{M0}$ and $\Omega_{\Lambda}$ denote the present critical fractions of matter and dark energy; it is actually well known that the dominance of dark energy suppresses the spectrum by a factor $(\Omega_{M0}/\Omega_{\Lambda})^2 = {\mathcal O}(0.1)$ (see, for instance, \cite{LIGO3}). In Eq. (\ref{STHR7}) $\nu_{\ast}$ denotes the frequency of the spectrum associated 
with $\eta_{\ast}$ and since $k_{\ast} = 1/\eta_{\ast}$ the corresponding 
comoving frequency is:
\begin{equation}
\nu < \nu_{\ast} = \biggl(1 + \frac{\alpha}{1 - \epsilon_{k}}\biggr) e^{\alpha N_{\ast} - \Delta N} \, \overline{\nu}_{max},\qquad 
\Delta N= N_{t} - N_{\ast}.
\label{STHR8}
\end{equation}
In Eq. (\ref{STHR8}) $N_{\ast} = \ln{(a_{\ast}/a_{i})}$ is the number 
of $e$-folds during the refractive stage while $N_{t} = \ln{(a_{1}/a_{i})}$ denotes the {\em total} number of $e$-folds; finally, as before,  $\overline{\nu}_{max}$ indicates the maximal frequency of the spectrum and it coincides with Eq. (\ref{NK6}) since, so far, the radiation dominance starts right after the end of inflation.  The tensor spectral index of Eq. (\ref{STHR9}) applies in the low and intermediate frequency ranges when the corresponding wavelengths exit during inflation and reenter in the radiation phase; in Eq. (\ref{STHR9}) $\alpha$ is always much larger than $\epsilon_{k}\simeq r_{T}/16 \leq 0.03/16 \ll 1$ so that the exact result can be accurately evaluated in the limit $\epsilon_{k} \ll 1$. While Eqs. (\ref{STHR9}) and (\ref{STHR7}) hold for $\nu<\nu_{\ast}$, 
the spectral energy density can also be evaluated in the range $\nu_{\ast} < \nu < \overline{\nu}_{max}$ (i.e.  $a_{\ast} \, H_{\ast} < k < a_{1}\, H_{1}$) corresponding to wavelengths that exited the comoving horizon when the refractive index was already constant (i.e. $n\to n_{\ast}$ and  $\eta_{ex} = \tau_{\ast}/n_{\ast}$); in this case the spectral energy density becomes:
 \begin{equation}
h_{0}^2 \Omega_{gw}(\nu,\tau_{0}) = \biggl(\frac{H_{1}}{M_{P}}\biggr)^2\, {\mathcal D}_{max}(\alpha, n^{(high)}_{T}) \, \, \biggl(\frac{\nu}{\overline{\nu}_{max}}\biggr)^{n^{(high)}_{T}}, \qquad \qquad \nu_{\ast}< \nu< \overline{\nu}_{max},
\label{STHR10}
\end{equation}
where the spectral index is given by $n^{(high)}_{T} = - 2 \epsilon_{k} = - r_{T}/8$ and 
\begin{equation}
{\mathcal D}_{max}(\alpha, n^{(high)}_{T}) =
\frac{4 \,h_{0}^2 \Omega_{R0} }{3 \pi} \, e^{(3 -n^{(high)}_{T})\alpha\,N_{\ast}} e^{ n^{(high)}_{T}\,\Delta N} \biggl(1 + \frac{\alpha}{1-\epsilon_{k}}\biggr)^{2-n^{(high)}_{T}} \, \biggl(\frac{g_{\rho, \, r}}{g_{\rho, \, eq}}\biggr) \biggl(\frac{g_{s,\, eq}}{g_{s,\, r}}\biggr)^{4/3} \, 
\biggl(\frac{\Omega_{M0}}{\Omega_{\Lambda}}\biggr)^2.
\label{dmax}
\end{equation}
Equation (\ref{STHR10}) evaluated for $\nu = \nu_{\ast}$ reproduces Eq. (\ref{STHR7}) computed at the same reference frequency and the equivalence of the two expressions ultimately follows from Eq. (\ref{STHR8}). Furtheremore, in Eqs. (\ref{STHR7}) and (\ref{STHR9}) $(H_{1}/M_{P})^2$ can be traded for $\pi \, \epsilon_{k}\, {\mathcal A}_{{\mathcal R}}$ where 
${\mathcal A}_{{\mathcal R}}$ is the amplitude of curvature inhomogeneities at the pivot scale $k_{p}$. It is finally worth recalling that, for a standard thermal history, $g_{s,\,eq} = 3.94$ while $g_{\rho,\, r} = g_{s,\, r}=106.75$ in Eqs. (\ref{dstar})--(\ref{dmax}). In Eqs. (\ref{STHR8}) and (\ref{dmax}) $N_{\ast}$ measures the range 
of variation of the refractive index during inflation and, for this reason, $N_{\ast} < N_{t}$. As we shall see in a moment, the relatively short inflationary stages (where $N_{t} \leq {\mathcal O}(61)$) seem to be preferred for a potential 
explanations of the PTA excesses.
\begin{figure}[!ht]
\centering
\includegraphics[height=5.7cm]{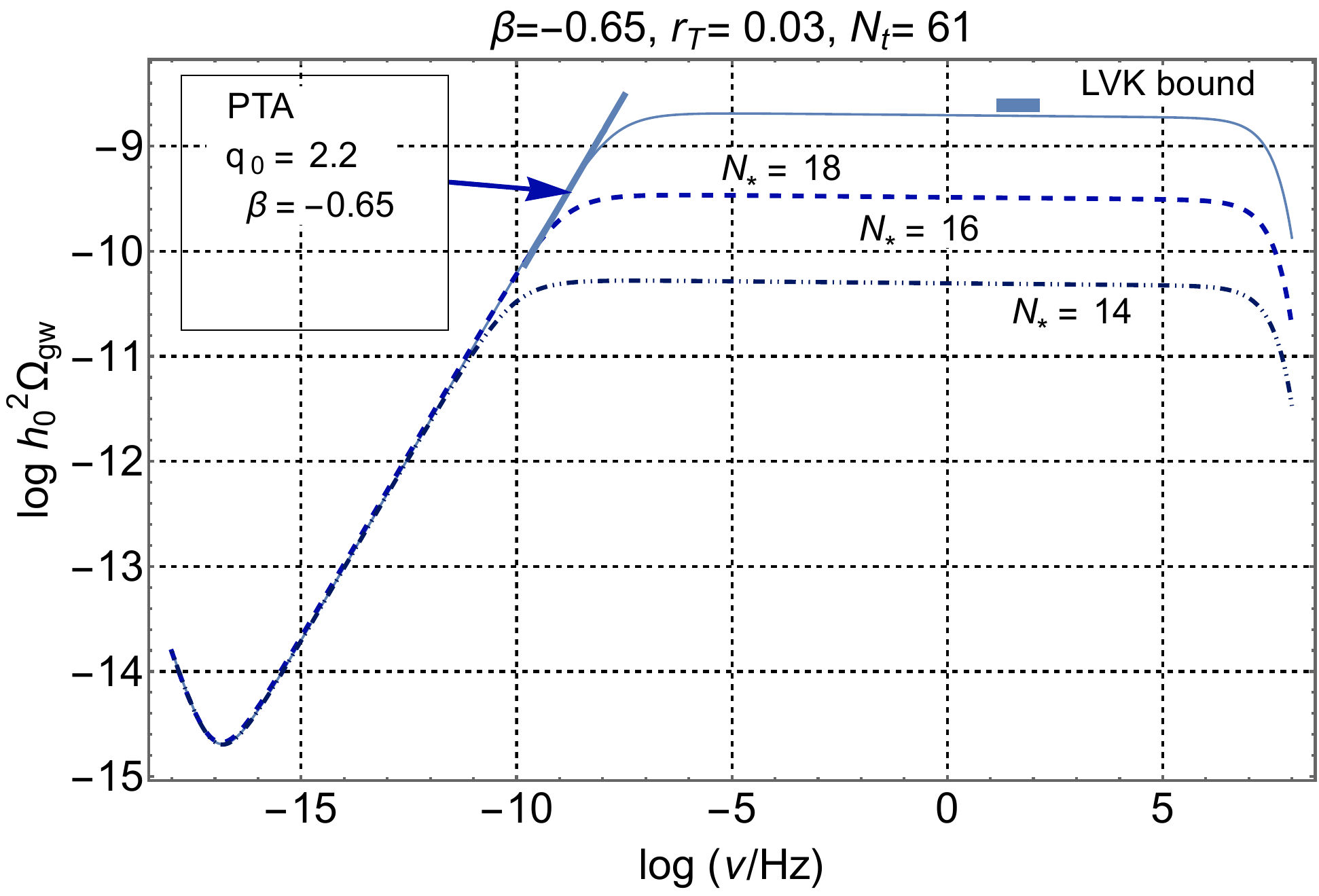}
\includegraphics[height=5.7cm]{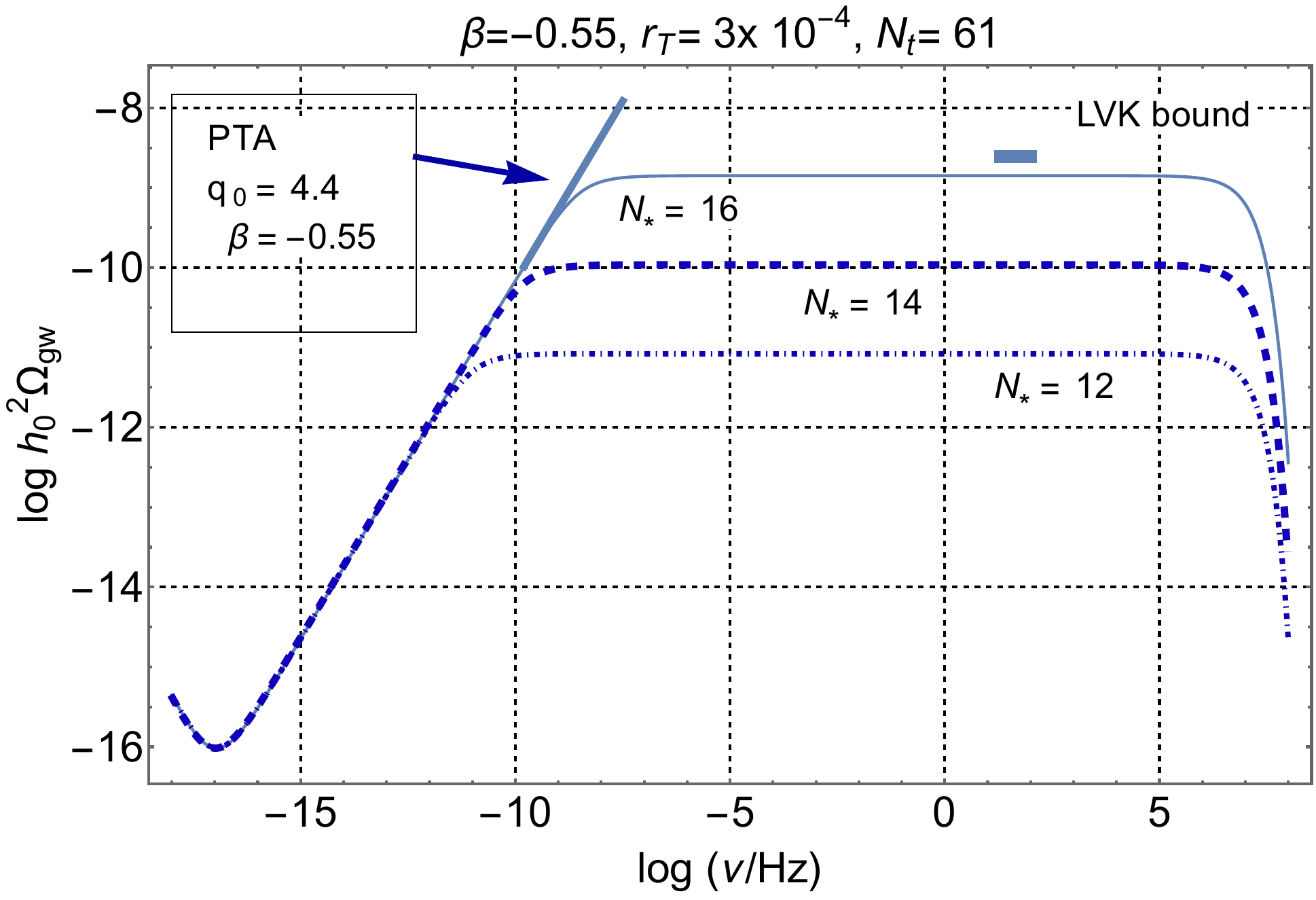}
\caption[a]{We illustrate the common logarithm of the spectral energy density in critical units as a function of the common logarithm of the comoving frequency. In both plots $N_{t} =61$ but the values of $\beta$ and $r_{T}$ do not coincide and they are indicated above each of the two cartoons. The arrows indicate the PTA signal for the spectral indices corresponding 
to the ones selected in each of the plots. The high frequency region labeled by LVK refers to the Ligo-Virgo-Kagra bound that applies in the audio band. The increasing branch and the flat plateau corresponds, respectively, to the analytic estimate of Eqs. (\ref{STHR5}) and (\ref{STHR7}).}
\label{FIG8a}      
\end{figure}
Equations (\ref{STHR5}), (\ref{STHR7}) and (\ref{STHR10}) are now compared with the parametrizations of the PTA signal given in Eqs.  (\ref{PTAR3}) and (\ref{PTAR5}). Since, by definition, the intermediate spectral index is given as 
$2 + 2\beta = n^{(low)}_{T}$  Eq. (\ref{STHR9}) implies a relation that determines 
$\alpha$ as a function of $\epsilon_{k}$ (or $r_{T}$) and $\beta$:
\begin{equation} 
\alpha = \frac{2 [\beta(\epsilon_{k} -1)-1]}{2 \beta -1}.
\label{STHR11}
\end{equation}
Moreover, given that $q_{0}$ depends on all the other parameters
determining the amplitude of $\Omega_{gw}(\nu,\tau_{0})$ (see Eqs. (\ref{STHR7}) and (\ref{STHR10})), we can demand that $\beta$ and $q_{0}$ fall within the phenomenologically 
allowed ranges and check if the results of Eqs. (\ref{STHR7})--(\ref{STHR10}) 
are compatible with the empirical determinations of the PTA. According to the Parkes PTA  the values of $\beta$ and $q_{0}$ fall, respectively,  in the following intervals:
\begin{equation}
-0.65 \leq \beta \leq -0.25, \qquad \qquad 2.2 < q_{0} < 4.4.
\label{STHR12}
\end{equation}
Equation (\ref{STHR12}) constrains the spectral energy density and 
the corresponding region of the theoretical parameters is illustrated in the left plot of Fig. where we report $q_{0}(\beta,\, N_{\ast})$ for different values of $N_{t}$; the shape of each shaded region  directly follows by requiring $ 2.2 < q_{0}(\beta,\,N_{\ast}) < 4.4$ for the various $N_{t}$ mentioned in the plot. On a technical side we note that Eq. (\ref{STHR11}) has been used with the purpose of trading directly $\alpha$ for $\beta$ at a fixed value of $\epsilon_{k}$. The same analysis illustrated in the case of the Parkes PTA can be repeated for the NANOgrav determinations with slightly different results; the analog of Eq. (\ref{STHR12}) is now given by \cite{NANO2}
\begin{equation}
-0.40 \leq \beta \leq -0.20, \qquad \qquad 3.7 < q_{0} < 10.6.
\label{STHR13}
\end{equation}
While the range of $\beta$ given in Eq. (\ref{STHR13}) is narrower than in Eq. (\ref{STHR12}), in the case of $q_{0}$  we observe the opposite: the allowed values of $q_{0}$ of Eq. (\ref{STHR13}) are comparatively larger than the ones of Eq. (\ref{STHR12}). 
A second class of constraints determining the shaded allowed regions is related to the direct bounds from the operating wide-band detectors; in particular we remind that the LIGO, Virgo and Kagra collaborations (LVK) reported a constraint implying \cite{LIGO0,LIGO0a,LIGO0b,LIGO1,LIGO3}:
\begin{equation}
\Omega_{gw}(\nu, \tau_{0}) < 5.8 \times 10^{-9}, \qquad\qquad 20 \,\, \mathrm{Hz} < \nu_{LVK} < 76.6 \,\, \mathrm{Hz},
\label{CONS2}
\end{equation}
in the case of a flat spectral energy density; in the present notations $\nu_{L}$ indicates the LIGO-Virgo-Kagra frequency. The limit of Eq. (\ref{CONS2}) improves on a series of bounds previously deduced by the wide-band interferometers (see Ref. \cite{LIGO3} for a review of the older results). 
\begin{table}[!ht]
\centering
\caption{Selected limits on the relic gravitons obtained by wide-band interferometers.
These limits will be generically referred to as the LIGO-Virgo-Kagra (LVK) bounds.}
\vskip 0.4 cm
\begin{tabular}{||c|c|c||}
\hline
\rule{0pt}{4ex}  $\sigma$ & frequency range if $\nu_{ref}$ [Hz] & Bound \\
\hline
\hline
$0$ &  $20-81.9$ & $\overline{\Omega}_{0} < 6 \times 10^{-8}$ Ref. \cite{LIGO0b}\\
$2/3$ & $20-95.2$ & $\overline{\Omega}_{2/3} < 4.8 \times 10^{-8}$ Ref. \cite{LIGO0b}\\
$3$ & $20-301$ & $\overline{\Omega}_{3} < 7.9 \times 10^{-9}$ Ref. \cite{LIGO0b}\\
$0$ & $20-76.6$ & $ \overline{\Omega}_{0} <    5.8\times10^{-9}$  Ref. \cite{LIGO1}\\
$2/3$ & $20-90.6$ & $  \overline{\Omega}_{2/3} < 3.4\times 10^{-9}$  Ref. \cite{LIGO1}\\
$3$ & $20-291.6$ & $\overline{\Omega}_{3} < 3.9\times 10^{-10}$ Ref. \cite{LIGO1}\\
\hline
\hline
\end{tabular}
\label{TABLE1}
\end{table}
In the present notations the parametrization of $\Omega_{gw}(\nu, \tau_{0})$ adopted by Ref. \cite{LIGO1} reads
\begin{equation}
\Omega_{gw}(\nu, \tau_{0}) = \overline{\Omega}(\sigma) \biggl(\frac{\nu}{\nu_{LVK}}\biggr)^{\sigma}, \qquad \qquad \nu_{LVK} = 25 \,\, \mathrm{Hz},
\label{NOT1}
\end{equation}
and the three specific cases constrained in Refs. \cite{LIGO0b,LIGO1} are reminded in Tab. \ref{TABLE1} 
As the value of $\sigma$ increases the bound becomes more restrictive for a fixed reference frequency and the three previous results are summarized by the following interpolating formula:
\begin{equation}
\log{\overline{\Omega}}(\sigma) < -\,8.236 -\, 0.335\, \sigma- 0.018\, \sigma^2.
\label{NOT2}
\end{equation}
 Since in the present case the bound (\ref{NOT2}) should be applied at high-frequencies we will have $ \sigma =- 2 \epsilon_{k}/ (1- \epsilon_{k})$ with $\epsilon \ll 0.1$; to leading order in $\epsilon_{k}$, 
Eq. (\ref{NOT2}) implies that $\log{\overline{\Omega}}(\epsilon_{k}) < -\,8.236 -\, 0.335\, \epsilon_{k} -0.393 \epsilon_{k}^2$. As an example in the two plots of Fig.  \ref{FIG8a} we considered two different values of $\beta$ (i.e. $\beta = -0.65$ and $\beta= -0.55$). If $N_{\ast}$ and $N_{t}$ are of the 
same order the refractive index stops evolving when inflation approximately ends and,
in this case, it is impossible to get a 
large signal in the nHz range without jeopardizing the big bang nuclosynthesis constraint.  
\begin{figure}[!ht]
\centering
\includegraphics[height=5.4cm]{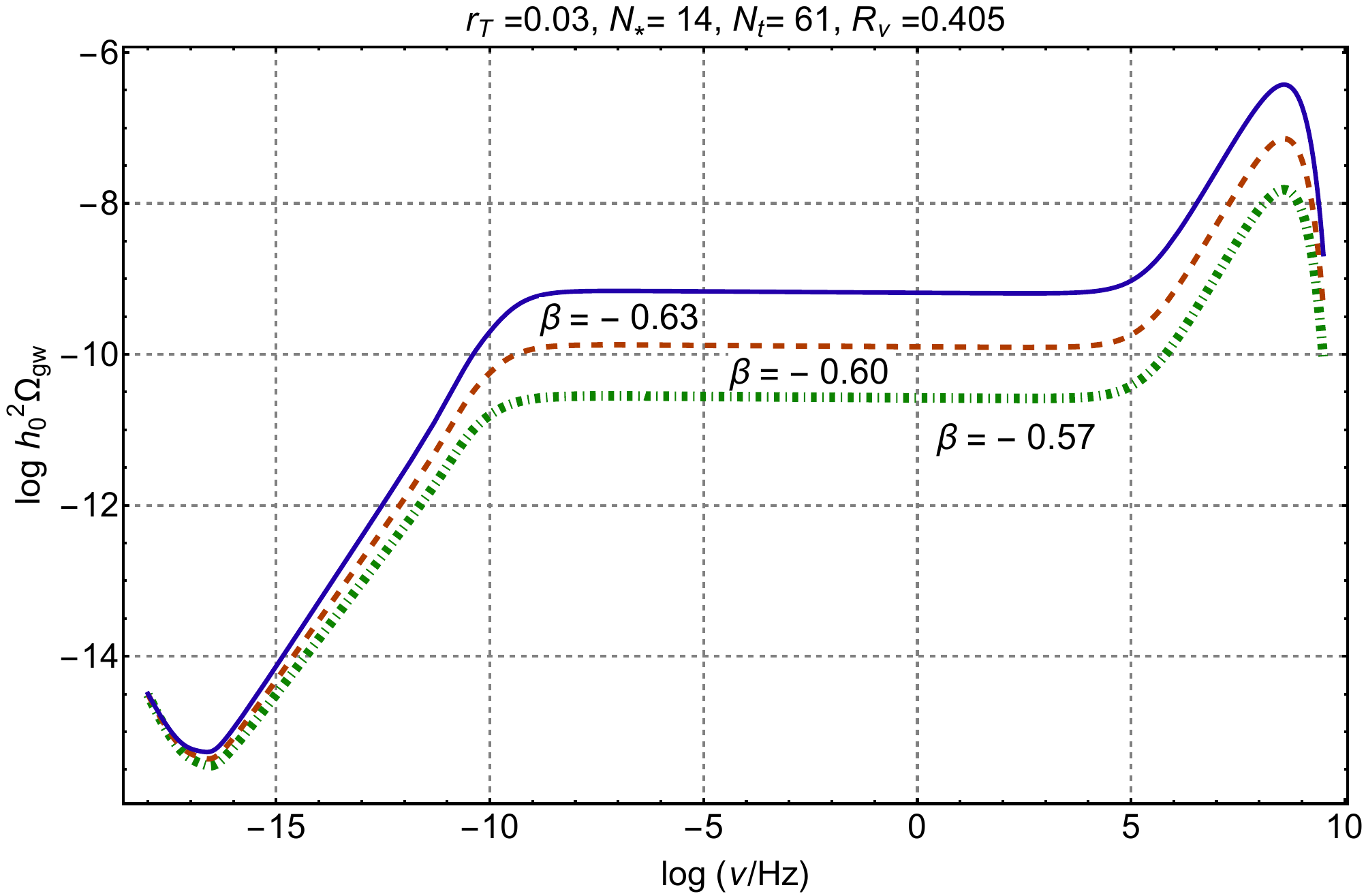}
\includegraphics[height=5.4cm]{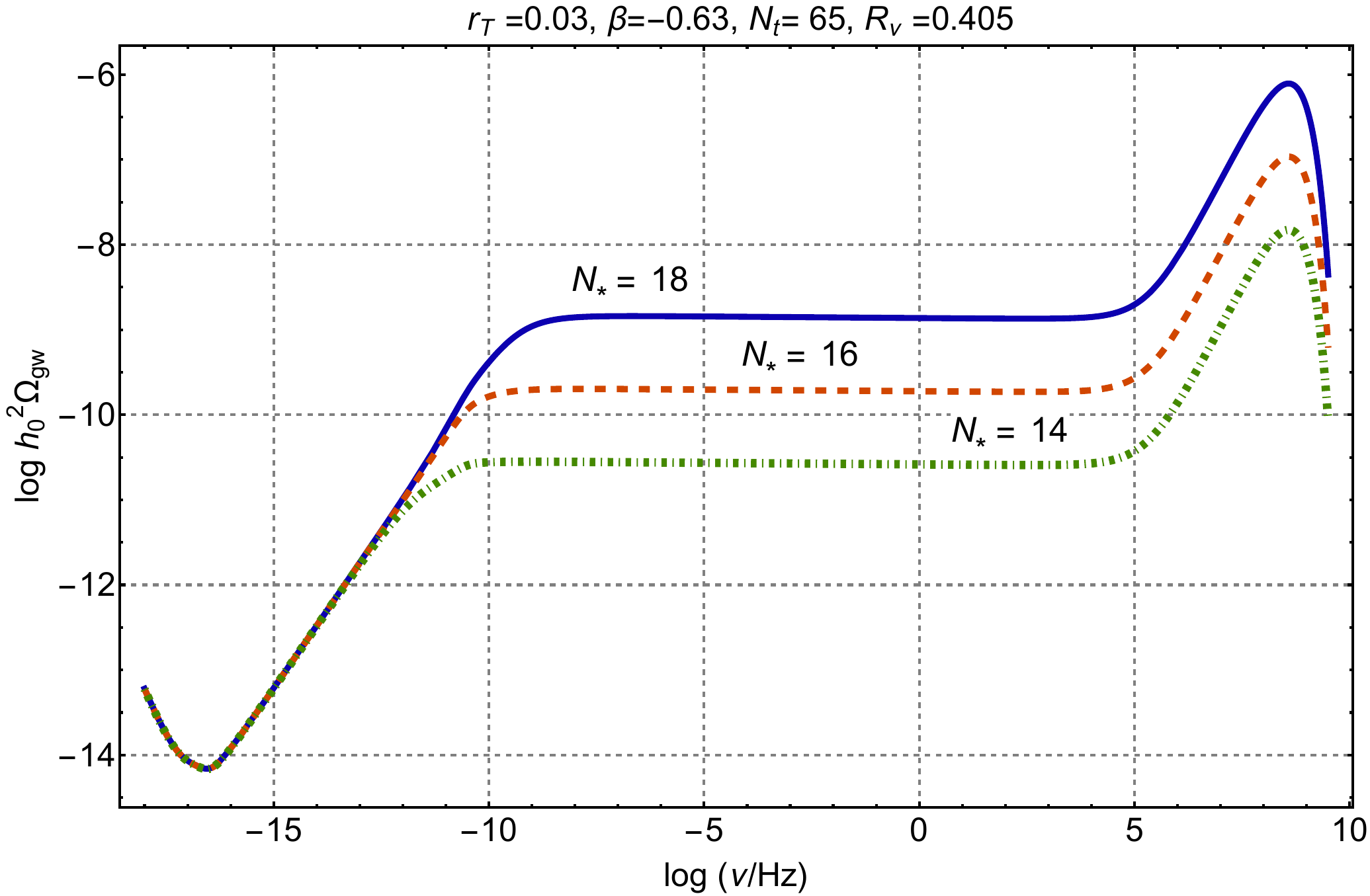}
\caption[a]{As in Fig. \ref{FIG8a} we illustrate the common logarithm of the spectral energy density as a function of the common logarithm of the 
comoving frequency. In the two plots the value of $r_{T}$ is the same but the values of $N_{t}$
are slightly dissimilar. In the plot at the left $N_{\ast}= 14$ while the three spectra correspond 
slightly different values of $\beta$. In the plot at the right $\beta = -0.63$ and the three curves 
illustrate the variation of $N_{\ast}$. Since the effect of neutrino free-streaming has been included, in both plots $R_{\nu}$ denotes the neutrino 
fraction.} 
\label{FIG8b}      
\end{figure}
Conversely,  when $N_{*} <N_{t}$ the refractive index stops evolving well before the onset of the post-inflationary stage, i.e. when the background is still inflating deep inside the quasi-de Sitter stage of expansion. In both plots of Fig.  \ref{FIG8a}, to ease the comparison, we selected $N_{t} = 61$ while different values of $N_{\ast}$ are illustrated. In both plots, for the same choice of the parameters, we also illustrated (with an arrow) the PTA excess and the Ligo-Virgo-Kagra bound. The PTA signal occurs for typical frequencies 
${\mathcal O}(\nu_{ref})$ while the LVK bound applies approximately between
$25$ and $100$ Hz. 

The previous discussion does not exclude the possibility of two 
concurrent modifications of the comoving horizon operating {\em before and after }
the end of inflation. This viewpoint is explored in Fig. \ref{FIG8b}
 where we consider the possibility that the refractive index 
stops its evolution well before the end of inflation (i.e. $N_{\ast} \ll N_{t}$);
The spectral energy density in critical units will therefore have three 
different slopes for $\nu > \nu_{eq}$. In both plots of Fig. \ref{FIG8b}, at intermediate 
frequencies $h_{0}^2 \Omega_{gw}(\nu,\tau_{0})$ 
has the same intermediate slopes appearing in Fig. \ref{FIG8a}  (see also Eqs. (\ref{STHR9})--
and (\ref{STHR11})). However, after the quasi-flat plateau, the spectral energy density 
exhibits a further increasing branch before the maximal frequency. The corresponding 
wavelengths left the comoving Hubble radius during inflation and reentered
in the post-inflationary stage before radiation dominance. In Fig. \ref{FIG8b}  the high frequency 
spectral slope is ${\mathcal O}(1)$ since during the post-inflationary stage 
the evolution is described by a stiff fluid with $\delta \simeq 1/2$ implying 
that $(a\, H)^{-1} \propto a^2$. The main difference between the plots of Figs. \ref{FIG8a} and \ref{FIG8b} comes from the high frequency shape where the bounds coming from BBN must be taken into account (see Eq. (\ref{BBN1}) and discussion therein). The theoretical perspective explored in this discussion strongly suggests that the problem is not yet to fit (more or less reliably) the existing data in terms of a series of preferred scenarios but to understand preliminarily whether or not the observed excesses in the nHz range are compatible with a modified evolution of the comoving horizon since this is the only way the spectrum of relic gravitons at intermediate frequencies can be affected. 
The most conventional option stipulates that the timeline of the comoving horizon is not modified during inflation so that the nHz excess is caused by the drastic change of the post-inflationary expansion rate prior to big bang nucleosynthesis.  This possibility can be 
safely ruled out. A second alternative implies a modified evolution of the tensor modes during a conventional inflationary stage as it happens, for instance, when the gravitons inherit an effective refractive index from the interactions with the geometry. This explanation seems viable in the light of the current observations. We may finally consider the possibility of an epoch of increasing curvature prior to the conventional decelerated stage of expansion and argue that this option is only reconcilable with the observed excesses provided the wavelengths crossing the comoving horizon at early times do not reenter in an epoch dominated by radiation. This option may also be viable with some caveats and has been explored in \cite{MGpuls}.

\subsection{Space-borne interferometers and the expansion history}
The direct measurements in the range $\nu_{ew} \leq \nu < \mathrm{Hz}$
may primarily clarify the nature of the post-inflationary expansion rate. Indeed, after inflation, the expanding stage could include a sequence of stages expanding either faster or slower than radiation; in this situation a hump in $h_{0}^2 \Omega_{gw}(\nu,\tau_{0})$ is generically expected below the a fraction of the Hz where the relic gravitons may exceed (even by eight orders of magnitude) the signals obtained under the hypothesis of radiation dominance throughout the whole expansion history prior to the formation of light nuclei. 

\subsubsection{The conventional wisdom}
An old and conventional viewpoint stipulates that between a fraction of the mHz and few Hz the spectral energy density of the inflationary gravitons can be disregarded even assuming the most optimistic sensitivities of the space-borne detectors. On the contrary, always within the standard lore, in the region between the $\mu$Hz and few Hz  the signals coming from the electroweak physics (or from some other phase transition) should represent the dominant contribution of cosmological origin. This perspective is not completely consistent for (at least) two independent reasons. 
\begin{itemize}
\item{} The first one (already mentioned earlier on in this 
section) is that the electroweak phase transition {\em does not} proceed through the formation 
of bubbles of the new phase and {\em does not} imply large deviations from homogeneity 
as required for the formation of a diffuse secondary background of gravitational radiation.
This statement hods given the measured values of the Higgs and $W$ masses. 
\item{} The usual counterargument  
is that we might expect strongly first-order phase transitions from new physics which 
did not show up so far from collider searches. This assumption is however ad hoc since there are no tangible signals 
of a new electroweak physics from colliders; it is therefore not clear why 
the purported new physics should always lead to a burst of gravitational radiation in a range 
compatible with $\nu_{ew}$.
\end{itemize}
We are now going to discuss how a modified expansion history may lead to a hump in the frequency domain compatible with $\nu_{ew}$. This is why any limit on the spectral energy density of the relic gravitons between few $\mu$Hz and the Hz indirectly constrains the timeline of the post-inflationary expansion rate.

\subsubsection{Chirp amplitudes and frequency dependence}
The direct bounds on the relic gravitons from the audio band ultimately depend upon the spectrum of the signal. For a nearly scale-invariant spectrum,  $h_{0}^2\Omega_{gw}(\nu, \tau_{0}) < {\mathcal O}(10^{-9})$  between $10$ Hz and $80$ Hz \cite{LIGO0,LIGO0a,LIGO0b,LIGO1} (see also \cite{LIGO3} for a recent review including earlier bounds).  To compare the ground-based detectors and the space-borne interferometers it is useful to express the spectral energy density in terms of the chirp amplitude $h_{c}(\nu,\tau_{0})$ \cite{LIGO3} when the typical frequencies fall in the audio band:
 \begin{equation}
h_{0}^2 \Omega_{gw}(\nu,\tau_{0}) = 6. 26 \times 10^{-9} \biggl(\frac{\nu}{0.1 \, \, \mathrm{kHz}}\biggr)^2 
\biggl[\frac{h_{c}(\nu,\tau_{0})}{10^{-24}} \biggr]^2.
\label{ONEeq}
\end{equation}
From left to right Eq. (\ref{ONEeq}) implies that to probe
$h_{0}^2 \Omega_{gw}(\nu,\tau_{0}) = {\mathcal O}(10^{-9})$ we should have 
 a sensitivity in the chirp amplitude ${\mathcal O}(10^{-24})$ for a typical frequency $\nu = {\mathcal O}(100)$ Hz.  From right to left the same relation suggests instead that, for comparable sensitivities 
in $h_{c}(\nu,\tau_{0})$, the minimal detectable $h_{0}^2 \Omega_{gw}(\nu,\tau_{0})$ gets 
comparatively smaller with the frequency; besides the absence of seismic noise this is probably one of strongest arguments  in favour of space-borne detectors 
for typical frequencies ranging between a fraction of the mHz and the Hz. This is why the 
minimal detectable spectral energy density could be $h_{0}^2 \Omega_{gw}(\nu,\tau_{0}) = {\mathcal O}(10^{-11})$ or even $h_{0}^2 \Omega_{gw}(\nu, \tau_{0}) = {\mathcal O}(10^{-15})$ under the hypothesis that the same sensitivity reached in the audio band for the chirp amplitude can also be achieved in the mHz range. With this great hope, various space-borne 
detectors have been proposed so far: the Laser Interferometric Space Antenna (LISA) \cite{LISA1,LISA2}, 
 the Deci-Hertz Interferometer Gravitational Wave Observatory (DECIGO) \cite{DECIGO1,DECIGO2},  the Ultimate-DECIGO \cite{UDECIGO} (conventionally 
 referred to as U-DECIGO), the Big Bang Observer (BBO) \cite{BBO}. This list has been recently enriched by the Taiji \cite{TAIJI1,TAIJI2} and by the TianQin \cite{TIANQIN1,TIANQIN2} experiments. Since these instruments are not yet operational (but might come into operation within the next twenty years) their actual sensitivities are difficult to assess, at the moment. However, without dwelling on the specific nature of the noise power spectra, Eq. (\ref{ONEeq}) shows that, as long as $h_{c} = {\mathcal O}(10^{-23})$ the space-borne detectors might probe $h_{0}^2 \Omega_{gw}(\nu, \tau_{0}) = {\mathcal O}(10^{-14})$ for $\nu_{S} = {\mathcal O}(0.01)$ Hz and this is, roughly 
 speaking, the daring expectation of DECIGO \cite{DECIGO1,DECIGO2} and of U-DECIGO \cite{UDECIGO}.

The fiducial frequency interval of space-borne interferometers ranges from a fraction 
 of the mHz to the Hz and, within this interval,  the minimal detectable spectral energy density (denoted hereunder by $h_{0}^2 \Omega_{gw}^{(min)}(\nu, \tau_{0})$) defines 
 the potential sensitivity of the hypothetical instrument. The LISA interferometers might hopefully probe the following region of the parameter space:
\begin{equation}
h_{0}^2 \Omega_{gw}^{(min)}(\nu, \tau_{0}) = {\mathcal O}(10^{-11.2}), \qquad \qquad 10^{-4} \mathrm{Hz} < \nu \leq 0.1 \, \mathrm{Hz}.
\label{SB1}
\end{equation}
In the case of the Deci-Hertz Interferometer Gravitational Wave Observatory (DECIGO)  the minimal 
detectable spectral energy density could be smaller 
\begin{equation}
10^{-17.5} \leq h_{0}^2 \Omega_{gw}^{(min)}(\nu, \tau_{0}) \leq {\mathcal O}(10^{-13.1}), \qquad \qquad 10^{-3} \mathrm{Hz} < \nu \leq 0.1 \, \mathrm{Hz}.
\label{SB2}
\end{equation}
The values of Eq. (\ref{SB2}) are still quite hypothetical so that it is prudent to choose $h_{0}^2 \Omega_{gw}^{(min)}(\nu, \tau_{0})$
between the standard values of the hoped sensitivity of the DECIGO project 
  (suggesting $h_{0}^2 \Omega_{gw}^{(min)}(\nu, \tau_{0}) = {\mathcal O}(10^{-13.1})$) 
 and the optimistic figure reachable by the Ultimate-DECIGO \cite{UDECIGO} (conventionally 
 referred to as U-DECIGO) where  $h_{0}^2 \Omega_{gw}^{(min)}(\nu, \tau_{0}) = {\mathcal O}(10^{-17.5})$. For the record, the Big Bang Observer (BBO) \cite{BBO} might reach sensitivities 
 \begin{equation}
 h_{0}^2 \Omega_{gw}^{(min)}(\nu, \tau_{0})=  {\mathcal O}(10^{-14.2}), \qquad  10^{-3} \mathrm{Hz} < \nu \leq 0.1 \, \mathrm{Hz}.
\label{SB3}
\end{equation}
There finally exist also recent proposals such as Taiji \cite{TAIJI1,TAIJI2} and  TianQin \cite{TIANQIN1,TIANQIN2}
leading to figures that are roughly comparable with the LISA values. In summary for the typical frequency of the space-borne detectors we consider the broad range $0.1\, \mathrm{mHz} <\nu_{S} < 0.1 \, \mathrm{Hz}$ and suppose that in this range
$h_{0}^2 \Omega_{gw}^{(min)}(\nu, \tau_{0})$  may take the following two extreme 
values:
\begin{equation}
h_{0}^2 \Omega_{gw}^{(min)}(\nu_{S}, \tau_{0}) = {\mathcal O}(10^{-11}), \qquad \qquad h_{0}^2 \Omega_{gw}^{(min)}(\nu_{S}, \tau_{0}) = {\mathcal O}(10^{-14}).
\label{SB4}
\end{equation}
While the two values of Eq. (\ref{SB4}) are both quite optimistic,  they are customarily assumed by the observational proposals and, for this reason, they are used here only for illustration.

\subsubsection{Humps in the spectra from the modified expansion rate}
\begin{figure}[!ht]
\centering
\includegraphics[height=5.5cm]{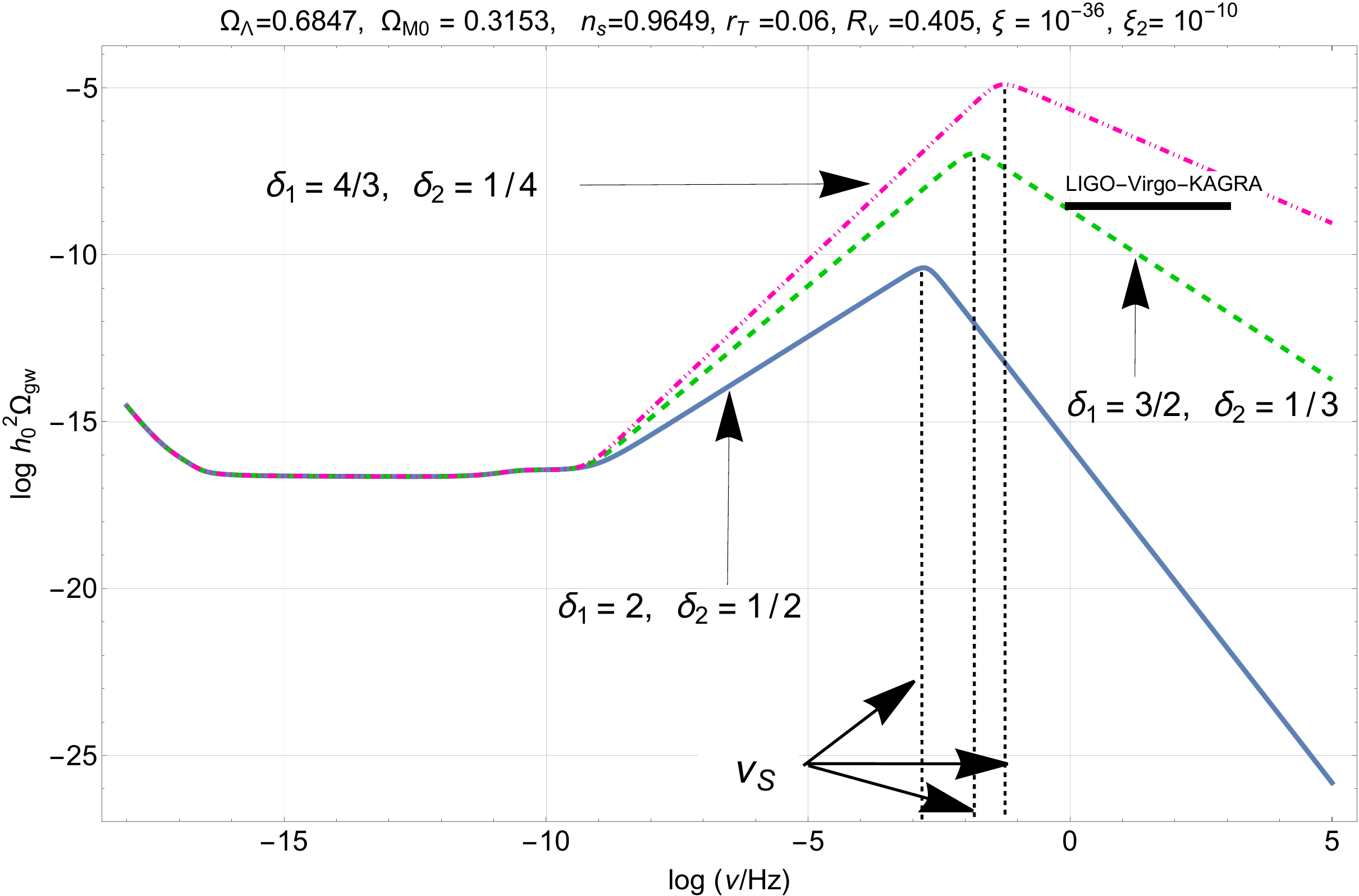}
\includegraphics[height=5.5cm]{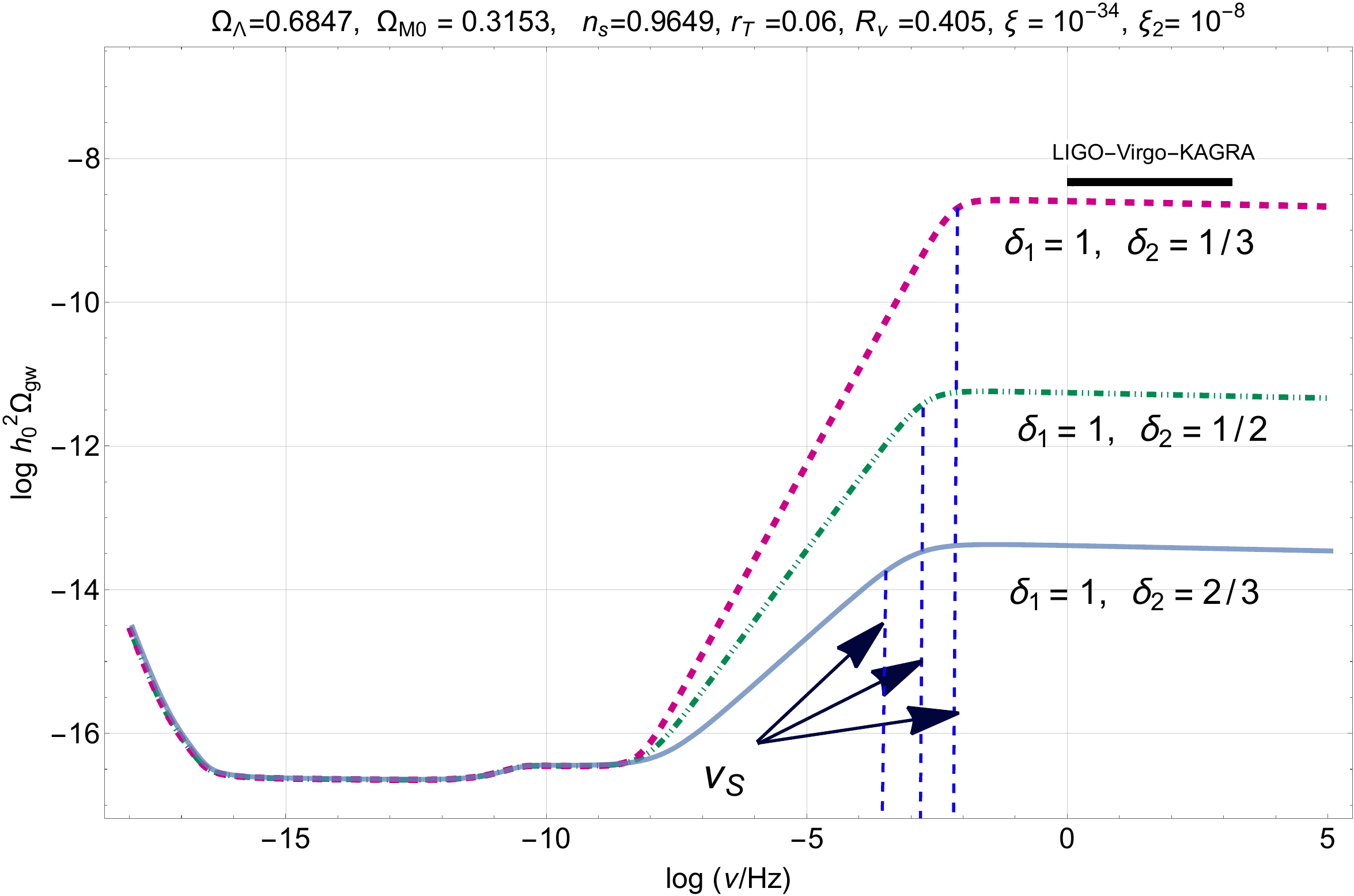}
\caption[a]{The common logarithm of $h_{0}^2 \, \Omega_{gw}(\nu,\tau_{0})$ is illustrated as a function of the common logarithm of the frequency expressed in Hz. In the left plot the dashed and the dot-dashed curves illustrate two models that are only marginally compatible with the big bang 
nucleosynthesis constraint  and with the LIGO-Virgo-Kagra limit; the parameters of the curve at the bottom (full line) are instead drawn 
from the allowed region of the parameter space. While in all the examples of the left plot $\delta_{1}> 1$ (and $h_{0}^2 \Omega_{gw}(\nu,\tau_{0})$  decreases for $\nu> \nu_{2}$), in the right plot $\delta_{1} \to 1$ and  the limits from the audio band are the most relevant ones. }
\label{FIG4}      
\end{figure}

The expansion rates can be bounded by requiring that for frequencies of the order 
of $\nu_{S}$ the corresponding spectral energy density exceeds $h_{0}^2 \Omega_{gw}^{(min)}(\nu_{S}, \tau_{0})$; all the other constraints on the diffuse backgrounds of gravitational radiation must also be satisfied. With this strategy it is possible to constrain the unconventional post-inflationary expansion histories by simultaneously obtaining a large signal for frequencies ${\mathcal O}(\nu_{S})$.
The spectral energy density of the relic gravitons might exhibit various 
successive local maxima but the simplest case consists in a single hump
for frequencies comparable with $\nu_{S}$. The $h_{0}^2\, \Omega_{gw}(\nu, \tau_{0})$ can be expressed in this case as:
\begin{equation}
h_{0}^2 \, \Omega_{gw}(\nu, \tau_{0}) = {\mathcal N}_{\rho} \, r_{T}(\nu_{p}) \biggl(\frac{\nu}{\nu_{p}}\biggr)^{n^{(low)}_{T}(r_{T})} \,\, {\mathcal T}^2_{low}(\nu/\nu_{eq}) \,\,{\mathcal T}^2_{high}(\nu, \nu_{2}, \nu_{r}, n_{T}^{(1)}, n_{T}^{(2)}),
\label{FOUR2}
\end{equation}
where, as usual, $n^{(low)}_{T}$ is the spectral index associated with the wavelengths leaving the Hubble radius during the inflationary phase and reentering during the radiation stage. In Eq. (\ref{FOUR2}) $\nu_{p}$ and $\nu_{eq}$ define the lowest frequency range of the spectral energy density:
\begin{eqnarray}
\nu_{p} &=&  3.092 \biggl(\frac{k_{p}}{0.002 \,\, \mathrm{Mpc}^{-1}}\biggr) \, \mathrm{aHz},
\nonumber\\
\nu_{eq} &=&   15.97 \biggl(\frac{h_{0}^2\,\Omega_{M0}}{0.1411}\biggr) \biggl(\frac{h_{0}^2\,\Omega_{R0}}{4.15 \times 10^{-5}}\biggr)^{-1/2}\,\, \mathrm{aHz},
\label{FOUR4}
\end{eqnarray}
The transfer function of Eq. (\ref{FOUR2}) also includes the dependence on the spectral slopes $n_{T}^{(1)}$ and $n_{T}^{(2)}$; up to corrections ${\mathcal O}(r_{T})$ the depend directly on the expansion rate expressed in the conformal time parametrization:
\begin{equation}
n_{T}^{(1)} = 2 ( 1 - \delta_{1}) + {\mathcal O}(r_{T}), \qquad\qquad 
n_{T}^{(2)} = 2 ( 1 - \delta_{2}) + {\mathcal O}(r_{T}),
\end{equation}
where $n_{T}^{(1)} < 0$ and $n_{T}^{(2)} >0$ since we consider the situation where 
during the first stage the Universe expands faster than radiation 
(i.e. $\delta_{1}>1$) while in the second stage it is 
slower than radiation (i.e. $\delta_{2}<1$). In the simplest 
case where the consistency relations are enforced we have
\begin{equation}
n^{(low)}_{T}(r_{T}) = - \frac{r_{T}}{8} + {\mathcal O}(r_{T}^2), \qquad\qquad {\mathcal N}_{\rho} = 4.165 \times 10^{-15} \biggl(\frac{h_{0}^2\,\Omega_{R0}}{4.15\times 10^{-5}}\biggr).
\label{FOUR3}
\end{equation}
In Eq. (\ref{FOUR4}) ${\mathcal T}_{low}(\nu/\nu_{eq})$ is the low-frequency 
transfer function of the spectral energy density \cite{LIGO3}:
\begin{equation}
{\mathcal T}_{low}(\nu, \nu_{eq}) = \sqrt{1 + c_{1}\biggl(\frac{\nu_{eq}}{\nu}\biggr) + c_{2}\biggl(\frac{\nu_{eq}}{\nu}\biggr)^2},\qquad c_{1}= 0.5238, \qquad c_{2}=0.3537.
\label{FOUR5}
\end{equation}
The high frequency transfer function ${\mathcal T}_{high}(\nu, \nu_{2}, \nu_{r}, \delta_{1}, \delta_{2})$ appearing in Eq. (\ref{FOUR2}) is specifically discussed in Ref. \cite{MGPT}.
In Figs. \ref{FIG4}  the spectral energy density has been explicitly illustrated for a selection of the 
parameters.  
In the left plot of Fig. \ref{FIG4} we selected $\xi = 10^{-36}$ and $\xi_{2} = 10^{-10}$ for different values of
$\delta_{1} > 1$ and $\delta_{2} < 1$. We recall that, by definition, 
\begin{equation}
\xi = \xi_{1}\, \xi_{2} = H_{r}/H_{1}, \qquad \xi_{1} = H_{2}/H_{1}, \qquad \xi_{2} = H_{r}/H_{2}.
\end{equation}
As expected the value of $\nu_{r}$ is always larger than $10^{-10}$. The parameters of the dot-dashed and of the dashed curves of the left plot in Fig. \ref{FIG4} have been selected in order to get an artificially large signal that is in fact excluded both by the BBN constraint and by the limit of ground-based detectors. The results of the right plot in Fig. \ref{FIG4} correspond instead to a slightly different choice of the parameters, namely 
$\xi = 10^{-34}$ and $\xi_{2} =10^{-8}$. For illustration we have chosen $\delta_{1} \to 1$ implying 
that between $\nu_{max}$ and $\nu_{2}$ the spectral energy density is quasi-flat. This is the most constraining case from the viewpoint of the limits coming from wide-band detectors \cite{LIGO0,LIGO0a,LIGO0b,LIGO1,LIGO3}.

\subsubsection{Complementary considerations}
So far we saw that different frameworks motivate the presence of post-inflationary stages expanding at rate either faster or slower 
than radiation and this is why the model independent perspective of Ref. \cite{EXP1} (see also \cite{ST3,ST3a}) is, in our opinion, 
the most useful.  We remind here that stiff post-inflationary phase is dynamically realized in different situations 
 and the first speculations along this direction probably date back to the ideas Zeldovich \cite{ZELD}, Sakharov \cite{BB1} and Grishchuk \cite{AA2}. After the formulation of conventional inflationary models Ford \cite{FORD1} noted that gravitational particle production at the end of inflation could account for the entropy of the present Universe and observed that the backreaction effects of the created quanta constrain the length of a stiff post-inflationary phase by making the expansion dominated by radiation. These effects typically lead, in our notations, to 
a pivotal frequency $\nu_{r}$ of the order of the mHz. It has been later argued by  Spokoiny \cite{spok} that various classes of scalar field potentials exhibit a transition from inflation to a stiff phase dominated by the kinetic energy of the inflaton. In more recent times it became increasingly plausible to have a single scalar field acting as inflaton in the early Universe and as quintessence field in the late Universe \cite{HIGHFF7,HIGHFF8}. A generic signature of a post-inflationary phase stiffer than radiation is the production of relic gravitons with increasing spectral energy density \cite{EXP1}.  In quintessential inflationary models the inflaton and the quintessence field are identified in a single scalar degree of freedom \cite{EXP4} and various concrete forms of the inflaton-quintessence potential $V(\varphi)$ have been proposed and scrutinized through the years. The transition between an inflationary phase and a kinetic phase can be realized both with power-law potentials and with exponential potentials. See also Refs. \cite{ST4,ST5,ST6} for further applications.
We pointed out so far that the expected signal coming from the phase transitions is probably rather small; however, as suggested in the past, a strong hypermagnetic background may be present in the symmetric phase of the electroweak theory \cite{MGew1,MGew2,MGew3} because 
of the symmetries of the plasma at finite density and finite conductivity.  The overall magnitude of the spectra of gravitational radiation induced by a hypermagnetic background have been estimated, for the first time, in \cite{MGew1,MGew2} and turn out to be generally different from the ones associated 
with a modified post-inflationary evolution \cite{MGew3}. 

\renewcommand{\theequation}{6.\arabic{equation}}
\setcounter{equation}{0}
\section{The expansion history and the high frequency gravitons}
\label{sec6}
The high frequency region of the spectrum ranges between few Hz and the THz since, as already discussed in section \ref{sec3}, their maximal frequency cannot exceed  the $\mathrm{THz}$ domain. Only for practical reasons, in this broad region we distinguish the {\em ultra-high frequency domain} (between the MHz and the THz) and the {\em high frequency band} ranging from the Hz to the MHz. To analyze the bounds on the post-inflationary expansion rate it is simpler to address first the THz domain and then focus on the MHz region that also includes the operating window of ground based interferometers. 
\subsection{Spikes in the GHz domain}
If the post-inflationary evolution consists of a single stage, the results of Eqs. (\ref{HH3}) and (\ref{SSTTnew3})--(\ref{SSTTnew3a}) suggest that the maximal signal should always be concentrated between the GHz and the THz. This happens when the expansion rate is slower than radiation (i.e. $\delta < 1$, see section \ref{sec2} and notations therein). If the expansion rate is instead faster than radiation (i.e. $\delta > 1$) the high frequency slope is negative (i.e. $n_{T}^{(high)}<0$) so that the spectral energy density is ultimately decreasing and potentially even smaller than the signal of the concordance paradigm (i.e. $h_{0}^2 \, \Omega_{gw}(\nu,\tau_{0}) \leq {\mathcal O}(10^{-17})$) in the same range of frequencies. 

\subsubsection{General considerations}
\begin{figure}[!ht]
\centering
\includegraphics[height=6.8cm]{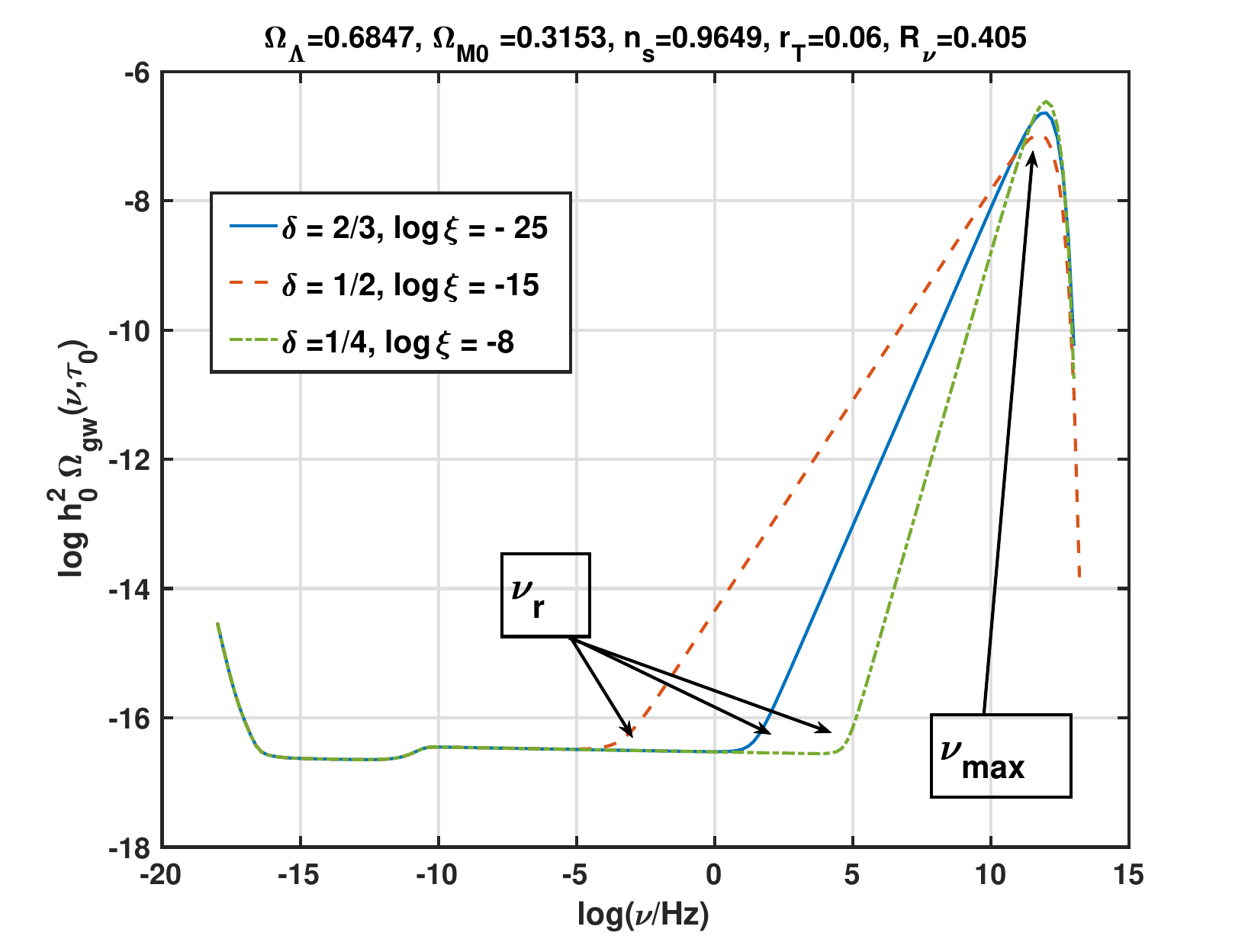}
\includegraphics[height=6.5cm]{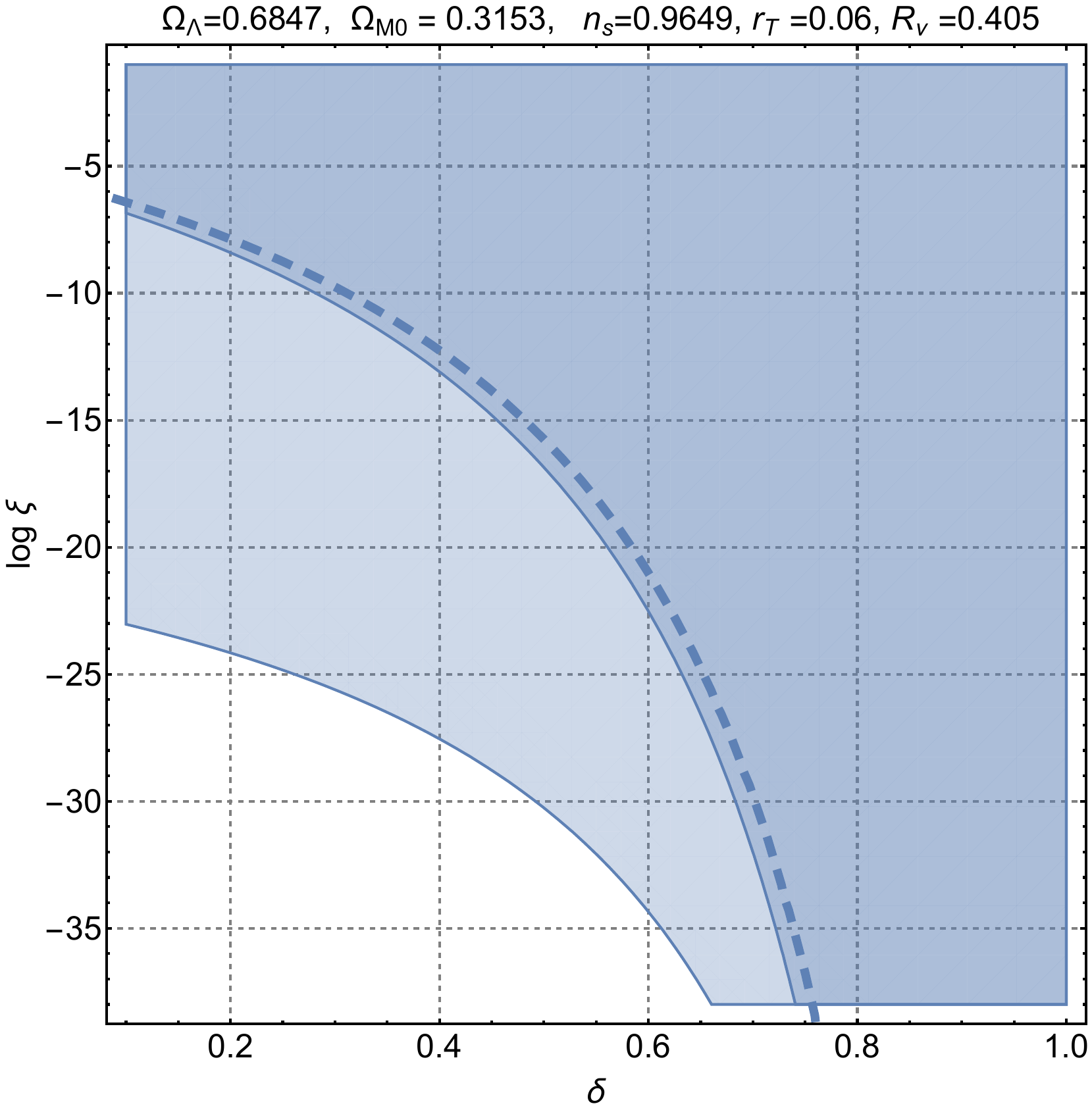}
\caption[a]{In the left plot the common logarithm of $h_{0}^2 \Omega_{gw}(\nu,\tau_{0})$ is illustrated as a function of the common logarithm of the frequency in the case $n_{T}^{(high)} >0$ (i.e. $\delta <1$). In the plot at the right a general bounds on the expansion rate are derived in the plane $(\delta, \log{\xi})$. The late-time parameters on top of the plots correspond to the last Planck release supplemented by the more constraining bounds on $r_{T}$ obtained later on \cite{RR1,RR2,RR3}. }
\label{FIGsingle1}      
\end{figure}

The high frequency branch of the spectrum bears the mark of the post-inflationary expansion rate and from the frequency profile of the spectral energy density 
we can directly infer the post-inflationary expansion rate, the maximal frequency and the other pivotal frequencies of the spectrum (including 
the approximate curvature scale of radiation dominance). 
In the left plot of  Fig. \ref{FIGsingle1} we report the spectral energy density in critical units as a function  of the frequency for a selection of examples (common logarithms are employed on both axes);  note that, for the reported spectra, the post-inflationary expansion rate is slower than radiation. In the right plot of the same figure the parameter space is illustrated in the plane $(\log{\xi}, \, \delta)$ where $\xi= H_{r}/H_{1}$ estimates the overall duration of the post-inflationary stage of expansion.  The shaded region in the right plot of Fig. \ref{FIGsingle1} denotes instead the allowed portion of the parameter space; in particular, the darker sector above the dashed curve accounts for the bounds coming 
from BBN which are ultimately the most constraining. 
The region with lighter 
shading {\em below} the dashed curve corresponds to the current 
limits coming from wide-band interferometers. Finally the dashed curve itself 
is deduced through a semianalytic approximation discussed hereunder. The specific features of $\Omega_{gw}(\nu,\tau_{0})$ illustrated Fig. \ref{FIGsingle1} allow for a quantitative reconstruction of the expansion rate if and when the sensitivities 
of the dedicated detectors (both in the audio band and in the GHz region) will be able to resolve the class of signals suggested here. Along this perspective the main features of Fig. \ref{FIGsingle1} motivate, in short, the following observations.
\begin{itemize}
\item{} In the aHz region the spectral energy density decreases as $\nu^{-2}$ while we can appreciate the suppression due to the neutrino free streaming close to $\nu_{bbn}$ \cite{STRNU0,STRNU1,STRNU2,STRNU3,STRNU4}. Other sources of suppression taken into account in Fig. \ref{FIGsingle1} and in the remaining plots include the late-time dominance of dark energy and the evolution of relativistic species. The spectra of Fig. \ref{FIGsingle1} have been deduced by using 
for the fiducial parameters the last Planck data release in the case of three massless neutrinos where $R_{\nu} = \rho_{\nu}/(\rho_{\gamma} + \rho_{\nu}) =0.405$, as indicated on top of each plots; this is the choice of the minimal $\Lambda$CDM scenario. If the radiation would dominate the whole post-inflationary evolution the quasi-flat plateau (decreasing because of the slow-roll corrections) would last up to frequencies ${\mathcal O}(300)$ MHz. 
\item{} When the expansion rate is faster than radiation (i.e. $\delta > 1$ in the 
notations of Fig. \ref{FIGsingle1}) the spectral energy density further 
{\em decreases} between $\nu_{r}$ and $\nu_{max}$: this timeline 
implies that $h_{0}^2 \Omega_{gw}(\nu,\tau_{0}) \ll {\mathcal O}(10^{-17})$ (in particular in the audio band). No further constraints (besides the low-frequency limits that translate into the upper bound on $r_{T}$ \cite{RR1,RR2,RR3}) appear 
when $\delta > 1$.  
\item{} When the post-inflationary expansion rate is slower than radiation 
(i.e. $\delta <1$ in Fig. \ref{FIGsingle1}) the spectral energy density grows for
$ \nu> \nu_{r}$ and eventually reaches a maximum that roughly 
corresponds to the onset of the exponential suppression taking place for $\nu>\nu_{max}$.
\end{itemize}
To trace the origin of the high frequency spike we remark that $h_{0}^2 \, \Omega_{gw}(\nu,\tau_{0})$ can be written, with compact notations, as:
\begin{equation}
h_{0}^2 \, \Omega_{gw}(\nu, \tau_{0}) = {\mathcal N}_{\rho} \, r_{T} \biggl(\frac{\nu}{\nu_{p}}\biggr)^{n^{(low)}_{T}} \, \, {\mathcal T}^2_{low}(\nu/\nu_{eq}) \,  {\mathcal T}^2_{high}(\nu/\nu_{r}, \delta),
\label{SING6}
\end{equation}
where $\nu_{p}$ and $\nu_{eq}$ are, respectively, the pivot and the 
equality frequencies already introduced in Eq. (\ref{FOUR4}). As usual $r_{T}= r_{T}(\nu_{p})$ is the tensor to scalar ratio evaluated at the pivot scale whereas ${\mathcal T}^2_{low}(\nu/\nu_{eq})$ and ${\mathcal T}^2_{high}(\nu/\nu_{r}, \delta)$ denote the transfer functions directly computed for the spectral energy density \cite{ST3,ST3a}; the value of ${\mathcal N}_{\rho} = {\mathcal O}(4)\times 10^{-15}$ can be accurately fixed both analytically and numerically\footnote{ The low-frequency transfer function ${\mathcal T}_{low}(\nu/\nu_{eq})$ has a definite form \cite{ST3,ST3a} the high frequency transfer function ${\mathcal T}_{high}(\nu/\nu_{r}, \delta)$ depends on the value of $\delta$ so that it does not have a general expression \cite{MGshift}. It should be stressed that we refer here to the transfer function {\em of the spectral energy density} \cite{MGshift,ST3,ST3a} which is numerically more accurate (when estimating $\Omega_{gw}(k,\tau_{0})$) than the transfer function for the amplitude \cite{CHten5,CHten6,CHten7,CHten8,CHten9}. }.
Since for $\nu> \nu_{r}$ the high-energy transfer function has the slope $n^{(high)}_{T}$ (i.e. ${\mathcal T}_{high}^{2} \to (\nu/\nu_{r})^{n^{(high)}_{T}}$) for the analytic estimates of the limits imposed on the spectral energy density we can express $h_{0}^2\,\Omega_{gw}(\nu,\tau_{0})$ 
in the following approximate form: 
\begin{equation}
h_{0}^2 \, \Omega_{gw}(\nu, \tau_{0})  = {\mathcal N}_{\rho} \, r_{T} \biggl(\frac{\nu}{\nu_{p}}\biggr)^{n^{(low)}_{T}} \, \, {\mathcal T}^2_{low}(\nu_{r}/\nu_{eq}) \, \biggl(\frac{\nu}{\nu_{r}}\biggr)^{n^{(high)}_{T}}, \qquad  \nu_{r} \leq \nu \leq \nu_{max}. 
\label{SING9}
\end{equation}
Equation (\ref{SING9}) rests on the observation that ${\mathcal T}_{low}(\nu_{r}/\nu_{eq}) \to 1$ for $\nu \geq \nu_{r}$ while, in the same limit, it is also true that $\overline{n}_{T} \ll 1$. Since the overall normalization mildly depends on $\nu$ and $r_{T}$ we can express the spectral energy density as:
\begin{equation}
h_{0}^2 \, \Omega_{gw}(\nu, \tau_{0})  = {\mathcal N}_{\rho}(r_{T}, \nu)  \biggl(\frac{\nu}{\nu_{r}}\biggr)^{n^{(high)}_{T}}, \qquad\qquad \nu > \nu_{r},
\label{SING9a}
\end{equation}
where ${\mathcal N}_{\rho}(r_{T}, \nu)$ is 
\begin{equation}
 {\mathcal N}_{\rho}(r_{T}, \nu) = {\mathcal N}_{\rho} \, r_{T} \biggl(\frac{\nu}{\nu_{p}}\biggr)^{n^{(low)}_{T}} \, \, {\mathcal T}^2_{low}(\nu_{r}/\nu_{eq}), \qquad \qquad \frac{ d \ln{{\mathcal N}_{\rho}}}{ d \ln{\nu}} = 
 - \frac{r_{T}}{8} \ll 1.
 \label{SING9b}
 \end{equation}
 \begin{figure}[!ht]
\centering
\includegraphics[height=7cm]{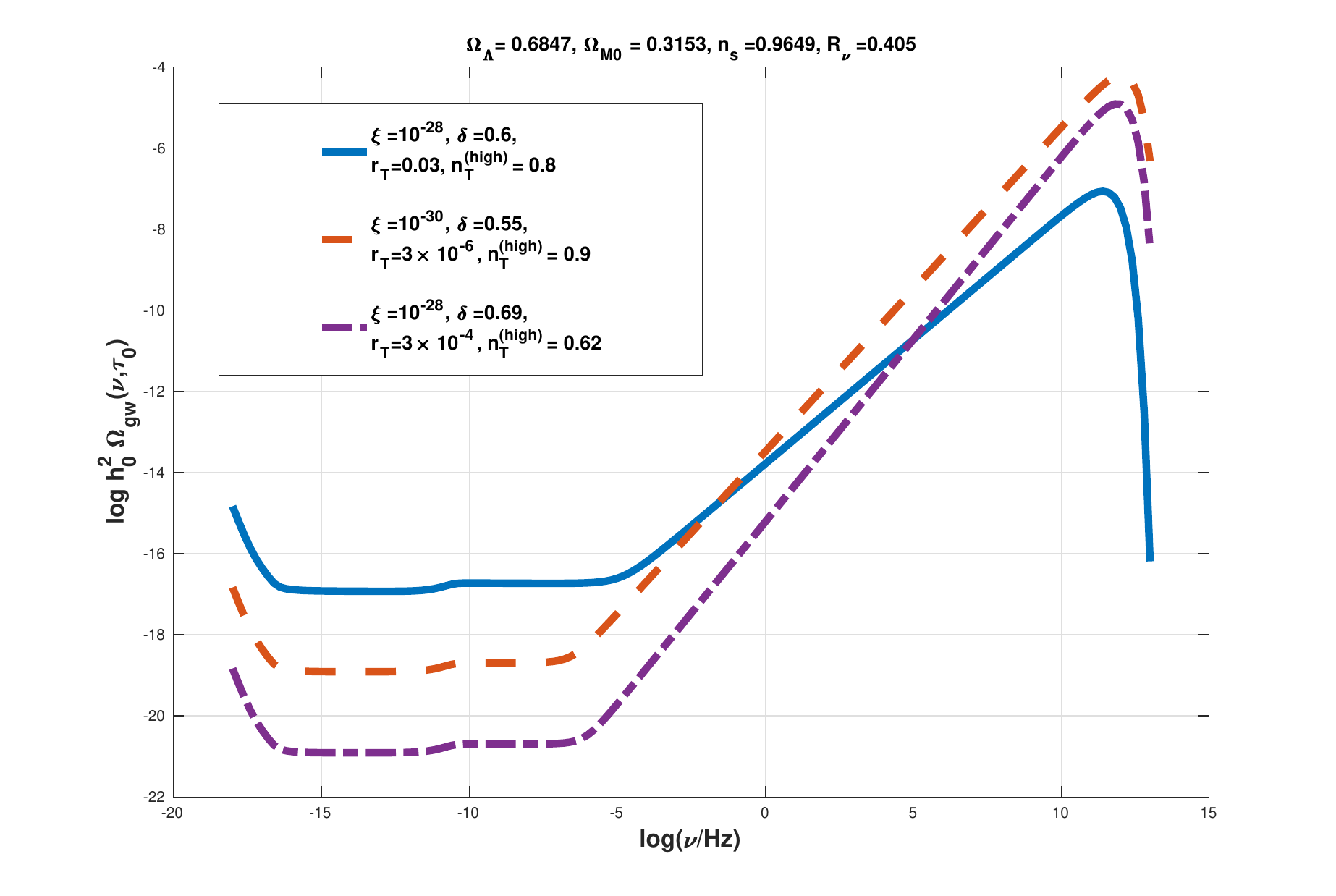}
\includegraphics[height=6.8cm]{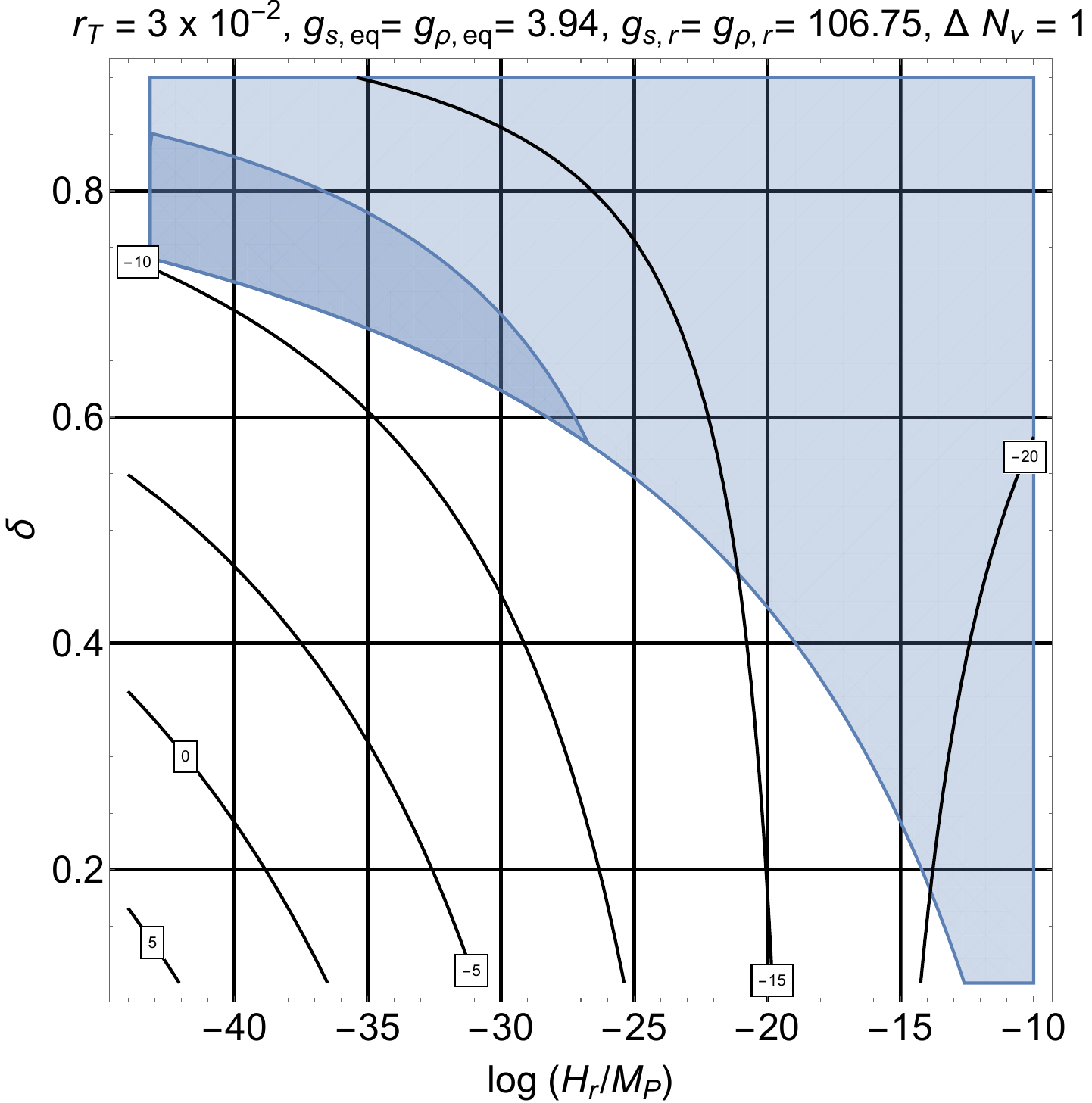}
\caption[a]{In the left plot  $h_{0}^2 \Omega_{gw}(\nu,\tau_{0})$ is illustrated as a function of the comoving frequency 
for three choices of  $r_{T}$; common logarithms are employed on both axes. In the plot at the right the shaded area denotes the region compatible with the BBN limit while darker shading follows by requiring that the resulting signal is ultimately detectable in the audio band; this is achieved by requiring, for instance, $10^{-13} \leq h_{0}^2 \, \Omega_{gw}(\nu_{LVK}, \tau_{0}) < 10^{-10}$ where $\nu_{LVK} = {\mathcal O}(100)$ is the frequency at which the sensitivity of wide-band detectors to diffuse backgrounds of gravitational radiation is maximal. In the complementary area of the shaded region the BBN is satisfied while $h_{0}^2 \Omega_{gw}(\nu,\tau_{0})< 10^{-13}$. The dashed curve in the left plot is barely compatible with the BBN bound although, overall, a reduction in $r_{T}$ does not necessarily entail a corresponding 
reduction of the maximum in the GHz region.}
\label{FIGsingle2}      
\end{figure}
Although ${\mathcal N}_{\rho}(r_{T}, \nu)$ exhibits a mild frequency dependence (mainly coming from neutrino free-streaming), for simplified analytic estimates this 
dependence can be approximately ignored. Along this perspective we may estimate ${\mathcal N}_{\rho} = {\mathcal O}(10^{-16.5})$ for $r_{T}\leq {\mathcal O}(0.06)$. In case a spectral energy density compatible with the one of Fig. \ref{FIGsingle1} we may deduce various pieces of information on the early expansion rate and on the various transitions that occurred throughout the evolution of the plasma. 

 \subsubsection{Invisible gravitons in the aHz region}
The results of Fig. \ref{FIGsingle1} do not rely on the specific value of $r_{T}$ and when 
$r_{T} \ll {\mathcal O}(0.06)$ the high frequency spike gets modified 
but does not disappear while the large-scale limits applicable to $\Omega_{gw}(\nu,\tau_{0})$
in the aHz region also affect the small-scale constraint as suggested for the first 
time in Ref. \cite{ST3,ST3a} (see also the discussion of section \ref{sec4}).
In Fig. \ref{FIGsingle2} we illustrate a sharp reduction of $r_{T}$ and a consequent suppression 
in the aHz region. When $r_{T}$ is reduced also the high frequency signal gets suppressed 
although this effect is easily counterbalanced by a smaller value of $H_{r}$. To clarify 
this point we first observe  that the values of $\xi = H_{r}/H_{1}$ (illustrated 
both in Fig. \ref{FIGsingle1} and \ref{FIGsingle2}) ultimately depend upon the assumed 
values of $r_{T}$: this happens since  $H_{1}$ is sensitive 
to the inflationary expansion rate so that eventually $\xi$ scales as 
$\xi \propto r_{T}^{-1/2}$ and it increases when $r_{T}$ gets progressively reduced. 
But $\nu_{max} = \xi^{(\delta-1)/[2 (\delta +1)]} \, \overline{\nu}_{max}$ depends 
on $\xi$ also because $\overline{\nu}_{max}$ itself scales as $r_{T}^{1/4}$ 
(see Eqs. (\ref{NK6})--(\ref{APA8}) and discussion therein). 
These different effects can be combined with the purpose 
of deducing the scaling of $h_{0}^2 \Omega_{gw}(\nu_{max}, \tau_{0})$ with $r_{T}$;
up to a numerical factor that depends on $\delta$ the result is:  
\begin{equation}
h_{0}^2 \Omega_{gw}(\nu_{max}, \tau_{0}) = {\mathcal B}(\delta)\, h_{0}^2 
\Omega_{R\,0} \,( r_{T}\, {\mathcal A}_{{\mathcal R}})^{\frac{2}{\delta +1}}\,
\biggl(\frac{H_{r}}{M_{P}}\biggr)^{2 \frac{\delta-1}{\delta+1}}, 
 \label{scalingrt}
\end{equation}
where ${\mathcal B}(\delta)= {\mathcal C}^4(g_{\rho}, g_{s}, \tau_{r}, \tau_{eq}) (16/\pi)^{(\delta-1)/(\delta +1)}/3$ is just a numerical factor that is not strictly essential in the forthcoming considerations. According to Eq. (\ref{scalingrt}), 
 for the same $H_{r}$ a reduction of $r_{T}$ entails an overall suppression of $h_{0}^2 \Omega_{gw}(\nu_{max}, \tau_{0})$. Conversely, when $r_{T}$ is kept fixed, a reduction of $H_{r}$ increases $h_{0}^2 \Omega_{gw}(\nu_{max}, \tau_{0})$ when $\delta < 1$;  when $\delta > 1$ a reduction 
of $H_{r}$ (for fixed $r_{T}$) further suppresses the spectral energy density.
This means, as anticipated,  that a reduction of $r_{T}$ may be compensated 
by an appropriate reduction of $H_{r}$ in the case when the post-inflationary expansion rate is slower than radiation. In the left plot of Fig. \ref{FIGsingle2} the high frequency spectral indices have been chosen exactly with the purpose of demonstrating that lower values of $r_{T}$ do not necessarily suppress the high-frequency signal that remains exceedingly large in comparison 
with ${\mathcal O}(10^{-17})$. 
In the right plot of Fig. \ref{FIGsingle2} we illustrate the parameter space in 
the plane defined by $H_{r}$ and $\delta$. As in Fig. \ref{FIGsingle1} the shaded area corresponds 
to the region allowed by the constraints stemming from BBN while 
the darker region comes from the bounds of the wide-band detectors 
operating in the audo band.
\begin{figure}[!ht]
\centering
\includegraphics[height=7cm]{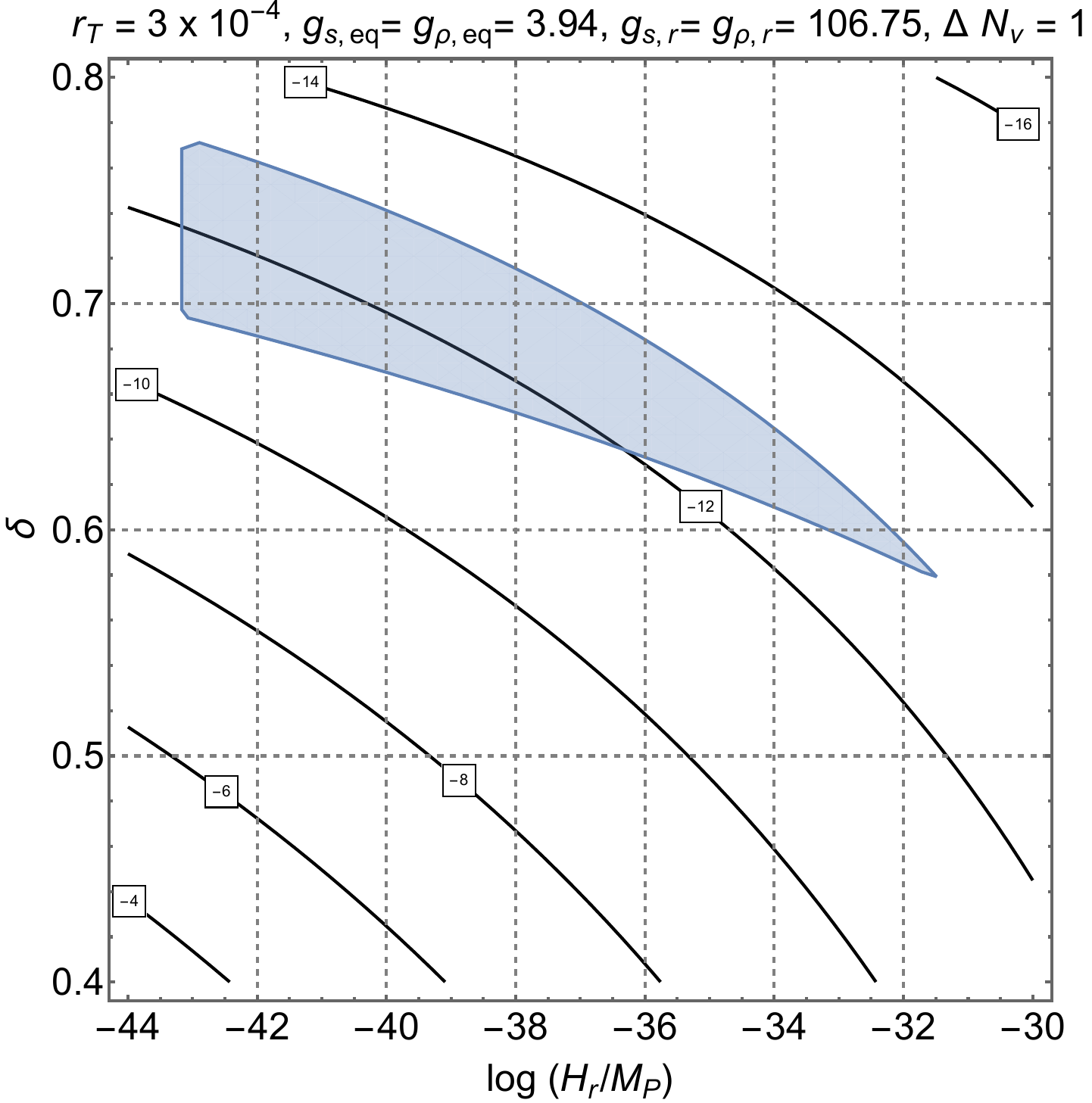}
\includegraphics[height=7cm]{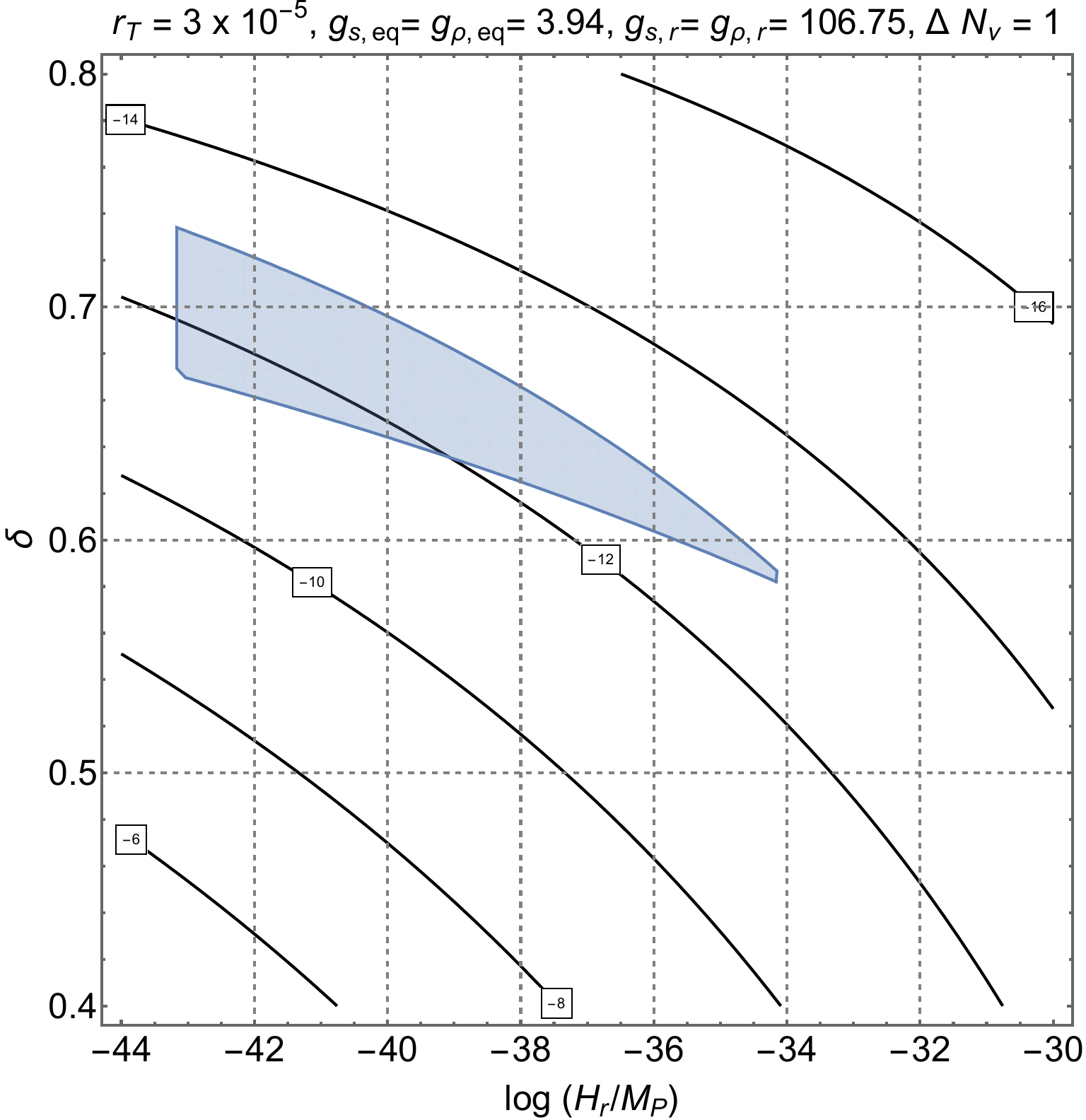}
\caption[a]{In the plane $(\log{H_{r}/M_{P}}$, $\delta$) we illustrate the allowed 
region of the parameter space where the BBN limit is enforced and the resulting signal is, in principle, 
detectable in the future by the wide-band detectors (see Eq. (\ref{BB3}) and discussion therein). The two plots correspond to different values of $r_{T}$ and are the ideal prosecution of Fig. \ref{FIGsingle2} (see, in particular, the right plot).}
\label{FIGsingle3}      
\end{figure}
 In the right plot Fig. \ref{FIGsingle2}  $r_{T} = 3 \times 10^{-2}$ while in the two plots of Fig. \ref{FIGsingle3} 
we selected instead $r_{T} = 3 \times 10^{-4}$ and $r_{T} = 3 \times 10^{-6}$ respectively. 
The shaded regions in Fig.  \ref{FIGsingle2} illustrate the intersection between the BBN bounds 
and the limits following from wide-band interferometers.
Indeed, a general requirement determining  the lowest value of $r_{T}$ is obtained from the current limits (summarized in Tab. \ref{TABLE1}) on the presence of relic graviton backgrounds in the audio band \cite{LIGO0,LIGO0a,LIGO0b,LIGO1,LIGO3}.
By following here this 
approach we adopted the condition 
\begin{equation}
10^{-13} \leq h_{0}^2 \, \Omega_{gw}(\nu_{LVK}, \tau_{0}) < 10^{-10}, \qquad\qquad \nu_{LVK} \leq {\mathcal O}(100)\,\, \mathrm{Hz},
\label{BB3}
\end{equation}
where $\nu_{LVK}$ denotes the Ligo-Virgo-Kagra frequency which can be estimated in terms of $\nu_{ref}$. The most sensitive region for the detection 
of relic gravitons in the audio band is, grossly speaking, below $0.1$ kHz since, in this band, the overlap reduction function has its first zero \cite{LIGO3}.
Equation (\ref{BB3}) requires, in practice, that the bounds coming from wide-band interferometers 
are satisfied while, in the same frequency range, $h_{0}^2 \, \Omega_{gw}(\nu,\tau_{0})$ is larger than ${\mathcal O}(10^{-13})$. We cannot foresee when 
the corresponding sensitivity will be reached by wide-band detectors 
but the requirement of Eq. (\ref{BB3}) follows from some of the optimistic claims suggested by the observational 
collaborations\footnote{Alternatively we may suppose that the relic gravitons backgrounds 
will not be accessible in the audio band;  in what follows we shall entertain a less pessimistic attitude which is mainly motivated by the steady increase of the sensitivity to relic gravitons in the last $20$ years. We must actually recall that in $2004$ wide-band detectors gave limits implying $h_{0}^2 \Omega_{gw}(\nu, \tau_{0}) < {\mathcal O}(1)$ \cite{LIGO0} while today the same limits improved by roughly $10$ orders of magnitude \cite{LIGO0a,LIGO0b}. } \cite{LIGO1}. 

\subsubsection{Bounds on the expansion rate}
In terms of Eqs. (\ref{SING9a})--(\ref{SING9b}) the BBN constraint  assumes a particularly simple analytical form and since the largest contribution to the integral 
comes from the bunch of frequencies ${\mathcal O}(\nu_{max})$, Eqs. (\ref{SING9a})--(\ref{SING9b}) can be used to set a limit on the integral of Eq. (\ref{HH2}); 
if we require, for instance, $h_{0}^2\,\Omega_{gw}(\nu_{max}, \tau_{0}) < 10^{-6}$
we obtain the following constraint in the $(\xi, \, \delta)$ plane:
\begin{equation}
\log{\xi} > \frac{(1+ \delta) (16 - r_{T})}{2[ 16 ( 1 - \delta) - r_{T} ( 2 - \delta)]} \bigl[ 6 + \log{{\mathcal N}_{\rho}(r_{T})}\bigr],
\label{SING12a}
\end{equation}
where we used the scaling of $(\nu_{max}/\nu_{r})$ with $\xi$, i.e.  $(\nu_{max}/\nu_{r}) \propto \xi^{- 1/(\delta +1)}$. As in Eq. (\ref{SING12a}) it is always true that $r_{T} \ll \delta$, 
Eq. (\ref{SING12a}) translates into
$\log{\xi} > -5.25 (1 + \delta)/(1 -\delta)$ (where we took $r_{T}=0.06$ and consequently estimated $\log{{\mathcal N}_{\rho}} = - 16.5$). Equation (\ref{SING12a}) implies then a lower bound on $\xi$. Indeed it can be argued that $\delta$ cannot get smaller than $1/2$ (see below Eq. (\ref{CRC8}) and discussion thereafter) and, in this case, we would have $\log{\xi} > -15.75$. If $\delta$ would decrease below $1/2$ the lower bound on $\xi$ would get larger: when $\delta=1/3$  the lower bound is given by  $\xi > 10^{-10.5}$, and so on. We finally remind, as already pointed out in section \ref{sec5}, that a further lower bound on $\xi$ is obtained by requiring that $\nu_{r} > \nu_{bbn}$; but then the bound is much less restrictive and it only demands $\xi > 10^{-38}$.

The limits obtained from Eqs. (\ref{SING9a})--(\ref{SING9b}) and (\ref{SING12a}) can be checked by 
direct numerical evaluation of the integral appearing in Eq. (\ref{HH2}). In the right plot of Fig. \ref{FIGsingle1} the shaded region illustrates the BBN constraint directly computed from Eq. (\ref{HH2}) and, in the same plot, the dashed curve describes the analytic bound coming from Eq. (\ref{SING12a}) for  $r_{T} \to 0.03$. The two determinations compare quite well and corroborate the approximation schemes of Eqs. (\ref{SING9a})--(\ref{SING9b}). 
We point out that in the right plot of Fig. \ref{FIGsingle1} the darker region corresponds to the BBN whereas the area defined by the lighter shading accounts for the LVK bounds of Tab. \ref{TABLE1} . 
 \begin{figure}[!ht]
\centering
\includegraphics[height=6.5cm]{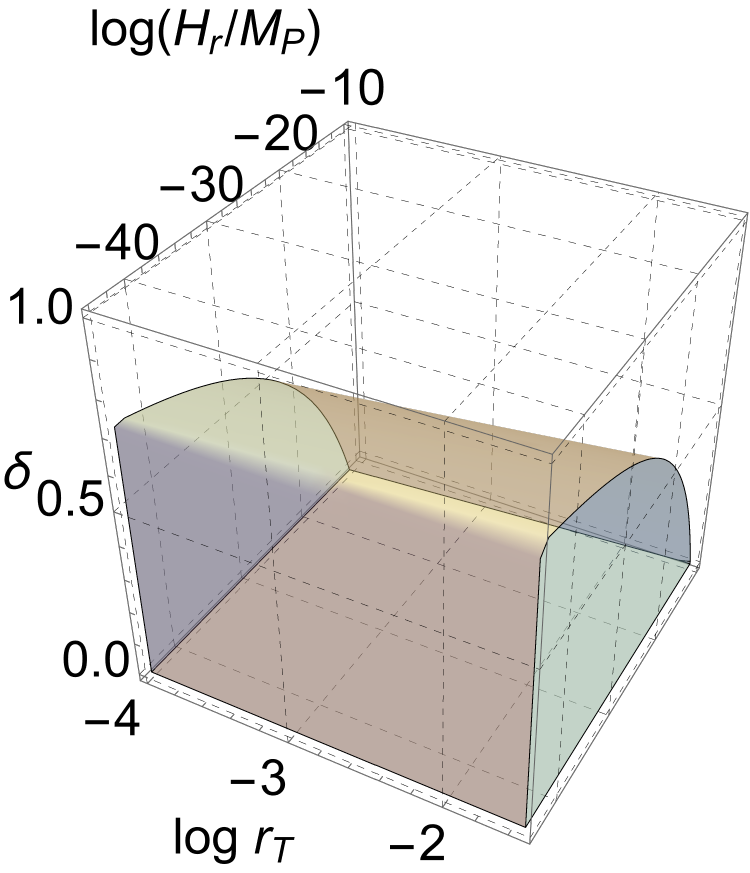}
\includegraphics[height=6.5cm]{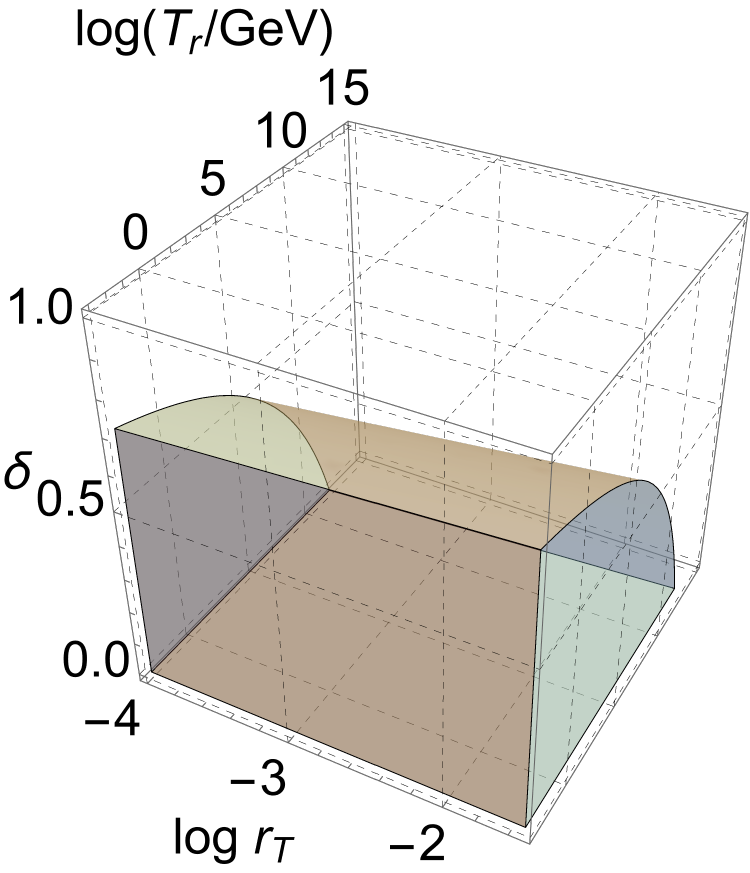}
\caption[a]{We illustrate the three-dimensional parameter space both in terms 
of $H_{r}$ and in terms of $T_{r}$. The limits illustrated in some of the previous plots are in fact 
two-dimensional slices of the three-dimensional parameter space illustrated in this figure.}
\label{FIGsingle5}      
\end{figure}
The limits illustrated in Figs. \ref{FIGsingle1}, \ref{FIGsingle2} and \ref{FIGsingle3} 
are two-dimensional slices of a three-dimensional parameter space where the values of $r_{T}$ are consistently reduced. The allowed 
region in three-dimensions is represented by volume in the space $(\delta,\, r_{T},\, H_{r})$.
To deduce the three-dimensional bounds we first observe, once more, that 
in spite of the complicated expansion timeline, the frequency $\nu_{r}$ is always 
related to $\overline{\nu}_{max}$ as $\nu_{r} = \sqrt{\xi} \,\,\overline{\nu}_{max}$. If we now require 
\begin{eqnarray}
\nu_{r} \geq \nu_{bbn} &=& \frac{\sqrt{H_{0} \, H_{bbn}}}{2 \pi} \, ( 2 \Omega_{R\,0})^{1/4} {\mathcal C}(g_{\rho}, g_{s}, \tau_{bbn}, \tau_{eq}), 
\nonumber\\
{\mathcal C}(g_{\rho}, g_{s}, \tau_{bbn}, \tau_{eq}) &=& (g_{\rho, \, bbn}/g_{\rho,\, eq})^{1/4}\, (g_{s, \, eq}/g_{s,\, bbn})^{1/3}, 
\end{eqnarray}
we obtain, in practice, that $H_{r} \geq H_{bbn}$. Recalling the considerations of Eqs. (\ref{QB27})--(\ref{QB28}) the simplest way of obtaining a bound on the expansion rate is to appreciate 
\begin{equation}
h_{0}^2 \Omega_{gw}(\nu_{max}, \tau_{0}) = \frac{128 \, \pi^3}{3 \, H_{0}^2 \, M_{P}^2} \nu_{max}^4 \, \overline{n}(\nu_{max}, \tau_{0}) \to  \frac{128 \, \pi^3}{3 \, H_{0}^2 \, M_{P}^2} \nu_{max}^4,
\label{PPEQ}
\end{equation}
since, by definition, $\overline{n}(\nu_{max}, \tau_{0}) = {\mathcal O}(1)$. From Eq. (\ref{PPEQ}) 
we can  write
\begin{equation}
h_{0}^2 \Omega_{R\,0} \,\, r_{T}\,\, {\mathcal A}_{{\mathcal R}}\,\, {\mathcal C}(g_{\rho}, g_{s}, \tau_{r}, \tau_{eq})
\, \xi^{2 (\delta-1)/(\delta+1)}.
\end{equation}
Because we are always requiring that $\nu_{r} \geq \nu_{bbn}$  the integral of Eq. (\ref{HH2})
can be approximated as follows
\begin{equation}
h_{0}^2 \, \int_{\nu_{r}}^{\nu_{max}} \Omega_{gw}(\nu,\tau_{0}) \,\frac{d \nu}{\nu} = 
\frac{{\mathcal N}_{\rho} \, r_{T}\, }{n_{T}^{(high)}} \biggl[ \biggl(\frac{\nu_{max}}{\nu_{r}}\biggr)^{n_{T}^{(high)}} - \biggl(\frac{\nu_{bbn}}{\nu_{r}}\biggr)^{n_{T}^{(high)}} \biggr],
\end{equation}
so that the BBN bound is now compactly expressed as:
\begin{equation}
r_{T}\, {\mathcal N}_{\rho} \, \biggl(\frac{\nu_{max}}{\nu_{r}}\biggr)^{n_{T}^{(high)}} \leq 5.61 \times 10^{-6}\, n_{T}^{(high)}\, \Delta N_{\nu}.
\end{equation}
A single post-inflationary stage expanding at a rate slower 
than radiation has fewer parameters in comparison with multiple stages of expansion (see e.g. Figs. \ref{FIGU0c} and \ref{FIGU0d}). In the simplest situation of a single post-inflationary stage,
the previous discussion clarifies that the three relevant parameters are: the tensor to scalar ratio $r_{T}$, 
the expansion rate during the post-inflationary evolution (related to $\delta$) and the 
Hubble rate at the onset of the radiation stage (i.e. $H_{r}/M_{P}$). We can eventually
trade $(H_{r}/M_{P})$ for the reheating temperature; the relation between the two 
quantities is obtained by assuming complete thermal equilibrium at $T_{r}$ and it is: 
\begin{equation}
\frac{H_{r}}{M_{P}} = \sqrt{ \frac{4 \pi^3 \, g_{\rho,\, r}}{45}} \biggl(\frac{T_{r}}{M_{P}}\biggr)^2.
\label{relT}
\end{equation}
The full three-dimensional parameter space is illustrated in Fig. \ref{FIGsingle5}: if the parameters 
fall within the shaded volume of Fig. \ref{FIGsingle5} all the relevant constraints 
are satisfied. The illustrative examples reported in Figs. \ref{FIGsingle3} and \ref{FIGsingle4}
can be viewed as two-dimensional projections of the three-dimensional parameter 
space of Fig. \ref{FIGsingle5}. From the shape of the spectral energy density it is then possible to infer the post-inflationary expansion rate and for a single post-inflationary phase the maximum of $h_{0}^2 \Omega_{gw}(\nu,\tau_{0})$ falls in the GHz region. If the expansion rate is more complicated the maximum can be from the GHz region to the audio band and this is 
the possibility examined in the following subsection.

\subsection{Spikes in the kHz domain}
When a single post-inflationary stage precedes the radiation epoch, $h_{0}^2\,\Omega_{gw}(\nu,\tau_{0})$ consists of three separate branches. If the timeline of the expansion rate contains different stages of expansion the spectral energy density may include multiple frequency domains and a maximum also develops below the MHz.
In the simplest situation there are two intermediate stages preceding the radiation-dominated phase. Besides  the standard aHz region and part of the intermediate branch (for $ \nu_{eq} < \nu < \nu_{r}$), the slopes in the two supplementary ranges (i.e. $\nu_{r} < \nu < \nu_{2}$ and  $\nu_{2} < \nu < \nu_{max}$) depend on the values of the expansion rates  (i.e. $\delta_{1}$ and  $\delta_{2}$) well before the electroweak epoch. 

\subsubsection{Maxima in the audio band}
\begin{figure}[!ht]
\centering
\includegraphics[height=6.3cm]{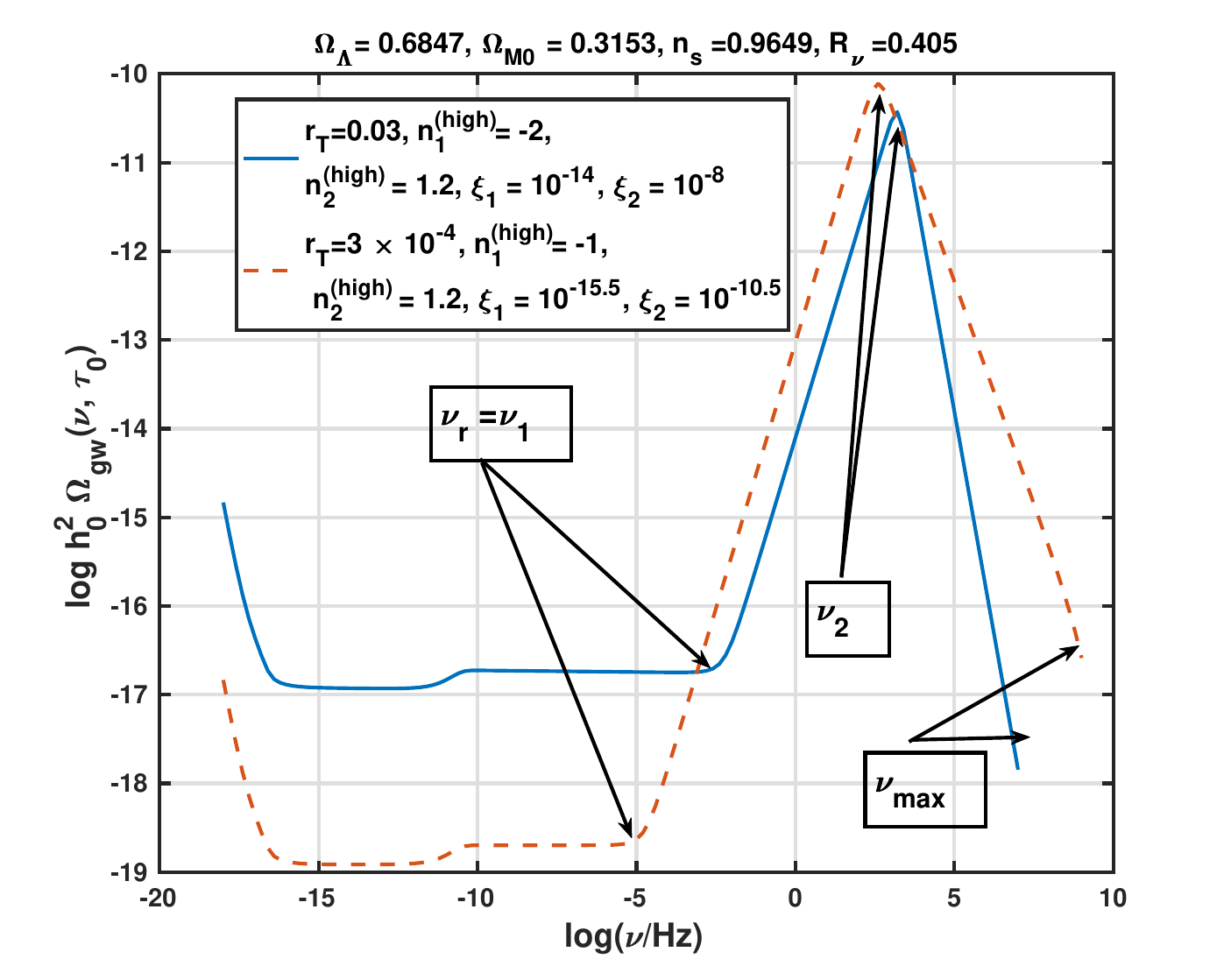}
\includegraphics[height=6.4cm]{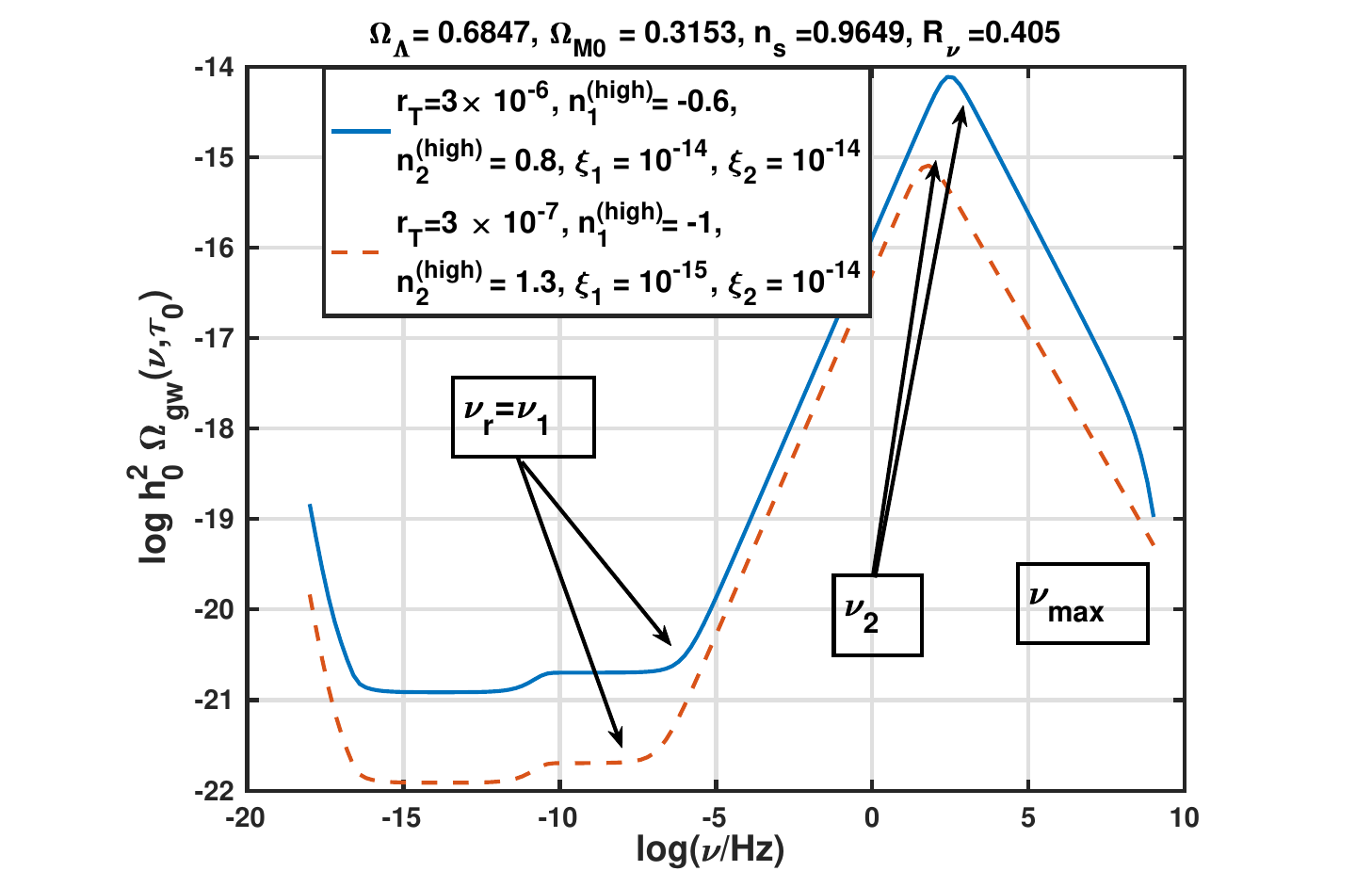}
\caption[a]{We illustrate the  peaks of the spectral energy density in the audio band.  The values of $r_{T}$ are similar to the ones of the previous plots although the spike of $h_{0}^2 \, \Omega_{gw}(\nu,\tau_{0})$now  falls in the audio band. The various parameters have been chosen by requiring that $\nu_{2}$ (i.e. the frequency of the spike) is such that $\nu_{2} = {\mathcal O}(\nu_{audio})$. This is one of the most constraining cases 
since the direct bounds of wide-band detectors fall into the audio band. Note that the maximum corresponds to frequencies $\nu = {\mathcal O}(\nu_{2})$ and not to $\nu_{max}$. Typical frequencies $\nu = {\mathcal O}(\nu_{max})$ are barely visible 
in rightmost region of the plot (see, in particular, the final part of the dot-dashed curve).}
\label{FIGsingle4}      
\end{figure}

In Fig. \ref{FIGsingle4} we illustrated few examples and the selected parameters also account for possible reductions of $r_{T}$. With a unified notation the spectral slopes (denoted in Fig. \ref{FIGsingle4} by $n^{(high)}_{1}$ and $n^{(high)}_{2}$) are:
\begin{equation}
n^{(high)}_{i} = \frac{32 - 4 r_{T}}{16 - r_{T}} - 2 \delta_{i}, \qquad \qquad r_{T} \ll 1, \qquad\qquad i = 1,\,\,2.
\label{AU1}
\end{equation}
The profiles of $h_{0}^2 \Omega_{gw}(\nu,\tau_{0})$ given in Fig. \ref{FIGsingle4} 
follow from the shape of the comoving horizon where, prior to radiation dominance, the post-inflationary evolution 
consists of two successive stages where the background first expands faster than radiation (i.e. $\delta_{1}>1$) and then slows down (i.e. $\delta_{2} < 1$). We have from Eq. (\ref{AU1}) 
that the spectral energy density decreases for $\nu> \nu_{2}$ (i.e. 
$n_{1}^{(high)} <0$) while it increases at lower frequencies 
(i.e. $n_{2}^{(high)} >0$ for $\nu< \nu_{2}$).  If $r_{T}\ll 0.03$ \cite{RR1,RR2,RR3}  Eq. (\ref{AU1}) 
reduces to:
\begin{equation}
n^{(high)}_{i} = 2 ( 1 - \delta_{i}) + {\mathcal O}(r_{T}), \qquad\qquad i = 1,\,\,2,
\label{AU2}
\end{equation}
 and  $n_{1}^{(high)} = 2( 1 -\delta_{1})<0$ for the wavelengths reentering before $a_{2}$ while for the wavelengths reentering between $a_{2}$ and $a_{r}$ we would have $n_{2}^{(high)} = 2( 1 -\delta_{2})>0$. In case the timeline 
 is reversed (and $\delta_{1}<1$ while $\delta_{2} >2$) 
instead of a spike $h_{0}^2 \Omega_{gw}(\nu,\tau_{0})$ exhibits a 
trough but this timeline would be comparatively less constrained than the one of Fig. \ref{FIGsingle4}.  
All in all, recalling the parametrization of Eqs. (\ref{SING9a})--(\ref{SING9b}),  the two high frequency 
branches of the spectral energy density can be parametrized as:
\begin{eqnarray}
h_{0}^2\,\Omega(\nu,\tau_{0}) &=& {\mathcal N}_{\rho}(r_{T},\nu) \biggl(\frac{\nu}{\nu_{r}}\biggr)^{n^{(high)}_{2}}, \qquad\qquad \nu_{r} < \nu < \nu_{2}, 
\label{AU3a}\\
h_{0}^2\,\Omega(\nu,\tau_{0}) &=& {\mathcal N}_{\rho}(r_{T},\nu) \biggl(\frac{\nu_{2}}{\nu_{r}}\biggr)^{n_{2}^{(high)}}\biggl(\frac{\nu}{\nu_{2}}\biggr)^{-|n^{(high)}_{1}|}, \qquad\qquad \nu_{2} < \nu < \nu_{max},
\label{AU3b}
\end{eqnarray}
where we are implicitly assuming that $n^{(high)}_{1}<0$ and $n^{(high)}_{2}>0$. The spectral energy density given of Eqs. (\ref{AU3a})--(\ref{AU3b}) exhibits a maximum for $\nu= {\mathcal O}(\nu_{2}$) but when $\delta_{1} \to 1$ the maximum is replaced by a plateau since $h_{0}^2\,\Omega_{gw}(\nu, \tau_{0})$ flattens out (i.e. $n^{(high)}_{1} \to 0$  for $\nu > \nu_{2}$) \cite{MGshift}.  We then illustrated the situations that are phenomenologically more constraining; on this basis it is now possible to derive further limits on $r_{T}$ under the hypothesis of an expansion history including at least two different post-inflationary stages different from radiation (i.e. $\delta_{i} \neq 1$).

\subsubsection{Again on the maximal frequency}
The maximal frequency of the relic gravitons depends on $H_{1}$ but a modified 
post-inflationary evolution may artificially increase the value of $\nu_{max}$ by few orders of magnitude and potentially contradict the quantum bound of Eq. (\ref{HH3}). The expansion histories leading to $\nu_{max} \gg \mathrm{THz}$ 
must then be rejected since the violations of the quantum bound also entail 
a violation of the limits set by BBN in the vicinity of $\nu_{max}$.
To be more specific we now assume that between the 
end of inflation and the dominance of radiation there are $n$ different stages of expansion 
that are arbitrarily different from radiation; this is, again, the 
general case illustrated in Fig. \ref{FIGU0c} and \ref{FIGU0d}.  We know from Eq. (\ref{APA8a}) that the value of the maximal frequency becomes, in this case:
\begin{equation}
\nu_{max} = \overline{\nu}_{max}\, \prod_{i=1}^{n-1} \,\, \xi_{i}^{\beta_{i}},\qquad \xi_{i} = H_{i+1}/H_{i} < 1,
\label{CRC4}
\end{equation}
where $\xi_{i}$ and $\beta_{i}  = (\delta_{i} -1)/[2\, (\delta_{i} + 1)]$ measure, respectively, the duration of each of the post-inflationary stages and the corresponding expansion rate. When all the $\beta_{i} \to 0$ (i.e. $\delta_{i} \to 1$),  the evolution is dominated by radiation from $H_{1}$ down to $H_{eq}$ and $\nu_{max} \to \overline{\nu}_{max}= {\mathcal O}(300)\, \mathrm{MHz}$. Conversely the value of $\nu_{max}$ given in Eq. (\ref{CRC4}) may exceed $300$ MHz provided at least one of the various $\delta_{i}$ gets smaller than $1$.  
In the case of $n$ intermediate stages preceding  the dominance of radiation at $a_{r}$, between $\nu_{max}$ and $\nu_{r}$ there will be $n-2$ intermediate frequencies corresponding to specific breaks in the spectral energy density. The post-inflationary contribution to $\nu_{max}$ is then maximized 
when the $\delta_{i}$ and $\xi_{i}$ take their minimal values:
\begin{equation}
\delta_{1} = \delta_{2}=\, .\,.\,.= \delta_{n-1} =  \overline{\delta} = 1/2, \qquad\qquad \xi_{1} \, \xi_{2}  \, .\,.\,. \xi_{n-1} =\xi_{r} = H_{bbn}/H_{1}.
\label{CRC8}
\end{equation}
The common value of the various $\delta_{i}$ corresponds to the slowest expansion rate 
of the primeval plasma. For instance in a perfect fluid the maximal value of the barotropic index (be it $w_{max}$) corresponds to the expansion rate, i.e.  $\delta_{min} = 2/(3 w_{max} +1)$ and since, at most, $w_{max} \to 1$ we obtain, as suggested in Eq. (\ref{CRC8}) $\delta_{min} \to 1/2$. The expansion rate can also be slower than radiation when the energy-momentum tensor is dominated by the oscillations of the inflaton and 
if assume that the minimum of the potential is located in $\varphi =0$,
$V(\varphi)$ can be parametrized as $V(\varphi)\simeq V_{1} \Phi^{2 q}$ (where, as usual, $\Phi = \varphi/M_{P}$). The averaged evolution of the comoving horizon can then mimic the timeline of a stiff epoch and the graviton spectra. Recalling Eqs. (\ref{TS22a})--(\ref{TS22b}), during the coherent oscillations of $\varphi$ the energy density of the scalar field is roughly constant \cite{turn1,turn2,turn3,turn4} and, in average, the expansion rate is $\delta=  (q+1)/(2 q-1)$. Thus $\delta_{min}$ is still ${\mathcal O}(1/2)$ and this happens when $q\gg 1$. When all the $\delta_{i}$ are equal the product of all the $\xi_{i}$ (denoted by $\xi_{r}$ in Eq. (\ref{CRC8})) is ultimately raised to the same common power implying that the contribution of the whole decelerated stage of expansion of Eq. (\ref{CRC4}) is maximized by a single expanding stage characterized by $\overline{\delta} = \delta_{min} < 1$.
Thanks to Eqs. (\ref{CRC8}) we therefore obtain the following bound on $\nu_{max}$
\begin{equation}
\nu_{max} < 10^{6} \, \biggl(\frac{H_{max}}{M_{P}}\biggr)^{2/3}\,\, \biggl(\frac{h_{0}^2 \Omega_{R0}}{4.15\times 10^{-5}}\biggr)^{1/4} \, \, \mathrm{THz},
\label{CRC13}
\end{equation}
where it has been assumed that $H_{r} \to H_{bbn} = 10^{-42} \, M_{P}$. If we now consider together Eqs. (\ref{CRC13}) and (\ref{HH3}) we must conclude that the quantum bound of Eq. (\ref{HH3}) is always more constraining \cite{MGF}. 

\subsection{Interplay between low-frequency and high frequency constraints}
The previous considerations suggest an interplay between the 
low-frequency constraints and 
the high frequency bound. We are going to examine first the
bounds on the inflationary potential coming from the high-frequency region and 
their connection with the low-frequency limits of section \ref{sec4}.
In the second part of this discussion we swiftly describe some notable quantum mechanical aspects of the relic gravitons at high frequencies. 

\subsubsection{General bounds on the inflationary potential}
\begin{figure}[!ht]
\centering
\includegraphics[height=6cm]{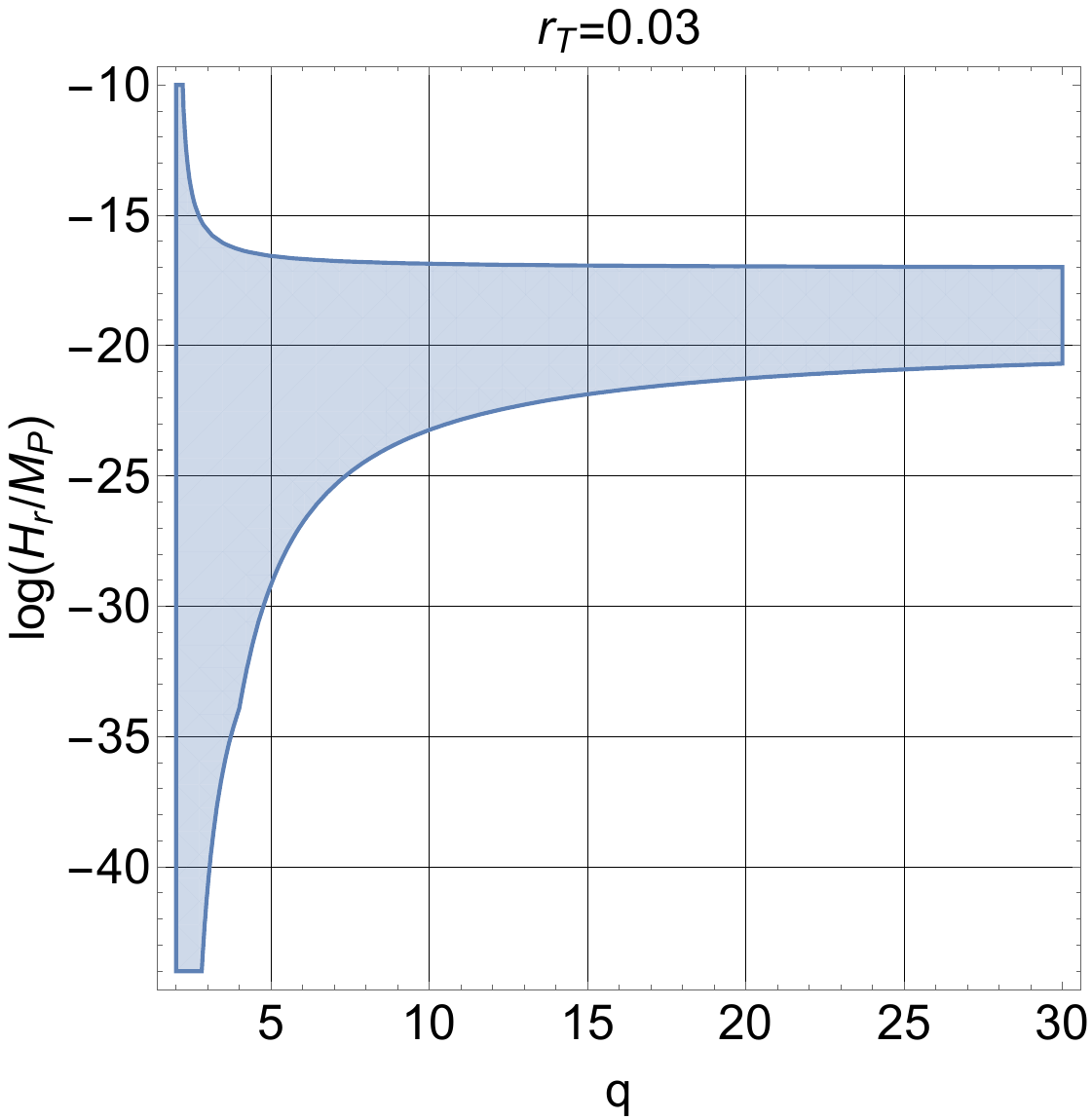}
\includegraphics[height=6cm]{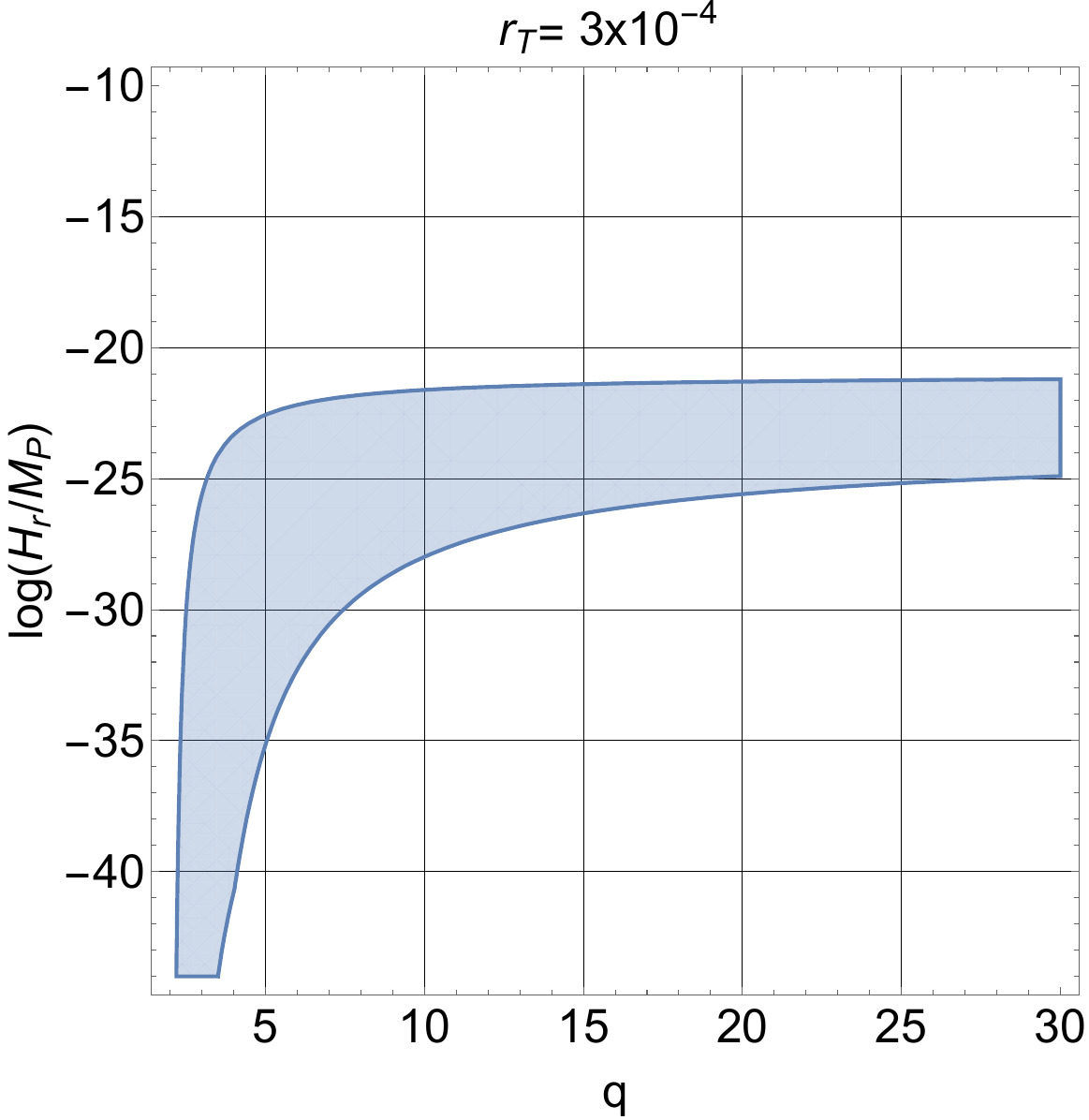}
\caption[a]{We illustrate the bounds on $q$ by using the results 
of Eqs. (\ref{BOUND1})--(\ref{BOUND2}) together with the BBN 
bound. We are here assuming an inflationary potential characterized by a flat 
plateau for $\Phi = \varphi/\overline{M}_{P} \gg 1$ and by an oscillating 
stage for $\Phi < 1$ where $V(\Phi) = V_{0} \Phi^{2 q}$.  }
\label{FIGsingle6}      
\end{figure}
Let us suppose, as suggested in section \ref{sec4}, that the inflationary 
potential interpolates between two complementary regimes:
it is inflationary for $\Phi = \varphi/\overline{M}_{P} \gg 1$ while it oscillates 
as $V(\Phi) = V_{0} \Phi^{2 \, q}$ in the limit $\Phi \ll 1$. 
Few examples of this class of potentials have been illustrated in 
section \ref{sec4} (see, in particular, Eqs. (\ref{POTEX1})--(\ref{POTEX2}) and (\ref{POTEX3})--(\ref{POTEX4})).
In this situation there is no absolute bound on the value of $q$ but the parameter space 
of the model is effectively three-dimensional: $r_{T}$ controls 
the low-frequency normalization, $H_{r}/M_{P}$ determines the reheating 
scale and $q$ fixes the high frequency spectral index 
of $h_{0}^2 \Omega_{gw}(k, \tau_{0})$ according to 
\begin{equation}
n_{T}^{(high)} = \frac{32 - 4 r_{T}}{16 - r_{T}} - \frac{ 2 ( q +1)}{2 q -1}.
\label{BOUND1}
\end{equation}
From the specific form of the spectral energy density at high-frequencies 
we may require that the BBN constraint is satisfied 
while, in the audio band we may require 
\begin{equation}
10^{-15} \leq h_{0}^2 \Omega_{gw}(\nu_{LVK}, \tau_{0}) \leq 10^{-10}, \qquad \nu_{LVK} \leq {\mathcal O}(100)\, \mathrm{Hz}.
\label{BOUND2}
\end{equation}
This condition roughly guarantees the enforcement of the constraints of Tab. \ref{TABLE1} together with a potentially detectable signal (in the far future); Eq. (\ref{BOUND2}) 
has the same content of Eq. (\ref{BB3}) with the difference that we consider here a slightly 
larger interval in the spectral energy density. For differnet values of $r_{T}$ the bounds are modified and this aspect illustrates once more the interplay between the constraints coming from different frequency regimes, as originally suggested in \cite{ST3,ST3a}. The bounds on $r_{T}$  appear in the aHz region whereas the bounds related to $n_{T}^{(high)}$ come from the high frequency range. In Fig. \ref{FIGsingle6} we consider two different values of $r_{T}$ and draw the allowed region in the plane $(q,\,H_{r}/M_{P})$. The value of $H_{r}$ must always be larger than $10^{-44}\, M_{P}$ roughly 
corresponding to the BBN scale. In Fig. \ref{FIGsingle6} we considered 
$r_{T}$ as a free parameter even though its potential suppression occurs 
via the total number of $e$-folds which is always larger than $60$ as long as $q >1$.
This result should compared also with the low-frequency limits on $q$, $r_{T}(k)$ and $n_{s}(k)$ discussed in section \ref{sec4}; see also, in this respect, the analysis of Ref. \cite{MGshift2}.

\subsubsection{Quantum sensing and the relic gravitons}
We already established that the quantum bound is more constraining than the classical limit of Eq. (\ref{CRC13}) and this is true in general terms since Eq. (\ref{HH3}) does not depend on the specific timeline of the post-inflationary evolution but just on the observation that at the maximal frequency only one graviton pair is produced. It makes then sense to normalize the chirp amplitude directly in the THz domain\footnote{The spectral energy density in critical units at the present 
time and the chirp amplitude $h_{c}(\nu, \tau_{0})$ are 
related as $\Omega_{gw}(\nu,\, \tau_{0}) = 2\, \pi^2 \, \nu^2 h_{c}^2/( 3 \, H_{0}^2)$.}; with this logic the bound on $\nu_{max}$ of Eq. (\ref{HH3}) can be converted into a limit on $h_{c}$. If the spectral energy is normalized in the THz domain with a putative high frequency slope $\nu^{n^{(high)}_{T}}$,  the minimal chirp amplitude required for the direct detection of cosmic gravitons must comply with the following limit
\begin{equation}
h^{(min)}_{c}(\nu, \tau_{0}) < 8.13 \times 10^{-32} \biggl(\frac{\nu}{0.1\, \mathrm{THz}}\biggr)^{-1 +n^{(high)}_{T}/2}.
\label{MM1}
\end{equation}
 This means that a sensitivity ${\mathcal O}(10^{-20})$ or even ${\mathcal O}(10^{-24})$ in the chirp amplitude for frequencies in the MHz or GHz regions is irrelevant for a direct or indirect detection of high frequency gravitons. It has been suggested long ago that microwave cavities \cite{CAV1,CAV2,CAV3,CAV4,CAV5,CAV6} operating in the MHz and GHz regions could be employed for the detection of relic gravitons \cite{EXP1,EXP2,EXP3}. The same class of instruments has been also invoked in \cite{FA1,FA2,FA3,FA4} with the difference that, unlike previous studies (more aware of the potential sources and of the instrumental noises), the required chirp amplitudes are now optimistically set in the range $h_{c}^{(min)} = {\mathcal O}(10^{-20})$ for arbitrarily high frequencies\footnote{To achieve $h^{(min)}_{c} = {\mathcal O}(10^{-20})$ is technologically interesting; from the physical viewpoint this minimal sensitivity is more than $10$ orders magnitude larger than the requirements associated with the direct detection of cosmic gravitons.}. Equation (\ref{MM1}) also clarifies why $h_{c}^{(min)}$ must be at least ${\mathcal O}(10^{-32})$ (or smaller) for a potential detection of cosmic gravitons in the THz domain.  In a more optimistic perspective,  for $n^{(high)}_{T} > 2$ the largest signal occurs at the largest frequency, for $n^{(high)}_{T} \leq 2$ the frequencies smaller than the THz are observationally convenient. If we consider, for instance, the case $n^{(high)}_{T} \to 1$ (which is, incidentally, typical of a post-inflationary stiff phase when we neglect here all the possible logarithmic enhancements) we would have that the chirp amplitude at in the MHz range could be ${\mathcal O}(10^{-28})$ (as also proposed in Refs. \cite{CAV5,CAV6} on the basis of more experimental considerations). Furthermore, when $n_{T}^{(high)} \to 2$ (typical of the ekpyrotic scenario) we would have instead that $h_{c}(\nu,\tau_{0})$ is the same at higher and smaller frequencies  \cite{EK1,EK2}. Finally for $n^{(high)}_{T} \to 3$ (as it happens in the case of the pre-big bang scenario \cite{EK3,EK3a}) the chirp amplitude at lower frequencies gets even smaller. We have therefore 
a trade-off between the optimal frequency, the features of the signal and the 
noises (especially the thermal one) indicating that the highest possible frequency 
(close to $\nu_{max}$) is not always the most convenient. Also this aspect should be 
taken into account if the goal is really an accurate assessment of the required sensitivities of high frequency instruments.

The limits following from Eq. (\ref{HH3})  are also relevant for the analysis of the statistical properties of the relic gravitons and, in particular, of their degrees of first- and second-order coherence. These observables follow by generalizing the appropriate Glauber correlators \cite{HBT0a,HBT0b} to the expectation values of tensor fields (see Refs. \cite{HBT1} and discussions therein); besides the physical aspects (discussed over a decade ago \cite{HBT2}) the main technical difference between the gravitons and the photons involves the polarization structure of the correlation functions. Mutatis mutandis the physical idea is however similar: if cosmic gravitons are detected by independent interferometers the correlated outputs are employed to estimate the degrees of second-order coherence. The analysis of the interplay between the Hanbury Brown-Twiss (HBT) interferometry and the high frequency gravitons has been recently discussed in Ref.  (see also \cite{HBT1,HBT2}); for the present purposes we avoid the polarization dependence and introduce 
 the {\em single-particle} (inclusive) density \cite{HBT3a,HBT3b}
\begin{equation}
\rho_{1}(\vec{k}) = \langle \widehat{A}^{\,\dagger}(\vec{k})\,\,\widehat{A}(\vec{k})\rangle, \qquad \widehat{A}(\vec{k}) 
=\int d^{3} p \,\, \widehat{a}_{\vec{p}} \,\, {\mathcal W} (\vec{k} - \vec{p}),
\label{HH1a}
\end{equation}
where $\widehat{A}(\vec{k},\tau)$ (and its Hermitian conjugate) are just a set of creation and annihilation operators that are non-zero inside the volume of the particle source associated with the three-dimensional integral (in real space)
of an appropriate window function ${\mathcal W}(\vec{x})$. By definition, $[\widehat{A}(\vec{k}),\,\,\widehat{A}^{\dagger}(\vec{k})] = \int d^{3} x |{\mathcal W}(\vec{x})|^2$. In the theory of Bose-Einstein interference \cite{HBT3a,HBT3b}
 \begin{equation}
 \rho_{2}(\vec{k}_{1}, \vec{k}_{2}) = \langle \widehat{A}^{\,\dagger}(\vec{k}_{1})\,\,\widehat{A}^{\,\dagger}(\vec{k}_{2})\,\,\widehat{A}(\vec{k}_{2})\,\,\widehat{A}(\vec{k}_{1})\rangle,
 \label{HH2a}
 \end{equation}
 is the {\em two-particle} inclusive density and according to Eqs. (\ref{HH1a})--(\ref{HH2a}) the normalized second-order correlation function
 \begin{equation}
 C_{2}(\vec{k}_{1}, \vec{k}_{2}) = \frac{\rho_{2}(\vec{k}_{1}, \vec{k}_{2})}{\rho_{1}(\vec{k}_{1})\,\,\rho_{1}(\vec{k}_{2})} \to 3 + {\mathcal O}\biggl(\frac{1}{\sqrt{\overline{n}(k_{1})\, \overline{n}(k_{2})}}\biggr),
 \label{HH3a}
 \end{equation}
 estimates the degree of second-order coherence \cite{HBT1,HBT2}. The value of $C_{2}(\vec{k}_{1}, \vec{k}_{2})$ is  always enhanced in comparison with so-called Poissonian limit so that the statistics of the relic gravitons is always super-Poissonian and generally super-chaotic. Indeed in the limit of a large number of graviton pairs $C_{2}(\vec{k}_{1}, \vec{k}_{2}) \to 3$ whereas $C_{2}(\vec{k}_{1}, \vec{k}_{2}) \to 2$ in the case of a chaotic mixture. This result is slightly refined by taking into account the polarisation structure of the correlators, as already discussed in the past \cite{HBT1,HBT2}; in this case $C_{2}(\vec{k}_{1}, \vec{k}_{2}) \leq 3$ but the statistics always remains super-Poissonian.  While the statistical properties of the relic gravitons determine the degrees of first- and second-order coherence, their potential detection depends from the achievable $h_{c}^{(min)}$ which is not the same in different ranges of comoving frequency. It is therefore not surprising that the analyses of the Bose-Einstein correlations overlooking the physical properties of the cosmic gravitons are inconclusive and often superficial. All in all we demonstrated, both at the classical and quantum level, that the largest frequency of the relic gravitons never exceeds the THz band while the minimal detectable chirp amplitude should be at least ${\mathcal O}(10^{-32})$ (or smaller) if the (hypothetical) detectors in the THz domain could claim (even in principle) the detection of a relic signal. However, if the pivotal frequencies of the instruments are reduced from the THz to the GHz (or even MHz) band the minimal required chirp amplitude may increase.

\subsubsection{The quantumness of relic gravitons}
The relic gravitons  are characterized by autocorrelation functions that are not invariant under a shift of the time coordinate (see section \ref{sec3} and discussion therein); 
this is why Eqs. (\ref{QB13})--(\ref{QB13}) do not only depend 
upon $\tau_{1} - \tau_{2}$ but also upon $\tau_{1} + \tau_{2}$. This property is 
 rooted into the quantum mechanical origin or the corresponding particles: the initial travelling waves 
associated with the quantum fluctuations turn eventually into a collection of standing 
waves because of the evolution of the underlying background geometry. 
The formation of standing waves (also called Sakharov oscillations) 
simply means that relic gravitons are produced in entangled states of opposite 
(comoving) three-momenta  according to the unitary process summarized by Eqs. (\ref{QB23})--(\ref{QB24}).
Although the field is initially in a pure state its entropy may increase if some information is lost and, for this reason, quantum measurements are somehow intrinsically associated with a loss 
of information. When observations are performed (for instance by means of HBT interferometry \cite{HBT1,HBT2}) the 
sign of the three-momentum cannot be determined; in other words only one of the members of pair is observed 
while the other one is in practice unobservable. 
The operators associated with the opposite momenta of a graviton pair effectively act on separated subspaces of the total Hilbert space of the problem. We can then focus on a single pair of gravitons so that the associated operators will be $\widehat{b}_{+}$ and $\widehat{b}_{-}$ (i.e. the signal and the idler mode in a quantum optical context \cite{MANDL,HBT4}). In this two-mode approximation the final state of the particle production process schematically corresponds to
 \begin{equation}
| z \rangle = \Sigma(z) | 0_{+} \, 0_{-} \rangle, \qquad \Sigma(z) = e^{z^{\ast} \, \widehat{b}_{+} \, \widehat{b}_{-} - 
z\widehat{b}_{+}^{\dagger} \, \widehat{b}_{-}^{\dagger}},
\label{SQ1}
\end{equation}
where $[\widehat{b}_{I}, \widehat{b}_{J}^{\dagger}] = \delta_{I,\,J}$; here $I, \, J = +,\,-$ and the $\pm$ are related to the sign of a (single) comoving three-momentum. The operator $\Sigma(z)$ can be factorized as the product of the exponentials of $L_{0}$ and $L_{\pm}$
 \begin{equation}
\Sigma(z) = \exp{\biggl[ -\frac{z}{|z|} \tanh{|z|} \,\,\, L_{+}\biggr]}\times \exp{[- 2 \ln{\cosh|z|}\,\,\, L_{0}]}\times \exp{\biggl[ \frac{z^{*}}{|z|} \, \tanh{|z|} \,\,\,L_{-}\biggr]},
\label{ENTR1}
\end{equation}
where $L_{0}$ and $L_{\pm}$ are the generators of the $SU(1,1)$ Lie algebra: 
\begin{equation}
L_{+} = \widehat{b}_{+}^{\,\dagger}\,\widehat{b}_{-}^{\,\dagger}, \qquad\qquad L_{-} = \widehat{b}_{+}\,\widehat{b}_{-},
\qquad\qquad L_{0} = \frac{1}{2} ( \widehat{b}_{+}^{\,\dagger} \,\widehat{b}_{+} + \widehat{b}_{-}\,\widehat{b}_{-}^{\,\dagger}),
\label{ENTR2}
\end{equation}
obeying the corresponding commutation relations $[L_{+}, \, L_{-}] = - 2 \, L_{0}$ and $[L_{0}, \, L_{\pm}] = \pm \, L_{\pm}$. The operator $\widehat{b}_{+}^{\, \dagger}$ creates a graviton of momentum $+\vec{k}$ while $\widehat{b}_{-}^{\,\dagger}$ creates a graviton with momentum $- \vec{k}$; the Fock states are an appropriate basis for the irreducible representations\footnote{An equivalent basis for the irreducible representations of $SU(1,1)$ is provided by the vectors  $|Q\,n_{t}\rangle$ where 
$Q = n_{+} - n_{-}$ is the total charge and $n_{t} = n_{+} + n_{-}$
is the total number of charged species. The vectors  $|Q\,n_{t}\rangle$  are the standard basis of the irreducible representations $T^{+k}$ of $SU(1,1)$ where $k$ is the principal quantum number and $m$ is the magnetic quantum number, i.e. the eigenvalue of $L_{0}$. The Casimir operator of the $SU(1,1)$ group can be notoriously written as $C = L_{0}\, (L_{0} -1) - L_{+} \, L_{-}$ so that, eventually, $C | k\, m \rangle = k(k-1)  | k\, m \rangle $. The commuting set of observables is formed in this case by the Casimir operator and by $L_{0}$; $k$ is usually 
referred to as the Bargmann parameter \cite{BARG}. 
The negative series $T^{-k}$ is symmetric under the exchange $n_{+} \to n_{-}$ while the principal (continuous) series will not play a specific role in the present considerations. In terms of $k$ and $m$ we have that the total charge and the total number of particles are given, respectively, by 
$Q= 2 k -1$ and by $n_{t} = 2m -1$. Note finally that the Bargmann parameter \cite{BARG} should not be confused with the modulus of the comoving three-momentum; this is actually impossible since the basis of the irreducible representations employed here is the one given in Eq. (\ref{ENTR3}) and not the Bargmann basis. } of the $SU(1,1)$
\begin{equation}
| n_{+}\,\, n_{-} \rangle = \frac{(\widehat{b}_{+}^{\,\dagger})^{n_{+}}}{\sqrt{n_{+}\, !}}\,\, \frac{(\widehat{b}_{-}^{\,\dagger})^{n_{-}}}{\sqrt{n_{+}\, !}}\,\, |0_{+}\,\, 0_{-}\rangle. 
\label{ENTR3}
\end{equation}
When the relic gravitons are produced in pairs of opposite three-momenta we have that $n_{+} = n_{-}$; furthermore the action of the group generators  on the two-mode vacuum is given by $L_{-} \, | 0_{+} \, 0_{-}\rangle =0$  while for $L_{0}$ we have instead $L_{0} \, | 0_{+} \, 0_{-}\rangle = | 0_{+} \, 0_{-}\rangle/2$. The density matrix associated with the state given in  Eq. (\ref{ENTR1}) 
is $\widehat{\rho} = | \,z \,\rangle \langle \,z\,|$  and since 
 the states are correctly normalized we can conclude that $\mathrm{Tr} \widehat{\rho}^2 = \mathrm{Tr} \widehat{\rho} =1$. In the Fock basis of Eq. (\ref{ENTR3}) the explicit form of $| z\rangle$ is:
\begin{equation}
| z \rangle = \frac{1}{\cosh{r}}  \sum_{n_{\pm}=0}^{\infty}\,\, \bigl(- e^{i \theta } \,\,\tanh{r}\bigr)^{(n_{+}+ n_{-})/2}\, \, \delta_{n_{+} \, n_{-}} |n_{-}\, n_{+} \rangle.
\label{ENTR4}
\end{equation}
Equation (\ref{ENTR4}) seems unnecessarily complicated: on the one hand, we summed over $n_{+}$ and $n_{-}$ while, on the other hand, we included the $\delta_{n_{+}\, n_{-}}$ that effectively cancels one of the two sums and enforces the conservation of the three-momentum. The redundant form of Eq. (\ref{ENTR4}) is however convenient in what follows since  the action of the operators acting over the different subspaces of the total Hilbert space is immediately clear. Indeed, the signal and the idler modes (i.e. $\widehat{b}_{+}$ and $\widehat{b}_{-}$ respectively) act on two different Hilbert subspaces; they may actually arise as ingredients of a quantum measurement but they correspond here to gravitons with opposite three-momenta. We can always construct a set of Hermitian observables acting on one of the two Hilbert subspaces; for instance $\widehat{N}_{+} = \widehat{b}_{+}^{\dagger} \widehat{b}_{+}$ is the averaged multiplicity of the signal whereas 
 $\widehat{I}_{+} =  \widehat{b}_{+}^{\dagger}\, \widehat{b}_{+}^{\dagger} \widehat{b}_{+} \widehat{b}_{+}$ measures the intensity of the signal. Similar operators can be introduced for the 
 idler mode by replacing $+ \to -$. If we eventually average these operators and take their 
 ratio we obtain, always in the case of the signal, $g_{+}^{(2)}= \langle \widehat{I}_{+} \rangle/ \langle \widehat{N}_{+} \rangle^2$; $g_{+}^{(2)}$ which is the degree of second-order coherence appearing in the analysis of the Hanbury-Brown Twiss correlations \cite{MANDL,HBT4} (see also \cite{HBBT1,HBBT2}). 
Let us now pretend to measure $\langle z| \widehat{N}_{+}| z \rangle$; we have, from Eq. (\ref{ENTR4}), that 
\begin{equation}
\langle z| \widehat{N}_{+}| z \rangle = \sum_{n_{\pm}=0}^{\infty}\,\sum_{m_{\pm}=0}^{\infty}
n_{+} \,  \delta_{n_{+} \, n_{-}}\, \delta_{m_{+} \, m_{-}}\,  \delta_{n_{+} \, m_{+}}\, \delta_{n_{-} \, m_{-}} \frac{ \bigl(\tanh{r}\bigr)^{m_{+} + n_{+}}}{\cosh^2{r}}. 
\label{ENTR4a}
\end{equation}
The result of Eq. (\ref{ENTR4a}) can also be expressed in a more transparent form way by introducing  the averaged multiplicity $\overline{n} = \sinh^2{r}$
\begin{equation}
\langle z| \widehat{N}_{+}| z \rangle =  \sum_{n=0}^{\infty} \, n\, p_{n} = \overline{n}, \qquad p_{n} = \frac{\overline{n}^{n}}{(\overline{n}+1)^{n +1}},
\label{ENTR4ab}
\end{equation}
 where $p_{n}$ now denotes the Bose-Einstein (geometric) distribution.  
The same discussion of Eq. (\ref{ENTR4a}) can be generalized to $\widehat{I}_{+}$  implying that 
the analog of Eq. (\ref{ENTR4a}) becomes
\begin{equation}
\langle z\, | \widehat{I}_{+}| z \rangle = \sum_{n=0}^{\infty} p_{n} \,\langle \, n\, | \widehat{I}_{+} | n \rangle, \qquad\qquad p_{n}=  \frac{\tanh^{2\,n}{r}}{\cosh^2{r}} = \frac{\overline{n}^{n}}{(\overline{n} + 1)^{n +1}}.
\label{ENTR5}
\end{equation}

\subsubsection{The entanglement entropy}
Equations (\ref{ENTR4a})--(\ref{ENTR5}) ultimately suggest that from the total density matrix $\widehat{\rho}$ a reduced density matrix can be obtained by tracing over the idler mode. To simplify the phases we can introduce
\begin{equation}
q_{-} = (n_{+} - m_{+})/2 + (n_{-} - m_{-})/2, \qquad q_{+} =(m_{+} + m_{-})/2 + (n_{+} + n_{-})/2,
\label{ENTR4bbb}
\end{equation}
so that the total density matrix in the Fock basis reads:
\begin{equation}
\widehat{\rho} =  \sum_{n_{\pm}=0}^{\infty}\,\sum_{m_{\pm}=0}^{\infty} e^{i q_{-} \theta} \,\frac{ \bigl(\tanh{r}\bigr)^{q_{+}}}{\cosh^2{r}} \delta_{m_{+}\,m_{-}}\, \delta_{n_{+}\,n_{-}}
|m_{-}\, m_{+} \rangle \langle n_{+}\, n_{-}|.
\label{ENTR4b}
\end{equation}
Equation (\ref{ENTR4b}) still describes a pure quantum state but if we now trace over the idler oscillator we obtain a reduced density operator that only depends on the signal:
\begin{eqnarray}
\widehat{\rho}_{\mathrm{red}} &=& \mathrm{Tr}_{-} \bigl[ \widehat{\rho}\bigr] 
\nonumber\\
&=& \sum_{k_{-} =0}^{\infty}\sum_{n_{\pm}=0}^{\infty}\,\sum_{m_{\pm}=0}^{\infty} e^{i q_{-} \theta} \,\frac{ \bigl(\tanh{r}\bigr)^{q_{+}}}{\cosh^2{r}} \delta_{m_{+}\,m_{-}}\, \delta_{n_{+}\,n_{-}}
\langle k_{-} |m_{-}\, m_{+} \rangle \langle n_{+}\, n_{-}|  k_{-} \rangle,
\label{ENTR4c}
\end{eqnarray}
where by definition $\mathrm{Tr}_{-}[.\,.\,.]$ denotes the trace over the idler mode. The final result for the reduced density operator becomes therefore 
\begin{equation}
\widehat{\rho}_{\mathrm{red}} = \sum_{n=0}^{\infty} \, p_{n} \, |\,n\, \rangle \langle\, n\,|, \qquad\qquad 
\mathrm{Tr} \widehat{\rho}^2_{\mathrm{red}} = \frac{1}{2 \overline{n} +1} < \mathrm{Tr} \widehat{\rho}_{\mathrm{red}}.
\label{ENTR6}
\end{equation}
All the Kr\"oneker deltas appearing in Eq. 
(\ref{ENTR4c}) can be used by recalling that $\langle k_{-} |m_{-}\, m_{+}\rangle = 
\delta_{k_{-}\, m_{-}} |m_{+}\rangle $ and that, similarly, 
$ \langle n_{+}\, n_{-}|  k_{-} \rangle = \delta_{n_{-}\, k_{-}} \langle n_{+} |$. Note also that in Eqs. (\ref{ENTR4c})--(\ref{ENTR6}) the summation index (i.e. $n_{+}\to n$) has been renamed. 
As in Eq. (\ref{ENTR4a}) the statistical weights of Eq. (\ref{ENTR6}) are $p_{n} = \overline{n}^{n}/(\overline{n} +1)^{n+1}$ and they correspond to the Bose-Einstein probability distribution even if the averaged multiplicity $\overline{n} = \sinh^2{r}$ is  non-thermal.  The reduced density matrix can be used to compute all the correlation functions relevant for the description of the signal but it carries no information of the idler mode. The loss of information associated with the trace over 
the idler mode is measured by the von Neumann entropy computed from Eq. (\ref{ENTR4}):
\begin{eqnarray}
s = - \mathrm{Tr} \bigl[ \widehat{\rho}_{\mathrm{red}} \, \ln{\widehat{\rho}_{\mathrm{red}} }\bigr]
= - \sum_{n=0}^{\infty} p_{n} \, \ln{p_{n}} = \ln{(\overline{n}+1)} - \overline{n} \ln{\biggl(\frac{\overline{n}}{\overline{n} +1}\biggr)}.
\label{ENTR7}
\end{eqnarray}
When a portion of the system is unobservable (or when observations are confined to a subset of the  degrees of freedom) information is lost and the total density matrix can then be reduced.
Equation (\ref{ENTR7}) quantitatively describes the loss of information associated with the trace over the idler mode.  Furthermore, from the result of Eq. (\ref{ENTR7}) we see that in the limit $\overline{n}\gg 1$ the entropy gets proportional to $\ln{\overline{n}}$, in other words 
\begin{equation}
\lim_{\overline{n} \gg1} \, s(\overline{n}) = \ln{\overline{n}}.
\label{ENTR8}
\end{equation}
We may now recall that in the process of particle production described by Eqs. (\ref{QB23})--(\ref{QB24}) the averaged multiplicity for each $k$ mode is proportional to $|\beta_{k}(\tau)|^2$
which we can estimate from Eq. (\ref{NNNew6}); this means that the result of Eq. (\ref{ENTR8}) can also be expressed as
 \begin{equation}
\ln{\overline{n}} = 2 \, r = 2 \ln{\biggl(\frac{a_{re}}{a_{ex}}\biggr)}.
\label{ENTR8a}
\end{equation}
The result of Eqs. (\ref{ENTR8})--(\ref{ENTR8a}) hold, strictly speaking, for two oscillators 
with opposite three-momenta;  the previous 
results are however valid for each pair of $\vec{k}$-modes of the field so that the density matrix of Eq. (\ref{ENTR4}) can also be written as
\begin{equation}
\widehat{\rho}_{\vec{k}} = \frac{1}{\cosh^2{r_{k}}} \sum_{n_{\vec{k}}=0}^{\infty} \sum_{m_{\vec{k}}=0}^{\infty} e^{- i \alpha_{k}(n_{\vec{k}} - m_{\vec{k}})} 
(\tanh{r_{k}})^{n_{\vec{k}} + m_{\vec{k}}} |n_{\vec{k}}\,\, n_{- \vec{k}}\rangle \langle m_{-\vec{k}}\,\,m_{\vec{k}}|,
\label{ENTR9}
\end{equation}
where, for completeness,  we have also considered the contribution of a further phase  $\alpha_{k}$ that is different for each $k$-mode. The density matrix can be written, in the Fock basis, as: 
\begin{equation}
\hat{\rho} = \sum_{\{n\}} \, P_{\{n\}} \, | \{n\} \rangle \langle \{n\}|,\qquad \sum_{\{n\}} \, P_{\{n\}} =1.
\label{ENTR10} 
\end{equation}
The multimode probability distribution appearing in Eq. (\ref{ENTR10}) is given by:
\begin{equation}
P_{\{n_{\vec{k}}\}} = \prod_{\vec{k}} P_{n_{\vec{k}}} , \qquad P_{n_{\vec{k}}}(\overline{n}_{k})= 
\frac{\overline{n}_{k}^{n_{\vec{k}}}}{( 1 + \overline{n}_{k})^{n_{\vec{k}} + 1 }},
\label{ENTR11}
\end{equation}
where $\overline{n}_{k}$ is the average multiplicity of each Fourier mode. Furthermore, following the standard notation, $ |\{n \}\rangle = |n_{\vec{k}_{1}} \rangle \, | n_{\vec{k}_{2}} \rangle \, | n_{\vec{k}_{3}} \rangle...$ where the ellipses stand for all the occupied modes of the field.  We also note that the density matrix can be reduced by considering the phases of the final multiparticle state to be 
unobservable. In this case the right hand side of Eq. (\ref{ENTR9}) can be averaged averaging over $\alpha_{k}$. The reduced density matrix would be given, in this case, by 
\begin{equation}
\hat{\rho}^{\mathrm{red}}_{\vec{k}} = \frac{1}{2\pi} \int_{0}^{2\pi} d \alpha_{k}\,\, \hat{\rho}_{\vec{k}}= \frac{1}{\cosh^2{r_{k}}} \sum_{n_{\vec{k}}=0}^{\infty} 
(\tanh{r_{k}})^{2 n_{\vec{k}}} |n_{\vec{k}}\,\, n_{- \vec{k}}\rangle \langle n_{-\vec{k}}\,\,n_{\vec{k}}|.
\label{ENTR12}
\end{equation}
This observation represents a further reduction scheme of the density matrix that ultimately 
leads to the same entropy of Eq. (\ref{ENTR7}) in the limit of the large averaged multiplicities.
To obtain the total entropy we must integrate over the whole spectrum for a fiducial Hubble volume at the present time; in this case the total entanglement entropy of the gravitons becomes:
\begin{equation}
S_{g \, 0} = \frac{8}{3} \pi H_{0}^{-3} \int_{k_{min}}^{k_{max}} \frac{d^{3} k}{(2 \pi)^3} \ln{\overline{n}_{k}}.
\label{ENTR13}
\end{equation}
The integral appearing in Eq. (\ref{ENTR13}) can be estimated by using either the general 
form of the averaged multiplicity deduced in in Eq. (\ref{FFF6}) or the result of Eq. (\ref{NNNew6}). For The power-law parametrization of Eq. (\ref{FFF6}) 
is actually compatible with Eqs. (\ref{NNNew6}) where the averaged multiplcity for $k \ll k_{max}$ 
depends on $(a_{re}/a_{ex}) \gg 1 $. If assume that between $a_{ex}$ and $a_{re}$ the background expands in this case the averaged multiplicity is given by\footnote{The explicit form of Eq. (\ref{ENTR14}) follows by assuming that the relevant wavelengths 
cross the Hubble radius for the first time during inflation  (i.e. $ k \tau_{ex} = k/{\mathcal H}_{ex}= {\mathcal O}(1)$) when the scale factor is given, approximately, by $a_{ex} = (- \tau_{1}/\tau_{ex})^{\beta_{ex}} \simeq |k \tau_{1}|^{\beta_{ex}}$; the reentry takes place instead when $a_{re} = (\tau_{re}/\tau_{1})^{\beta_{re}} \simeq |k \tau_{1}|^{-\beta_{re}}$.
In the conventional situation $\beta_{ex} = 1/(1- \epsilon)$ (where $\epsilon$ is the slow-roll parameter) and $\beta_{ex} = {\mathcal O}(1)$: this means that $\overline{n}_{k}\simeq (k/k_{max})^{-4}$.  If  the wavelengths reenters the Hubble radius during a maximally stiff phase we have instead 
that $\overline{n}_{k}\simeq (k/k_{max})^{-3}$ since $\beta_{re} = {\mathcal O}(1/2)$. The considerations based on Eq. (\ref{ENTR14}) are then consistent with Eq. (\ref{FFF6}).}:
\begin{equation} 
\overline{n}_{k} = \frac{1}{4}\biggl(\frac{a_{re}}{a_{ex}}\biggr)^2 \simeq \biggl(\frac{k}{k_{max}}\biggr)^{ - 2 \beta_{re} - 2\beta_{ex}},
\label{ENTR14}
\end{equation}
where the subleading contributions have been neglected since they do not affect the final 
value of the integral.  Thanks to the explicit form of Eq. (\ref{ENTR4}) the integration variable appearing in Eq. (\ref{ENTR13}) can be rescaled and the total entropy of the gravitons is 
\begin{equation}
S_{g\, 0}= \frac{32}{3} \,\pi^2\, \biggl(\frac{\nu_{max}}{H_{0}}\biggr)^3 {\mathcal K}(m_{T}), 
\label{ENTR15}
\end{equation}
where ${\mathcal K}(m_{T})(4 - m_{T}) \int_{\nu_{min}/\nu_{max}}^{1} x^2 \, \ln{x} \, d \, x = {\mathcal O}(1) $ is a numerical factor that depends on the spectral slope 
$m_{T} = n_{T}^{(high)}$. Since in the integral ${\mathcal K}(m_{T})$ the value of $m_{T}$ is not essential and the 
lower limit of integration goes to zero (at least in practice since $\nu_{p} \simeq \nu_{min} =
3\, 10^{-18} \mathrm{Hz}$) we have, from Eq. (\ref{ENTR15}), that $S_{g\, 0} ={\mathcal O}(10) (\nu_{max}/H_{0})^{3}$. The result for $S_{g\, 0}$ can now be compared with the well known result of the thermal entropy of the cosmic microwave background computed within the same fiducial Hubble volume 
\begin{equation}
S_{\gamma\, 0} = \frac{4}{3} \pi \, H_{0}^{-3}\, s_{\gamma}, \qquad \qquad s_{\gamma}= 
\frac{4}{45} \pi^2 T_{\gamma\, 0}^3.
\label{ENTR16}
\end{equation}
Since $T_{\gamma\, 0} = T_{\gamma 0} = (2.72548 \pm 0.00057) \, \mathrm{K}$  we also have that 
\begin{equation}
T_{\gamma 0} = 356.802 \biggl(\frac{T_{\gamma 0}}{2.72548\,\, \mathrm{K}}\biggr) \, \, \mathrm{GHz}.
\label{ENTR17}
\end{equation}
If we now require that $S_{g\, 0} \leq S_{\gamma\, 0} = {\mathcal O}(10^{90})$ we have from Eqs. (\ref{ENTR15})--(\ref{ENTR16}) and (\ref{ENTR17}) that 
\begin{equation}
S_{g\, 0} \leq S_{\gamma\, 0}\qquad \Rightarrow \qquad \nu_{max} \leq \mathrm{THz}.
\label{ENTR18}
\end{equation}
The maximal frequency of the spectrum deduced in Eq. (\ref{HH3}) implies that the entanglement entropy of the gravitons cannot exceed the total entropy of the 
cosmic microwave background. The quantum theory of parametric amplification however provides a natural cosmological arrow associated with an entanglement entropy \cite{ST8,ST9,ST9a,ST9b,ST10,ST11}. In our context the growth of the averaged multiplicity of the produced gravitons is naturally associated with the increase of the entanglement entropy of the gravitational field. In an idealized experiment based, for instance, on the analysis of the gravitational analog of the Hanbury Brown-Twiss correlations 
only one of the two gravitons of the pair is typically detected. In this situation the resulting entanglement entropy is proportional to the logarithm of the averaged multiplicity. Since the final multiplicity is always large,  the entropy is effectively proportional to the squeezing parameter $r$ \cite{ST10,ST11}.  The reduction of the density matrix can be performed in different bases \cite{ST9a,ST9b,ST10,ST11};
the results depend on the basis but the asymptotic limit when the averaged multiplicity is large  
is generally proportional to $2\,r$ and hence to the logarithm of $\overline{n}$. The 
coarse grained entropy employed here is quite close to the notion of information theoretic 
entropy introduced many years ago \cite{DDP1,DDP2} with the purpose of reformulating the 
indetermination relations. This idea actually inspired the reduction scheme discussed in Refs. \cite{ST9a,ST9b}. 
We finally stress that the value of the entropy of the gravitons 
for each single mode is equal to $2 r$ and this result follows by considering the 
entropy as a measure of uncertainty in the result of a measurement or 
preparation of a given quantum mechanical observable. Let us consider 
the following canonical transformation of a two-mode harmonic oscillator
\begin{equation}
x \to x_{r} = e^{-r} x, \qquad \widetilde{x} \to \widetilde{x}_{r} = e^{-r}\, \widetilde{x},
\label{cantrans1}
\end{equation}
with $r>0$. 
Since the transformation is canonical the conjugate momenta 
will transform $p \to p_{r} = e^{r} p$ and as $\widetilde{p} \to \widetilde{p}_{r} = e^{r} \, \widetilde{p}$.  It is clear from Eq. (\ref{cantrans1}) that operators 
$x$ and $\widetilde{x}$ fluctuate above the quantum noise 
since for a quantum state $| \psi \rangle$ the corresponding normalized wavefunction reads 
\begin{equation}
\langle x \, \widetilde{x}| \psi \rangle = \psi(x, \widetilde{x}) = \sqrt{\frac{\sigma}{\pi}}\, e^{- \sigma (x^2 + \widetilde{x}^2)/2} 
, \qquad \sigma = e^{- 2 r}.
\label{cantrans2}
\end{equation}
From Eq. (\ref{cantrans2}) it is simple to compute $(\Delta x)^2 = 
\langle x^2 \rangle - \langle x\rangle^2$ (and similarly for $(\Delta \widetilde{x)}^2$);
thanks to Eq. (\ref{cantrans2}) $\Delta x = \sqrt{(\Delta x)^2 } = e^{r}/\sqrt{2}$ 
and $\Delta \widetilde{x} = \sqrt{(\Delta \widetilde{x})^2 } = e^{r}/\sqrt{2}$.
Both operators $x$ and $\widetilde{x}$ fluctuate above the quantum noise and 
a natural measure of uncertainty in the result of a measurement of the 
superfluctuant operators is given by 
\begin{equation}
s_{super} = -  \int \, d x \int d\widetilde{x} |\langle x \, \widetilde{x}| \psi \rangle|^2 \ln  |\langle x \, \widetilde{x}| \psi \rangle|^2.
\label{cantrans3} 
\end{equation}
Inserting Eq. (\ref{cantrans2}) into Eq. (\ref{cantrans3}) we obtain 
$s_{super} = 2 \, r + \ln{(e \, \pi)}$ which coincides with $2\, r$ in the limit 
of large produced particles. The 
coarse grained entropy analyzed here is then quite close to the notion of information theoretic 
entropy introduced many years ago \cite{DDP1,DDP2} with the purpose of reformulating the 
indetermination relations. This idea actually inspired the reduction scheme discussed in Refs. \cite{ST9a,ST9b}. 
\newpage
\renewcommand{\theequation}{7.\arabic{equation}}
\setcounter{equation}{0}
\section{Concluding Remarks}
\label{sec7}
Since the Universe became transparent to the propagation of electromagnetic 
disturbances only after matter-radiation equality, the photons coming from the 
primeval stages of the evolution of the plasma cannot be detected so that 
earlier tests on the expansion history are actually related to the remarkable successes of big bang nucleosynthesis taking place when the expansion rate was of the order of $10^{-44}\, M_{P}$. This figure should be compared with the  approximate value of the inflationary expansion rate (i.e. ${\mathcal O}(10^{-6})\, M_{P}$) inferred from the amplitudes of the curvature inhomogeneities that affect the CMB temperature and polarization anisotropies. Between these two scales the expansion rate spanned $38$ orders of magnitude where the evolution of the plasma could have been rather different from radiation.

During the last fifty years  the interplay between high-energy physics and cosmology has been guided by the assumption that radiation should be the dominant component of the plasma well before the onset of big bang nucleosynthesis and immediately after the end of inflation.  This conventional wisdom is consistent both with an early stage of inflationary  expansion and with the concordance scenario at late times but it is not unique. The physical foundations of this paradigm are not corroborated by direct observations and they could be either partially or totally refuted in the years to come.  Since during inflation the particle horizon diverges (while the event horizon is finite) all the wavelengths that are currently shorter than the Hubble 
radius were in causal contact during inflation provided the overall duration of inflation was sufficiently long. The length of the inflationary stage is customarily 
assessed in terms of the number of $e$-folds which should be ${\mathcal O}(60)$ if the post-inflationary expansion rate is dominated by radiation. This estimate can be either reduced (down to ${\mathcal O} (45)$) or increased (up to ${\mathcal O}(75)$) depending on the post-inflationary expansion rate that may become either faster or slower than radiation, respectively. 

Any presumption about the timeline of the expansion rate should necessarily acknowledge that every variation of the space-time curvature  produces shots of gravitons with specific averaged multiplicities. After the actual detection of gravitational radiation there are no direct physical limitations forbidding the empirical scrutiny of the spectra of the relic gravitons (either in the audio band or in higher frequency domains) within the following score year. Since different timelines ultimately correspond to specific profiles of $h_{0}^2 \, \Omega_{gw}(\nu,\tau_{0})$ (for frequencies ranging between the aHz and the THz),  the expansion rate can be systematically inferred from the slopes of the observed spectra and from their pivotal frequencies.  The results outlined here specifically address the interplay between the expansion history of the plasma and the spectral energy density of the relic gravitons in the concrete situations inspired by the current phenomenological lore at low, intermediate and high frequencies.
\begin{itemize}
\item{} The inflationary observables in the aHz region depend on the timeline of the post-inflationary evolution. In  single-filed inflationary scenarios this means, in particular, that the tensor to scalar ratio and the scalar spectral index are more or less
suppressed if the timeline of the expansion rate is either slower or faster than radiation respectively. At higher frequencies the pulsar timing arrays (operating in the nHz range) are now setting interesting bounds on the post-inflationary expansion rate. The apparent  excesses appearing in the last data releases of two pulsar timing arrays could actually come from an increasing spectrum of relic gravitons at intermediate frequencies.
\item{} Between the $\mu\mathrm{Hz}$ and the Hz various space-borne detectors might be operational in the far future although the signals expected in the mHz region are dominated by astrophysical sources (e.g. galactic white dwarves, solar-mass black holes, supermassive black holes coming from galaxy mergers). The only cosmological sources customarily considered in this framework are associated with the phase transitions at the TeV scale although perturbative and non-perturbative estimates consistently suggest that the standard electroweak theory leads to a cross-over regime where drastic deviations from homogeneity (and the consequent bursts of gravitational radiation) should not be expected. The inflationary signal (often regarded as irrelevant between few $\mu$Hz and the Hz) could be in fact much larger 
that the purported signal coming from a realistic dynamics at the electroweak scale. 
Moreover, since the slopes of $h_{0}^2 \Omega_{gw}(\nu,\tau_{0})$ obtained in the case of a putative strongly first-order phase transition are much steeper than the ones associated with a modified expansion history, the most severe phenomenological bounds on the relic gravitons between the $\mu\mathrm{Hz}$ and the Hz arise (by continuity in frequency) from the audio band 
and from the operating ground-based detectors. 
\item{} The window of wide-band detectors notoriously ranges between few Hz and $10$ kHz. The current limits imply that the sensitivity of correlated interferometers for the detection of a flat spectral energy density of relic gravitons is approximately $h_{c}^{(min)}= {\mathcal O}(10^{-24})$ for typical frequencies in the audio band. Sharp deviations from scale-invariance lead to similar orders of magnitude and while these figures may improve in the years to come, the frequency domain of ground-based interferometers will remain the same. For this reason it is important to promote new instruments operating in higher frequency domains where the potential signals coming from the past history of the plasma are dominant. More than twenty years ago it was suggested that microwave cavities (operating between the MHz and the GHz regions) 
could be used for the detection of relic gravitons associated with post-inflationary phases stiffer than radiation. While forty years ago the typical sensitivities of these instruments were $h_{c}^{(min)} = {\mathcal O}(10^{-17})$ they improved later on and reached $h_{c}^{(min)} = {\mathcal O}(10^{-20})$ in the early 2000s. Similar prototypes aimed at the detection of dark matter could be used as high frequency detectors of gravitational waves. The target sensitivities of these instruments are often set by requiring in the MHz (or even GHz regions) the same sensitivities reached (today) by the interferometers in the audio band. This means that the features of the instruments are not guided by the signals of the available sources in the corresponding frequency domain. To detect directly relic gravitons with high frequency instruments operating between the MHz
and the GHz the minimal detectable chirp amplitude should be $h_{c}^{(min)} = {\mathcal O}(10^{-32})$ (or smaller). However, if the pivotal frequencies of the instruments are reduced from the THz to the GHz (or even MHz) band the minimal required chirp amplitude may increase. With these specifications, the detectors in the MHz and GHz domains may be able to probe directly the relic gravitons and their quantumness.
\end{itemize}
Both at the classical and quantum level,  the largest frequency of the relic gravitons never exceeds the THz band and above the maximal frequency the averaged multiplicity is exponentially suppressed so that $\nu_{max}$ ultimately corresponds to the production of a single graviton pair. Since the relic gravitons are inherently quantum mechanical,  their quantumness can be measured in terms of an entanglement entropy that is caused by the loss of the complete information on the underlying quantum field. The reduction of the density matrix in different bases leads to the same von Neumann entropy whose integral over all the modes of the spectrum is dominated again by the maximal frequency. Whenever the THz bound is applied, it turns out that the total integrated entropy of the relic gravitons is comparable with the entropy of the cosmic microwave background but not larger.  A potential detection of relic gravitons both at low and high frequencies may therefore represent a direct evidence of macroscopic quantum states associated with the gravitational field.  For this reason the detectors operating in the MHz and GHz regions are quantum sensitive to the second-order interference effects. 
As in the case of optical photons, the interferometric techniques 
pioneered by Hanbury-Brown and Twiss in the 1950s could be applied to high-frequency gravitons with the purpose of distinguishing the statistical properties of thermal and non-thermal gravitons. 

\section*{Acknowledgements}
I wish to acknowledge relevant discussions with the late Ph. Bernard, G. Cocconi and E. Picasso on high frequency gravitons and microwave cavites. It is a pleasure to thank A. Gentil-Beccot, P. Birtwistle,  A. Kohls,  L. Pieper, S. Rohr and J. Vigen of the CERN Scientific Information Service for their kind help along the different stages of this investigation. Some of the discussions presented here have been developed on the 
occasion of few seminars and of a set of lectures; I thank the questions and the remarks of students and colleagues.

\newpage 

\begin{appendix}
\renewcommand{\theequation}{A.\arabic{equation}}
\setcounter{equation}{0}
\section{Complements on the curvature inhomogeneities}
\label{APPA}
\subsection{General considerations}
The evolution of curvature inhomogeneities appears in various discussions throughout this 
article and this is why it is useful to present a self-contained account of the problem. We recall that the action of the scalar modes of the geometry can be 
expressed as 
\begin{equation}
S_{{\mathcal R}} = \frac{1}{2} \int \, d^{3} x\, \int d \tau \, z_{\varphi}^2(\tau) \biggl[ \partial_{\tau} {\mathcal R} \, \partial_{\tau} {\mathcal R} - \partial_{k} {\mathcal R} \, \partial_{k} {\mathcal R}\biggr],
\label{APPB1}
\end{equation}
where ${\mathcal R}$ is the gauge-invariant variable denoting the curvature inhomogeneities 
on comoving orthogonal hypersurfaces and $z_{\varphi} = a \, \varphi^{\prime}/{\mathcal H}$; the prime indicates, throughout this first appendix, a derivation with respect to the conformal time coordinate $\tau$. From Eq. (\ref{APPB1}) the canonical momenta are $\pi_{{\mathcal R}} = z_{\varphi}^2 \partial_{\tau} {\mathcal R}$ and the associated classical
Hamiltonian is:
\begin{equation}
H_{{\mathcal R}}(\tau) = \frac{1}{2} \int \, d^{3}x \, \biggl[ \frac{\pi_{{\mathcal R}}^2}{z_{\varphi}^2} + 
z_{\varphi}^2 \partial_{k}{\mathcal R}\, \partial_{k} {\mathcal R}\biggr].
\label{APPB2}
\end{equation}
From Eq. (\ref{APPB2}) the corresponding Hamilton's equations read $\partial_{\tau} \pi_{{\mathcal R}} = z_{\varphi}^2 \nabla^2 {\mathcal R}$ and $\partial_{\tau} {\mathcal R} = \pi_{{\mathcal R}}/z_{\varphi}^2$. The scalar modes of the geometry are quantized by promoting the classical variables to the status
of field operators as:
\begin{equation}
{\mathcal R}(\vec{x}, \tau) \to \widehat{{\mathcal R}}, \qquad \pi_{{\mathcal R}}(\vec{x}, \tau) \to \widehat{\pi}_{{\mathcal R}},\qquad H_{{\mathcal R}}(\tau) \to \widehat{H}_{{\mathcal R}}.
\label{APPB3}
\end{equation}
The field operators obey the canonical commutation relations at equal time, i.e. 
\begin{equation}
[\widehat{{\mathcal R}}(\vec{x}, \tau),\, \widehat{\pi}_{{\mathcal R}}(\vec{y}, \tau) ] = \, i\, \delta^{(3)}(\vec{x} -\vec{y}).
\end{equation}
The explicit form of the field operators can then be written as:
\begin{eqnarray}
 \widehat{{\mathcal R}}(\vec{x}, \tau) &=& \frac{1}{(2\pi)^{3/2}} \int d^{3} k\, \biggl[ \widehat{a}_{\vec{k}} \, F_{k}^{(s)}(\tau) \, e^{- i \, \vec{k}\cdot\vec{x}} + \mathrm{H.\, c.}\biggr],
\label{APPB4a}\\
\widehat{\pi}_{{\mathcal R}}(\vec{x}, \tau) &=& \frac{z_{\varphi}^2}{(2\pi)^{3/2}} \int d^{3} k\, \biggl[ \widehat{a}_{\vec{k}} \, G_{k}^{(s)}(\tau) \, e^{- i \, \vec{k}\cdot\vec{x}} + \mathrm{H.\, c.}\biggr],
\label{APPB4b}
\end{eqnarray}
where $F_{k}^{(s)}(\tau)$ and $G_{k}^{(s)}(\tau)= F_{k}^{(s)\, \prime}(\tau)$ are the associated mode functions. From Eqs. (\ref{APPB4a})--(\ref{APPB4b}) the commutation relations at equal times remain canonical throughout the dynamical evolution  provided 
$F_{k}^{(s)}(\tau)$ and $G_{k}^{(s)}(\tau)$ obey the Wronskian normalization condition
\begin{equation}
F_{k}^{(s)}(\tau) \, G_{k}^{(s)\,\ast}(\tau) - F_{k}^{(s)\, \ast}(\tau) \, G_{k}^{(s)}(\tau) = \frac{i}{z_{\varphi}^2(\tau)}.
\label{APPB4c}
\end{equation}
Together with this normalization condition (that preserves the canonical commutation relation) the evolution of the mode functions can be written as 
\begin{equation}
F_{k}^{(s)\prime\prime} + 2 \frac{z_{\varphi}^{\prime}}{z_{\varphi}}  F_{k}^{(s)\prime} + k^2 F_{k}^{(s)} =0, \qquad\qquad 
G_{k}^{(s)} = F_{k}^{(s)\, \prime}.
\label{APPB5}
\end{equation}
In terms of the evolution of the underlying fields of the background sources we have that 
\begin{equation}
\frac{z_{\varphi}^{\prime}}{z_{\varphi}} = a\, H( 1 + \eta + \epsilon) = a\, H(1 + 2 \epsilon - \overline{\eta}), 
\label{APPB6}
\end{equation}
where $\eta = \ddot{\varphi}/( H\, \dot{\varphi})$ and $\epsilon= - \dot{H}/H^2$ are the usual slow-roll
parameters; we can also redefine $\eta$ as $\eta = \eta - \overline{\eta}$ where, as already mentioned 
in the main text, $\overline{\eta} = \overline{M}_{P}^2 (V_{,\varphi\varphi}/V)$.   
The evolution of the mode functions can be rescaled by defining $f_{k}^{(s)} = z_{\varphi} \, F_{k}^{(s)}$ 
and $g_{k}^{(s)} = z_{\varphi} \, G_{k}^{(s)}$; in this parametrization the evolution of $f_{k}^{(s)}$ and 
$g_{k}^{(s)}$ is given by  $f_{k}^{(s)\,\prime\prime} + [k^2 - z_{\varphi}^{\prime\prime}/z_{\varphi}] f_{k}^{(s)} =0$ 
with $g_{k}^{(s)} = f_{k}^{(s)\,\prime} - (z_{\varphi}^{\prime}/z_{\varphi}) f_{k}^{(s)}$.

\subsection{The scalar power spectra}
The power spectrum $P_{{\mathcal R}}(k,\tau)$ of curvature inhomogeneities is defined in the following manner:
\begin{equation}
\langle 0 | \widehat{{\mathcal R}}(\vec{x}, \tau) \, \widehat{{\mathcal R}}(\vec{x} + \vec{r}, \tau) |0 \rangle 
= \int \, d \ln{k} P_{{\mathcal R}}(k,\tau)\,\, j_{0}(k\, r),
\label{APPB7}
\end{equation}
where $j_{0}(x) = \sin{x}/x$ is the zeroth order spherical Bessel function \cite{abr1,abr2} and $|0 \rangle$ is the state annihilated by $\widehat{a}_{\vec{k}}$. Using the explicit form of the field operators given in Eq. (\ref{APPB4a}) the scalar power spectrum reduces to:
\begin{equation}
P_{{\mathcal R}}(k,\tau) = \frac{k^3}{2 \pi^2} \bigl| F_{k}^{(s)}(\tau) \bigr|^2 = \frac{k^3}{ 2 \,\pi^2 \, z_{\varphi}^2}  \bigl| f_{k}^{(s)}(\tau) \bigr|^2.
\label{APPB8}
\end{equation}
We can also represent the field operator in Fourier space: 
\begin{equation}
\widehat{{\mathcal R}}(\vec{x}, \tau) = \frac{1}{(2\pi)^{3/2}} \int \widehat{{\mathcal R}}_{\vec{k}}(\tau) 
\, e^{- i \vec{k}\cdot\vec{x}} \, d^{3} k, 
\label{APPB9}
\end{equation}
so that, eventually, the expectation value of the Fourier amplitudes evaluated for different three-momenta is
\begin{equation}
\langle \widehat{{\mathcal R}}_{\vec{k}}(\tau)\,\widehat{{\mathcal R}}_{\vec{p}}(\tau) \rangle = 
\frac{2 \pi^2}{k^3} \, P_{{\mathcal R}}(k,\tau) \, \delta^{(3)}(\vec{k}+\vec{p}).
\label{APPB10}
\end{equation}
Recalling now the result of Eq. (\ref{APPB6}) it is easy to obtain the 
explicit expression for $z_{\varphi}^{\prime\prime}/z_{\varphi}$:
\begin{equation}
\frac{z_{\varphi}^{\prime\prime}}{z_{\varphi}} = a^2 \, H^2 ( 1 + 2 \epsilon - \overline{\eta}) (2 + \epsilon - \overline{\eta})
= \frac{ ( 1 + 2 \epsilon - \overline{\eta}) (2 + \epsilon - \overline{\eta})}{ ( 1 - \epsilon)^2 \tau^2},
\label{APPB11}
\end{equation}
where the second equality follows by appreciating that, during slow-roll, $(1 - \epsilon) a\, H= - 1/\tau$. From the second equality of Eq. (\ref{APPB11}) it also follows that the evolution of the mode function can be expressed as 
\begin{equation}
f_{k}^{(s)\prime\prime} + \biggl[ k^2 - \frac{\nu_{s}^2 - 1/4}{\tau^2} \biggr] f_{k}^{(s)} =0, \qquad \nu_{s} = \frac{3 + 3 \epsilon - 2 \overline{\eta}}{ 2 ( 1 - \epsilon)}.
\label{APPB12}
\end{equation}
The solution of the evolution of the mode functions with the correct boundary conditions is finally: 
\begin{equation}
F_{k}^{(s)}(\tau) =\frac{f_{k}^{(s)}}{z_{\varphi}(\tau)}=  \frac{{\mathcal N}_{s}}{z_{\varphi} \, \sqrt{2 k}} \sqrt{- k\tau} H_{\nu_{s}}^{(1)}(- k\tau), \qquad 
{\mathcal N}_{s}= \sqrt{\frac{\pi}{2}}\, e^{i \pi(2 \nu_{s} +1)/4},
\label{APPB13}
\end{equation}
where $H^{(1)}_{\nu_{s}}(x)$ are the Hankel functions of first kind with index $\nu_{s}$ and generic argument $x$ \cite{abr1,abr2}.
Equation (\ref{APPB13}) leads therefore to the following explicit expression of the scalar power 
spectrum:
\begin{equation}
P_{{\mathcal R}}(k,\tau) = \frac{k^2}{8\pi \, z_{\varphi}(\tau)} (-k\tau) \bigl| H_{\nu_{s}}^{(1)}(-k\tau)\bigr|^2, \qquad k\tau = 
\frac{k}{(1 - \epsilon) a\, H}.
\label{APPB14}
\end{equation}
The limit $k < a\, H$ coincides with $|k \, \tau| < 1$ and $\epsilon< 1$. The small argument limit of the Hankel functions 
together with the explicit form of $z_{\varphi}(\tau)$ lead to the following explicit form of the scalar 
power spectrum in the long-wavelength limit
\begin{equation}
P_{{\mathcal R}}(k,\tau) = {\mathcal C}_{s}(\nu_{s}, \epsilon) \biggl(\frac{H^4}{\dot{\varphi}^2}\biggr) \biggl| \frac{k}{a\, H}\biggr|^{3 - 2 \nu_{s}}, \qquad {\mathcal C}_{s}(\nu_{s}, \epsilon) = \frac{2^{2 \nu_{s} -3}}{\pi^3} \Gamma^2(\nu_{s}) (1 - \epsilon)^{2 \nu_{s} -1}.
\label{APPB15}
\end{equation}
The scalar power spectrum of Eq. (\ref{APPB15}) is usually evaluated in the long wavelength limit since the initial conditions 
for the CMB anisotropies are usually set when the relevant wavelengths are larger than the Hubble radius 
before matter-radiation equality. From Eq. (\ref{APPB15}) we can also deduce the dependence of the spectral index upon the slow-roll parameters. Recalling the explicit form of $\nu_{s}$ given in Eq. (\ref{APPB12}) we have that, by definition, $n_{s} -1 = 3 - 2 \nu_{s}$ which means that 
\begin{equation}
n_{s} = \frac{1 - 7 \epsilon + 2 \overline{\eta}}{1 - \epsilon} = 1 - 6 \epsilon + 2 \overline{\eta} + {\mathcal O}(\epsilon^2).
\label{APPB16}
\end{equation}
Equation (\ref{APPB15}) can be written in analogous forms. As 
emphasized in section \ref{sec2}, according to some the scalar power spectrum is viewed as a tool for the reconstruction of the inflaton potential. Along this perspective the slow-roll approximation of Eq. (\ref{SINGLE4}) can be used with the purpose of eliminating 
the expansion rate;  two equivalent forms of Eq. (\ref{APPB15}) are then
\begin{equation}
P_{{\mathcal R}}(k,\tau) = \frac{{\mathcal C}_{s}(\nu_{s}, \epsilon)}{6 \, \epsilon} \biggl(\frac{V}{\overline{M}_{P}^4}\biggr) \, \biggl| \frac{k}{a\, H}\biggr|^{n_{s} -1} = \frac{32\,\pi^2 \,{\mathcal C}_{s}(\nu_{s}, \epsilon)}{3 \,\epsilon} \biggl(\frac{V}{M_{P}^4}\biggr) \, \biggl| \frac{k}{a\, H}\biggr|^{n_{s} -1}.
\label{APPB17}
\end{equation}
The difference between the two expressions of Eq. (\ref{APPB17}) follows by recalling that, within the present conventions, 
the reduced Planck mass is given by $\overline{M}_{P} = M_{P}/\sqrt{8\,\pi}$. The perspective of section \ref{sec2} is slightly more general; instead of using the scalar power spectrum to reconstruct the potential it seems more appropriate, for the present purposes, to phrase Eq. (\ref{APPB15}) in terms of the expansion rate; in this way we obtain:
\begin{equation}
P_{{\mathcal R}}(k,\tau) = \frac{{\mathcal C}_{s}(\nu_{s}, \epsilon)}{2 \, \epsilon} \biggl(\frac{H}{\overline{M}_{P}}\biggr)^2 \, \biggl| \frac{k}{a\, H}\biggr|^{n_{s} -1} = \frac{4\,\pi \,{\mathcal C}_{s}(\nu_{s}, \epsilon)}{\epsilon}  \biggl(\frac{H}{M_{P}}\biggr)^2 \, \biggl| \frac{k}{a\, H}\biggr|^{n_{s} -1}.
\label{APPB18}
\end{equation}
Recalling that, to leading-order in the slow-roll parameters, $\Gamma(\nu_{s}) \simeq \sqrt{\pi}/2$ we have that ${\mathcal C}_{s}(\nu_{s}, \epsilon) \to (4 \pi^2)^{-1}$. This means that, in the same approximation, when a given scale crosses 
the Hubble radius $P_{{\mathcal R}}(k,1/k) \simeq (\pi \epsilon_{k})^{-1} (H_{k}/M_{P})^2$ where, as already explained 
in section \ref{sec2}, the time dependent factors are evaluated for $\tau =1/k$. 

\subsection{The tensor to scalar ratio}
While in the bulk of the article we preferred to employ the WKB approximation, we report here the derivation of $r_{T}(k,\tau)$ in terms of the expressions of the inflationary mode functions. 
Since $r_{T}(k,\tau) = P_{T}(k,\tau)/P_{{\mathcal R}}(k,\tau)$ we just need to express the tensor power spectrum within the same notations of the previous subsection. From the results of section \ref{sec3} the tensor mode functions during the inflationary stage can be given in full analogy with the scalar result of Eq. (\ref{APPB13}) 
\begin{equation}
F_{k}^{(t)}(\tau) = \frac{{\mathcal N}_{t}}{a \, \sqrt{2 k}} \sqrt{- k\tau} H_{\nu_{t}}^{(1)}(- k\tau), \qquad 
{\mathcal N}_{t}= \sqrt{\frac{\pi}{2}}\, e^{i \pi(2 \nu_{t} +1)/4},
\label{APPB19}
\end{equation}
where $\nu_{t} = (3 - \epsilon)/[2 (1-\epsilon)]$. Within this notation we have that in the limit $k < a\, H <1$ 
the tensor power spectrum can be written as:
\begin{equation}
P_{T}(k,\tau) = {\mathcal C}_{t}(\nu_{t}, \epsilon)\,\, \biggl(\frac{H}{\overline{M}_{P}}\biggr)^2 \,\, \biggl| \frac{k}{a\, H}\biggr|^{n_{T}}, \qquad {\mathcal C}_{t}(\nu_{t}, \epsilon) = \frac{2^{2\,\nu_{t}}}{\pi^3} (1 - \epsilon)^{2 \nu_{t} -1}\, \Gamma^2(\nu_{t}),
\label{APPB20}
\end{equation}
where $n_{T} = 3 - 2 \nu_{t}$ is, by definition, the tensor spectral index. 
With these notations the tensor-to-scalar ratio can be written as
\begin{equation}
r_{T}(k,\tau) = 16 \, \epsilon\, \frac{2^{2(\nu_{t} -\nu_{s})}}{(1- \epsilon)^{2(\nu_{t} - \nu_{s})}} \frac{\Gamma^2(\nu_{t})}{\Gamma^2(\nu_{s})} \,\,  \biggl| \frac{k}{a\, H}\biggr|^{2 (\nu_{s} - \nu_{t})}.
\label{APPB21}
\end{equation}
From this expression it is clear that, to leading order in the slow-roll parameters, $\nu_{s} \simeq \nu_{t}$ so that 
$r_{T}(k,1/k) \to 16 \epsilon_{k}$, as repeatedly discussed in sections \ref{sec3} and \ref{sec4}.

\renewcommand{\theequation}{B.\arabic{equation}}
\setcounter{equation}{0}
\section{The action and the energy density of the relic gravitons}
\label{APPB}
The evolution of gravitational waves in curved backgrounds is ultimately gauge invariant and frame-invariant. This means that the early expansion history of the background has a well defined meaning not only in general relativity but also in its extensions.  The evolution can be always treated in the most convenient frame but the spectral energy density will always be the same in spite of the frame employed in the description of the dynamical evolution.
\subsection{Generalities}
Every discussion on gravitational radiation involves, as a first step,  the evolution of general relativistic disturbances in flat-space time with the aim of showing that only two degrees of freedom propagate, at least in the case of Einsteinian theories of gravity. Since the ideas analyzed here suggest a direct connection between the spectrum of the relic gravitons and the early expansion history of the Universe, it is more appropriate to consider the propagation of weak disturbances in general background geometries that do not necessarily coincide with the conventional Minkowski space-time. For this purpose the full metric $g_{\mu\nu}(x)$ (where $x$ denotes the space-time point)  is separated into a background value $\overline{g}_{\mu\nu}(x)$ supplemented by the corresponding disturbance $\delta^{(1)} g_{\mu\nu}(x)$:
\begin{equation}
g_{\mu\nu}(x) = \overline{g}_{\mu\nu}(x) + \delta^{(1)} g_{\mu\nu}(x),\qquad \delta^{(1)}g_{\mu\nu}(x) = f_{\mu\nu}(x),  
\label{APPA1}
\end{equation}
where $|f_{\mu\nu}(x)| \ll 1$ denotes the whole metric fluctuation that also encompasses the tensor modes 
of the geometry. The fluctuations of all the interesting geometric quantities can be obtained in terms of $f_{\mu\nu}$; for instance the fluctuations 
of the Christoffel connection to can be compactly expressed as:
\begin{eqnarray}
\delta^{(1)} \Gamma_{\mu\nu}^{\,\,\,\,\,\,\alpha} = \frac{1}{2} \overline{g}^{\alpha\beta} \biggl[ -  \overline{\nabla}_{\beta} f_{\mu\nu} 
+ \overline{\nabla}_{\nu} f_{\beta\mu} +  \overline{\nabla}_{\mu} f_{\nu\beta} \biggr], \qquad
\delta^{(2)} \Gamma_{\mu\nu}^{\,\,\,\,\,\,\alpha} = \frac{1}{2}\,f^{\alpha\beta} \biggl[ \overline{\nabla}_{\beta} f_{\mu\nu} 
- \overline{\nabla}_{\nu} f_{\beta\mu} -  \overline{\nabla}_{\mu} f_{\nu\beta}  \biggr],
\label{APPA2}
\end{eqnarray}
where  $\overline{\nabla}_{\nu}$ denotes the covariant derivative with respect to the background metric $\overline{g}_{\mu\nu}(x)$. Thanks to the well known Palatini identity stipulating that $\delta^{(1)} R^{\alpha}_{\,\,\,\,\,\,\,\,\mu\beta\nu} = \overline{\nabla}_{\beta} \delta^{(1)} \Gamma_{\mu\nu}^{\,\,\,\,\,\,\alpha}- \overline{\nabla}_{\nu} \delta^{(1)} \Gamma_{\mu\beta}^{\,\,\,\,\,\,\alpha}$ the first-order fluctuations 
of the Riemann tensor become:
\begin{eqnarray}
\delta^{(1)} R^{\alpha}_{\,\,\,\,\,\,\,\,\mu\beta\nu} &=& \frac{1}{2} \biggl[ - \overline{\nabla}_{\beta} \overline{\nabla}^{\alpha} f_{\mu\nu}  - 
\overline{\nabla}_{\nu} \overline{\nabla}_{\mu} \, f_{\beta}^{\,\,\,\, \alpha} + 
\overline{\nabla}_{\beta} \overline{\nabla}_{\nu} f^{\alpha}_{\,\,\,\,\mu} 
\nonumber\\
&+& \overline{\nabla}_{\beta} \overline{\nabla}_{\mu} f^{\alpha}_{\,\,\,\,\nu}
+ \overline{\nabla}_{\nu} \overline{\nabla}^{\alpha} \, f_{\mu\beta} - \overline{\nabla}_{\nu} \overline{\nabla}_{\beta} f^{\alpha}_{\,\,\,\,\mu}\biggr].
\label{APPA3}
\end{eqnarray}
From Eqs.(\ref{APPA2})--(\ref{APPA3}) it is straightforward to obtain the first-order fluctuations of the Ricci tensor, of the scalar 
curvature and of all the other quantities arising in the  effective evolution of the four-dimensional space-time geometry. When the gravitational waves propagate far from the sources the background equations imply that $2 \overline{R}_{\alpha\beta} = \overline{g}_{\alpha\beta} \overline{R}$ and in this approximation the evolution of the disturbances can be expressed in terms of a linear combination of $f_{\mu}^{\,\,\,\nu}$ and of its trace, i.e. 
$ \psi_{\mu}^{\,\,\,\,\nu} = f_{\mu}^{\,\,\,\,\nu} - f \,\delta_{\mu}^{\,\,\,\,\nu}/2$ where $f = \overline{g}^{\alpha\beta} \, f_{\alpha\beta}$. The 
equation obeyed by $\psi_{\mu}^{\,\,\nu}$ reads 
\begin{equation}
\square \psi_{\mu}^{\,\,\,\,\nu} - 2 \psi^{\alpha\lambda} \overline{R}^{\nu}_{\,\,\,\,\,\alpha\lambda\mu} - \overline{\nabla}_{\mu} \overline{\nabla}_{\alpha} \psi^{\nu\alpha} 
- \overline{\nabla}^{\nu} \overline{\nabla}_{\alpha} \psi^{\alpha}_{\,\,\,\,\mu} + \delta_{\mu}^{\,\,\,\,\nu} \overline{\nabla}_{\alpha} 
\overline{\nabla}_{\beta} \psi^{\alpha\beta} =0.
\label{APPA4}
\end{equation}
Both $f_{\mu}^{\,\,\,\nu}$ and $\psi_{\mu}^{\,\,\,\nu}$ change for infinitesimal coordinate transformations of the type  $x^{\mu} \to \widetilde{\,x\,}^{\mu} = x^{\mu} + \epsilon^{\mu}$. In particular we have that  $f_{\mu\nu}$ changes according to the Lie derivative in the direction $\epsilon^{\mu}$, i.e. $\widetilde{\,\,f\,\,}_{\mu\nu}= 
f_{\mu\nu} -  \overline{g}_{\beta\nu}\, \,\overline{\nabla}_{\mu} \epsilon^{\beta}- \overline{g}_{\beta\mu} \,\,\overline{\nabla}_{\nu} \epsilon^{\beta} $. For the same infinitesimal coordinate shift the transformation of $\psi_{\mu\nu}$ can be written as
$ \widetilde{\,\,\psi\,\,}_{\mu\nu}= 
\psi_{\mu\nu} - \overline{\nabla}_{\mu} \epsilon_{\nu} - \overline{\nabla}_{\nu} \epsilon_{\mu}
+ \overline{g}_{\mu\nu} \overline{\nabla}_{\alpha} \, \epsilon^{\alpha}$. By looking at Eq. (\ref{APPA4}) it is clear that 
in the coordinate system where $\overline{\nabla}_{\mu} \psi^{\mu}_{\,\,\,\,\,\nu} =0$ the evolution of $\psi_{\mu\nu}$ eventually 
becomes 
\begin{equation}
 \square \psi_{\mu\nu} - 2 \,\psi^{\alpha\lambda}\, \overline{R}_{\lambda\mu\nu\alpha} =0, \qquad \overline{\nabla}_{\mu} \psi^{\mu}_{\,\,\,\,\,\nu} =0.
\label{APPA5}
\end{equation}
Since  $\psi_{\mu\nu}$ is modified under infinitesimal coordinate shifts 
also the condition $\overline{\nabla}_{\mu} \psi^{\mu}_{\,\,\,\,\,\nu}$  is altered: 
\begin{equation}
\overline{\nabla}_{\mu} \psi^{\mu}_{\,\,\,\,\,\nu} \to \widetilde{\,\,\overline{\nabla}_{\mu} \psi^{\mu}_{\,\,\,\,\,\nu}\,\,} = \overline{\nabla}_{\mu} \psi^{\mu}_{\,\,\,\,\,\nu} - \overline{\nabla}_{\alpha} \overline{\nabla}^{\alpha} \epsilon_{\mu} - \epsilon^{\gamma} \, \overline{R}_{\gamma\mu},
\label{APPA6}
\end{equation}
where $\overline{R}_{\gamma\nu}$ denotes the background Ricci tensor; this term is a consequence 
of the observation that the covariant derivatives in Riemannian and pseudo-Riemannian space-times 
do not commute; in particular Eq. (\ref{APPA6}) can be easily derived by recalling that 
$\epsilon_{\alpha\,;\mu;\,\nu} - \epsilon_{\alpha\,;\nu;\,\mu} = \overline{R}^{\lambda}_{\,\,\,\mu\alpha\beta} \, \epsilon_{\lambda}$. 
Therefore, whenever $\overline{\nabla}_{\nu} \psi^{\nu\mu}\neq 0$ we can always 
perform a gauge transformation (\ref{APPA6}) and select a coordinate system where 
$\widetilde{\,\,\overline{\nabla}_{\mu} \psi^{\mu}_{\,\,\,\,\,\nu}\,\,} =0$.
This condition can always be imposed provided the infinitesimal shift 
 $\epsilon_{\mu}$ obeys $ \overline{\nabla}_{\alpha} \overline{\nabla}^{\alpha} \epsilon_{\mu} + \epsilon^{\gamma} \, \overline{R}_{\gamma\mu}
 = \overline{\nabla}_{\mu} \psi^{\mu}_{\,\,\,\,\,\nu}$ where $\overline{\nabla}_{\mu} \psi^{\mu}_{\,\,\,\,\,\nu}$ is evaluated in the original coordinate 
system and, by assumption, it does not vanish.  For a further infinitesimal coordinate transformation of the type $x^{\mu} \to \widetilde{\,x\,}^{\mu} = x^{\mu} + \epsilon^{\mu}$ the gauge condition remains unaltered provided $\epsilon_{\mu}$ satisfies the equation:
\begin{equation}
\square \epsilon_{\mu} + \epsilon^{\gamma} \overline{R}_{\gamma\mu} =0\qquad \Rightarrow\qquad \square \epsilon_{\mu} + \frac{1}{2}  \overline{R}\, \epsilon_{\mu}=0,
 \label{APPA7}
\end{equation}
which is valid far from the background sources.
 It follows that also in curved backgrounds and in the absence of sources the gauge
freedom can be completely removed by simultaneously enforcing the following three conditions:
\begin{equation}
\overline{\nabla}_{\mu} \psi^{\mu}_{\,\,\,\,\,\nu} =0, \qquad\qquad \overline{g}^{\mu\nu} \psi_{\mu\nu} =0, \qquad\qquad 
\psi^{\mu\nu}\, u_{\nu} =0,
\label{APPA8}
\end{equation}
where $u_{\nu}$ is a unit time-like vector associated with the observer detecting the gravitational radiation. Equation (\ref{APPA8}) amounts, overall, to $8$ independent conditions and the counting goes, in short, as follows. If we choose the vector $u^{\nu}$ to coincide with $(1,\, 0,\, 0,\, 0)$
we have that the requirement $\psi_{\mu\nu} \, u^{\nu} =0$
imposes the independent conditions $\psi_{0\,0} =0$ and $\psi_{i\,0}=0$ (with $i=1, \, 2,\, 3$); 
consequently $\psi_{\mu\nu} \, u^{\nu} =0$ leads overall to $4$ independent conditions.
The condition of vanishing trace (i.e. $\psi =0$) implies $\psi_{0\,0} - \psi_{i\, i} =0$ (sum over the repeated indices is understood); 
but since $\psi_{0\,0}=0$ (as a consequence of the requirement $\psi_{\mu\nu} \, u^{\nu} =0$) we have that 
$\psi=0$ implies $\psi_{i\, i} =0$ (i.e. one independent condition). Finally the gauge choice $\partial_{\nu} \psi^{\nu}_{\,\,\,\mu} =0$ for $\nu=0$ and $\nu=i$  imposes 
two separate requirements:
\begin{eqnarray}
\partial_{\nu} \psi^{\nu}_{\,\,\,0} = \partial_{0} \psi^{0}_{\,\,\,0} +  \partial_{i} \psi^{i}_{\,\,\,0}=0,\qquad \partial_{\nu} \psi^{\nu}_{\,\,\,i} = \partial_{0} \psi^{0}_{\,\,\,i} + \partial_{k} \psi^{k}_{\,\,\, i} =0.
\label{CONDGW2} 
\end{eqnarray}
Since the first equation of (\ref{CONDGW2}) is trivially satisfied 
as a consequence of $\psi_{\mu\nu} \, u^{\nu} =0$, only the second equation of (\ref{CONDGW2})  
is independent and it imposes $3$ independent conditions for $i = 1,\,2,\, 3$.
In summary, from Eq. (\ref{APPA8}) we have that $\psi_{\mu\nu} \, u^{\nu} =0$ amounts to $4$ independent conditions, $\psi=0$ requires $1$ independent condition and $\partial_{\nu} \psi^{\nu}_{\,\,\,\mu} =0$ corresponds to $3$ conditions. The independent conditions are therefore $8$, as anticipated after Eq. (\ref{APPA8}). The conditions expressed in Eq. (\ref{APPA8}) are sometimes referred to as {\em transverse traceless gauge} and depend ultimately upon the choice of the observer. Since $\psi_{\mu\nu}$ contains $10$ independent components only $2$ out of $10$ degrees of freedom are dynamical, exactly as in the case of flat space-time. Note, finally, that in the case where the Riemann tensor of the 
background vanishes consistently Eq. (\ref{APPA5}) coincides with the result of flat space-time.
Since the condition $\overline{g}^{\alpha\beta} \psi_{\alpha\beta} =0$ implies that $f =0$, we can also 
write Eq. (\ref{APPA5}) as
\begin{equation}
 \square f_{\mu\nu} - 2 \,f^{\alpha\lambda}\, \overline{R}_{\lambda\mu\nu\alpha} =0, \qquad \overline{\nabla}_{\mu} f^{\mu}_{\,\,\,\,\,\nu} =0.
\label{APPA9}
\end{equation}
If the contribution of the matter sources is included the form of Eq. (\ref{APPA9}) 
may be different. However the result of Eq. (\ref{APPA9}) holds in a variety of physical situations.
For instance, in the case of perfect fluid sources the analog of Eq. (\ref{APPA9}) becomes
\begin{eqnarray}
&& \Box f_{\mu}^{\,\,\,\,\nu} +  \overline{\nabla}_{\mu} \overline{\nabla}^{\nu} f 
- 2 f^{\alpha \lambda} \overline{R}^{\nu}_{\,\,\,\,\, \alpha\lambda\mu} - \overline{\nabla}_{\mu} \overline{\nabla}_{\alpha} f^{\nu\alpha} 
- \overline{\nabla}^{\nu} \overline{\nabla}_{\alpha} \, f^{\alpha}_{\,\,\,\, \mu}
\nonumber\\
&&+ \delta_{\mu}^{\nu} \biggl\{ -\frac{1}{2} f \overline{R} - \Box f + \overline{\nabla}_{\alpha} \overline{\nabla}_{\beta} f^{\alpha\beta} - \ell_{P}^2 \biggl[(\overline{p}_{t} + \overline{\rho}_{t}) u_{\alpha} u_{\beta} f^{\alpha\beta} - \overline{p}_{t}\, f\biggr] \biggr\} 
\nonumber\\
&&= 2 \ell_{P}^2 (\overline{p}_{t} + \overline{\rho}_{t}) u_{\mu} \, u_{\alpha} \, f^{\alpha\nu}.
\label{APPA9a}
\end{eqnarray}
In the gauge $u^{\mu} f_{\mu\nu} =0$, $\overline{\nabla}_{\mu} f^{\mu\nu}=0$ and $f = \overline{g}^{\mu\nu} \, f_{\mu\nu} =0$ Eq. (\ref{APPA9a}) takes again the form (\ref{APPA9}).

\subsection{Second-order action in the Einstein frame}
Equation (\ref{APPA9}) also follows from the second-order action for the tensor modes of the geometry. There are different ways in which the second-order action can be derived but the first step is to observe 
that the Einstein-Hilbert action can be written in explicit terms by 
isolating the contribution of the total derivatives; more specifically we have that
the sum of the gravity action and of a generic matter contribution $S_{m}$ becomes:
\begin{eqnarray}
S = \frac{1}{2 \ell_{P}^2} \int d^{4} x \sqrt{-g} \, g^{\alpha\beta}\biggl[ \Gamma_{\alpha\beta}^{\,\,\,\,\,\mu} \Gamma_{\mu\nu}^{\,\,\,\,\,\nu} - \Gamma_{\alpha\nu}^{\,\,\,\,\,\mu} \Gamma_{\mu\beta}^{\,\,\,\,\,\nu} \biggr] +  \frac{1}{2\ell_{P}^2} \int d^{4} x \sqrt{-g}\, \, g^{\alpha\beta} \biggl(\nabla_{\beta} \,\Gamma_{\alpha\lambda}^{\,\,\,\,\,\lambda} - \nabla_{\lambda}  \,\Gamma_{\alpha\beta}^{\,\,\,\,\,\lambda}\biggr).
\label{APPA10}
\end{eqnarray}
The third term of Eq. (\ref{APPA10}) combine in a single total derivative that does not contribute to the second-order action which can be derived in, at least, two different ways. Within the covariant approach the second-order action 
\begin{eqnarray}
S_{g} &=& \frac{1}{8 \ell_{P}^2} \int d^{4} x\, \sqrt{-\overline{g}} \biggl[ \overline{\nabla}_{\rho} f_{\mu\nu} \, 
\overline{\nabla}^{\rho} \,f^{\,\,\mu\nu}
+ 2 \overline{R}_{\rho}^{\,\,\,\,\sigma} \,\, f_{\alpha\sigma} \, f^{\alpha\rho} + 2 \, \overline{R}^{\,\mu\,\,\,\,\,\,\,\,\,\,\,\nu}_{\,\,\,\,\,\rho\sigma}\, f_{\mu\nu} \,f^{\rho \sigma}
\nonumber\\
&+& \ell_{P}^2 (\rho_{t} - p_{t}) f_{\mu\nu} \,f^{\mu\nu} \biggr].
\label{APPA11}
\end{eqnarray}
The result of  Eq. (\ref{APPA9a}), originally obtained from the first-order fluctuations 
of the Einstein's equations, follows now by extremizing the action (\ref{APPA11}) with respect to the variation of  $ f^{\mu\nu}$. 
The background Ricci tensor appearing in the first line of Eq. (\ref{APPA11}) can be eliminated 
by using the background Einstein's equations written in the form $\overline{R}_{\mu\nu} = \ell_{P}^2 [ (p_{t} + \rho_{t}) u_{\mu} u_{\nu} + \overline{g}_{\mu\nu} ( p_{t} - \rho_{t})/2]$. The explicit form of Eq. (\ref{APPA11}) then becomes :
\begin{equation}
S_{g} = \frac{1}{8 \ell_{P}^2} \int d^{4} x\, \sqrt{-\overline{g}} \biggl[ \overline{\nabla}_{\rho} f_{\mu\nu} \,
 \overline{\nabla}^{\rho} \,f^{\,\,\mu\nu} + 2 \, \overline{R}^{\mu\,\,\,\,\,\,\,\,\,\,\,\nu}_{\,\,\,\,\,\rho\sigma}\, f_{\mu\nu} \, f^{\rho \sigma}
\biggr].
\label{APPA12}
\end{equation}
In the case of a conformally flat background geometry $\overline{g}_{\mu\nu} = a^2(\tau) \, \eta_{\mu\nu}$ (where $\eta_{\mu\nu}$ 
is the Minkowski metric and $a(\tau)$ is the scale factor) Eq. (\ref{APPA12}) reduces to 
\begin{eqnarray}
S_{g} &=& \frac{1}{8 \ell_{P}^2} \int d^{4} x \, \sqrt{- \overline{g}} \,\, \overline{g}^{\mu\nu} \, \partial_{\mu} h_{i\, j} \, \partial_{\nu} h^{i\, j} = \int d^{3}x\, \int d\tau \, {\mathcal L}_{g}(\vec{x},\tau), 
\nonumber\\
{\mathcal L}_{g}(\vec{x},\tau) &=& \frac{a^2}{8 \ell_{P}^2} \biggl[ \partial_{\tau} h_{i\,j}\partial_{\tau} h^{i\,j} -\partial_{k} h_{i\,j}\partial^{k} h^{i\,j} \biggr].
\label{APPA13}
\end{eqnarray}
where the amplitude has been redefined as $f_{i\, j} = - a^2(\tau) \, h_{i\,j}$;
note that ${\mathcal L}_{g}(\vec{x},\tau)$ denotes the Lagrangian density. From Eq. (\ref{APPA13}) we can deduce 
the energy-momentum pseudo-tensor by taking the variation of $S_{g}$ with respect to $\overline{g}_{\mu\nu}$ and 
the result is 
\begin{equation}
T^{(gw)}_{\mu\nu} = \frac{1}{4 \ell_{P}^2} \biggl[ \partial_{\mu} h_{i\,j} \partial_{\nu} h^{i\,j} 
- \frac{1}{2} \overline{g}_{\mu \nu} \,\biggl(\overline{g}^{\alpha\beta}\, \partial_{\alpha} h_{i\,j}\, \partial_{\beta} h^{i\,j} \biggr)\biggr].
\label{APPA14}
\end{equation}
The most sound prescription for the energy-momentum pseudo-tensor of the relic gravitons follows from the variation of the second-order action with respect to the background metric. The other approaches are fully equivalent in the high frequency limit (i.e. inside the Hubble radius) but lead to various drawbacks when the wavelengths exceed the Hubble radius (i.e. in the low-frequency regime)\cite{effective}. Since the rate of variation of the space-time curvature can be both larger and smaller than the typical frequencies  of the relic gravitons, it is desirable to adopt a definition for the energy-momentum pseudo-tensor that is well defined in spite of of the frequency of the gravitons. In this respect the most plausible definition is the one following from the functional derivative of the effective action with respect to the background metric. 

\subsection{Second-order action in the Jordan frame}
The actions of Eqs. (\ref{APPA12})--(\ref{APPA13}) have been derived in the Einstein frame. The evolution 
of the tensor modes of the geometry could be studied in any action conformally related to the Einstein 
frame. The evolution will clearly be the same and by changing frame nothing dramatic should happen. 
This means, broadly speaking, that the spectrum of the relic gravitons is ultimately the same in all 
frames that are conformally related to the Einstein frame. To clarify this statement we consider here the scalar-tensor action
written in a  generalized Jordan frame 
\begin{equation}
S_{J} = \int d^4 x\, \sqrt{ - G}\,\biggl[ -  \frac{A(\overline{\varphi})}{ 2 \ell_{P}^2}\, R_{J} + \frac{B(\overline{\varphi})}{2} G^{\alpha\beta} \partial_{\alpha} \,\overline{\varphi} \partial_{\beta} \overline{\varphi} 
- W(\overline{\varphi}) \biggr],
\label{APPA15}
\end{equation}
where $A(\overline{\varphi}) $ and $B(\overline{\varphi})$ are both dimensionless;  $\overline{\varphi}$ denotes the scalar 
field written in the Jordan frame. Equation (\ref{APPA15}) corresponds to a canonical action in the Einstein frame 
\begin{equation}
S_{E} = \int d^4 x\, \sqrt{ - g}\,\biggl[ - \frac{R}{ 2 \ell_{P}^2} \,+ \,\frac{1}{2} g^{\alpha\beta} \partial_{\alpha} \varphi \partial_{\beta} \varphi 
- V(\varphi) \biggr].
\label{APPA16}
\end{equation}
The metrics, the scalar fields and the potentials in the two frames are in fact related as
\begin{equation}
A \,\, G_{\alpha\beta} = g_{\alpha\beta}, \qquad d \overline{\varphi} \,\,\sqrt{ \frac{B}{A} + \frac{3}{2 \ell_{P}^2} \biggl(\frac{d \ln{A}}{d \overline{\varphi}}\biggr)^2 } = d \varphi, \qquad V = \frac{W}{A^2}.
\label{APPA17}
\end{equation}
The metric rescaling of Eq. (\ref{APPA17}) becomes more explicit if it is separately written for the background 
and for the first-order fluctuations. The scale factors are then related as $a_{J}^2 \,\,A = a^2$ while 
the connection between the first-order (tensor) fluctuations in the two frames is $a^2_{J} \,\,h^{(J)}_{ij} = a^2\,\, h_{ij}$.
It is already apparent that the connection between the two frames provided by Eq. (\ref{APPA17}) 
must have a direct counterpart in the second-order action. We can then expect that the Jordan frame 
action perturbed to second-order must be directly equivalent to the second-order action in the Einstein frame.
The second-order action in the Jordan frame can be written in terms of ${\mathcal Z}_{\alpha\beta}$ whose explicit form 
is given by 
\begin{equation}
{\mathcal Z}_{\alpha\beta} = \Gamma_{\alpha\beta}^{\,\,\,\,\,\mu} \Gamma_{\mu\nu}^{\,\,\,\,\,\nu} - \Gamma_{\alpha\nu}^{\,\,\,\,\,\mu} \Gamma_{\mu\beta}^{\,\,\,\,\,\nu},
\label{APPA18}
\end{equation}
where the Christoffel symbols are now computed in the Jordan frame. We can therefore denote 
$\overline{{\mathcal Z}}_{\alpha\beta}$ as the background value, $\delta_{t}^{(1)} {\mathcal Z}_{\alpha\beta}$ as 
the first-order tensor fluctuations, $\delta_{t}^{(2)} {\mathcal Z}_{\alpha\beta}$ as 
the second-order tensor fluctuations. In this manner we can then obtain the second-order fluctuations of the Jordan frame action:
\begin{eqnarray}
 \delta^{(2)}_{t} S_{J} &=&\int d^{4} x  \, \biggl\{ \frac{1}{2 \ell_{P}^2}\,\biggl[ A(\overline{\varphi}) \, \overline{G}^{\alpha\beta} \, \overline{{\mathcal Z}}_{\alpha\beta} \,\,\delta^{(2)}_{t} \sqrt{-G} 
  \nonumber\\
 &+& A(\overline{\varphi})\, \sqrt{ -\overline{G}} \biggl( \delta^{(2)}_{t} G^{\alpha\beta} \,\overline{{\mathcal Z}}_{\alpha\beta} +
 \delta^{(1)}_{t} G^{\alpha\beta} \,  \delta^{(1)}_{t} {\mathcal Z}_{\alpha\beta} 
 + \overline{G}^{\alpha\beta}\,\,\delta^{(2)}_{t} {\mathcal Z}_{\alpha\beta} \biggr)
 \nonumber\\
&-& \delta^{(2)}_{t}\biggl( \sqrt{-G} \, G^{\alpha\beta} \, \Gamma_{\alpha\lambda}^{\,\,\,\,\,\lambda} \, \partial_{\beta} A(\overline{\varphi})\biggr) 
+  \delta^{(2)}_{t}\biggl( \sqrt{-G} \, G^{\alpha\beta} \, \Gamma_{\alpha\beta}^{\,\,\,\,\,\lambda} \, \partial_{\lambda} A(\overline{\varphi})\biggr) \biggr]
\nonumber\\
&+&  \delta_{t}^{(2)} \sqrt{-G} \,\biggl(\frac{B(\overline{\varphi})}{2} \overline{G}^{\alpha\beta} \partial_{\alpha} \overline{\varphi} \,\partial_{\beta}\overline{\varphi} 
- W(\overline{\varphi})\biggr) 
\nonumber\\
&+& \sqrt{-\overline{G}} \,\frac{B(\overline{\varphi})}{2} \delta_{t}^{(2)} G^{\alpha\beta} \,\partial_{\alpha} \overline{\varphi} \partial_{\beta}\overline{\varphi}\biggr\}.
\label{APPA19}
\end{eqnarray}
 After some lengthy algebra the explicit expression of Eq. (\ref{APPA19}) assumes a more readable form 
\begin{eqnarray}
 \delta^{(2)} S_{J} &=& \frac{1}{8 \ell_{P}^2} \int d^{4}x \sqrt{-\overline{G}} \, \overline{G}^{\alpha\beta} \, \, A(\overline{\varphi}) \, \partial_{\alpha} h^{\,\,\,\,(J)}_{i\,j}
\partial_{\beta} h^{(J)\,\,\,i\,j} 
\nonumber\\
&-& \frac{1}{8\ell_{P}^2} \int d^{4} x \, a^2_{J} A(\overline{\varphi}) h^{\,\,\,\,(J)}_{k\ell}  h^{(J)\,\,k \ell} \bigg[ 4 {\mathcal H}_{J}^{\prime} 
+ 2 {\mathcal M}^{\prime}
\nonumber\\
&+& 2 ({\mathcal H}_{J}^2 + {\mathcal H}_{J} {\mathcal M} + {\mathcal M}^2) 
+ \frac{ 2 \ell_{P}^2}{A} \biggl( \frac{B}{2} \overline{\varphi}^{\prime \, \,2} - W\, a_{J}^2\biggr)\biggr],
\label{APPA20}
\end{eqnarray}
where the auxiliary quantities ${\mathcal M} = A^{\prime}/A$ and 
${\mathcal H}_{J} = a_{J}^{\prime}/a_{J}$ have been introduced.
 The tensor amplitude $h^{(J)}_{ij}$ entering Eq. (\ref{APPA20}) is defined directly
 in the Jordan frame, i.e.  $\delta_{t}^{(1)} G_{ij} = -a_{J}^2 \, h^{\,(J)}_{ij}$, where, as already mentioned,  
 $a_{J}$ is the scale factor appearing in  the $J$-frame, i.e. $\overline{G}_{\alpha\beta} = a^2_{J} \,\,\eta_{\alpha\beta}$. The expression inside the squared bracket of Eq. (\ref{APPA20}) vanishes identically since it corresponds to the $(ij)$ component of the  background equations derived from the extremization of the action (\ref{APPA15}) with 
respect to the variation of the metric. Therefore the final result is
\begin{equation}
\delta^{(2)} S_{J} = \frac{1}{8 \ell_{P}^2} \int d^{4}x \sqrt{-\overline{G}} \, \overline{G}^{\alpha\beta} \, \, A(\overline{\varphi}) \, \partial_{\alpha} h^{\,\,\,\,(J)}_{i\,j}
\partial_{\beta} h^{(J)\,\,\,i\,j}.
\end{equation}
If we now consider the case of a conformally 
flat background in the Jordan frame we obtain, from the previous 
equation,
\begin{equation}
\delta^{(2)} S_{J} =  \frac{1}{8 \ell_{P}^2} \int d^{4}x \, a_{J}^2\, A(\overline{\varphi}) \, \eta^{\alpha\beta}\,\partial_{\alpha} h^{\,\,\,\,(J)}_{i\,j}\partial_{\beta} h^{(J)\,\,\,i\,j}.
\label{APPA21}
\end{equation}
We may now insert the two conditions $a_{J}^2 \,\,A = a^2$ and $a^2_{J} \,\,h^{(J)}_{ij} = a^2\,\, h_{ij}$
so that the explicit expression of the action becomes:
\begin{equation}
\delta^{(2)} S_{J} =
 \frac{1}{8 \ell_{P}^2} \int d^{4}x \, a^2 \biggl[ \partial_{\tau} h_{i\,j}\partial_{\tau} h^{i\,j} -\partial_{k} h_{i\,j}\partial^{k} h^{i\,j} \biggr].
\label{APPA22}
\end{equation}
 This result shows that Eqs. (\ref{APPA13}) and (\ref{APPA22}) are ultimately one and the same equation even if the evolution of the backgrounds and of the related fluctuations in the two frames may look different:
the equivalence of the two actions guarantees that all the relevant observables must coincide. This property can be directly verified in the case of the energy density and of the spectral energy density \cite{MGB}.

\subsection{More general form of the effective action}
The effective action of the tensor modes of the geometry may be written in a form that is more general than the one of Eq. (\ref{APPA13}):
\begin{equation}
S_{g} = \frac{1}{8\ell_{P}^2} \int \, d^{3} x \int \, d \tau \biggl[ d_{1}(\tau) \,\partial_{\tau} h_{i\, j}\, \partial_{\tau} h^{i\, j} - d_{2}(\tau) \, \partial_{k} h_{i \, j} \,\partial^{k}h^{i\, j} 
- d_{3}(\tau)\, m_{c}^2\, h_{i \, j} \, \partial h^{i \, j} \biggr].
\label{APPG1}
\end{equation}
The parity-breaking terms associated 
with quadratic combinations involving either the dual Riemann or the dual Weyl tensors have been neglected; both terms would appear in the effective action and can polarize the backgrounds of relic gravitons. 
While $d_{1}(\tau)$ and $d_{2}(\tau)$ are related to the expanding 
dimensions while $d_{3}(\tau)$ may appear in the case of compact extra-dimensions. We can always factor one of the coefficients; if $d_{1}(\tau)$ is factored the resulting expression can be written as:
\begin{equation}
S_{g} = \frac{1}{8\ell_{P}^2} \int \, d^{3} x \int \, d \tau \,d_{1}(\tau) \biggl[\partial_{\tau} h_{i\, j} \partial_{\tau} h^{i\, j} - \frac{1}{n^2(\tau)} \,\partial_{k} h_{i \, j} \partial^{k} h^{i\, j} - \frac{1}{\overline{n}^2(\tau)} m_{c}^2  h_{i \, j} h^{i \, j} \biggr],
\label{APPG2}
\end{equation}
where $n(\tau)$ and $\overline{n}(\tau)$ are, respectively, the refractive indices associated with the expanding and with the compact dimensions  $n(\tau) = \sqrt{d_{1}(\tau)/d_{2}(\tau)}$ and $\overline{n}(\tau) = \sqrt{d_{1}(\tau)/d_{3}(\tau)}$. The final form of (\ref{APPG2}) can be 
simplified even further by introducing $b(\eta) = \sqrt{d_{1}(\eta)/n(\eta)}$
and $r_{c}(\eta) = n(\eta)/\overline{n}(\eta)$:
\begin{equation}
S_{g} = \frac{1}{8\, \ell_{P}^2} \int \, d^{3} x \int \, d \eta \, \, b^2(\eta) \biggl[\partial_{\eta} h_{i\, j} \partial_{\eta} h^{i\, j} -  \, \, \partial_{k} h_{i \, j} \partial^{k} h^{i\, j} 
- r_{c}^2(\eta)  m_{c}^2 \, h_{i \, j} h^{i \, j} \biggr].
\label{APPG3}
\end{equation}
In the absence of a contribution from the internal dimensions (i.e. $m_{c} \to 0$)
Eq. (\ref{APPG3}) reproduces exactly Eq.  when $n \to 1$ and $d_{1}(\tau) = a^2(\tau)$. Equation (\ref{APPG3}) follows from Eq. (\ref{APPG2}) by first changing the time 
parametrization from $\tau$ (the conformal time coordinate) to $\eta$ 
according to $n(\eta) d\eta = d\tau$.  Let us therefore consider the simplest situation 
where the refractive index increases during inflation as suggested in Eq. (\ref{STHR3}); in this case for $a< a_{\ast}$ we would have $n(a) = n_{\ast} (a/a_{\ast})^{\alpha}$ (with $\alpha >0$) so that the relation between the conformal time coordinate $\tau$ and the $\eta$-time can be swiftly worked since 
$d \eta = d\tau/n(a)$. From the definition of $\eta$ we therefore have:
\begin{equation}
\eta = \int \frac{da}{a^2 \, H\, n} = - \frac{1}{a\, H\, n} + (\epsilon - \alpha)\int \frac{d a}{a^2 \, H\, n},
\label{APPG4}
\end{equation}
where, as in Eq., $H = \dot{a}/a$ and the overdot denotes a derivation 
with respect to the cosmic time coordinate. The second equality in Eq. (\ref{APPG4}) follows after integration by parts since $\dot{\epsilon} \ll 1$ and $\dot{\alpha} =0$. Equation 
(\ref{APPG4}) also implies that $a \, H\, n= - 1/[(1 -\epsilon + \alpha) \eta]$; note 
once more that when $n \to 1$ we also have $\alpha \to 0$ and the standard 
relation $a \, H= -1/[(1-\epsilon)\tau]$ is immediately recovered.
In the $\eta$-time parametrization the evolution of the mode functions simplifies 
and it is given by
\begin{equation}
\partial_{\eta}^2 f_{k} + \biggl[ k^2 - \frac{\partial_{\eta}^2 b}{b}\biggr] f_{k} =0, \qquad 
g_{k} = \partial_{\eta} f_{k} - \frac{\partial_{\eta}b}{b} f_{k}.
\label{APPG5}
\end{equation}
From Eq. (\ref{APPG5}) it follows that the crossing of a given 
wavelength occurs when $k^2 = (\partial_{\eta}^2 b)/b$. This expression 
generalizes therefore the notion of the comoving horizon 
during the refractive phase. 
\end{appendix}


\newpage

\begin{thebibliography}{99}

\itemsep -4pt

\bibitem{AE1} A. Einstein, Preuss. Akad. Wiss. Berlin {\bf 1},  688 (1916).

\bibitem{AE2} A. Einstein,  Preuss. Akad. Wiss. Berlin {\bf 1}, 154 (1918).

\bibitem{AE2a}  A. Einstein and N. Rosen, Journal of the Franklin Institute {\bf 223}, 43 (1937).

\bibitem{AE2b}  J.~Weber and J.~A.~Wheeler,  Rev.\ Mod.\ Phys.\  {\bf 29},  509 (1957).

\bibitem{AE2c}  F.~A.~E.~Pirani, Acta Phys.\ Polon.\  {\bf 15}, 389 (1956)  [Gen.\ Rel.\ Grav.\  {\bf 41}, 1215 (2009)].

\bibitem{WEB1} J.~Weber, Phys. Rev. Lett. {\bf 18}, 498 (1967).

\bibitem{WEB2} J.~Weber,  Phys.\ Rev.\ Lett.\  {\bf 24}, 276 (1970).

\bibitem{AE3}  J. H. Taylor snd J.~M. Weisberg, Astrophys.~J. {\bf 253}, 908 (1982).

\bibitem{AE4} B. P.  Abbott  {\it et~al.},  Phys. Rev. Lett. {\bf 116}, 061102 (2016).

\bibitem{AE5} B. P.  Abbott  {\it et~al.},  Phys. Rev. Lett. {\bf 116}, 241103 (2016).

\bibitem{AE6} B. P.  Abbott  {\it et~al.},  Phys. Rev. Lett. {\bf 118}, 221101 (2017).

\bibitem{AA01}  L.~Parker,  Phys.\ Rev.\ Lett.\  {\bf 21}, 562 (1968).

\bibitem{AA02}  L.~Parker,  Phys.\ Rev.\  {\bf 183}, 1057 (1969).

\bibitem{AA1} L.~P.~Grishchuk,  Sov.\ Phys.\ JETP {\bf 40}, 409 (1975)   [Zh.\ Eksp.\ Teor.\ Fiz.\  {\bf 67}, 825 (1974)].

\bibitem{AA2} L.~P.~Grishchuk,  Annals N.\ Y.\ Acad.\ Sci.\  {\bf 302}, 439 (1977).

\bibitem{AA3}  L. H. Ford and L. Parker, Phys.\ Rev.\ D {\bf 16}, 245 (1977).

\bibitem{AA4} B. L. Hu and L. Parker, Phys. Lett. A {\bf 63}, 217 (1977).

\bibitem{INFL1}  A.~A.~Starobinsky,  Phys.\ Lett.\ B {\bf 91}, 99 (1980)  [Adv.\ Ser.\ Astrophys.\ Cosmol.\  {\bf 3}, 130 (1987)].

\bibitem{INFL2}  A.~H.~Guth,  Phys.\ Rev.\ D {\bf 23}, 347 (1981) [Adv.\ Ser.\ Astrophys.\ Cosmol.\  {\bf 3}, 139 (1987)].
  
\bibitem{INFL3} A.~D.~Linde,  Phys.\ Lett.\  {\bf 108B}, 389 (1982) [Adv.\ Ser.\ Astrophys.\ Cosmol.\  {\bf 3}, 149 (1987)].  
  
\bibitem{INFL4} A.~Albrecht and P.~J.~Steinhardt,  Phys.\ Rev.\ Lett.\  {\bf 48}, 1220 (1982)  
[Adv.\ Ser.\ Astrophys.\ Cosmol.\  {\bf 3}, 158 (1987)]. 

\bibitem{AA7}  A. A. Starobinsky, JETP Lett. {\bf 30}, 682 (1979) [Pis'ma Zh. Eksp. Teor. Fiz. {\bf 30}, 719 (1979)].

\bibitem{AA8} L.~F.~Abbott and M.~B.~Wise, Nucl.\ Phys.\ B {\bf 244}, 541 (1984).
   
\bibitem{AA9}  S.~W.~Hawking,  Phys.\ Lett.\  {\bf 150B}, 339 (1985).

\bibitem{AA10}  V. A. Rubakov, M. V. Sazhin, and A. V. Veryaskin, Phys. Lett. B {\bf 115}, 189 (1982).

\bibitem{HBB1} G. Gamow, Phys. Rev. {\bf 70}, 572 (1946).

\bibitem{HBB2} R. Alpher, H. Bethe, and G. Gamow, Phys. Rev. {\bf 73}, 803 (1948).

\bibitem{HBB3} R. Alpher and R. Herman, Rev. Mod. Phys. {\bf 22}, 153 (1950).

\bibitem{HBB4} A.~A.~Penzias and R.~W.~Wilson,  Astrophys.\ J.\  {\bf 142}, 419 (1965).

\bibitem{HBB4a} P. J. E. Peebles,  Astrophys. J. {\bf 142}, 1317 (1965).

\bibitem{HBB4b} P. J. E. Peebles,  Phys. Rev. Lett.  {\bf 16},  410 (1966).

\bibitem{HBB5} P. J. E. Peebles, L. A. Page  Jr., R. B. Partridge, {\it Finding the Big Bang}, (Cambridge University Press, Cambridge, UK, 2009).

\bibitem{LIGO0}  B.~Abbott {\it et al.} [LIGO Collaboration], Phys.\ Rev.\ D {\bf 69}, 122004 (2004); Phys.\ Rev.\ Lett.\  {\bf 95}, 221101 (2005).  

\bibitem{LIGO0a} J.~Aasi {\it et al.} [LIGO/Virgo Collaboration], Phys.\ Rev.\ Lett.\  {\bf 113}, 231101 (2014); Phys.\ Rev.\ D {\bf 91},  022003 (2015).

\bibitem{LIGO0b} B.~P.~Abbott {\it et al.} [LIGO/Virgo Collaboration],  Phys.\ Rev.\ Lett.\  {\bf 118}, 121101 (2017)  Erratum: [Phys.\ Rev.\ Lett.\  {\bf 119}, 029901 (2017)]; Phys.\ Rev.\ D {\bf 100}, 061101(R) (2019).

\bibitem{LIGO1} R.~Abbott {\it et al.} [KAGRA, Virgo and LIGO Scientific], Phys. Rev. D {\bf 104}, 022004 (2021).

\bibitem{LIGO3} M.~Giovannini, Prog. Part. Nucl. Phys. \textbf{112}, 103774 (2020).

\bibitem{NANO1} Z. Arzoumanian {\it et al.}, Astrophys. J. Lett. {\bf 905}, L34 (2020).

\bibitem{NANO2} G.~Agazie \textit{et al.} [NANOGrav], Astrophys. J. Lett. {\bf 951},  L8 (2023).

\bibitem{PPTA1} B. Goncharov {\it et al.} Astrophys. J. Lett. {\bf 917}, L19 (2021).

\bibitem{PPTA2} D.~J.~Reardon, {\it et al.}, Astrophys. J. Lett. \textbf{951},  L6 (2023).

\bibitem{RR1}  Y.~Akrami {\it et al.} [Planck Collaboration], Astron. Astrophys. {\bf 641}, A10 (2020).

\bibitem{RR2}  N.~Aghanim {\it et al.} [Planck Collaboration], Astron. Astrophys. {\bf 641}, A6 (2020).

\bibitem{RR3} P.~Ade {\it et al.} [BICEP and Keck], Phys. Rev. Lett. {\bf 127}, 151301 (2021).

\bibitem{EXP1} M.~Giovannini, Phys. Rev. D \textbf{58}, 083504 (1998).

\bibitem{EXP2} M.~Giovannini, Phys. Rev. D \textbf{60}, 123511 (1999).

\bibitem{EXP3} M.~Giovannini, Class. Quant. Grav. \textbf{16}, 2905 (1999).

\bibitem{EXP4} P.~J.~E.~Peebles and A.~Vilenkin,  Phys.\ Rev.\ D {\bf 59}, 063505 (1999).

\bibitem{MGB} M. Giovannini, {\it Relic Gravitons}, (World Scientific, Singapore, 2024).

\bibitem{PJE} P. J. E. Peebles, {\it Physical Cosmology}, (Princeton University Press, Princeton NJ, 1971).

\bibitem{SW1} S. Weinberg, {\it Gravitation and Cosmology}, (Wiley, New York, 1972).

\bibitem{SW2} S. Weinberg, {\it The first three minutes}, (basic Books, New York, 1977).

\bibitem{BBK1}  A.R. Liddle, D. Lyth, {\it Cosmological Inflation and Large-Scale Structure},  (Cambridge University Press, Cambridge, UK, 2000).

\bibitem{BBK2} M. Giovannini, {\it A primer on the Physics of the Cosmic Microwave Background}, (World Scientific, Singapore, 2008). 

\bibitem{BBK3} S. Weinberg, {\it Cosmology} (Oxford University Press, Oxford, UK, 2008).

\bibitem{EFFT} S.~Weinberg, Phys.\ Rev.\ D {\bf 77}, 123541 (2008).

\bibitem{bardeen}  J. M. Bardeen, {\it Phys. Rev. D }  {\bf 22}, 1882 (1980).

\bibitem{bardeen2} J. Bardeen, P. Steinhardt, M. Turner, {\it Phys. Rev. D }  {\bf 28}, 679 (1983).

\bibitem{liddleetal} D. Wands {\it et al.}, {\it Phys. Rev. D} {\bf 62}, 043527 (2000).

\bibitem{wein1} S.~Weinberg, {\it Phys.\ Rev.\ D}  {\bf 67}, 123504 (2003).

\bibitem{wein2}  S.~Weinberg, {\it Phys.\ Rev.\ D} {\bf 70}, 043541 (2004).

\bibitem{wein3} S.~Weinberg, {\it Phys.Rev.D } {\bf 70}, 083522 (2004). 

\bibitem{MB} C.-P. Ma and E. Bertschinger, {\it Astrophys. J.}{\bf 455}, 7 (1995).

\bibitem{LID1}  A.~R.~Liddle and S.~M.~Leach, Phys. Rev. D {\bf 68}, 103503 (2003).

\bibitem{MGshift} M.~Giovannini, Phys. Rev. D \textbf{105}, 103524 (2022).

\bibitem{turn1} M. S. Turner, Phys. Rev. D \textbf{28}, 1243 (1983).

\bibitem{turn2} C. Pathinayake and L. H. Ford, Phys. Rev. D \textbf{35}, 3709 (1987).

\bibitem{turn3}  L. H. Ford, Phys. Rev. D \textbf{35}, 2955 (1987).

\bibitem{turn4} J. D. Barrow, Phys. Rev. D \textbf{48}, 1585 (1993).

\bibitem{MGshift2} M.~Giovannini, Phys. Rev. D \textbf{108},  123508 (2023).

\bibitem{BB1}  A. D. Sakharov,  Sov. Phys. JETP {\bf 22}, 241 (1966) [Zh. Eksp. Teor. Fiz. {\bf 49}, 345 (1965)].

\bibitem{BB2} P.~J.~E.~Peebles and J.~T.~Yu,  Astrophys.\ J.\  {\bf 162} 815 (1970).

\bibitem{BB3} R.~A.~Sunyaev and Y.~B.~Zeldovich,  Astrophys.\ Space Sci.\  {\bf 7}, 3 (1970).

\bibitem{FLS1} V.~Sahni, Phys.\ Rev.\ D {\bf 42}, 453 (1990).

\bibitem{FLS2} L.~P.~Grishchuk and M.~Solokhin,  Phys.\ Rev.\ D {\bf 43}, 2566 (1991).

\bibitem{MGSTOC} M. Giovannini, JCAP {\bf 11}, 027 (2024).

\bibitem{CORR6} L.~P.~Grishchuk, Usp. Fiz. Nauk \textbf{182}, 222 (2012).

\bibitem{abr1}  A. Erdelyi, W. Magnus, F. Oberhettinger, and F. R. Tricomi {\it Higher Trascendental Functions} (Mc Graw-Hill, New York, 1953).

\bibitem{abr2} M. Abramowitz and I. A. Stegun, {\it Handbook of Mathematical Functions} (Dover, New York, 1972).

\bibitem{STOC1} C. W. Gardiner,  {\it Handbook of stochastic methods}, (Springer-Verlag, Berlin, 1987).

\bibitem{STOC2} S. Karlin and H. M. Taylor, {\it A first course in stochastic processes} (Academic Press, New York, 1975).

\bibitem{WK1}  N. Wiener, Acta Math. {\bf 55}, 117 (1930).

\bibitem{WK2}  A. Khintchine, Math. Ann. {\bf 104}, 415  (1931).

\bibitem{CORR1} P. Michelson,  Mon. Not. Roy. Astron. Soc. {\bf 227}, 933 (1987).

\bibitem{CORR2} N. Christensen,  Phys. Rev. D {\bf 46}, 5250 (1992).

\bibitem{CORR3} E.  Flanagan ,Phys. Rev. D {\bf 48},  2389 (1993)

\bibitem{CORR4} D.  Babusci and M. Giovannini, Phys. Rev. D {\bf 60},  083511 (1999).

\bibitem{CORR7}  N.  Christensen, Rep. Prog. Phys. {\bf 82}, 016903 (2019).

\bibitem{PARKTH} L. Parker, Nature {\bf 261}, 20 (1976).

\bibitem{BIRREL}  N. D. Birrel and P. C. W. Davies, {\it Quantum fields in curved spaces} (Cambridge Univ. Press, Cambridge, England, 1982).

\bibitem{PTOMS}  L. Parker and D. Toms, {\it Quantum Field Theory in Curved Space-time}, (Cambridge University Press, Cambridge 2009).

\bibitem{MANDL} L. Mandel and E. Wolf, {\it Optical Coherence and Quantum Optics} (Cambridge University Press, Cambridge, 1995).

\bibitem{HBT4}  R. Loudon, {\it The Quantum Theory of Light} (Clarendon Press, Oxford, 1983).

\bibitem{louisell} W. H. Louisell, A. Yariv, and A. E. Siegman Phys. Rev. {\bf 124}, 1646 (1961).

\bibitem{mollow1}  B. L. Mollow and R. J. Glauber, Phys. Rev. {\bf 160}, 1076 (1967).

\bibitem{mollow2}   B. L. Mollow and R. J. Glauber, Phys. Rev. {\bf 160}, 1097 (1967).

\bibitem{RGS0} L. P. Grishchuk and Y. V. Sidorov, Phys. Rev. D {\bf 42}, 3413 (1990).

\bibitem{HBTG1} M.~Giovannini, Phys. Rev. D \textbf{83}, 023515 (2011).

\bibitem{WK}  L.~M.~Krauss and F.~Wilczek, Phys. Rev. D \textbf{89},  047501 (2014).

\bibitem{MGF} M.~Giovannini, Phys. Lett. B \textbf{854}, 138769 (2024).

\bibitem{MGQ} M.~Giovannini, Phys. Rev. D {\bf 110}, 123520 (2024).

\bibitem{bbn1}  V.F. Schwartzman, Pis'ma Zh. Eksp. Teor. Fiz. {\bf 9}, 315 (1969) [JETP Lett. {\bf 9}, 184 (1969)].
  
\bibitem{bbn2} M.~Giovannini, H.~Kurki-Suonio and E.~Sihvola, Phys.\ Rev.\  D {\bf 66}, 043504 (2002).

\bibitem{bbn3} R. Cyburt, B.~D.~Fields, K.~A.~Olive, and E.~Skillman, Astropart.\ Phys.\ {\bf 23}, 313 (2005).

\bibitem{bbn4} D.~M.~Siegel and M.~Roth, Astrophys. J. \textbf{784}, 88 (2014).

\bibitem{bbn5} I.~Lopes and J.~Silk, Astrophys. J. \textbf{794}, 32 (2014).

\bibitem{ST3} M.~Giovannini, Phys. Lett. B \textbf{668}, 44 (2008).

\bibitem{ST3a} M.~Giovannini, Class. Quant. Grav. \textbf{26}, 045004 (2009).

\bibitem{DD5} T.~L.~Smith, E.~Pierpaoli and M.~Kamionkowski, Phys. Rev. Lett. \textbf{97}, 021301 (2006).

\bibitem{DD6} D.~M.~Siegel and M.~Roth, Astrophys. J. \textbf{784}, 88 (2014).

\bibitem{MGHF} M.~Giovannini, JCAP \textbf{05}, 056 (2023).

\bibitem{FA} F.~Dyson, Int. J. Mod. Phys. A \textbf{28}, 1330041 (2013).

\bibitem{FB} M. E. Gertsenshtein,  Sov. Phys. JETP {\bf 14}, 84 (1962) [Zh.\ Eksp.\ Teor.\ Fiz.\  {\bf 41}, 113 (1961)].

\bibitem{FC} V.~Anastassopoulos \textit{et al.} [CAST], Nature Phys. \textbf{13}, 584-590 (2017).

\bibitem{FA1} Y. Kahn, B. R. Safdi, and J. Thaler, Phys. Rev. Lett. {\bf 117}, 141801 (2016).

\bibitem{FA2} S.~Chaudhuri, P.~W.~Graham, K.~Irwin, J.~Mardon, S.~Rajendran and Y.~Zhao,
Phys. Rev. D \textbf{92}, 075012 (2015).

\bibitem{FA3} J. L. Ouellet {\it et al.}, Phys. Rev. Lett. {\bf 122} , 121802 (2019).

\bibitem{FA4} R. Lasenby,  Phys. Rev. D {\bf 102}, 015008 (2020).

\bibitem{FD} A.~Arvanitaki and A.~A.~Geraci, Phys. Rev. Lett. \textbf{113}, 161801 (2014).

\bibitem{CAV1} F. Pegoraro, L.  Radicati, Ph. Bernard, and E. Picasso, Phys. Lett. A {\bf 68}, 165 (1978).

\bibitem{CAV2} F. Pegoraro, E. Picasso, and L. Radicati, J. Phys. A {\bf 11}, 1949 (1978).

\bibitem{CAV3} C. M. Caves, Phys. Lett. B {\bf 80}, 323 (1979).

\bibitem{CAV4} C.  Reece, P. Reiner, and A.  Melissinos, Nucl. Inst. and Methods, A {\bf 245}, 299 (1986).

\bibitem{CAV5} Ph. Bernard, G. Gemme, R. Parodi and E. Picasso, Rev. Sci. Instrum. {\bf 72}, 2428 (2001).

\bibitem{CAV6} R. Ballantini, P. Bernard, A. Chincarini, G. Gemme, R. Parodi and E. Picasso, Class. Quant. Grav. {\bf 21}, S1241 (2004).

\bibitem{WG1} V. B. Braginsky and M. B. Menskii,  Pis'ma Zh. Eksp. Teor. Fiz. {\bf 13}, 585 (1971) 
[JETP Lett. {\bf 13}, 417 (1971)].

\bibitem{WG2} V. B. Braginsky, L.P. Grishchuk, A. G. Doroshkevich, Ya. B. Zeldovich, I. D. Novikov and M. Sazhin,  Sov. Phys. JETP {\bf 38}, 865 (1974) [Zh. Eksp. Teor. Fiz. {\bf 65}, 1729 (1973)].

\bibitem{WG3} A. M. Cruise, Class. Quantum Grav. {\bf 17} , 2525 (2000).

\bibitem{WG4} A. M. Cruise and R. M. Ingley, Class. \ Quantum \ Grav. {\bf 23},  6185 (2006).

\bibitem{WG5}  F.~Y.~Li, M.~X.~Tang and D.~P.~Shi,  Phys.\ Rev.\  D {\bf 67}, 104008 (2003).

\bibitem{SI1} A.~Nishizawa, \textit{et al.} Phys. Rev. D \textbf{77}, 022002 (2008).

\bibitem{SI2} S. Dimopoulos, P. Graham, J.  Hogan, M. Kasevich, S. Rajendran Phys.Rev.D {\bf 78} 122002 (2008).

\bibitem{SI3} S.~Dimopoulos, P.~W.~Graham, J.~M.~Hogan, M.~A.~Kasevich and S.~Rajendran, Phys. Lett. B \textbf{678}, 37 (2009).

\bibitem{ST11} J.~T.~Hsiang and B.~L.~Hu, Universe \textbf{8}, 27 (2022).

\bibitem{RR1a}  L. Dai, M. Kamionkowski, J. Wang Phys. Rev. Lett.  {\bf 113}, 041302 (2014).

\bibitem{RR2a} J.~B.~Munoz and M.~Kamionkowski, Phys. Rev. D {\bf 91}, 043521 (2015). 

\bibitem{RR3a}  J.~L.~Cook, E.~Dimastrogiovanni, D.~A.~Easson and L.~M.~Krauss, JCAP \textbf{04}, 047 (2015).

\bibitem{REF3a} P. A. R. Ade et al. (BICEP2 Collaboration), Phys. Rev. Lett. {\bf 112}, 241101 (2014).

\bibitem{fluid1} M.~Giovannini, Phys. Rev. D \textbf{88}, 021301 (2013).

\bibitem{REF3b} M.~Giovannini, Phys. Rev. D \textbf{89}, 123517 (2014).

\bibitem{HT3} R. Easther, B. Bahr-Kalus, and D. Parkinson, Phys. Rev. D {\bf 106}, L061301 (2022).

\bibitem{HT1} L.~Boubekeur and D.~H.~Lyth, JCAP \textbf{07}, 010 (2005)

\bibitem{HT2} N.~K.~Stein and W.~H.~Kinney, JCAP \textbf{03}, 027 (2023).

\bibitem{HT2aa} W. J. Wolf, Phys. Rev. D {\bf 110}, 043521 (2024).

\bibitem{FR1} H. Motohashi and A. A. Starobinsky, JCAP {\bf 11}, 025 (2019).

\bibitem{FR2}  M. Guerrero, D. Rubiera-Garcia and D. Saez-Chillon Gomez, Phys. Rev. D {\bf 102}, 123528 (2020). 

\bibitem{FR3} A. Mohammadi, T. Golanbari, S. Nasri and K. Saaidi, Phys. Rev. D {\bf 101}, 123537 (2020).

\bibitem{HT2a} A.~Mohammadi, N.~Ahmadi and M.~Shokri, JCAP \textbf{06}, 058 (2023).

\bibitem{effective} M.~Giovannini, Phys. Rev. D \textbf{100}, 083531 (2019).

\bibitem{ZELD} Ya. Zeldovich, Sov. Phys. Usp. {\bf 6}, 475 (1964) [Usp. Fiz. Nauk. {\bf 80}, 357 (1963)].

\bibitem{FORD1} L. H. Ford, Phys. Rev. D {\bf 35}, 2955  (1987).

\bibitem{spok} B.~Spokoiny, Phys. Lett. B \textbf{315}, 40 (1993).

\bibitem{HIGHFF7} P. J. E. Peebles and B. Ratra, Astrophys. J. {\bf 325}, L17 (1988).

\bibitem{HIGHFF8} R. R. Caldwell, R. Dave, and P. J. Steinhardt, Phys. Rev. Lett. {\bf 80}, 1582 (1998).

\bibitem{ST4} J.~Haro, W.~Yang and S.~Pan, JCAP {\bf 01}, 023 (2019).

\bibitem{ST5} M.~Gorghetto, E.~Hardy and H.~Nicolaescu, JCAP {\bf 06}, 034 (2021).

\bibitem{ST6} B.~Li and P.~R.~Shapiro, JCAP {\bf 10}, 024 (2021).

\bibitem{MGew1} M.~Giovannini, Phys. Rev. D \textbf{61}, 063004 (2000).

\bibitem{MGew2} M.~Giovannini, Phys. Rev. D \textbf{61}, 063502 (2000).

\bibitem{MGew3} M.~Giovannini, Class. Quant. Grav. \textbf{34}, 135010 (2017).

\bibitem{STRNU0}  S. Weinberg, Phys. Rev. D {\bf 69}, 023503 (2004).

\bibitem{STRNU1} D.~A.~Dicus and W.~W.~Repko,  Phys.\ Rev.\  D {\bf 72}, 088302 (2005).
 
\bibitem{STRNU2} H.~X.~Miao and Y.~Zhang, Phys.\ Rev.\ D {\bf 75}, 104009 (2007).

\bibitem{STRNU3}  B.~A.~Stefanek and W.~W.~Repko, Phys.\ Rev.\ D {\bf 88},  083536 (2013).

\bibitem{STRNU4}  K.~W.~Ng, Phys.\ Rev.\ D {\bf 86}, 103510 (2012).

\bibitem{Dn1} C.J. Copi, D.N. Schramm, and M.S. Turner, Phys. Rev. D {\bf 55}, 3389 (1997).

\bibitem{Dn2}  S. Burles, K.M. Nollett, J.W. Truran, and M.S. Turner, Phys. Rev. Lett. {\bf 82}, 4176 (1999).

\bibitem{Dn3} R. Cyburt, B. Fields, and K. Olive, Astropart. Phys. {\bf 17}, 87 (2002).

\bibitem{Dn4} R. Cyburt, B.~D.~Fields, K.~A.~Olive and T.~H.~Yeh,  Rev.\ Mod.\ Phys.\  {\bf 88}, 015004 (2016).

\bibitem{PTFIRST1} D.~A.~Kirzhnits and A.~D.~Linde, Phys. Lett. B \textbf{42}, 471 (1972).

\bibitem{PTFIRST2} D.~A.~Kirzhnits and A.~D.~Linde, Annals Phys. \textbf{101}, 195-238 (1976).

\bibitem{PTFIRST3} A.~D.~Linde, Rept. Prog. Phys. \textbf{42}, 389 (1979).

\bibitem{PTFIRST4} A.~D.~Linde, Phys. Lett. B \textbf{96}, 289 (1980).

\bibitem{PTFIRST5} K.~Kajantie {\it et al}., Nucl. Phys. B \textbf{458}, 90 (1996).

\bibitem{PTFIRST6} K.~Kajantie {\it et al}., Nucl. Phys. B \textbf{466}, 189 (1996).

\bibitem{PTFIRST7} K.~Kajantie,  {\it et al}., Phys. Rev. Lett. \textbf{77}, 2887 (1996).

\bibitem{PTFIRST8} F.~Csikor, Z.~Fodor and J.~Heitger, Phys. Lett. B \textbf{441}, 354 (1998).

\bibitem{PTFIRST9} F.~Csikor, Z.~Fodor and J.~Heitger, Phys. Rev. Lett. \textbf{82}, 21 (1999).

\bibitem{MGpuls}  M.~Giovannini, Eur. Phys. J. C \textbf{84}, 67 (2024).

\bibitem{EPTA1} S.~Chen, {\it et al.} Mon. Not. Roy. Astron. Soc. {\bf 508},  4970 (2021). 

\bibitem{EPTA2} J.~Antoniadis, {\it et al.} Astron. Astrophys.  {\bf 678}, 50 (2023).

\bibitem{IPTA1} J.~Antoniadis, {\it et al.} Mon. Not. Roy. Astron. Soc. \textbf{510},  4873 (2022).

\bibitem{CPTA} H.~Xu, S.~Chen,  {\it et al.} Res. Astron. Astrophys. \textbf{23},  075024 (2023).

\bibitem{PP1a} M. V. Sazhin,  Sov. Astron.  {\bf 22},  36 (1978) [Astron. Zh. {\bf 55}, 65 (1979)].

\bibitem{PP1b} S. Detweiler, Astrophys. J. {\bf 234},  1100 (1979).

\bibitem{PP1c} R. W. Hellings and G. S. Downs, Astrophys. J. Lett. {\bf  265} L39 (1983).

\bibitem{PP2a} V.~M.~Kaspi, J.~H.~Taylor, and M.~F.~Ryba,   Astrophys.\ J.\ {\bf 428}, 713 (1994).

\bibitem{PP2b} F.~A.~Jenet {\it et al.},  Astrophys.\ J.\  {\bf 653}, 1571 (2006).

\bibitem{PP2c} W.~Zhao,  Phys.\ Rev.\ D {\bf 83}, 104021 (2011).

\bibitem{PP2d} P.~B.~Demorest {\it et al.},  Astrophys.\ J.\  {\bf 762}, 94 (2013).

\bibitem{CC2}  M.~Giovannini, Class.\ Quant.\ Grav.\  {\bf 33}, 125002 (2016).

\bibitem{CC1} P.~Szekeres,  Annals Phys.\  {\bf 64}, 599 (1971).

\bibitem{CC1a} P.~C.~Peters,  Phys.\ Rev.\ D {\bf 9}, 2207 (1974).

\bibitem{CC3} M.~Giovannini, Eur. Phys. J. C \textbf{82}, 117 (2022).

\bibitem{CC4} M.~Giovannini, Phys. Rev. D \textbf{98}, 103509 (2018).

\bibitem{CC5} M.~Giovannini, Phys. Lett. B \textbf{789}, 502 (2019).

\bibitem{LISA1} P.~Amaro-Seoane \textit{et al.} [LISA], [arXiv:1702.00786 [astro-ph.IM]].

\bibitem{LISA2} LISA documents webpage, https://www.cosmos.esa.int/web/lisa/lisa-documents.

\bibitem{DECIGO1} N. Seto, S. Kawamura and T. Nakamura,  Phys. Rev. Lett. {\bf 87},  221103 (2001).

\bibitem{DECIGO2} S. Kawamura {\it et al.},  Class. Quant. Grav. {\bf 28}, 094011 (2011).

\bibitem{UDECIGO} H. Kudoh, A. Taruya, T. Hiramatsu and Y. Himemoto, Phys. Rev. D {\bf 73}, 064006 (2006).

\bibitem{BBO}  G. M. Harry {\it et al.}, Class. Quant. Grav. {\bf 23}, 4887 (2006).

\bibitem{TAIJI1} W.-R. Hu and Y.-L. Wu,  Natl. Sci. Rev. {\bf 4}, 685 (2017).

\bibitem{TAIJI2} W.-H. Ruan, Z.-K. Guo, R.-G. Cai and Y.-Z. Zhang, arXiv:1807.09495.

\bibitem{TIANQIN1} T.J.~Luo \textit{et al.} [TianQin], Class. Quant. Grav. \textbf{33}, 035010 (2016).

\bibitem{TIANQIN2} X.~C.~Hu {\it et al.}, Class. Quant. Grav. \textbf{35}, 095008 (2018).

\bibitem{MGPT} M.~Giovannini, Eur. Phys. J. C \textbf{82}, 828 (2022).

\bibitem{CHten5}  M.~S.~Turner, M.~J.~White and J.~E.~Lidsey,  Phys.\ Rev.\  D {\bf 48}, 4613 (1993).

\bibitem{CHten6} L.~M.~Krauss and M.~J.~White,  Phys.\ Rev.\ Lett.\  {\bf 69}, 869 (1992).

\bibitem{CHten7} B.~Allen and S.~Koranda,  Phys.\ Rev.\ D {\bf 50}, 3713 (1994).

\bibitem{CHten8} K.~w.~Ng and A.~D.~Speliotopoulos,  Phys.\ Rev.\ D {\bf 52}, 2112 (1995).
  
\bibitem{CHten9}  L.~Knox,  Phys.\ Rev.\ D {\bf 52}, 4307 (1995). 

\bibitem{EK1} J.~Khoury, B.~A.~Ovrut, P.~J.~Steinhardt and N.~Turok,  Phys.\ Rev.\ D {\bf 64}, 123522 (2001).
  
\bibitem{EK2} L.~A.~Boyle, P.~J.~Steinhardt and N.~Turok,  Phys.\ Rev.\ D {\bf 69}, 127302 (2004).  

\bibitem{EK3}  M.~Gasperini and M.~Giovannini,  Phys.\ Rev.\ D {\bf 47}, 1519 (1993).

\bibitem{EK3a} R.~Brustein, M.~Gasperini, M.~Giovannini and G.~Veneziano,  Phys.\ Lett.\ B {\bf 361}, 45 (1995).

\bibitem{HBT0a} R.~J.~Glauber,  Phys.\ Rev.\ Lett.\  {\bf 10}, 84 (1963); Phys.\ Rev.\  {\bf 130}, 2529 (1963);  Phys.\ Rev.\  {\bf 131}, 2766 (1963).

\bibitem{HBT0b}  E.~C.~C.~Sudarshan, Phys.\ Rev.\ Lett.\  {\bf 10}, 277 (1963).

\bibitem{HBT1} M.~Giovannini, Phys. Rev. D \textbf{99}, 123507 (2019); Class. Quant. Grav. \textbf{34}, 035019 (2017).

\bibitem{HBT2} M.~Giovannini, Phys. Rev. D \textbf{83}, 023515 (2011).

\bibitem{HBT3a} G. Cocconi, Phys. Lett. B {\bf 49}, 459 (1974).

\bibitem{HBT3b} D.~H.~Boal, C.~K.~Gelbke, B.~K.~Jennings, Rev.\ Mod.\ Phys.\  {\bf 62}, 553 (1990); 
G. Baym, Acta Phys.\ Polon.\  B {\bf 29}, 1839 (1998).

\bibitem{HBBT1} R. Hanbury Brown and R. Q. Twiss, Nature {\bf 178}, 1046 (1956).

\bibitem{HBBT2} R. Hanbury Brown and R. Q. Twiss, Proc. Roy. Soc. (London) {\bf A242}, 300 (1957); 
Proc. Roy. Soc. (London)  {\bf A243}, 291 (1958).

\bibitem{BARG} V. Bargmann, Ann. Math. {\bf 48},  568 (1947).

\bibitem{ST8} B.~L.~Hu and H.~E.~Kandrup, Phys. Rev. D \textbf{35}, 1776 (1987).

\bibitem{ST9} H. E. Kandrup, Phys. Rev. D \textbf{37}, 3505 (1988).

\bibitem{ST9a} M.~Gasperini and M.~Giovannini, Phys. Lett. B \textbf{301}, 334 (1993).

\bibitem{ST9b} M.~Gasperini and M.~Giovannini, Class. Quant. Grav. \textbf{10}, L133 (1993).

\bibitem{ST10} B. L. Hu, G. Kang, and A. Matacz, Int. J. Mod. Phys. A {\bf 9}, 991 (1994).

\bibitem{DDP1} D. Deutsch, Phys. Rev. Lett. {\bf 50}, 631 (1983).

\bibitem{DDP2} M. H. Partovi, Phys. Rev. Lett. {\bf 50}, 1883 (1983).

\end{thebibliography}
\end{document}